\documentclass[12pt,letterpaper,twoside,openright]{book}

\usepackage[utf8]{inputenc}
\usepackage{graphicx,amssymb,amsmath,amsthm,amsfonts}
\numberwithin{equation}{section}
\usepackage{float}
\usepackage[lmargin=3.5cm,rmargin=2.5cm,tmargin=2.5cm,bmargin=2.5cm]{geometry} 
\usepackage{makeidx}
\usepackage{bm}
\usepackage{enumerate}
\usepackage[usenames,dvipsnames]{color}
\usepackage{relsize}
\usepackage{empheq}
\usepackage{lscape}
\usepackage[square,colon]{natbib}
\usepackage{fancyhdr}  
\graphicspath{{Graphs2/}{Graphs3/}{Graphs4/}{Graphs5/}{Graphs6/}}

\providecommand{\norm}[1]{\lVert#1\rVert}
\providecommand{\inner}[2]{\langle #1 | #2 \rangle}
\providecommand{\ket}[1]{\lvert #1 \rangle}
\providecommand{\bra}[1]{\langle #1 \lvert}
\providecommand{\conm}[2]{\left[ #1 , #2 \right]}
\providecommand{\esp}[1]{\langle #1 \rangle}
\newcommand{\MeijerG}[7]{G \begin{smallmatrix} #1\! & #2 \\ #3\! & #4 \end{smallmatrix}\!\! \left( \begin{smallmatrix} #5 \\ #6 \end{smallmatrix} \middle\vert #7 \right) }

\usepackage[pdftex, plainpages = false, pdfpagelabels, 
                 pdfpagelayout = SinglePage,
                 bookmarks,
                 bookmarksopen = true,
                 bookmarksnumbered = true,
                 breaklinks = true,
                 linktocpage,
                 colorlinks = true,
                 linkcolor = blue,
                 urlcolor  = blue,
                 citecolor = red,
                 anchorcolor = green,
                 hyperindex = true,
                 hyperfigures
                 ]{hyperref} 
\usepackage{hyperref}
\hypersetup{
    bookmarks=true,         
    unicode=false,          
    pdftoolbar=true,        
    pdfmenubar=true,        
    pdffitwindow=false,     
    pdfstartview={FitH},    
    pdftitle={Polynomial Heisenberg algebras and Painleve equations},    
    pdfauthor={David Bermudez},     
    pdfsubject={Mathematical Physics},   
    pdfcreator={David Bermudez},   
    pdfproducer={David Bermudez}, 
    pdfkeywords={supersymmetry, SUSY QM, Painleve, quantum, algebras, Heisenberg, PhD, thesis, physics, mathematics}, 
    pdfnewwindow=true,      
    colorlinks=true,       	
    linkcolor=magenta,      
    citecolor=MidnightBlue, 
    filecolor=red,      	
    urlcolor=red           	
}

\makeatletter
\def\thickhrulefill{\leavevmode \leaders \hrule height 1ex \hfill \kern \z@}
\def\@makechapterhead#1{%
  \vspace*{10\p@}%
  {\parindent \z@ \raggedleft \reset@font
            \scshape \@chapapp{} \thechapter
        \par\nobreak
        \interlinepenalty\@M
    \Huge \bfseries #1\par\nobreak
    \hrulefill
    \par\nobreak
    \vskip 100\p@
  }}
\def\@makeschapterhead#1{%
  \vspace*{10\p@}%
  {\parindent \z@ \raggedleft \reset@font
            \scshape \vphantom{\@chapapp{} \thechapter}
        \par\nobreak
        \interlinepenalty\@M
    \Huge \bfseries #1\par\nobreak
    \hrulefill
    \par\nobreak
    \vskip 100\p@
  }}
\title{\textbf{Polynomial Heisenberg algebras\\
and Painlev\'e equations}}
\author{
A thesis by\\
\textbf{David Berm\'udez Rosales}\\ \\ \\
In partial fulfillment of the requirements\\
for the degree of\\ \\
\textit{Doctor of Philosophy}\\
\textit{in Physics}\\ \\
Advised by\\
\textbf{David Jos\'e Fern\'andez Cabrera} \\ \\ \\
\includegraphics[scale=0.32]{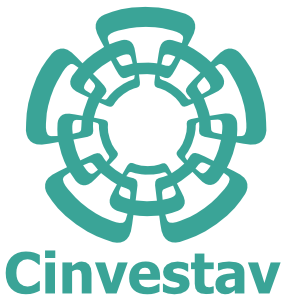}\\
Department of Physics \\
\textit{Center for Research and Advanced Studies} \\ 
\textit{of the National Polytechnic Institute} \\
\date{July, 2013}
}

\begin{document}
\maketitle

\frontmatter
\thispagestyle{empty}
\newenvironment{dedication} {\cleardoublepage
\thispagestyle{empty}
\vspace*{\stretch{1}} \begin{center} \em} {\end{center}
\vspace*{\stretch{3}} \clearpage}
\begin{dedication}
\hfill \
\parbox{10cm}{
\begin{verse}
A mis padres y a mi hermana...
\end{verse}
}
\end{dedication}
\chapter*{Agradecimientos
\markboth{Agradecimientos}{Agradecimientos}}
En este espacio quiero agradecer a varias personas e instituciones que me han ayudado durante estos cinco a\~nos de mi vida, en los que he aprendido mucho, de f\'isica y de muchas otras cosas m\'as.

A mis pap\'as, mi hermana y Oliver, porque ellos me han ense\~nado lo m\'as importante de la vida y gracias a ellos he tenido la fuerza para seguir adelante. Espero que sepan que siempre pienso en ellos aunque estemos lejos.

A mis t\'ios Pepe y Paty y a mis primos Jos\'e Eduardo y Regina porque se convirtieron en mi segunda familia. Estoy feliz de haber contado con ellos en los momentos dif\'iciles y que me hayan dado ese contacto familiar que muchas veces se pierde durante el posgrado. Al resto de mi familia, porque siempre han sido ejemplo de dedicaci\'on y de valores. 

A mis amigos del Departamento de F\'isica, algunos se fueron, otros se quedaron por m\'as tiempo, pero siempre compartiremos experiencias \'unicas. Abraham, Ariadna, Abril, Alonso, Fercho, Juan Carlos, Maly, Manuel, Alfonso, Marco, Iraq, Roger, Lily, Jorge, Pedro, L\'azaro, Jes\'us, Caro, Juli\'an y Hugo. A mis amigas del Cinvestav: Itzel, Derly, Ale, Alma y Lore, de ustedes aprend\'i lo emocionante que son tambi\'en el resto de las ciencias. A mis amigas Jan y Vane, por compartir su vida conmigo y mostrarme ese otro lado de la vida durante mi estancia en la ciudad de M\'exico.

Un agradecimiento especial a mi asesor David por su gu\'ia durante el desarrollo de mi investigaci\'on, sin duda alguna, una de las personas que m\'as admiro tanto a nivel personal como acad\'emico. Reconozco que me dio toda su confianza y apoyo durante los tres a\~nos y medio que trabajamos juntos, espero haber cumplido con sus expectativas.

A Nicol\'as Fern\'andez Garc\'ia, Javier Negro y Alonso Contreras Astorga con quienes colabor\'e durante mi doctorado y desarrollamos juntos temas muy interesantes que finalmente forman parte de esta tesis. Espero que volvamos a hacerlo.

A profesores y personal del Departamento de F\'isica, que me ense\~naron y me apoyaron durante mi estancia. En especial a Bogdan Mielnik, \'Oscar Rosas Ort\'iz, Nora Bret\'on, Sara Cruz Cruz, Eduard de la Cruz, Pepe M\'endez y Mim\'i. Al resto de los estudiantes del {\it Bar Quantum} con quienes aprend\'i mucho en discusiones y compart\'i cursos y charlas.

A Bernardo Wolf, Alexander Turbiner, Gabriel L\'opez Castro y \'Oscar Rosas Ort\'iz por fungir como sinodales de esta tesis y por sus valiosos comentarios durante la revisi\'on del escrito.

Tambi\'en reconozco el apoyo de Conacyt a trav\'es de la beca 219665 que hizo posible mi dedicaci\'on plena a este trabajo.

\phantom{.}\vspace{1mm}
\hfill \emph{David Berm\'udez Rosales}
\chapter*{Resumen\markboth{Resumen}{Resumen}}
Primero estudiaremos la mec\'anica cu\'antica supersim\'etrica (SUSY QM), con un \'enfasis especial en los osciladores arm\'onico y radial. Despu\'es, mostraremos dos contribuciones originales de esta tesis en el \'area: una nueva f\'ormula del Wronskiano para la transformaci\'on SUSY confluente \citep{BFF12} y la aplicaci\'on de la SUSY QM al potencial del {\it oscilador invertido} \citep{BF13b}.

Posteriormente, presentaremos las definiciones del \'algebra de Heisenberg-Weyl y de las {\it \'alge\-bras de Heisenberg polinomiales} (PHA). Estudiaremos los sistemas generales descritos por PHA: para \'ordenes cero y uno obtenemos a los osciladores arm\'onico y radial, respectivamente; para segundo y tercer orden, el potencial queda determinado por soluciones de las ecuaciones de Painlev\'e IV ($P_{IV}$) y Painlev\'e V ($P_{V}$), respectivamente.

M\'as tarde, haremos una breve revisi\'on general de las seis ecuaciones de Painlev\'e y estudiaremos espec\'ificamente los casos de $P_{IV}$ y $P_V$. Probaremos un {\it teorema de reducci\'on} \citep{Ber10,BF11a}, para que PHA de orden $2k$ se reduzcan a \'algebras de segundo orden. Demostraremos tambi\'en un teorema an\'alogo para que PHA de orden $2k+1$ se reduzcan a \'algebras de tercer orden. Mediante estos teoremas encontraremos soluciones a $P_{IV}$ y $P_V$ dadas en t\'erminos de funciones hipergeom\'etricas confluentes ${}_1F_1$ \citep{CFNN04,BF11b}. Para algunos casos especiales, \'estas pueden clasificarse en diversas {\it jerarqu\'ias de soluciones} \citep{BF11a,BF13a}. De esta manera, encontraremos soluciones reales con par\'ametros reales y soluciones complejas con par\'ametros reales y complejos de ambas ecuaciones \citep{Ber12,BF12}.

Finalmente, estudiaremos los {\it estados coherentes} (CS) para los socios SUSY del oscilador arm\'onico que se conectan con $P_{IV}$ a trav\'es del teorema de reducci\'on, a los cuales llamaremos {\it estados coherentes tipo Painlev\'e IV}. Ya que estos sistemas siempre tienen operadores de escalera de tercer orden $l_k^\pm$, buscamos primero los CS como eigenestados del operador de aniquilaci\'on $l_k^-$. Definimos tambi\'en operadores an\'alogos al operador de desplazamiento $D(z)$ y obtenemos CS a partir de los estados extremales de cada subespacio del espacio de Hilbert $\mathcal{H}$ en el que se descompone el sistema. Concluimos el tratamiento aplicando un proceso de {\it linealizaci\'on} de los operadores de escalera para definir un nuevo operador de desplazamiento $\mathcal{D}(z)$ con el cual obtener CS que involucren a todo $\mathcal{H}$.
\chapter*{Abstract\markboth{Abstract}{Abstract}}
We shall study first the supersymmetric quantum mechanics (SUSY QM), specially the cases of the harmonic and radial oscillators. Then, we will show two original contributions of this thesis in the area: a new Wronskian formula for the confluent SUSY transformation \citep{BFF12} and the application of SUSY QM to the {\it inverted oscillator} potential \citep{BF13b}.

After that, we will present the definitions of the Heisenberg-Weyl algebra and the so called {\it polynomial Heisenberg algebras} (PHA). We will study the general systems described by PHA: for zeroth- and first-order we obtain the harmonic and radial oscillators, respectively; for second- and third-order PHA, the potential is determined in terms of solutions to Painlev\'e IV ($P_{IV}$) and Painlev\'e V ($P_{V}$) equations, respectively.

Later on, we will give a brief general review of the six Painlev\'e equations and we will study specifically the cases of $P_{IV}$ and $P_V$. We will prove a {\it reduction theorem} \citep{Ber10,BF11a} for $2k$th-order PHA to be reduced to second-order algebras. We will also prove an analogous theorem for the $(2k+1)$th-order PHA to be reduced to third-order ones. Through these theorems we will find solutions to $P_{IV}$ and $P_V$ given in terms of confluent hypergeometric functions ${}_1F_1$ \citep{CFNN04,BF11b}. For some special cases, those can be classified in several {\it solution hierarchies} \citep{BF11a,BF13a}. In this way, we will find real solutions with real parameters and complex solutions with real and complex parameters for both equations \citep{Ber12,BF12}.

Finally, we will study the {\it coherent states} (CS) for the specific SUSY partners of the harmonic oscillator that are connected with $P_{IV}$ through the reduction theorem, which we will call {\it Painlev\'e IV coherent states}. Since these systems always have third-order ladder operators $l_k^\pm$, we will seek first the CS as eigenstates of the annihilation operator $l_k^-$. We will also define operators which are analogous to the displacement operator $D(z)$ and we will get CS departing from the extremal states in each subspace in which the Hilbert space $\mathcal{H}$ is decomposed. We conclude our treatment applying a {\it linearization} process to the ladder operators in order to define a new displacement operator $\mathcal{D}(z)$ to obtain CS involving the entire $\mathcal{H}$.
\tableofcontents
\listoffigures
\listoftables
\newpage

\mainmatter
\chapter{Introduction}

\section{Background} 

At the end of the 19th century, many scientists believed that all principles of physics had already been found and that from that moment, they had only left the solution of practical problems. Nevertheless, time showed them wrong with a pile of experiments that either did not have theoretical explanation or even worst, the explanation was completely wrong. Some of these problems were the {\it black-body radiation}, {\it the photoelectric effect}, {\it the magnetic moment of the electron}, among others.

During these uncertain times, the {\it quantum hypothesis} appeared in an insightful work of \citet{Pla00} to explain the black-body radiation, electromagnetic radiation emitted by a body in thermodynamic equilibrium with its environment. This radiation has a very specific pattern that could not be explained by classical electrodynamics. In his work, Planck suggested that the radiation emitted by a specific system can be {\it divided} into {\it discrete} elements of energy $E$ and not only that, but that the energy of these elements would depend on its frequency $\nu$. This hypothesis seems bizarre even now and we can only imagine the resistance of the scientific community at the beginning of the 20th century to receive this idea. Even Planck himself did not like it and considered it as a {\it mathematical trick}. But then, why did he even proposed it? The reason is that it gives the correct answer, not only qualitatively but also quantitatively while the usual theory was completely wrong.

Later on, \citet{Ein05} took this idea even further. Einstein generalized the {\it discreteness} proposed by Planck to explain the photoelectric effect, which is caused when some material absorbs light, emits electrons, and produces an electric current. Based on the Planck's quantum hypothesis, Einstein postulated that matter does not only absorbs and emits light in a discrete way, but rather that light itself is made of discrete particles. This was a complete turn from the {\it classical} theories from 19th century that were so successful. Actually, Einstein was not the first to think of light as particles. It had also been proposed by Isaac Newton, but Maxwell's work proved that light was a special type of electromagnetic waves. Nevertheless, in the specific case of photoelectric effect, the classical theory could not give a correct answer, while Einstein theory not only correct it, but also simplify it. These light particles were later on called {\it photons}.

In the following years, a complete quantum theory was developed in Europe by both, young and old physicists. Some of the main contributors to the theory are Heisenberg, Born, Jordan, Dirac, Pauli, Schr\"{o}dinger, among others. They all built their names working in quantum mechanics. The fundamental topics in quantum theory are those of \emph{uncertainty} and \emph{discretization}.

The starting point was the work of \citet{Hei25} (translated in \citet{vW67}), where the basis for a fundamental quantum theory were laid on the notion that only those physical quantities that can be measured are important. The reinterpretation of Heisenberg's results by \citet*{BJ25}, and one more work from \citet*{BHJ26} gave rise to what is now known as the \emph{matrix formulation} of quantum mechanics.

Later on, in a series of works by Schr\"odinger published in 1926 (recompiled in \citep{Sch82}), an alternative presentation of quantum mechanics was born, called \emph{wave mechanics}. It is curious to note that Schr\"odinger himself proved in a following article \citep{Sch26a} that both formulations were equivalent.

During those years, in two consecutive works by \citet{Dir25,Dir26}, a further refinement of Heisenberg's formulation started. Dirac also obtained quantum theory from a process of \emph{algebraic quantization} from the Hamiltonian formulation of classical mechanics. Those advances were culminated by \citet{Hei27} in an elegant work where he formulated what is known today as the \emph{uncertainty principle}.

Basically the whole formalism of quantum mechanics was completed by that time, and from the publication of two books, \emph{``The principles of quantum mechanics''} by \citet{Dir30} and \emph{``Fundamentals of quantum mechanics''} by \citet{Foc31}, a new area of physics was born, quantum mechanics. Those were the first specialized books in the topic and they integrated the whole formalism of quantum theory.

What a remarkable 30 years journey for science! From the notions that physics was basically completed and that the basic principles were already in place, to laying ground for a whole new area of knowledge. Arguably, it has been the fastest development period in science. Two other important periods precede this one: the rapid progress in thermodynamics and chemistry during the industrial revolution; and the electric, magnetic, and light phenomena, explained by the electromagnetic theory in the 19th century. The knowledge gathered up during those scientific revolutions, together with the social principles and art concepts, form the complete body of knowledge and are the basis of today's Western civilization.

Nowadays, quantum theory keeps attracting both physicist and mathematicians, who are still making important contributions to the theory. For example, the supersymmetric quantum mechanics (SUSY QM), one of the specialties of the Department of Physics at Cinvestav, has received a lot of attention in the last years.

\section{Factorization method}
In order to describe a system in quantum mechanics one must solve an eigenvalue problem for the matrix formulation or a second-order differential equation with boundary conditions for the wave formulation. An elegant procedure to solve this problem in quantum mechanics consists in using the {\it factorization method}, where a certain differential operator is factorized in terms of other differential operators. The first ideas about this method were proposed by \citet{Dir30} and \citet{Foc31} in order to solve the one-dimensional harmonic oscillator and later on they were exploited by Schr\"{o}dinger \citep{Sch40,Sch40a,Sch41} to solve different problems.

The first generalization of this technique was given by \citet{Inf41} and then after several contributions from various scientists, it was finished ten years later with the seminal paper of \citet{IH51}, where they perform an exhaustive classification of all the systems solvable through factorization method. This includes the harmonic oscillator, the hydrogen atom potential, the free particle, the radial oscillator, some spin systems, the P\"oschl-Teller potential, Lam\'e potentials, among others. For many years this work was considered to be the culmination of the technique, i.e., if someone wanted to see the viability to use the factorization method, he or she simply checked the paper by Infeld and Hull. This also meant that people thought this method was essentially finished.

After many years, and contrary to the common belief that the factorization method was completely explored, \citet{Mie84} made an important contribution. In his work, Mielnik did not consider the particular solution used in the factorization method of Infeld and Hull, but rather the general solution and he used it to find a family of new factorizations of the harmonic oscillator that also lead to related new solvable potentials. In this way, after 33 years, not only one, but a whole family of new solvable potentials were obtained by the factorization method. In this classic work, Mielnik obtained a family of potentials isospectral to the harmonic oscillator.

Many years later to Infeld and Hull's article, and from a totally different area of physics, \citet{Wit81} proposed a mechanism to form hierarchies of isospectral Hamiltonians, which are now called {\it supersymmetric partners}. In this work, a toy model for the supersymmetry in quantum field theory is considered. It turns out that this technique is closely related with the generalization of the factorization method proposed by Mielnik. In terms of the now completely developed theory, we would say that Mielnik found the first-order SUSY partner potentials of the harmonic oscillator for the specific factorization energy $-1/2$. As a result, the study of analytically solvable Hamiltonians was reborn. This generalization of the factorization method or intertwining technique is gathered now in an area of science that is commonly called {\it supersymmetric quantum mechanics}, or SUSY QM, and there is a big community of scientists working on this topic nowadays.

Almost immediately after Mielnik's work, \citet{Fer84} applied the same technique to the hydrogen atom and he also obtained a new one-parameter family of potentials with the same spectrum. In the mean time, \citet{Nie84}; \citet*{ABI84}; and \citet{Suk85a} developed the formal connection between SUSY QM and the factorization method. They were the first to understand the full power of the technique in order to obtain new solvable potentials in quantum mechanics by generalizing the process used by Mielnik and Fern\'andez. Now we say that they generalized the factorization method to a general solvable potential with an arbitrary factorization energy. All these developments caused a new interest in the algebraic methods of solution in quantum mechanics and the search for new exactly-solvable potentials.

Until that moment, the factorization operators were always of first-order. This is natural, being the Hamiltonian a second-order differential operator, it is expected to be factorized in terms of lower-order operators. Nevertheless, \citet*{AIS93} proposed to use higher-order operators (see also \citet{AICD95}). An alternative point of view of this work was proposed by \citet{BS95}.

After many years away from these developments, it is worth to notice that the group of Cinvestav returned to the study of SUSY QM. In a remarkable work, \citet{FGN98} generalized the factorization method from Mielnik's point of view, in order to obtain new families of potentials isospectral to the harmonic oscillator, using second-order differential intertwining operators. Soon after, \citet{Ros98a,Ros98b} applied the same techniques to the hydrogen atom. This generalization was achieved using two iterative first-order transformations as viewed by Mielnik and Sukumar in the 1980's. With this theory, it was possible to obtain an energy spectrum with spectral gaps, i.e., the regularity of the spectrum was lost. A review of SUSY QM from the point of view of a general factorization method can be found in the works by \citet{MR04} and by \citet{FF05}.

Furthermore, it is important to mention that even when most of the papers on this theory are gathered under the keyword of SUSY QM, there is a lot of work on this topic under different points of view. We can mention for example, Darboux transformations \citep{MS91,FR08}, intertwining technique \citep*{CRF01}, factorization method \citep{MR04}, N-fold supersymmetry \citep{AST01,ST02,GLT04,BT09}, and non-linear hidden supersymmetry \citep{LP03,Ply04,CNP07,CJNP08}.

\section{Polynomial Heisenberg algebras}
Lie algebras and their deformations play an important role in several problems of physics, for example, Higgs algebra \citep{Hig79} is applied to several Hamiltonians with analytic solution \citep{BDK94}. For Lie algebras, the commutators are linear combinations of the generators. On the other hand, in deformed Lie algebras, the commutators are non-linear functions of the generators \citep{DGRS99}.

In this thesis we will study the polynomial Heisenberg algebras (PHA), i.e., systems for which the commutators of the Hamiltonian $H$ and the ladder operators $L^{\pm}$ (sometimes also known as creation and annihilation operators) are the same as for the harmonic oscillator, but the commutator $[L^{+},L^{-}]$ is a polynomial $P(H)$ of $H$. Some of these algebras are constructed taking $L^{\pm}$ as a $m$th-order differential operator \citep{Fer84D,DEK92,SRK97,FH99,ACIN00}.

Furthermore, it is important to study not only these specific algebras, but also the characterization of the general systems ruled by these PHA. We will see in this thesis that the difficulties in the study of this problem are dramatically increased with the order $m$ of the polynomial, for zeroth- and first-order PHA, the systems are the harmonic and the radial oscillators, respectively \citep{Fer84D,DEK92,Adl93,SRK97}. On the other hand, for second- and third-order PHA, the determination of the potentials is reduced to find solutions of Painlev\'e IV and V equations, $P_{IV}$ and $P_{V}$, respectively \citep{Adl93,WH03}.

This means that, in order to have a system described by these PHA, we need solutions of $P_{IV}$ and $P_V$. Nevertheless, in this thesis we will use this connection but in the opposite direction, i.e., we look for systems that we know before hand that are described by PHA and then we develop a method to find solutions of those Painlev\'e equations.

\section{Painlev\'e equations}

There has been different connections between quantum mechanics and non-linear differential equations. The simplest case was studied by \citet{Dir30} connecting the Schr\"odinger equation, a second-order linear differential equation, and the Riccati equation, a first-order non-linear differential equation. Further examples are the SUSY partners of the free particle potential, which lead to solutions of the Korteweg-de Vries (KdV) equation \citep{Mat92}.

In particular, in this work we will study the relation between SUSY QM, PHA, and Painlev\'e equations. When we started working on the topic it was already known that specific PHA were connected with solutions of some Painlev\'e equations. It was also known that the first-order SUSY partner potentials of the harmonic and radial oscillator were ruled by these algebras, and thus connected with solutions to some Painlev\'e equations. At this moment, several questions arise: can their higher-order SUSY partners lead to more solutions? And if so, which are the conditions on the quantum systems? What kind of solutions do they lead to? In this thesis we will answer these questions.

We will see that second-order PHA are related with the Painlev\'e IV equation ($P_{IV}$) and third-order ones with Painlev\'e V equation ($P_V$). Not only that, but we will use higher-order SUSY QM to obtain additional systems described by these two kind of algebras departing from the harmonic and the radial oscillators, which will allow us to find new solutions of $P_{IV}$ and $P_V$. After that, we will study and classify these solutions into the so called {\it solution hierarchies}.

Painlev\'e equations are non-linear second-order differential equations in the complex plane which pass the so called {\it Painlev\'e test}, i.e., their movable singularities are poles. In general, they are not solvable in terms of special functions, but rather they define new functions called {\it Painlev\'e trascendents}.

Painlev\'e equations received that name because they were developed by a method derived by \citet{Pai00,Pai02} at the beginning of the 20th century. Painlev\'e wanted to classify second-order non-linear differential equations that would define some new useful functions according to some mathematical properties, therefore, they should not be {\it reduced} to first-order differential equations or to {\it elliptic} functions. In this way, their general solutions, called {\it Painlev\'e trascendents}, cannot be given in terms of special functions, because that would mean that the equations are {\it reducible}. Painlev\'e found the first four equations through this method, then his school finished the job: \citet{Gam10} identified the fifth equation and \citet{Fuc07} found the last and most general one. The explicit form of the six Painlev\'e equations is presented in chapter \ref{cappain}.

Painlev\'e trascendents play an important role in the topic of non-linear ordinary differential equations. Some specialists \citep{IKSY91,CM08} consider that during the 21st century, Painlev\'e trascendents will be new members of the set of special functions. At the moment, both physicists and mathematicians are already employing these functions, for example, they have been used to describe a great variety of systems, as quantum gravity \citep{FIK91}, superconductivity \citep{KSSLW09}, and random matrix models \citep{ASM95}.

As far as we know, the first people who realized the connection between SUSY QM, second-order PHA, and Painlev\'e equations were \citet{VS93}, \citet{Adl93}, and \citet{DEK94}. This connection has been explored more thoroughly by \citet*{ACIN00,FNN04,CFNN04}; and \citet*{MN08}.

In chapter \ref{cappain} of this work we will make use of the factorization method to show by induction that if the factorization energies are given by an equidistant set connected by the annihilation and creation operators of the harmonic oscillator, then the {\it natural} ladder operators $L_k^\pm$ of $(2k+1)$th-order associated with the SUSY Hamiltonian $H_k$ will be factorized in terms of a polynomial of $H_k$ and a new pair of ladder operators $l_k^\pm$ which will always be of third-order.

Then, in the following chapter, we will do the analogous procedure for the SUSY partners of the  radial oscillator. In this case, the natural ladder operators are of $(2k+2)$th-order, and through the reduction process we will factorize it in terms of a polynomial of $H_k$ and new fourth-order ladder operators $\ell_k^\pm$, which will describe a third-order PHA.

This means that through this specific factorization we first obtain higher-order PHA which are {\it reducible} to second-order ones in the case of the harmonic oscillator and to third-order ones in the case of the radial oscillator. We will also prove in chapter \ref{pha} that these systems are connected with Painlev\'e IV ($P_{IV}$) and Painlev\'e V ($P_V$) equations, respectively. Thus, in order to obtain systems described by these algebras, we would need to solve $P_{IV}$ and $P_V$. What we will do is to use these ideas for the inverse problem, i.e., we will derive systems described by these PHA and then we will find solutions to the Painlev\'e equations.

Furthermore, we must remark that these solutions will be obtained by purely algebraic methods, although Painlev\'e equations are non-linear second-order differential equations that are difficult to solve, specifically, there is no general method to solve them.

Some of the work of this thesis is already published in the scientific literature, e.g., in \citet*{Ber10,Ber12}; \citet*{BF11a,BF11b,BF12,BF13a,BF13b}; and \citet*{BFF12}.
\chapter{Supersymmetric quantum mechanics}
\label{capsusyqm}

The supersymmetric quantum mechanics (SUSY QM), the factorization method, and the intertwining technique are closely related and their names will be used indistinctly in this work to characterize a specific method, through which it is possible to obtain new exactly-solvable quantum systems departing from known ones. These new systems are indeed solution families which can be manipulated to perform the spectral design of quantum systems.

In the first half of this chapter, sections \ref{secsusy1} to \ref{secHORO}, we will study the basis of SUSY QM; this material is not new and one can find several recent reviews in the scientific literature, e.g., \citet*{MR04}, \citet*{FF05}, \citet*{Fer10}, \citet*{AI12}, and references therein. We will review this method because it is the main tool we are going to use in the original research of this thesis. In the second half of this chapter, sections \ref{difconfluente} and \ref{inverted}, we shall show two new developments inside the general framework of SUSY QM, obtained as part of my PhD work.

To begin with, in section \ref{secsusy1} we will study the first-order SUSY QM, which contains the essence of the factorization method, since most higher-order cases are generalizations of this process. Next, higher-order SUSY QM will be studied by two different approaches. First, in section \ref{secite} we will review the {\it iterative approach}, in which several first-order SUSY transformation are performed one after the other. Second, in section \ref{secdir} we will study the {\it direct approach}, where we perform a global $k$th-order SUSY transformation in just one step. The direct approach becomes increasingly complicated as the order of the transformation grows up; therefore, we will focus mainly in the second-order case, where the resulting equations can still be solved using a simple ansatz and the results are satisfactory. Moreover, there is a theorem \citep{AS07,Sok08} stating that all non-singular SUSY transformations can be seen as compositions of non-singular first- and second-order ones, implying that the treatment presented here is the most general possible. Then, in section \ref{secHORO} we will apply the SUSY transformations to two interesting systems, the {\it harmonic} and the {\it radial oscillators} because we will use them later to generate solutions to Painlev\'e equations. After that, our new general developments will be shown, namely, in section \ref{difconfluente} a new differential formula for the confluent SUSY QM will be obtained and then, in section \ref{inverted} we will obtain for the first time the SUSY partner potentials associated with the inverted oscillator.

\section{First-order SUSY QM}
\label{secsusy1}
\subsection{Real first-order SUSY QM}
Let $H_0$ and $H_1$ be two Schr\"odinger Hamiltonians
\begin{equation}
H_i = -\frac{1}{2}\frac{\text{d}^2}{\text{d}x^2} + V_i (x) \text{, \ \ \ \ \ } i=0,1.
\end{equation}
For simplicity, we are taking {\it natural units}, i.e., $\hbar=m=1$. Next, let us suppose the existence of a first-order differential operator $A_1^{+}$ that {\it intertwines} the two Hamiltonians in the way
\begin{equation}
H_1 A_1^{+}=A_1^{+}H_0,\label{entre}
\end{equation}
with
\begin{equation}
A_1^{+}=\frac{1}{2^{1/2}}\left[-\frac{\text{d}}{\text{d}x} + \alpha_1 (x)\right],
\label{A1}
\end{equation}
where the \emph{superpotential} $\alpha_1 (x)$ is still to be determined. Equation \eqref{entre} is known as the {\it intertwining relation}.

On the other hand, we must remind that these equations involve operators, which means that in order to interchange the differential operator  $\text{d}^k/\text{d}x^k$ with any operator multiplicative function $f(x)$ we must use the following relations
\begin{subequations}
\begin{align}
\frac{\text{d}}{\text{d}x} f &=f\frac{\text{d}}{\text{d}x} + f',\\
\frac{\text{d}^2}{\text{d}x^2} f &=f\frac{\text{d}^2}{\text{d}x^2} +2f'\frac{\text{d}}{\text{d}x} +f'',\\
		&\ \ \ \ \ \vdots\nonumber\\
\frac{\text{d}^n}{\text{d}x^n} f &= \sum_{k=0}^{n} {n\choose k} f^{(k)}\frac{\text{d}^{n-k}}{\text{d}x^{n-k}},
\end{align}
\end{subequations}
where the {\it binomial coefficient} ${n\choose k}$ is defined as
\begin{equation}
{n\choose k} \equiv \frac{n!}{k!(n-k)!}.
\end{equation}

Then, for equation \eqref{entre} it is straightforward to show that
\begin{subequations}
\begin{align}
2^{1/2}H_1 A_1^{+} &=\frac{1}{2}\frac{\text{d}^3}{\text{d}x^3}-\frac{\alpha_1}{2}\frac{\text{d}^2}{\text{d}x^2}-(V_1 + \alpha'_1)\frac{\text{d}}{\text{d}x} + \alpha_1 V_1 - \frac{\alpha''_1}{2},\\
2^{1/2}A_1^{+}H_0 &= \frac{1}{2}\frac{\text{d}^3}{\text{d}x^3}-\frac{\alpha_1}{2}\frac{\text{d}^2}{\text{d}x^2}-V_0 \frac{\text{d}}{\text{d}x} +\alpha_1 V_0 -V'_0.
\end{align}\label{AH}
\end{subequations}
\hspace{-1.5mm}Matching the powers of the differential operator $\text{d}/\text{d}x$ of equations \eqref{AH} and solving the coefficients, we get
\begin{subequations}
\begin{align}
V_1 & = V_0 - \alpha'_1,\label{v1v0}\\
\alpha_1 V_1 - \frac{\alpha''_1}{2}&=\alpha_1 V_0 - V_{0}'. \label{v1}
\end{align}\label{v1p}
\end{subequations}
\hspace{-1mm}Substituting $V_1$ from equation \eqref{v1v0} into \eqref{v1} and integrating the result we obtain a Riccati equation
\begin{equation}
\alpha'_1+\alpha^{2}_1 = 2(V_0-\epsilon).
\end{equation}

From now on, we will explicitly express the superpotential dependence on the {\it factorization energy} $\epsilon$ as $\alpha_1(x,\epsilon)$:
\begin{subequations}
\begin{align}
V_1(x) &= V_0(x) - \alpha'_1(x,\epsilon),\label{Valfa}\\
\alpha'_1(x,\epsilon)+\alpha^{2}_1(x,\epsilon) &= 2[V_0(x)-\epsilon]. \label{alfa}
\end{align}\label{2alfas}
\end{subequations}

If we use a new function $u^{(0)}(x)$ such that $\alpha_1(x,\epsilon)=u^{(0)'}/u^{(0)}$, then equations \eqref{2alfas} are mapped into
\begin{subequations}
\begin{align}
V_1 &=  V_0- \left[\frac{u^{(0)'}}{u^{(0)}}\right]',\label{Vu}\\
-\frac{1}{2}u^{(0)''} + V_0 u^{(0)} &= \epsilon u^{(0)},\label{ucero}
\end{align}\end{subequations}
which means that $u^{(0)}$ is a solution of the initial stationary Schr\"odinger equation associated with $\epsilon$, although it may not have physical interpretation, i.e., $u^{(0)}$ might not fulfill any boundary condition.

Starting from equations \eqref{2alfas} we obtain that $H_0$ and $H_1$ can be factorized as
\begin{subequations}
\begin{align}
H_0 &= A_1^{-} A^{+}_1+\epsilon,\\
H_1 &= A^{+}_1 A_1^{-}+\epsilon, 
\end{align}\label{ffH}
\end{subequations}
\hspace{-1mm}where
\begin{equation}
A_1^{-}\equiv(A^{+}_1)^{\dagger}=\frac{1}{2^{1/2}}\left[\frac{\text{d}}{\text{d}x} + \alpha_1(x,\epsilon)\right],\label{amam}
\end{equation}
i.e., $A^{-}_1$ is the Hermitian conjugate operator of $A^{+}_1$. The constraint \eqref{amam} leads to a real $\alpha_1(x,\epsilon)$, but it can be generalized to the case where $A_1^{-}\neq (A_1^{+})^{\dagger}$ with $\alpha_1(x,\epsilon)$ being a complex function (associated with a complex $\epsilon$, a complex initial potential, or a complex linear combination of the solutions $u^{(0)}(x)$).

\begin{figure}\centering
\includegraphics[scale=0.35]{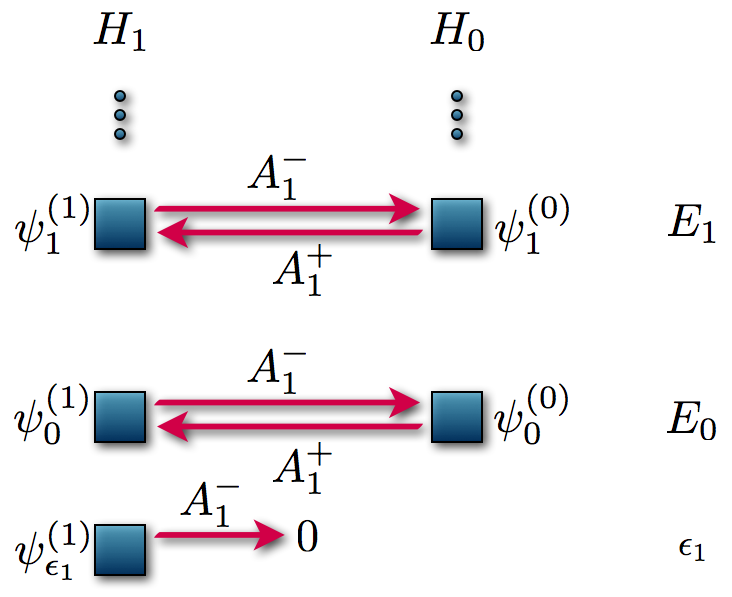}
\caption{\small{Representation of the first-order SUSY transformation. The final Hamiltonian $H_1$ has the same spectrum of the initial Hamiltonian $H_0$, but a new level at the factorization energy $\epsilon_1$ has been added.}}
\label{fig.susyqm1er}
\end{figure}

Let us assume that $V_0(x)$ is a solvable potential with normalized eigenfunctions $\psi^{(0)}_n (x)$ and eigenvalues such that Sp$(H_0)=\{E_n,n=0,1,2,\dots\}$. Besides, we know a non-singular solution $\alpha_1(x,\epsilon)$ [a $u^{(0)}(x)$ without zeroes] to the Riccati equation \eqref{alfa} [Schr\"odinger \eqref{ucero}] for a certain value of the factorization energy $\epsilon=\epsilon_1 \leq E_0$, where $E_0$ is the ground state energy for $H_0$. Then, the potential $V_1(x)$ given in equation \eqref{Valfa} [in \eqref{Vu}] is completely determined, its normalized eigenfunctions are expressed by
\begin{subequations}
\begin{align}
\psi^{(1)}_{\epsilon_1}(x) &\propto \exp\left(-\int^{x}_0 \alpha_1(y,\epsilon_1)\text{d}y\right)=\frac{1}{u^{(0)}_1(x)},\label{psimathcal}\\
\psi^{(1)}_{n}(x) &= \frac{A^{+}_{1} \psi_n^{(0)}(x)}{(E_n-\epsilon_1)^{1/2}},
\end{align}\label{psis}
\end{subequations}
\hspace{-1.5mm}while its eigenvalues are such that Sp$(H_1)=\{\epsilon_1,E_n;n=0,1,\dots\}$. An scheme of the way the first-order supersymmetric transformation works, and the resulting spectrum, is shown in figure \ref{fig.susyqm1er}.

We must emphasize the importance of the restriction $\epsilon_1 \leq E_0$ to avoid the existence of singularities in the superpotential $\alpha_1(x,\epsilon_1)$, in the potential $V_1(x)$, and in the eigenfunctions $\{\psi^{(1)}_{\epsilon_1},\psi^{(1)}_n\}$ given by \eqref{psis}. As a matter of fact, if $\epsilon_1 > E_0$, the transformation function $u^{(0)}_1(x)$ would have non-removable zeroes in the initial domain for $x$ and thus $\alpha_1(x,\epsilon_1)$ will have singularities at those points. On the other hand, if $\epsilon_1 \leq E_0$, then $u^{(0)}_1(x)$ could have at most one zero, but this can be moved to the boundary of the domain. In fact, by exploring the two-dimensional solution subspace associated with $\epsilon_1 \leq E_0$ it is possible to find solutions without zeroes \citep{Suk85a,Suk85b}.

\begin{figure}\centering
\includegraphics[scale=0.6]{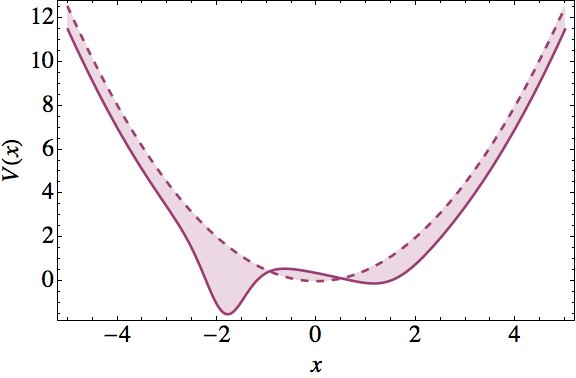}
\caption{\small{SUSY partner $V_1(x)$ (solid line) of the harmonic oscillator potential $V_0(x)$ (dashed line), generated using $u^{(0)}(x)$ with $\epsilon=0,\ \nu=0.9$. We remark the difference between the two potentials.}}\label{figsusy1}
\end{figure}

For now, let us assume that the factorization energy used to generate the new Hamiltonian is below the ground state of the initial Hamiltonian. Moreover, we suppose that for a given factorization energy $\epsilon_1$, the arbitrary parameter of the general solution of the Riccati equation has been adjusted to avoid singularities. An example of the generated potentials can be seen in figure \ref{figsusy1}.

\subsection{Complex first-order SUSY QM}
\label{complexsusy}
Let us begin here with a given Hamiltonian $H$ which has been completely solved, i.e., all its eigenvalues $E_n$ and eigenfunctions $\psi_n,\ n=0,1,\dots$ are known:
\begin{subequations}
\begin{align}
&H=-\frac{1}{2}\frac{\text{d}^2}{\text{d}x^2} +V(x),\label{defH}\\
&H\psi_n(x)=E_n\psi_n(x).\label{eigen}
\end{align}
\end{subequations}

Now, let us propose that $H$ is factorized as
\begin{equation}\label{facH}
H=A^{-}A^{+}+\epsilon.
\end{equation}
In the standard factorization method there is indeed an extra condition $A^{+}\equiv (A^{-})^{\dag}$ (see equation \eqref{amam}). In this section we will not use this constraint, but rather we simply ask that
\begin{equation}
A^{+} =\frac{1}{2^{1/2}}\left[-\frac{\text{d}}{\text{d}x} +\beta(x)\right],\quad A^{-} =\frac{1}{2^{1/2}}\left[\frac{\text{d}}{\text{d}x} +\beta(x)\right],\label{As}
\end{equation}
where $\beta(x)$ is a complex function to be found. This choice represents a more general factorization than the usual real one \citep{RM03}.

Working out the operations in equation~\eqref{facH}, using the definitions in \eqref{defH} and \eqref{As}, we obtain one condition for $\beta(x)$,
\begin{equation}
\beta'+\beta^{2}=2[V(x)-\epsilon],\label{riccatieq}
\end{equation}
which is a Riccati equation.

On the other hand, if we consider a similar factorization but in a reversed order and introduce a new Hamiltonian $\widetilde{H}$, defined by
\begin{equation}
\widetilde{H}=-\frac{1}{2}\frac{\text{d}^2}{\text{d}x^2} +\widetilde{V}
\end{equation}
and
\begin{equation}\label{facH2}
\widetilde{H}=A^{+}A^{-}+\epsilon,
\end{equation}
it turns out that
\begin{equation}
\widetilde{V}(x)=V(x)-\beta'(x).
\end{equation}

Besides, from equations \eqref{facH} and \eqref{facH2} it is straightforward to show that \citep{AICD99,RM03}
\begin{equation}
\widetilde{H}A^{+} =A^{+}H,\quad HA^{-} =A^{-}\widetilde{H},\label{entre1}
\end{equation}
which are the well known \textit{intertwining relations} with $A^{+}$, $A^{-}$ being the \textit{intertwining operators}. From equations \eqref{eigen} and \eqref{entre1} we can obtain the eigenvalues and eigenfunctions of the new Hamiltonian $\widetilde{H}$ as follows
\begin{subequations}
\begin{align}
\widetilde{H}A^{+}\psi_n(x) &=A^{+}H\psi_n(x)=E_n A^{+}\psi_n(x),\\
\widetilde{H}\left[A^{+}\psi_n(x)\right] &=E_n \left[A^{+}\psi_n(x)\right].
\end{align}
\end{subequations}
Therefore, the eigenfunctions $\widetilde{\psi}_n$ of $\widetilde{H}$ associated with the eigenvalues $E_n$ become
\begin{equation}
\widetilde{\psi}_n\propto A^{+}\psi_n(x)\propto \frac{W(u,\psi_n)}{u},
\end{equation}
where $W(f,g)\equiv fg'-gf'$ is the Wronskian of $f$ and $g$, $\beta(x)=[\ln\, u(x)]'$, and $u(x)$ is a nodeless seed solution of the stationary Schr\"odinger equation for $H$ associated with the complex {\it factorization energy} $\epsilon$, i.e.
\begin{equation}
Hu=\epsilon u.\label{scheq}
\end{equation}
Furthermore, the eigenstates $\widetilde{\psi}_n$ are not automatically normalized as in the real SUSY QM since now
\begin{equation}
\langle A^{+}\psi_n | A^{+}\psi_n \rangle = \langle \psi_n | (A^{+})^{\dag}A^{+} \psi_n \rangle,
\end{equation}
and in this case $(A^{+})^{\dag}A^{+}\neq (H-\epsilon)$. Nevertheless, since they can be normalized we introduce a normalizing constant $C_n$, chosen for simplicity as $C_n \in \mathbb{R}^{+}$, so that
\begin{equation}
\widetilde{\psi}_n(x)=C_n A^{+}\psi_n(x),\quad \langle \widetilde{\psi}_n|\widetilde{\psi}_n\rangle=1.
\end{equation}
Finally, there is a function
\begin{equation}
\widetilde{\psi}_{\epsilon}\propto \frac{1}{u},
\end{equation}
that is also an eigenfunction of $\widetilde{H}$
\begin{equation}
\widetilde{H}\widetilde{\psi}_{\epsilon}=\epsilon\widetilde{\psi}_{\epsilon}.
\end{equation}
If it is normalized, it turns out that $\widetilde{V}(x)$ is a complex potential which corresponding Hamiltonian $\widetilde{H}$ has the following spectrum
\begin{equation}
\text{Sp}(\widetilde{H})=\{\epsilon\}\cup\{E_n,n=0,1,\dots\},\label{spectrum}
\end{equation}
with $\epsilon\in\mathbb{C}$ although, in particular $\epsilon$ could also be real.

\subsection{First-order SUSY partners of the harmonic oscillator}\label{complexsusyHO}
Let us consider now the harmonic oscillator potential
\begin{equation}
V(x)=\frac{1}{2}x^2.\label{hopot}
\end{equation}
In order to apply the first-order SUSY transformation, we just need to supply either a solution of the Riccati equation \eqref{riccatieq} or one of the Schr\"odinger equation \eqref{scheq}. It turns out that the general solution of the Schr\"odinger equation for $V(x)$ in \eqref{hopot} and any $\epsilon\in\mathbb{C}$ is given by
\begin{equation}
u(x)=\exp(-x^2/2)\left[ {}_1F_1\left(\frac{1-2\epsilon}{4},\frac{1}{2};x^2\right)+(\lambda+i\kappa)\,x\, {}_1F_1\left(\frac{3-2\epsilon}{4},\frac{3}{2};x^2\right)\right],
\label{solu}
\end{equation}
with $\lambda,\kappa\in\mathbb{R}$. Thus, in this formalism the first-order SUSY partner potential $\widetilde{V}(x)$ of the harmonic oscillator is
\begin{equation}
\widetilde{V}(x)=\frac{1}{2}x^2-[\ln u(x)]''.
\end{equation}
The previously known result for the real case \citep{JR98} is obtained by taking $\epsilon\in\mathbb{R}$, $\epsilon\leq E_0=1/2$, $\kappa=0$, and expressing $\lambda$ as
\begin{equation}
\lambda=2\nu\frac{\Gamma\left(\frac{3-2\epsilon}{4}\right)}{\Gamma\left(\frac{1-2\epsilon}{4}\right)}.
\end{equation}
On the other hand, for $\epsilon\in\mathbb{C}$ the transformation function $u(x)$ is complex and so is $\widetilde{V}(x)$.

\begin{figure}[H]
\begin{center}
\includegraphics[scale=0.38]{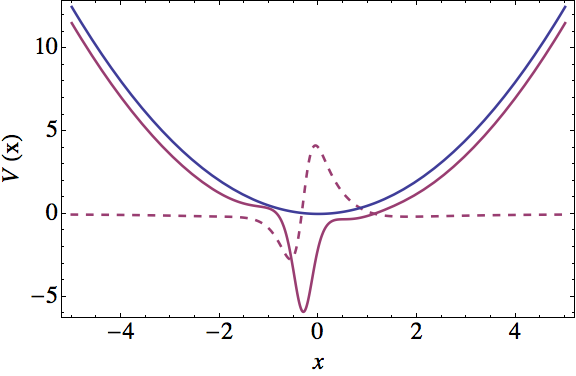}
\includegraphics[scale=0.38]{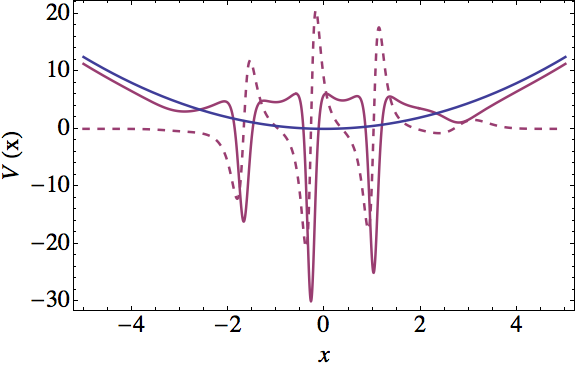}
\end{center}
\vspace{-5mm}
\caption{\small{Examples of SUSY partner potentials of the harmonic oscillator using the two complex factorization {\it energies} $\epsilon=-1+i$ with $\lambda=\kappa=1$ (left) and $\epsilon=3+i10^{-3}$ with $\lambda=\kappa=2$ (right). Its real (magenta solid line) and imaginary (magenta dashed line) parts are compared to the harmonic oscillator potential (blue line).}}\label{susypot}
\end{figure}

\begin{figure}[H]
\begin{center}
\includegraphics[scale=0.35]{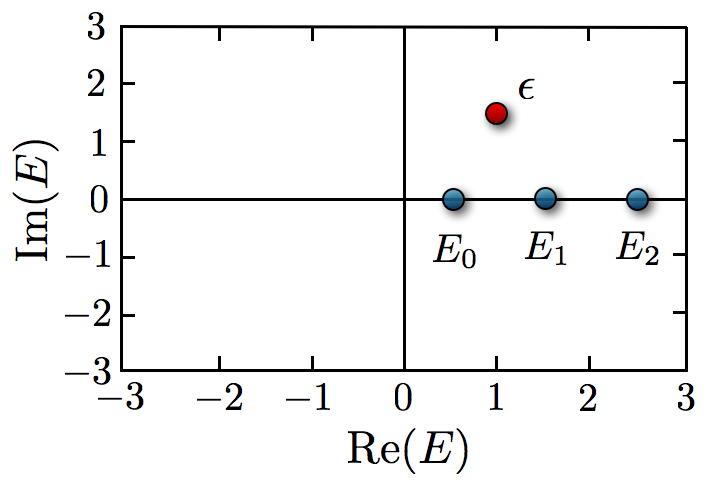}
\end{center}
\vspace{-5mm}
\caption{\small{The complex {\it energy} plane which contains the eigenvalues of the SUSY generated Hamiltonian $\widetilde{H}$. Along the real axis it is shown the usual ladder formed by the energy levels $E_n=n+1/2$, $n= 0,1,\dots $; out of the real line there is a new level at the complex value $\epsilon$.}}
\label{parspace}
\end{figure}

In figure~\ref{susypot} we present some examples of complex SUSY partner potentials of the harmonic oscillator generated for $\epsilon\in\mathbb{C}$ and compare them with the initial potential. Let us note that these new Hamiltonians have the same real spectra as the harmonic oscillator, except that they have one extra {\it energy} level, located at the complex value $\epsilon$. This kind of spectrum is represented in the diagram of figure~\ref{parspace}, which can be interpreted as the superposition of the two ladders shown in equation~\eqref{spectrum}.

\section{$k$th-order SUSY QM. Iterative approach}
\label{secite}

\subsection{Second-order SUSY QM}
Let us apply iteratively the technique discussed in section \ref{secsusy1}, i.e., taking now the resulting $V_1(x)$ as a solvable potential which is used to generate a new one $V_2(x)$ through another intertwining operator $A^{+}_2$ and a new factorization energy $\epsilon_2$, with the restriction $\epsilon_2<\epsilon_1\leq 1/2$ once again taken to avoid singularities in the new potentials and in their eigenfunctions. The corresponding intertwining relation reads
\begin{equation}
H_2A^{+}_2=A^{+}_2H_1,
\end{equation}
which leads to equations similar to \eqref{2alfas} for $V_2$ and $\alpha_2$:
\begin{subequations}
\begin{align}
V_2(x)&=V_1(x)-\alpha'_{2}(x,\epsilon_2),\\
\alpha'_{2}(x,\epsilon_2)+\alpha^2_2(x,\epsilon_2)&= 2[V_1(x)-\epsilon_2].\label{alfa22b}
\end{align}\label{alfa2}
\end{subequations}
\hspace{-1.5mm}In terms of $u^{(1)}_2(x)$ such that $\alpha_2 (x,\epsilon_2)=u_2^{(1)}{}'(x)/u^{(1)}_2(x)$ we have
\begin{subequations}
\begin{align}
V_2 & =  V_1 -\left[\frac{u_2^{(1)}{}'}{u^{(1)}_2}\right]',\\
-\frac{1}{2}u_2^{(1)}{}''+V_1 u^{(1)}_2 &= \epsilon_2 u^{(1)}_2.
\end{align}
\end{subequations}

An important result that will be proven next is that the solution of equation~\eqref{alfa22b} can be algebraically determined using the solutions of the initial Riccati equation \eqref{alfa} for the factorization energies $\epsilon_1$ and $\epsilon_2$ \citep{FHM98,Ros98a,Ros98b,FH99,MNR00}. First, let us take two solutions of the initial Riccati equation 
\begin{equation}
\alpha'_1(x,\epsilon_j)+\alpha^{2}_1(x,\epsilon_j) = 2[V_0(x)-\epsilon_j], \quad j=1,2. \label{alfas}
\end{equation}
Therefore, for the Schr\"odinger equation we have
\begin{equation}
H_0u^{(0)}_j(x)=\epsilon_j u^{(0)}_j(x),
\end{equation}
where
\begin{equation}
u^{(0)}_j(x) \propto \exp\left(\int^{x}_0 \alpha_1 (y,\epsilon_j)\text{d}y\right).
\end{equation}

Let us recall that $u^{(0)}_1(x)$ is used to implement the first SUSY transformation and the eigenfunction of $H_1$ associated with $\epsilon_1$ is given by equation~\eqref{psimathcal}. On the other hand, the eigenfunction of $H_1$ associated with $\epsilon_2$ is
\begin{equation}
u^{(1)}_2 \propto A^{+}_1 u^{(0)}_2 \propto - u_2^{(0)}{}'+\alpha_1 (x,\epsilon_1)u^{(0)}_2 \propto \frac{W\left(  u^{(0)}_1,u^{(0)}_2 \right) }{u^{(0)}_1},
\end{equation}
and taking into account that
\begin{equation}
u_2^{(0)}{}' = \alpha_1 (x,\epsilon_2)u^{(0)}_2,
\end{equation}
we have
\begin{equation}
u^{(1)}_2 \propto [\alpha_1(x,\epsilon_1)-\alpha_1(x,\epsilon_2)]u^{(0)}_2.\label{u2u1}
\end{equation}

To implement the second SUSY transformation we express $u^{(1)}_2$ in terms of the corresponding superpotential
\begin{equation}
u^{(1)}_2(x) \propto \exp\left(\int^{x}_0 \alpha_2 (y,\epsilon_2)\text{d}y\right). \label{u2alfa}
\end{equation}
Substituting \eqref{u2alfa} in equation~\eqref{u2u1}, we obtain
\begin{equation}
\exp\left(\int^{x}_0 \alpha_2 (y,\epsilon_2)\text{d}y\right) \propto [\alpha_1(x,\epsilon_1)-\alpha_1(x,\epsilon_2)]u^{(0)}_2(x).
\end{equation}
Taking the logarithm on both sides
\begin{equation}
\int^{x}_0 \alpha_2 (y,\epsilon_2)\text{d}y = \ln\left[u^{(0)}_2(x)\right] + \ln[\alpha_1(x,\epsilon_1)-\alpha_1(x,\epsilon_2)] +\text{constant},
\end{equation}
and applying the derivative with respect to $x$ it turns out that
\begin{equation}
\alpha_2 (x,\epsilon_2) = \alpha_1(x,\epsilon_2)+\frac{\alpha'_1 (x,\epsilon_1)-\alpha'_1(x,\epsilon_2)}{\alpha_1 (x,\epsilon_1)-\alpha_1(x,\epsilon_2)}.
\end{equation}
Using the initial Riccati equations \eqref{alfas} we obtain that
\begin{equation}
\frac{\alpha'_1 (x,\epsilon_1)-\alpha'_1(x,\epsilon_2)}{\alpha_1 (x,\epsilon_1)-\alpha_1(x,\epsilon_2)} = -\alpha_1(x,\epsilon_2)-\alpha_1(x,\epsilon_1)-\frac{2(\epsilon_1-\epsilon_2)}{\alpha_1(x,\epsilon_1)-\alpha_1(x,\epsilon_2)}.
\end{equation}
Therefore
\begin{equation} \alpha_2(x,\epsilon_2)=-\alpha_1(x,\epsilon_1)-\frac{2(\epsilon_1-\epsilon_2)}{\alpha_1(x,\epsilon_1)-\alpha_1(x,\epsilon_2)}.\label{alfa2b}
\end{equation}

This formula expresses the solution of equation~\eqref{alfa22b} with $V_1(x)=V_0(x)-\alpha'_1(x,\epsilon_1)$ as a {\it finite difference formula} that involves two solutions $\alpha_1(x,\epsilon_1)$ and $\alpha_1(x,\epsilon_2)$ of the Riccati equation \eqref{alfa} for the factorization energies $\epsilon_1,\epsilon_2$ \citep{FHM98}. Even more, a similar expression was found by \citet{Adl94} in order to discuss the B\"acklund transformations of the Painlev\'e equations.

\begin{figure}\centering
\includegraphics[scale=0.4]{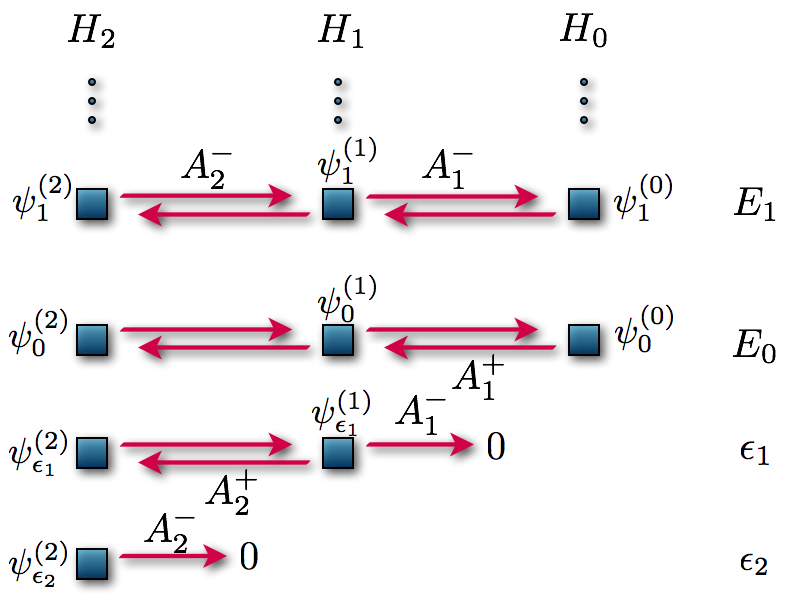}
\caption{\small{Iteration of two consecutive first-order SUSY transformations applied to $H_0$ and $H_1$ using as transformation functions two non-physical eigenfunctions of $H_0$ with factorization energies $\epsilon_1 \leq E_0$ and $\epsilon_2 < \epsilon_1$.}}\label{fig.susyqm2do}
\end{figure}

On the other hand, the potential $V_2(x)$ is expressed as
\begin{equation}
V_2(x)=V_1(x)-\alpha'_2 (x,\epsilon_2)=V_0(x) +\left[\frac{2(\epsilon_1-\epsilon_2)}{\alpha_1(x,\epsilon_1)-\alpha_1(x,\epsilon_2)}\right]',
\end{equation}
the eigenfunctions associated with $H_2$ are given by
\begin{subequations}
\begin{align}
\psi^{(2)}_{\epsilon_2}(x) &\propto \exp\left(-\int^{x}_0 \alpha_2(y,\epsilon_2)\text{d}y\right)=\frac{1}{u^{(1)}_2(x)}, \label{psi22a}\\
\psi^{(2)}_{\epsilon_1}(x) &= \frac{A^{+}_{2} \psi^{(1)}_{\epsilon_1}(x)}{(\epsilon_1-\epsilon_2)^{1/2}},\\
\psi^{(2)}_{n}(x) &= \frac{A^{+}_{2} \psi^{(1)}_{n}(x)}{(E_n-\epsilon_2)^{1/2}}= \frac{A^{+}_2A^{+}_1\psi^{(0)}_n(x)}{[(E_n-\epsilon_1)(E_n-\epsilon_2)]^{1/2}},
\end{align}
\end{subequations}
and the corresponding eigenvalues are such that Sp$(H_2)=\{\epsilon_2,\epsilon_1,E_n;n=0,1,\dots\}$. The scheme representing this transformation is shown in figure \ref{fig.susyqm2do}.

\subsection{$k$th-order SUSY QM}
This iterative process can be continued to higher orders using several solutions associated with different factorization energies $\epsilon_i$ of the initial Riccati equation \eqref{alfa}. Let us assume that we know $k$ of them, $\{ \alpha_1(x,\epsilon_j); j=1,\dots,k, \}$, where $\epsilon_{j+1}<\epsilon_{j}$ and the process is iterated $k$ times. Therefore, we obtain a new solvable Hamiltonian $H_k$, whose potential reads
\begin{equation}
V_k(x)=V_{k-1}(x)-\alpha'_k(x,\epsilon_k)=V_0(x) - \sum_{j=1}^{k}\alpha'_j(x,\epsilon_j),\label{Vk}
\end{equation}
where $\alpha_j(x,\epsilon_j)$ is given by a recursive finite difference formula which can be obtained as a generalization of equation \eqref{alfa2b}, i.e.,
\begin{equation}
\alpha_{j+1}(x,\epsilon_{j+1})=-\alpha_j(x,\epsilon_j)-\frac{2(\epsilon_j-\epsilon_{j+1})}{\alpha_j(x,\epsilon_j)-\alpha_j(x,\epsilon_{j+1})}, \ \ \ j=1,\dots,k-1. \label{alfai}
\end{equation}

The eigenfunctions of $H_k$ are given by
\begin{subequations}
\begin{align}
\psi^{(k)}_{\epsilon_k}(x) &\propto \exp\left(-\int^{x}_0 \alpha_k(y,\epsilon_k)\text{d}y\right),\\
\psi^{(k)}_{\epsilon_{k-1}}(x) &= \frac{A^{+}_{k}\psi^{(k-1)}_{\epsilon_{k-1}}(x)} {(\epsilon_{k-1}-\epsilon_k)^{1/2}},\label{psie}\\
				&\ \ \vdots \nonumber
\end{align}
\begin{align}
\psi^{(k)}_{\epsilon_1}(x) &= \frac{A^{+}_{k}\dots A^{+}_2 \psi^{(1)}_{\epsilon_1}(x)} {[(\epsilon_1-\epsilon_2)\dots (\epsilon_1-\epsilon_k)]^{1/2}},\\
\psi^{(k)}_{n}(x) &=\frac{A^{+}_k\dots A^{+}_1\psi^{(0)}_n(x)} {[(E_n-\epsilon_1)\dots(E_n-\epsilon_k)]^{1/2}}.\label{psin}
\end{align}\label{psisk}
\end{subequations}
\hspace{-1.5mm}The corresponding eigenvalues belong to the set Sp$(H_k)=\{ \epsilon_j, E_n; j=k,\dots,1;$ $n=0,1,\dots \}$.

In order to have a complete scheme, let us recall how the SUSY partners $H_j$ of the initial Hamiltonian $H_0$ are intertwined to each other
\begin{equation}
H_j A^{+}_j=A_j^{+}H_{j-1}, \quad j=1,\dots ,k. \label{HAAH}
\end{equation}
Then, starting from $H_0$ we have generated a chain of factorized Hamiltonians in the way
\begin{subequations}
\begin{align}
H_0&=A_1^{-}A_1^{+}+\epsilon_1,\\
H_j&=A^{+}_jA_j^{-}+\epsilon_j=A^{-}_{j+1}A^{+}_{j+1}+\epsilon_{j+1}, \quad j=1,\dots,k-1,\\
H_k&=A^{+}_kA^{-}_k+\epsilon_k,
\end{align}\label{HAA}
\end{subequations}
\hspace{-1mm}where the final potential $V_k(x)$ can be determined recursively using equations \eqref{Vk} and \eqref{alfai}.

In addition, since we are asking for $k$ solutions of the initial Riccati equation, $\{\alpha_1(x,\epsilon_i);\ i=1,\dots,k\}$, then we also obtain $k$ non-equivalent factorizations of the Hamiltonian $H_0$,
\begin{equation}
H_0 = \frac{1}{2}\left[\frac{\text{d}}{\text{d}x} + \alpha_1(x,\epsilon_i)\right]\left[-\frac{\text{d}}{\text{d}x}+\alpha_1(x,\epsilon_i)\right]+\epsilon_i, \quad i=1,\dots,k.
\end{equation}

We must note now that there exists a $k$th-order differential operator, $B_{k}^{+} \equiv A^{+}_k\dots A^{+}_1$, that intertwines the initial Hamiltonian $H_0$ with the final one $H_k$ as follows
\begin{equation}
H_k B^{+}_k = B^{+}_k H_0.
\label{Bdag}
\end{equation}

From equations \eqref{psisk} we obtain
\begin{equation}
B^{+}_k\psi^{(0)}_n = [(E_n-\epsilon_1)\dots(E_n-\epsilon_k)]^{1/2}\psi^{(k)}_n,
\label{Bdag2}
\end{equation}
meanwhile the equation adjoint to \eqref{Bdag} leads to
\begin{equation}
B_k^{-}\psi^{(k)}_n = [(E_n-\epsilon_1)\dots(E_n-\epsilon_k)]^{1/2}\psi^{(0)}_n.
\label{Bdag3}
\end{equation}

These equations immediately lead to the higher-order SUSY QM \citep{AIS93,AICD95,BS97,FGN98,FHM98,Ros98a, Ros98b,FH99,BGBM99,MNR00}. In this treatment, the standard SUSY algebra with two generators $\{Q_1,Q_2\}$ \citep{Wit81},
\begin{equation}
[Q_i,H_{ss}]=0, \ \ \ \{Q_i,Q_j\}=\delta_{ij}H_{ss}, \ \ \ i,j=1,2,
\label{Qis}
\end{equation}
it can be realized from $B_k^{-}$ and $B^{+}_k$ through the definitions
\begin{subequations}
\begin{align}
Q^{-}&=\left(\begin{array}{cc}
  0 &  0 \\
  B_k^{-} &  0 
  \end{array}\right),\\
Q^{+}&=\left(\begin{array}{cc}
  0 &  B^{+}_k \\
  0 &  0 
  \end{array}\right),\\
H_{ss}\equiv\{Q^{-},Q^{+}\}&=\left(\begin{array}{cc}
  B^{+}_kB_k^{-} &  0 \\
  0 &  B_k^{-}B^{+}_k 
  \end{array}\right),
  \label{Qis2}
\end{align}\end{subequations}
\hspace{-1mm}where $Q_1\equiv (Q^{+}+Q^{-})/2^{1/2}$ and $Q_2\equiv i(Q^{-}-Q^{+})/2^{1/2}$. Given that
\begin{subequations}
\begin{align}
B^{+}_kB_k^{-}&=A^{+}_k\dots A^{+}_1 A_1^{-}\dots A_k^{-}\nonumber\\
								&=A^{+}_k\dots A^{+}_2 (H_1-\epsilon_1)A_2^{-}\dots A_k^{-}\nonumber\\
								&=A^{+}_k\dots A^{+}_3 (H_2-\epsilon_1)A_2^{+}A_2^{-}\dots A_k^{-}\nonumber\\
								&=A^{+}_k\dots A^{+}_3 (H_2-\epsilon_1)(H_2-\epsilon_2)A_3^{-}\dots A_k^{-}\nonumber\\
								&=(H_k-\epsilon_1)\dots (H_k-\epsilon_k), \label{Bdag4}\\
B_k^{-}B^{+}_k			&=A_1^{-}\dots A_k^{-} A^{+}_k\dots A^{+}_1\nonumber\\
								&=A_1^{-}\dots A_{k-1}^{-}(H_{k-1}-\epsilon_k)A^{+}_{k-1}\dots A^{+}_1\nonumber\\
								&=A_1^{-}\dots A_{k-2}^{-}(H_{k-2}-\epsilon_k)A_{k-1}^{-}A^{+}_{k-1}\dots A^{+}_1\nonumber\\
								&=(H_0-\epsilon_1)\dots (H_0-\epsilon_k),\label{Bdag5}
\end{align}
\end{subequations}
it turns out that the SUSY generator ($H_{ss}$) is a $k$th-order polynomial of the Hamiltonian $H^p_s$ that involves the two intertwined Hamiltonians $H_0$ and $H_k$,
\begin{equation}
H_{ss}=(H^p_s -\epsilon_1)\dots(H^p_s -\epsilon_k),
\label{Hss}
\end{equation}
where
\begin{equation}
H^p_s=\left(\begin{array}{cc}
  H_k &  0 \\
  0 &  H_0 
  \end{array}\right).\\
  \label{Hps}
\end{equation}

In particular, for $k=1$ we obtain the standard SUSY QM for which $H_{ss}=(H^p_s-\epsilon_1)$, i.e., $H_{ss}$ is linear in $H^p_s$. For $k=2$ we get the quadratic superalgebra or SUSUSY QM due to $H_{ss}=(H^p_s-\epsilon_1)(H^p_s-\epsilon_2)$ \citep{AIS93,Fer97,FGN98}.

\section{Second-order SUSY QM. Direct approach}
\label{secdir}
For the direct approach to the $k$th-order SUSY QM we propose from the start that the intertwining operator in equation \eqref{Bdag} is of $k$th-order
\begin{equation}
B^{+}_k=2^{-k/2}\left[(-1)^{k}\frac{\text{d}^k}{\text{d}x^k} +f_{k-1}(x)\frac{\text{d}^{k-1}}{\text{d}x^{k-1}}+\dots + f_1(x)\frac{\text{d}}{\text{d}x} +f_0\right],
\label{Bk}
\end{equation}
where the real functions $\{f_j(x);j=0,\dots,k-1\}$ can, in principle, be determined through a similar approach to the first-order case. Furthermore, equations~(\ref{Bdag}--\ref{Hps}) are still valid. Next we present the simplest non-trivial case for $k=2$, which will clearly illustrate the advantages for the {\it spectral design} of the direct approach when compared with the iterative one.

The second-order SUSY QM \citep{AIS93,AICD95,BS97,Fer97,FR06} emerges when one considers a second-order intertwining operator such that
\begin{subequations}
\begin{align}
H_2B_2^{+}&=B_2^{+}H_0,\label{H2B2}\\
						 H_j&=-\frac{1}{2}\frac{\text{d}^{2}}{\text{d}x^2}+V_j(x), \quad j=0,2,\label{2S_Ht}\\
	 B^{+}_2&=\frac{1}{2}\left[\frac{\text{d}^2}{\text{d}x^2} - g(x)\frac{\text{d}}{\text{d}x} + h(x)\right].\label{2S_Bmas}
\end{align}
\end{subequations}

The calculation for the left hand side of equation \eqref{H2B2} leads to
\begin{equation}
2H_2B^{+}_2=-\frac{1}{2}\frac{\text{d}^4}{\text{d}x^4} + \frac{g}{2}\frac{\text{d}^3}{\text{d}x^3} +\left(g'-\frac{h}{2}+V_2\right)\frac{\text{d}^2}{\text{d}x^2} + \left(\frac{g''}{2}-h'- g V_2\right)\frac{\text{d}}{\text{d}x} +h V_2 - \frac{h''}{2},
\end{equation}
while from the right hand side it follows that
\begin{equation}
2B^{+}_2H_0=-\frac{1}{2}\frac{\text{d}^4}{\text{d}x^4} + \frac{g}{2}\frac{\text{d}^3}{\text{d}x^3} +\left(V_0-\frac{h}{2}\right)\frac{\text{d}^2}{\text{d}x^2} + \left(2V'_0 - g V_0\right)\frac{\text{d}}{\text{d}x} +V''_0 - g V'_0+h V_0.
\end{equation}

Matching the coefficients of the same powers of the differential operator $\text{d}/\text{d}x$, the following system of equations is found
\begin{subequations}
\begin{align}
V_2 &= V_0 - g',\label{gammaa}\\
\frac{g''}{2}-h'-g V_2 &=2V'_0-g V_0,\label{gammab}\\
h V_2-\frac{h''}{2}&= V''_0-g V'_0+h V_0.\label{gammac}
\end{align}
\end{subequations}
Substituting equation \eqref{gammaa} in \eqref{gammab} and solving for $h'$ we obtain
\begin{equation}
h'=\frac{g''}{2}+gg'-2V'_0.\label{hder}
\end{equation}
Integrating this equation with respect to $x$ it turns out that
\begin{equation}
h=\frac{g'}{2}+\frac{g^2}{2}-2V_0+d,
\label{gammai}
\end{equation}
where $d$ is a real constant. If we derive equation \eqref{hder} we arrive to
\begin{equation}
h''= \frac{g'''}{2}+gg''+g'^{2}-2V''_0.
\label{gamma2a}
\end{equation}

Now, solving equation \eqref{gammac} for $h''$ we get
\begin{equation}
h''=2(g V'_0 - hg' - V''_0).
\label{gamma2b}
\end{equation}

Substituting equation \eqref{gammai} in the right side of \eqref{gamma2b} and matching with the result from \eqref{gamma2a} we obtain
\begin{equation}
\frac{g'''}{2}+gg''+2g'^2=2(g V'_0+2g'V_0)-g^2g'-2dg'.
\end{equation}

If we multiply this equation by $g$ and we add and subtract $g'g''/2$ it turns out that
\begin{equation}
2(g^2 V_0)'=\frac{gg'''}{2}+\frac{g'g''}{2}+(g^2g')'+g^3g'+2dgg' -\frac{g'g''}{2}.
\end{equation}
Integrating with respect to $x$ and rearranging the terms it is found that
\begin{equation}
\frac{gg''}{2}-\frac{g'^{2}}{4}+g^2g'+\frac{g^4}{4}-2V_0g^2+dg^2+c=0,
\label{eta}
\end{equation}
where $c$ is a real constant. Therefore, given $V_0(x)$ we can obtain the new potential $V_2(x)$ and the function $h(x)$ from equations \eqref{gammaa} and \eqref{gammai} once we have the explicit solution for $g(x)$ of equation~\eqref{eta}. To find this solution, we make use of the following \emph{ansatz} \citep{FGN98,Ros98a,Ros98b,FH99}
\begin{equation}
g'=-g^2+2\gamma g+2\xi,
\label{etap}
\end{equation}
where $\gamma(x)$ and $\xi(x)$ are functions to be determined. With this assumption we obtain
\begin{subequations}
\begin{align}
g''&=-2gg'+2\gamma' g+2\gamma g'+2\xi',\\
\frac{gg''}{2}&=-g^2g'+\gamma' g^2+\gamma gg'+g\xi',\label{ans2}\\
\frac{g'^2}{4}&=\frac{g^4}{4}-\gamma g^3+(\gamma^2-\xi)g^2+2\gamma\xi g+\xi^2.\label{ans3}
\end{align}\label{ans}
\end{subequations}
\hspace{-1.5mm}The substitution of equations \eqref{ans2} and \eqref{ans3} in \eqref{eta} leads to
\begin{equation}
\gamma' g^2+\gamma gg'+g\xi' + \gamma g^3+(\xi-\gamma^2)g^2 - 2\gamma\xi g-\xi^2 -2V_0 g^2 +dg^2+c=0,
\end{equation}
and using again equation \eqref{etap} to eliminate $g'$ we get
\begin{equation}
(\gamma'+\gamma^2-2V_0+\xi+d)g^2+\xi' g-\xi^2+c=0. \label{beta}
\end{equation}
As this equation should be valid for any function $g$, the coefficient for each power of $g$ must be zero, which leads to $c \equiv \xi^2 $. Therefore
\begin{subequations}
\begin{align}
\gamma'(x)+\gamma^2(x)&=2[V_0(x)-\epsilon],\label{ecbeta}\\
\epsilon&\equiv \frac{d+\xi}{2}.
\end{align}
\end{subequations}

Alternatively, we can work with the Schr\"odinger equation related with \eqref{ecbeta} through the substitution $\gamma=u^{(0)'}/u^{(0)}$ \citep{CRF01}
\begin{equation}
-\frac{1}{2}u^{(0)''}+V_0u^{(0)}=\epsilon u^{(0)}. \label{riccati}
\end{equation}
According to whether $c$ is zero or not, $\xi$ can vanish or take two different values $\xi=\pm c^{1/2}$. For $c=0$, we need to solve \eqref{ecbeta} for $\gamma(x)$ and then the resulting equation \eqref{etap} for $g(x)$. For $c\neq 0$ we have two different equations \eqref{ecbeta} for $\gamma(x)$, with two factorization energies
\begin{subequations}
\begin{align}
\epsilon_1 & \equiv \frac{d+c^{1/2}}{2},\\
\epsilon_2 & \equiv \frac{d-c^{1/2}}{2}.
\end{align}\label{matEs}
\end{subequations}
\hspace{-1mm}Once these equations are solved, it is possible to construct a common algebraic solution $g(x)$ for the two equations \eqref{etap}. There is an essential difference between the cases for $c>0$ and $c<0$, since the first leads to real $\epsilon_1$, $\epsilon_2$ and the second to complex ones. Therefore, we obtain a natural classification of the solutions $g(x)$ based on the sign of $c$, namely, there are three different cases: real, confluent, and complex (see table~\ref{table1}).
\begin{table}\centering
\begin{tabular}{ll}
\hline
Value of $c$&Type of transformation\\
\hline
$c>0$&Real case\\
$c=0$&Confluent case\\
$c<0$&Complex case\\
\hline
\end{tabular}
\vspace{-3mm}
\caption{\small{The three types of second-order SUSY transformations.}} \label{table1}
\end{table}

\begin{figure}\centering
\includegraphics[scale=0.4]{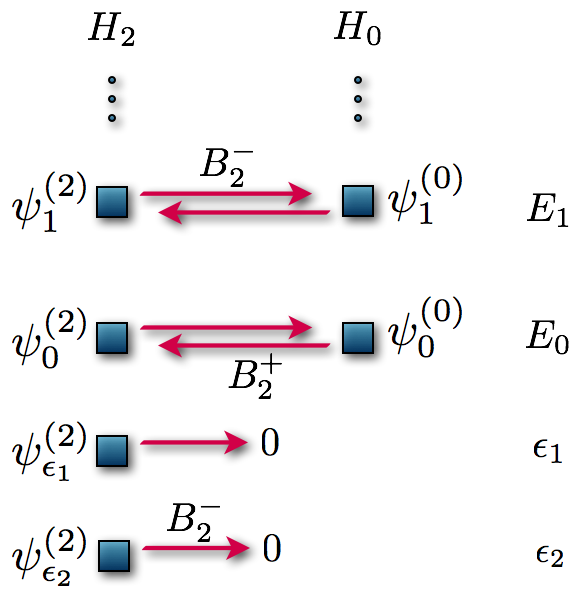}
\caption{\small{Diagram of the second-order SUSY transformation for the direct approach.}}\label{fig.susyqm2dod}
\end{figure}

A scheme representing the second-order SUSY transformation for the direct approach is shown in figure \ref{fig.susyqm2dod}.

\subsection{Real case ($c>0$)}
\label{realcase}
In this case we consider $\epsilon_1,\epsilon_2 \in \mathbb{R}$ such that $\epsilon_1 \neq \epsilon_2$ and the corresponding solutions of the Riccati equation \eqref{ecbeta} are denoted as $\gamma_1(x)$ and $\gamma_2(x)$, respectively. Each of them leads to a different expression for equation \eqref{etap}, namely,
\begin{subequations}
\begin{align}
g'(x)&=-g^2(x)+2\gamma_1(x)g(x)+2(\epsilon_1-\epsilon_2),\\
g'(x)&=-g^2(x)+2\gamma_2(x)g(x)+2(\epsilon_2-\epsilon_1).
\end{align}
\end{subequations}
When we subtract them, we obtain an algebraic solution for $g(x)$ in terms of $\epsilon_1$, $\epsilon_2$, $\gamma_1(x)$, and $\gamma_2(x)$
\begin{equation}
g(x)=-\frac{2(\epsilon_1-\epsilon_2)}{\gamma_1(x)-\gamma_2(x)}.
\end{equation}
If we use now the corresponding solutions of the Schr\"odinger equation we obtain
\begin{equation}
g(x)=\frac{2(\epsilon_1-\epsilon_2)u_1^{(0)}u_2^{(0)}}{W(u_1^{(0)},u_2^{(0)})}=\frac{W'(u_1^{(0)},u_2^{(0)})}{W(u_1^{(0)},u_2^{(0)})}.\label{wronsk}
\end{equation}
Therefore, the potentials $V_2(x)$ do not have singularities if $W(u_1^{(0)},u_2^{(0)})$ does not have zeroes, as can be seen from equation \eqref{gammaa}.

The spectrum of $H_2$, Sp($H_2$), can differ from Sp($H_0$) according to the normalization of the two mathematical eigenfunctions $\psi_{\epsilon_1}^{(2)},\psi_{\epsilon_2}^{(2)}$ of $H_2$ associated with $\epsilon_1$ and $\epsilon_2$, that belong to the kernel of $B_2^{-}$
\begin{subequations}
\begin{align}
B_2^{-} \psi_{\epsilon_j}^{(2)}&=0, \ \ \ \ j=1,2,\\
H_2\psi_{\epsilon_j}^{(2)}&=\epsilon_j \psi_{\epsilon_j}^{(2)}.
\end{align}
\end{subequations}

For the solution associated with $\epsilon_1$ the two explicit equations to solve are
\begin{subequations}
\begin{align}
\psi_{\epsilon_1}^{(2)}{}'' + g \psi_{\epsilon_1}^{(2)}{}'+(h+g')\psi_{\epsilon_1}^{(2)}&=0,\\
\psi_{\epsilon_1}^{(2)}{}''-2(V_2 - \epsilon_1)\psi_{\epsilon_1}^{(2)}&=0.
\end{align}\label{psi1}
\end{subequations}
\hspace{-1mm}Eliminating $\psi_{\epsilon_1}^{(2)}{}''$ from both equations we get
\begin{equation}
g\psi_{\epsilon_1}^{(2)}{}'+(g' +h+2V_2-2\epsilon_1)\psi_{\epsilon_1}^{(2)}=0,
\end{equation}
with the expressions for $V_2$ and $h$ given by equations \eqref{gammaa} and \eqref{gammai} with $d=\epsilon_1+\epsilon_2$, which can be obtained by adding equations \eqref{matEs}. Then
\begin{equation}
\frac{\psi_{\epsilon_1}^{(2)}{}'}{\psi_{\epsilon_1}^{(2)}}=\frac{g'-g^2+2(\epsilon_1-\epsilon_2)}{2g},
\end{equation}
and using the \emph{ansatz} proposed in \eqref{etap} we obtain
\begin{equation}
\frac{\psi_{\epsilon_1}^{(2)}{}'}{\psi_{\epsilon_1}^{(2)}}=\frac{g'}{g}-\gamma_1=\frac{g'}{g}-\frac{u_1^{(0)}{}'}{u_1^{(0)}},
\end{equation}
which can be easily integrated to obtain
\begin{equation}
\psi_{\epsilon_1}^{(2)}\propto \frac{g}{u_1^{(0)}}\propto \frac{u_2^{(0)}}{W(u_1^{(0)},u_2^{(0)})}.
\label{psi2}
\end{equation}
A similar procedure leads to
\begin{equation}
\psi_{\epsilon_2}^{(2)}\propto \frac{g}{u_2^{(0)}}\propto \frac{u_1^{(0)}}{W(u_1^{(0)},u_2^{(0)})}.
\end{equation}

The second-order SUSY QM offers wider possibilities of spectral manipulation. Indeed, it has been found an heuristic criterion that provides useful information about these possibilities \citep{Fer10}. Remember that the product
\begin{equation}
B_2^{-}B_2^{+}=(H_0-\epsilon_1)(H_0-\epsilon_2),
\end{equation}
is a positive definite operator in its domain. In particular, this is valid for the basis of energy eigenstates $|\psi_n^{(0)} \rangle$ of $H_0$, and therefore we have
\begin{equation}
\langle \psi_n^{(0)}|B_2^{-}B_2^{+}|\psi_n^{(0)}\rangle = (E_n-\epsilon_1)(E_n-\epsilon_2) \geq 0, \quad n=0,1,\dots \label{b2b2}
\end{equation}
In 1-SUSY the equivalent result is $(E_n-\epsilon_1)\geq 0$, and that is why we conclude that $\epsilon_1 \leq E_n$ for that case. But now equation \eqref{b2b2} opens up unexpected possibilities for the positions of the new levels $\epsilon_1,\epsilon_2$. A non-exhaustive list of different situations for the spectral design is shown next

\begin{enumerate}[(a)]

\item If $\epsilon_2 < \epsilon_1< E_0$.\\
The heuristic criterion suggests that it is possible to find $u_1^{(0)}$ and $u_2^{(0)}$ such that $W(u_1^{(0)},u_2^{(0)})$ has no zeroes and that $\psi_{\epsilon_1}^{(2)}$ and $\psi_{\epsilon_2}^{(2)}$ can be normalized. In fact, if we choose the ordering $\epsilon_2 <\epsilon_1$, one has to find a $u_1^{(0)}$ without zeroes and $u_2^{(0)}$ with only one. Let $x_0$ be such that $u_2^{(0)}(x_0)=0$. Due to $W'(u_1^{(0)},u_2^{(0)})=2(\epsilon_1-\epsilon_2)u_1^{(0)}u_2^{(0)}$, in this point $W''(u_1^{(0)},u_2^{(0)})/W(u_1^{(0)},u_2^{(0)})=2(\epsilon_1-\epsilon_2)>0$, which implies that the Wronskian acquires a minimum positive or a maximum negative in $x_0$; in such a case it turns out that $W(u_1^{(0)},u_2^{(0)})$ has no zeroes. The spectrum of the new Hamiltonian is Sp$(H_2)=\{ \epsilon_2,\epsilon_1,E_n;n=0,1,\dots \}$. An illustration of a potential $V_2(x)$ generated from the harmonic oscillator is presented in figure \ref{fig.hsusyqm1}, although the treatment presented here is valid for any potential with exact solution. This is the typical case obtained when we follow the iterative approach to obtain a second-order SUSY transformation from two first-order ones.

\begin{figure}\centering
\includegraphics[scale=0.6]{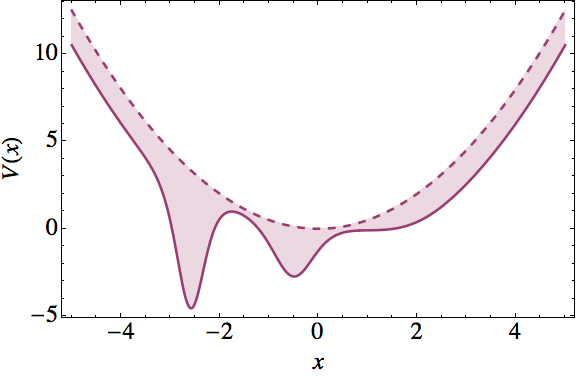}
\caption{\small{SUSY partner potential $V_2(x)$ (solid line) of the harmonic oscillator $V_0(x)$ (dashed line), generated using two seed solutions $u_1^{(0)}$ with $\epsilon_1=-1,\ \nu_1=0.99$ and $u_2^{(0)}$ with  $\epsilon_2=-1.2,\ \nu_1=1.01$. We mark in color the difference between the two potentials.}}\label{fig.hsusyqm1}
\end{figure}

\item If $E_j<\epsilon_2<\epsilon_1<E_{j+1}$.\\
The heuristic criterion suggests that we can find two solutions $u_1^{(0)}$ and $u_2^{(0)}$ such that $W(u_1^{(0)},u_2^{(0)})$ has no zeroes and that $\psi_{\epsilon_1}^{(2)}$ and $\psi_{\epsilon_2}^{(2)}$ can be normalized so that Sp$(H_2)=\{E_0,\dots,E_j,\epsilon_2,\epsilon_1,E_{j+1},\dots \}$ \citep{Sam99,FMRS02}. A plot for a potential $V_2(x)$ of this kind is shown in figure \ref{fig.hsusyqm2}. The possibility is fulfilled if we choose two solutions $u_2^{(0)}$ and $u_1^{(0)}$ with $j+2$ and $j+1$ zeroes, respectively. Taking into account the {\it oscillation theorem}, which states that between two zeroes of $u_2^{(0)}$ there is, at least, one zero of $u_1^{(0)}$, it turns out that the $2j+3$ zeroes are alternating. These zeroes, $x_0< x_1 <\dots< x_{2j+2}$, are also critical points of $W(u_1^{(0)},u_2^{(0)})\equiv W(x)$. Due to $W(x_j)/W(x_{j+1})>0$, thus $W(x)$ has no zeroes in the interval $(x_j,x_{j+1})$, then, it has no zeroes in the domain $(x_0,x_{2j+2})$. Finally, $x_0$ is a zero of $u_2^{(0)}$, therefore $W''(x_0)/W(x_0)=2(\epsilon_1-\epsilon_2)>0$. Therefore, $W(x)$ is a maximum negative or a minimum positive in $x_0$. In both cases $W(x)$ never crosses the $x$-axis in the interval $(-\infty,x_0)$. A similar consideration lead us to conclude that the Wronskian is not zero in $(x_{2j+2},\infty)$ and, therefore, it does not have zeroes in the full real axis.

\begin{figure}\centering
\includegraphics[scale=0.6]{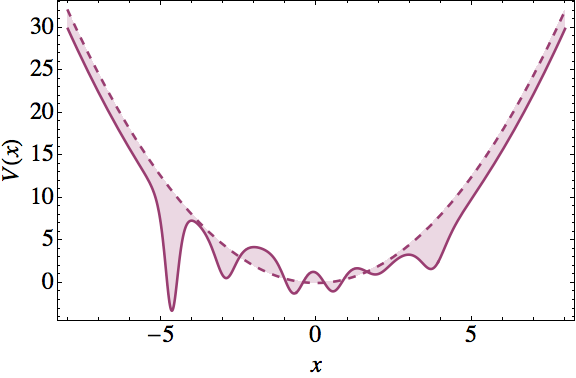}
\caption{\small{SUSY partner potential $V_2(x)$ (solid line) of the harmonic oscillator $V_0(x)$ (dashed line), generated with two seed solutions: $u_1^{(0)}$ with $\epsilon_1=3.2,\ \nu_1=1.01$ and $u_2^{(0)}$ with $\epsilon_2=3,\ \nu_1=0.99$. We remark in color the difference between the two potentials.}}\label{fig.hsusyqm2}
\end{figure}

\item For $\epsilon_2=E_j,\ \epsilon_1=E_{j+1}$ and $u_2^{(0)}=\psi_j^{(0)},\ u_1^{(0)}=\psi_{j+1}^{(0)}$.\\
Then $W(u_1^{(0)},u_2^{(0)})$ has no zeroes but $\psi_{\epsilon_1}^{(2)}$ and $\psi_{\epsilon_2}^{(2)}$ are not normalizable. To see this, we must note that $u_2^{(0)}$ and $u_1^{(0)}$ have $j$ and $j+1$ zeroes, respectively. Due to the null asymptotic behaviour for $x\rightarrow \pm \infty$ of both solutions, it turns out that these $2j+1$ zeroes, ordered as $x_0<x_1<\dots <x_{2j}$, are once again alternating but now $x_0$ and $x_{2j}$ are zeroes of $u_1^{(0)}$. Using an argument similar to the previous case, we obtain that $W(x)$ has no zeroes in $(x_0,x_{2j})$. On the other hand, $W(x)$ is monotonically increasing on the interval $(-\infty,x_0)$ and $W''(x_0)/W(x_0)=2(\epsilon_2-\epsilon_1)<0$, which implies that $W(x)$ reaches a positive maximum or a negative minimum in $x=x_0$. Since $\lim_{x\rightarrow-\infty} W(x)=0$, it turns out that the only zero for $W(x)$  in the interval $(-\infty,x_0)$ appears when $x\rightarrow -\infty$. A similar procedure shows that in $(x_{2j},\infty)$, $W(x)$ has a null asymptotic behaviour as $x\rightarrow\infty$. In conclusion, the Wronskian does not have zeroes in the full real axis, except for the null asymptotic behaviour when $x\rightarrow\pm\infty$. This implies that the second-order SUSY transformation is not singular in the initial domain, therefore the intertwining operator reproduces the same boundary conditions for the eigenfunctions of $H_0$, except that now the eigenfunctions of $H_2$ associated with $E_j$ and $E_{j+1}$ are no longer square-integrable. Therefore,  Sp$(H_2)=\{E_0,\dots,E_{j-1},E_{j+2},\dots \}$, i.e., in a way we have \emph{deleted} the levels $E_j,E_{j+1}$ to generate $V_2(x)$.
\end{enumerate}

According to the standard treatment of SUSY QM, the new levels will always be below the ground state of the initial Hamiltonian. Nevertheless, in cases $(b)$ and $(c)$ we have shown that this can be surpassed, gaining more freedom to manipulate the spectrum of the final potential. In principle, this atypical cases can also be obtained from two consecutive first-order SUSY transformations, but the corresponding interpretation would be strange, since in the first transformation we would generate a singular potential $V_1$, whose singularities are caused by the zeroes of the transformation function in use. Then, the second transformation would remove all those singularities to finally obtain a non-singular potential $V_2$. Next, we are going to explore the other cases of the classification induced by $c$ and we will expand the domain of the SUSY transformations to new potentials, which cannot be obtained by iterations of first-order transformations.

\subsection{Confluent case ($c=0$)}
\label{secconfluente}
In this case $c=0$, therefore $\epsilon_1=\epsilon_2 \equiv \epsilon \in \mathbb{R}$. Once we have found a solution $\gamma (x)$ of the Riccati equation \eqref{ecbeta}, we must solve the Bernoulli equation resulting from \eqref{etap}:
\begin{equation}
g'=-g^2+2\gamma g.
\end{equation}

To solve it, we make $g=1/y$, which implies that
\begin{equation}
y'+2\gamma y =1,
\end{equation}
whose general solution is given by
\begin{equation}
y=\left[ w_0 + \int \exp \left(2\int\gamma(x)\text{d}x\right)\text{d}x\right]\exp\left(-2\int\gamma(x)\text{d}x\right),
\end{equation}
$w_0$ being a real constant. Therefore, the general solution of $g$ is
\begin{equation}
g(x)=\frac{\exp\left(2\int\gamma(x)\text{d}x\right)}{w_0 + \int\exp\left(2\int\gamma(x)\text{d}x\right)\text{d}x}.
\end{equation}
In terms of the solution of the Schr\"odinger equation, $u^{(0)}(x)\propto \exp\left(\int\gamma(x)\text{d}x\right)$, we have
\begin{equation}
g(x)=\frac{[ u^{(0)}(x) ]^2}{w_0 + \int_{x_0}^{x}[ u^{(0)}(y) ]^2 \text{d}y}=\frac{w'(x)}{w(x)},\label{confformula}
\end{equation}
where $x_0$ is a fixed point in the domain of $V_0$ and
\begin{equation}
w(x)=w_0 + \int_{x_0}^{x} [ u^{(0)}(y) ]^2 \text{d}y.\label{intformula}
\end{equation}

\begin{figure}\centering
\includegraphics[scale=0.6]{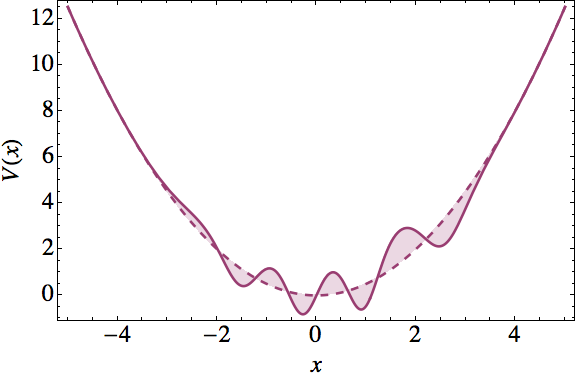}
\caption{\small{SUSY partner potential $V_2(x)$ (solid line) of the harmonic oscillator $V_0(x)$ (dashed line), generated through the confluent algorithm using the third excited state of $H_0$, $w_0=-0.51$ and $x_0=10$. We remark in color the difference between the two potentials.}}\label{fig.hsusyqm3}
\end{figure}

To accomplish that $V_2(x)$ has no singularities, $w(x)$ must not have zeroes and as $w(x)$ is a monotonically non-decreasing function, a simple solution \citep{FS03} consists in a using transformation function $u^{(0)}$ such that
\begin{equation}
\lim_{x\rightarrow\infty}u^{(0)}=0, \quad I_{+}=\int_{x_0}^{\infty} [ u^{(0)}(y) ]^2 \text{d}y < \infty,\label{imas}
\end{equation}
or
\begin{equation}
\lim_{x\rightarrow - \infty}u^{(0)}=0, \quad I_{-}=\int_{-\infty}^{x_0} [ u^{(0)}(y) ]^2 \text{d}y < \infty.\label{imenos}
\end{equation}
In both cases it is possible to find a domain of $w_0$ where $w(x)$ does not have zeroes \citep{FS03}. For example, if equation \eqref{imas} is fulfilled and $u^{(0)}$ is a non-physical eigenfunction of $H_0$ associated with $\epsilon$, it turns out that the domain in which $w(x)$ has no zeroes is $w_0 \leq -I_{+}$. A similar procedure implies that for the transformation functions that satisfy equation \eqref{imenos}, the domain becomes $w_0 \geq I_{-}$. An example of $V_2(x)$ generated through the confluent algorithm can be found in figure \ref{fig.hsusyqm3}.

Similarly to the case with $c>0$, now we can look for a function $\psi_{\epsilon}^{(2)}$ in the kernel of $B_2^{-}$ that is also an eigenfunction of $H_2$ with eigenvalue $\epsilon$. In fact, equations \eqref{psi1} and \eqref{psi2} are still valid in this case, we should only substitute $\epsilon_1$ and $\epsilon_2$ by $\epsilon$, $u_1^{(0)}$ by $u^{(0)}$, and $\gamma_1$ by $\gamma$. Then
\begin{equation}
\psi_{\epsilon}^{(2)}(x)\propto \frac{g(x)}{u^{(0)}(x)}\propto \frac{u^{(0)}(x)}{w(x)}.
\end{equation}

The spectrum of $H_2$ depends on whether $\psi_{\epsilon}^{(2)}$ is normalized or not. In particular, for $\epsilon \geq E_0$ it is possible to find solutions $u^{(0)}$ that satisfy equation \eqref{imas} or \eqref{imenos} such that $\psi_{\epsilon}^{(2)}$ is indeed normalized. This means that the confluent second-order SUSY transformations allow us to add only one energy state but in any arbitrary position we wish, even above the ground state of $H_0$. We must remember that this cannot be achieved by iterations of first-order SUSY transformations without introducing singularities. As we expected, these spectral possibilities are consistent with the heuristic criterion formulated previously.

As a part of the work of this thesis, we have developed a new alternative algorithm to calculate this confluent SUSY transformation. It uses parametric derivatives instead of integration, and for several potentials it is easier to calculate, as we will show in the examples. The results of this research are presented in section \ref{difconfluente}.

\subsection{Complex case ($c<0$)}\label{complexcase}
If $c<0$ we have that $\epsilon_1=\bar{\epsilon}_2\equiv \epsilon \in \mathbb{C}$. We must remark that the heuristic criterion allows this possibility without contradicting the positiveness of $B_2^{-}B_2^{+}$. Let us analyze here only the case when $V_2(x)$ is real, which implies that $\gamma_1(x)=\bar{\gamma}_2(x)\equiv \gamma(x)$. Following analogue steps as for the real case, one gets $g (x)$ in terms of the complex solution $\gamma(x)$ of the Riccati equation associated with $\epsilon$ \citep{FMR03}
\begin{equation}
g(x)=-\frac{2\, \text{Im}(\epsilon)}{\text{Im}[\gamma(x)]}.
\end{equation}

Using the complex solution of the Schr\"odinger equation $u^{(0)}(x)$ we also obtain that
\begin{equation}
g(x)=\frac{w'(x)}{w(x)}, \quad w(x)=\frac{W(u^{(0)},\bar{u}^{(0)})}{2(\epsilon-\bar{\epsilon})}.\label{gcomp2}
\end{equation}

To avoid singularities in $V_2(x)$, we know that $w(x)$ must not have zeroes. Due to $w'(x)= |u^{(0)}(x)|^2$, it turns out that $w(x)$ is a monotonically non-decreasing function; therefore, to assure that $w(x) \neq 0\ \forall \ x \in \mathbb{R}$ we can ask that
\begin{equation}
\lim_{x\rightarrow\infty}u_1^{(0)}(x)=0, \quad \text{or}\quad \lim_{x\rightarrow - \infty} u_1^{(0)}(x)=0.
\end{equation}

\begin{figure}\centering
\includegraphics[scale=0.6]{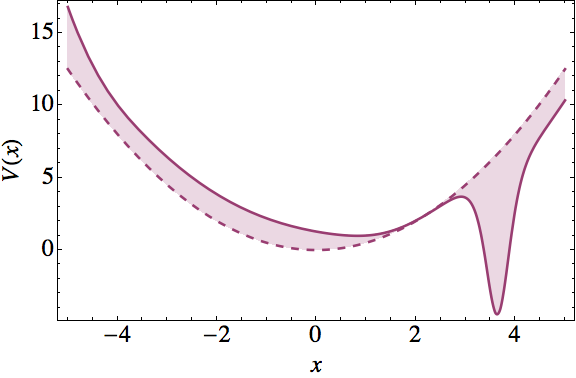}
\caption{\small{Real SUSY partner potential $V_2(x)$ (solid line) of the harmonic oscillator $V_0(x)$ (dashed line), generated using  a transformation function with {\it complex energy} $\epsilon_1=1/2 +i 10^{-2}$. We remark the difference between the two potentials.}}\label{fig.hsusyqm4}
\end{figure}

For transformation functions that fulfill one of these conditions, it turns out that $V_2(x)$ is a real potential which is isospectral to $V_0(x)$ (see figure \ref{fig.hsusyqm4}). This transformation can also be obtained by iterations of first-order SUSY transformations, but the intermediate potential will be complex.

\section{$k$th-order SUSY partners of the harmonic and radial oscillators}\label{secHORO}
\subsection{$k$-SUSY partners of the harmonic oscillator}

In this section we will apply the $k$th-order SUSY QM to the harmonic oscillator potential, because in chapter \ref{pha} we will use this Hamiltonian to obtain second-order PHA and in chapter \ref{cappain} to obtain solutions to $P_{IV}$. In order to accomplish this, once again we need the general solution $u(x,\epsilon)$ of the Schr\"odinger equation for $V_0(x) = x^2/2$ with an arbitrary factorization energy $\epsilon\in\mathbb{R}$ (see section \ref{complexsusyHO} and \citet{JR98}), which is given by
\begin{align}
u(x,\epsilon ) & = e^{-x^2/2}\left[ {}_1F_1\left(\frac{1-2\epsilon}{4},\frac12;x^2\right) + 2x\nu\frac{\Gamma(\frac{3 - 2\epsilon}{4})}{\Gamma(\frac{1-2\epsilon}{4})}\, {}_1F_1\left(\frac{3-2\epsilon}{4},\frac32;x^2\right)\right] \nonumber \\
& = e^{x^2/2}\left[ {}_1F_1\left(\frac{1+2\epsilon}{4},\frac12;-x^2\right) + 2x\nu\frac{\Gamma(\frac{3 - 2\epsilon}{4})}{\Gamma(\frac{1-2\epsilon}{4})}\, {}_1F_1\left(\frac{3+2\epsilon}{4},\frac32;-x^2\right)\right] , \label{hyper}
\end{align}
$_1F_1$ being the confluent hypergeometric (Kummer) function. Note that, for $\epsilon < 1/2$ this solution will not have zeroes for $\vert\nu\vert < 1$ while it will have only one for $\vert\nu\vert > 1$.

Let us perform now a non-singular $k$th-order SUSY transformation which creates precisely $k$ new
levels, additional to the standard ones $E_n = n + 1/2, \ n=0,1,2,\dots$ of $H_0$, in the way
\begin{equation}
\text{Sp}(H_k)  = \left\{ \epsilon_k, \dots,\epsilon_1, \frac12,\frac32,\dots \right\} , \label{spect}
\end{equation}
where $\epsilon_k < \cdots < \epsilon_1 < 1/2$. In order that the Wronskian $W(u_1,\dots,u_k)$ would be nodeless, the parameters $\nu_j$ have to be taken as $\vert\nu_j\vert < 1$ for $j$ odd and $\vert\nu_j\vert > 1$ for $j$ even, $j=1,\dots,k$. The corresponding potential turns out to be
\begin{equation}
V_k(x) = \frac{x^2}2 - \{\ln[W(u_1,\dots,u_k)]\}''.
\end{equation}

Now, it is important to note that there is a pair of natural ladder operators $L_k^\pm$ for $H_k$:
\begin{equation}
L_k^{\pm} = B_k^{+} a^{\pm} B_k^{-},
\end{equation}
which are differential operators of $(2k+1)$-th order such that
\begin{equation}
[H_k,L_k^\pm]= \pm L_k^\pm ,
\end{equation}
and $a^{-},\, a^{+}$ are the standard annihilation and creation operators of the harmonic oscillator.

From the intertwining relation of equation~\eqref{Bdag}, its Hermitian conjugate, and the factorizations in equations \eqref{Bdag4} and \eqref{Bdag5}, it is straightforward to show the following relation in terms of the {\it extremal energies} $\mathcal{E}_j$:
\begin{align}
N(H_k) =& L_k^+ L_k^- =\prod_{j=1}^{2k+1}(H_k-\mathcal{E}_j)\nonumber \\
=& \left(H_k - \frac12\right) \prod_{j=1}^k \left(H_k - \epsilon_j \right) \left(H_k - \epsilon_j - 1\right), \label{annum}
\end{align}
which will be quite useful later on.

\subsection{$k$-SUSY partners of the radial oscillator}\label{SUSYRO}
Now we apply $k$-SUSY QM to the radial oscillator potential. This will be used in chapter \ref{pha} to study third-order PHA with fourth-order ladder operators and in chapter \ref{5painleve} to obtain solutions to $P_V$.

If we start form the harmonic oscillator in three dimensions, whose potential will be $V(\mathbf{r})\propto r^2$ and perform the separation of variables we obtain a one dimensional reduced problem, described by an {\it effective potential}. This new potential is called the {\it radial oscillator}and is given by
\begin{equation}
V(x)=\frac{x^2}{8}+\frac{\ell(\ell+1)}{2x^2}, \quad \ell\geq 0,
\end{equation}
where $\ell$ characterizes the associated angular momentum and we have re-scaled the potential for reasons that we will explain later. Also, $x$ is the reduced variable from the three dimensional problem and in this case its domain is $[0,\infty)$. Then, the radial oscillator Hamiltonian takes the form
\begin{equation}
H_\ell=-\frac{1}{2}\frac{\text{d}^2}{\text{d}x^2}+\frac{x^2}{8}+\frac{\ell(\ell+1)}{2x^2},
\end{equation}
where we have added the subscript $\ell$ to denote the dependence of the Hamiltonian in the angular momentum.

To perform the higher-order SUSY transformations onto this potential we will explore first the usual factorization method \citep{FND96,CR08,CFNN04}. The radial oscillator Hamiltonian can be factorized in four different ways. The first two are
\begin{equation}
H_\ell=a_\ell^- a_\ell^+ +\frac{\ell}{2}-\frac{1}{4} = a_{\ell+1}^+a_{\ell+1}^-+\frac{\ell}{2}+\frac{3}{4},
\end{equation}
with
\begin{equation}
a_\ell^\pm \equiv \frac{1}{2^{1/2}}\left(\mp\frac{\text{d}}{\text{d}x}-\frac{\ell}{x}+\frac{x}{2}\right).
\end{equation}

The commutator of these operators is
\begin{equation}
[a_\ell^-,a_\ell^+]=\frac{\ell}{x^2}+\frac{1}{2},
\end{equation}
from which we can see that $a_\ell^\pm$ are not ladder operators but rather {\it shift operators}, i.e., they change the angular momentum of the original Hamiltonian and actually intertwine it with a different one $H_{\ell-1}$ to create a hierarchy of radial oscillator Hamiltonians with different angular momentum
\begin{subequations}
\begin{align}
H_\ell a_\ell^- &= a_\ell^-\left(H_{\ell-1}-\frac{1}{2}\right),\\
H_{\ell-1} a_\ell^+ &= a_\ell^+\left(H_{\ell}+\frac{1}{2}\right).
\end{align}\label{interRO}
\end{subequations}

Now, let $\psi_{n\ell}(x)$ be an eigenfunction of $H_\ell$ with eigenvalue $E_{n\ell}$
\begin{equation}
H_\ell \psi_{n\ell} = E_{n\ell}\psi_{n\ell}.
\end{equation}
Then, from equations~\eqref{interRO} we obtain that
\begin{subequations}
\begin{align}
H_{\ell+1} (a_{\ell+1}^- \psi_{n\ell}) &= \left(E_{n\ell}-\frac{1}{2}\right)(a_{\ell+1}^-\psi_{n\ell}),\\
H_{\ell-1} (a_{\ell}^+ \psi_{n\ell}) &= \left(E_{n\ell}+\frac{1}{2}\right)(a_{\ell}^+\psi_{n\ell}).
\end{align}
\end{subequations}

Moreover, the change $\ell\rightarrow -(\ell+1)$ produces the other two factorizations, although this causes small changes in the equations. For the factorizations we have
\begin{equation}
H_\ell=a_{-(\ell+1)}^- a_{-(\ell+1)}^+-\frac{\ell}{2}-\frac{3}{4}=a_{-\ell}^+ a_{-\ell}^- -\frac{\ell}{2}+\frac{1}{4},
\end{equation}
for the intertwinings
\begin{subequations}
\begin{align}
H_{\ell-1} a_{-\ell}^- &= a_{-\ell}^-\left(H_{\ell}-\frac{1}{2}\right),\\
H_{\ell} a_{-\ell}^+ &= a_{-\ell}^+\left(H_{\ell-1}+\frac{1}{2}\right),
\end{align}
\end{subequations}
and for the eigenvalue equations
\begin{subequations}
\begin{align}
H_{\ell-1} (a_{-\ell}^- \psi_{n\ell}) &= \left(E_{n\ell}- \frac{1}{2}\right)(a_{-\ell}^-\psi_{n\ell}),\\
H_{\ell+1} (a_{-(\ell+1)}^+ \psi_{n\ell}) &= \left(E_{n\ell}+\frac{1}{2}\right)(a_{-(\ell+1)}^+\psi_{n\ell}).
\end{align}
\end{subequations}

\begin{figure}\centering
\includegraphics[scale=0.4]{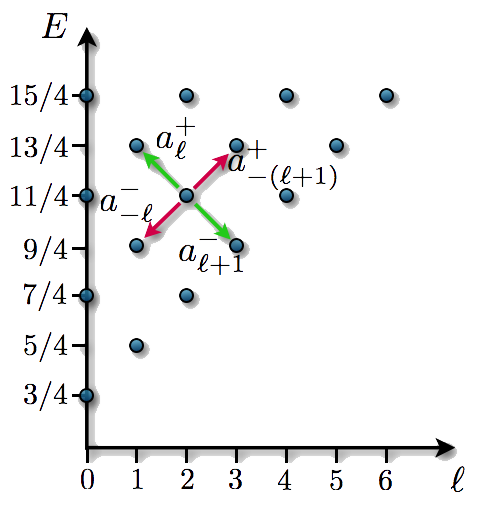}
\includegraphics[scale=0.4]{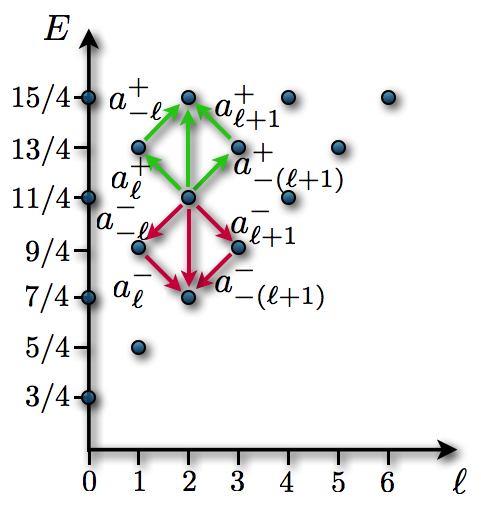}
\caption{\small{Diagram of the action of the first-order intertwining operators $a_\ell^\pm$. On the horizontal axis we have different values of the angular momentum index $\ell$, on the vertical one we have different energy values. We can see that the joint action of two first-order shift operators leads to the ladder operators $b_\ell^\pm$.}}\label{diaRO}
\end{figure}

As we stated before, neither of these are ladder operators. Nevertheless, through them we can define some ladder operators for the radial oscillator, but these will necessarily be of higher-order, in this case of second-order. From diagram in figure~\ref{diaRO} we can see that the joint action of two shift operators can lead to an effective definition of a ladder operator. Indeed, let us take $b_\ell^\pm$ such that

\begin{subequations}
\begin{align}
b_\ell^- &= a_{-(\ell+1)}^-a_{\ell+1}^-=a_\ell^-a_{-\ell}^-,\\
b_\ell^+ &= a_{\ell+1}^+a_{-(\ell+1)}^+=a_{-\ell}^+a_{\ell}^+.
\end{align}
\end{subequations}
Then we can easily show that
\begin{subequations}
\begin{align}
H_\ell b_\ell^- &= b_\ell^-(H_\ell-1),\\
H_\ell b_\ell^+ &= b_\ell^+(H_\ell + 1).
\end{align}
\end{subequations}

\begin{figure}\centering
\includegraphics[scale=0.5]{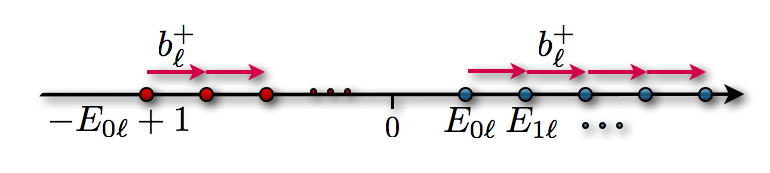}
\caption{\small{Spectrum of the radial oscillator. The blue circles represent the physical solutions starting from $E_0$ and the red circles represent the non-physical solutions departing from $-E_{0\ell}+1$.}}\label{speRO}
\end{figure}

Furthermore, we can simply obtain the commutator with the Hamiltonian as
\begin{equation}
[H_\ell,b_\ell^{\pm}]=\pm b_\ell^\pm,
\end{equation}
which proves that $b_\ell^\pm$ are ladder operators. Their explicit form is
\begin{equation}
b_\ell^\pm=\frac{1}{2}\left(\frac{\text{d}^2}{\text{d}x^2}\mp x\frac{\text{d}}{\text{d}x}+\frac{x^2}{4}-\frac{\ell(\ell+1)}{x^2}\mp\frac{1}{2}\right).
\end{equation}
On the other hand, we can obtain the eigenstates if we start from the {\it ground state} $\psi_{0\ell}$, defined as $b_\ell^- \psi_{0\ell}=0$. In this systems there are two states that are annihilated by $b_\ell^-$
\begin{subequations}
\begin{alignat}{3}
\psi_{\mathcal{E}_1} &\propto x^{\ell+1}\exp(-x^2/4), & \quad \mathcal{E}_1 & =\frac{\ell}{2}+\frac{3}{4}\equiv E_{0\ell},\\
\psi_{\mathcal{E}_2} &\propto x^{-\ell}\exp(-x^2/4), & \quad \mathcal{E}_2 & = -\frac{\ell}{2}+\frac{1}{4}= -E_{0\ell}+1,
\end{alignat}\label{solsRO}
\end{subequations}
\hspace{-1mm}but only the first one fulfills the boundary conditions and therefore leads to a ladder of {\it physical} eigenfunctions. The spectrum of the radial oscillator is therefore
\begin{equation}
\text{Sp}(H_\ell)=\{E_{n\ell}=n+\frac{\ell}{2}+\frac{3}{4}, n=0,1,\dots \}.
\end{equation}
We can see a diagram of this spectrum in figure~\ref{speRO} where we represent both the physical and non-physical solutions given by equations~\eqref{solsRO}.

An analogue of the number operator can now be defined for the radial oscillator as
\begin{equation}
b_\ell^+b_\ell^-=(H_\ell-\mathcal{E}_1)(H_\ell-\mathcal{E}_2)=\left(H_\ell-\frac{\ell}{2}-\frac{3}{4}\right)\left(H_\ell+\frac{\ell}{2}-\frac{1}{4}\right).
\end{equation}

In order to perform now the SUSY transformations, we need the general solution of the stationary Schr\"odinger equation for any factorization energy $\epsilon$, which is given by \citep{JR98,Car01,CFNN04}
\begin{align}
u(x,\epsilon)=\, & x^{-\ell}\text{e}^{-x^2/4}\left[{}_1F_1\left(\frac{1-2\ell-4\epsilon}{4},\frac{1-2\ell}{2};\frac{x^2}{2}\right)\right. \nonumber\\
& + \left. \nu \frac{\Gamma\left(\frac{3+2\ell-4\epsilon}{4}\right)}{\Gamma\left(\frac{3+2\ell}{2}\right)}\left(\frac{x^2}{2}\right)^{\ell+1/2}{}_1F_1\left(\frac{3+2\ell-4\epsilon}{4},\frac{3+2\ell}{2};\frac{x^2}{2}\right)\right].\label{solRO}
\end{align}
Three conditions must be fulfilled to avoid singularities in the transformation
\begin{equation}
x>0, \quad \epsilon<E_{0\ell}, \quad \nu\geq -\frac{\Gamma\left(\frac{1-2\ell}{2}\right)}{\Gamma\left(\frac{1-2\ell-4\epsilon}{4}\right)}.\label{condRO}
\end{equation}

We apply now the iterative approach of section \ref{secite} to the $k$th-order SUSY QM, where the Riccati equation reads
\begin{equation}
\alpha_j'+\alpha_j^2=\left(\frac{x^2}{4}+\frac{\ell(\ell+1)}{x^2}-2\epsilon_j\right)=2(V_\ell(x)-\epsilon_j ),
\end{equation}
which can be transformed into the Schr\"odinger equation
\begin{equation}
-\frac{1}{2}u_j''+V_\ell(x)u_j=\epsilon u_j.
\end{equation}

\begin{figure}\centering
\includegraphics[scale=0.37]{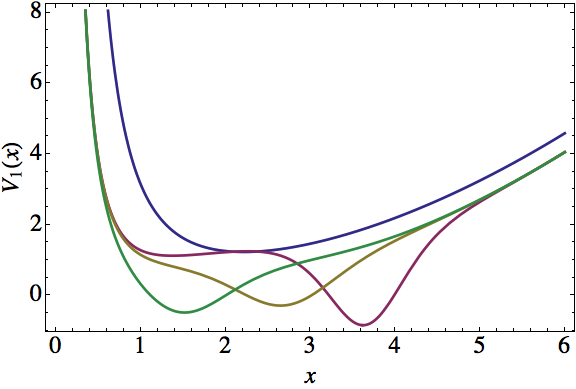}
\includegraphics[scale=0.37]{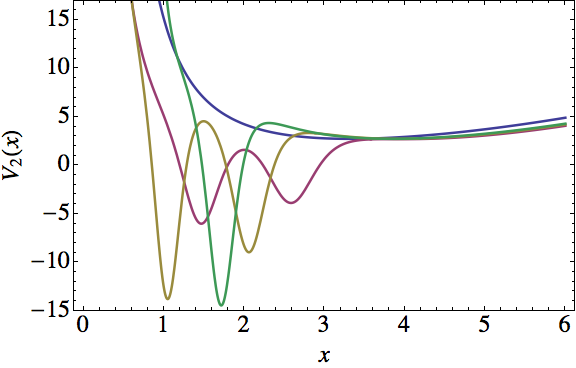}
\caption{\small{SUSY partner potentials of the radial oscillator (blue). The left plot is for $k=1$, $\ell=2$, $\epsilon=1/2$, and $\nu=\{-$0.59 (magenta), $-0.4$ (yellow), 1 (green)$\}$. The right one is for $k=2$, $\ell=5$, $\nu=1$, and $\epsilon=\{$0 (magenta), $-2$ (yellow), $-4$ (green)$\}$.}}\label{potRO}
\end{figure}

Then, the deformed potential is given by
\begin{equation}
V_k(x)=\frac{x^2}{8}+\frac{\ell(\ell+1)}{2x^2}-\{\ln[W(u_1,\dots , u_k)]\}'',
\end{equation}
with the spectrum 
\begin{equation}
\text{Sp}(H_k)=\{\epsilon_k,\dots,\epsilon_1,E_{0\ell},E_{1\ell},\dots \}.
\end{equation}
In figure~\ref{potRO} we show some examples of first- and second-order SUSY partner potentials of the radial oscillator. 

As in the harmonic oscillator case, we can define now a pair of natural ladder operators $L_k^{\pm}$ for the $k$-SUSY partners $H_k$ of the radial oscillator as
\begin{equation}
L_k^\pm=B_k^+b_\ell^\pm B_k^-,
\end{equation}
which are of $(2k+2)$th-order and fulfill
\begin{equation}
[H_k,L_k^\pm]=\pm L_k^\pm.
\end{equation}

From the intertwining relations we can obtain the analogue of the number operator for the radial oscillator as
\begin{align}
N(H_k)&=L_k^+L_k^-=\prod_{j=1}^{2k+2}(H_k-\mathcal{E}_j)\nonumber\\
		  &=\left(H_k-\frac{\ell}{2}-\frac{3}{4}\right)\left(H_k+\frac{\ell}{2}-\frac{1}{4}\right)\prod_{j=1}^{k}(H_k-\epsilon_j)(H_k-\epsilon_j-1).
\end{align}

This means that the $k$-SUSY partners of the radial oscillator have $(2k+2)$th-order differential ladder operators, e.g., the 1-SUSY partner has fourth-order ladder operators. We will show in chapter \ref{pha} that this system is ruled by third-order PHA and is connected to $P_V$. Moreover, in chapter~\ref{5painleve} we will show a method to generate new solutions to $P_V$ through this kind of systems.

\section{Differential formula for confluent SUSY}
\label{difconfluente}
In section \ref{secconfluente} we presented this special case of the second-order SUSY transformation, in which the two factorization energies tend to the same value. Taking this limit appropriately, this transformation leads to more flexibility on the spectral design compared to the first-order case. Nevertheless, the formula in section \ref{secconfluente} requires to solve an indefinite integral, which is sometimes difficult to accomplish. In this section we are going to derive a differential formula to calculate the confluent SUSY partners of an arbitrary potential. This algorithm includes derivatives of the transformation function with respect to the factorization energy. Here, we introduce a method to obtain the general formula and we apply it to two cases, the free particle, which has been already studied using an integral expression, and the Lam\'e potential, which has not been addressed before. This last example clearly shows the advantages of this method, because the confluent transformation has not been applied before to the Lam\'e potential since the integration is difficult to perform.

This research is part of the original contributions of this thesis and it is already published in \citet*{BFF12}.

\subsection{Introduction}
It was already shown that the confluent second-order SUSY QM \citep{MNR00,FS03}, for which the two involved factorization energies converge to the same value, increases the possibilities of spectral manipulation which are available. However, the main issue for implementing this method has to do with the difficulty to calculate the integral of equation \eqref{intformula}.

On the other hand, it has been shown \citep{FS05} that in the confluent case the Wronskian formula is preserved if solutions closing a Jordan chain of length two are used as seeds for implementing the algorithm. In this section we will take advantage of this fact by introducing a differential version of the technique which will preserve as well the general Wronskian formula and will avoid to evaluate the previously mentioned integrals. In this way, an alternative calculation tool will be available for implementing the confluent second-order SUSY QM.

We will present next the Wronskian formula derived by \citet*{FS05}. Then we will derive our own version of the algorithm in terms of parametric derivatives. After that, we will apply this alternative method to the free particle and to the single-gap Lam\'e potentials. A summary of our original results and some conclusions are presented in the last subsection.

\subsection{Confluent SUSY QM}
It was shown in section \ref{secconfluente} that the key function $w(x)$ to implement the confluent algorithm is given by equation \eqref{intformula}, namely,
\begin{equation}\label{wconfluente}
w(x)=w_0+\int_{x_0}^{x}u_1^2(y)\text{d}y,
\end{equation}
where $w_0$, $x_0$ are real constants which can be chosen at will in order to avoid singularities in the new potential and $u_1(x)$ is a solution of the initial stationary Schr\"odinger equation associated with the factorization energy $\epsilon$.

On the other hand, let us consider now the following pair of generalized eigenfunctions of $H$, of first and second rank, associated with $\epsilon$ \citep{DK67,FS05,EJM06},
\begin{subequations}
\begin{align}
(H-\epsilon)u_1 & = 0,\label{uv1}\\
(H-\epsilon)u_2 & = u_1, \label{uv2}
\end{align}
\end{subequations}
which is known as Jordan chain of length two. By solving equation \eqref{uv2} for $u_2$ through the method of variation of parameters, supposing that $u_1$ is given, we get
\begin{equation}
u_2=\left(k+\int \frac{w(x)}{u_1^2(x)}\text{d}x\right)u_1(x).\label{u2}
\end{equation}
Moreover, by using the following Wronskian identity
\begin{equation}
W(f,hf)=h'f^2,\label{wrons}
\end{equation}
which is valid for two differentiable but otherwise arbitrary functions $f$ and $h$, it is straightforward to show that
\begin{equation}
w(x)=W(u_1,u_2).\label{wW}
\end{equation}
Therefore, the Wronskian formula for the non-confluent second-order SUSY QM given by equation \eqref{wronsk} is preserved for the confluent case \citep{FS05}. Moreover, it can be used to construct a one-parameter family of exactly-solvable potentials for each solution $u_1$ of the initial stationary Schr\"odinger equation associated with $\epsilon$. However, if $u_1$ has an involved explicit form, the task of evaluating the corresponding integral is not simple. In the next section we shall present an alternative version of the Wronskian formula for the confluent case which will make unnecessary the evaluation of the integrals of equations \eqref{wconfluente} and \eqref{u2}.

\subsection{Wronskian differential formula for the confluent SUSY QM}\label{wrodiff}
Now, let us look for the general solution of equation \eqref{uv2} in a slightly different way. Let $u_1$ denote once again the given solution $u_1=u$ of \eqref{uv1}. It is well known that the general solution of the inhomogeneous second-order differential equation \eqref{uv2} takes the following form
\begin{equation}
u_2=u_2^{h}+u_2^{p},
\end{equation}
where $u_2^{h}$ is the general solution of the homogeneous equation and $u_2^{p}$ denotes a particular solution of the inhomogeneous one. Since the homogeneous equation is of second order, it has two linearly independent solutions. They can be taken as $u$ and its orthogonal function $u^\perp$ defined by $W(u,u^\perp) = 1$. The last equation can be immediately solved for $u^\perp$, yielding
\begin{equation}
u^\perp (x) = u(x)\int\frac{\text{d}x}{u^2(x)}.
\end{equation}
Then, it turns out that
\begin{equation}
u_2^{h}=Cu+Du^\perp,
\end{equation}
with $C,\,D \in \mathbb{R}$.

In order to find the particular solution $u_2^{p}$, let us suppose from now on that $u$ and its parametric derivative with respect to $\epsilon$, $\partial_\epsilon u$, are well defined continuous functions in a neighborhood of $\epsilon$. Hence, by deriving equation \eqref{uv1} with respect to $\epsilon$ we obtain
\begin{equation}
\left(H-\epsilon\right)\frac{\partial u}{\partial\epsilon} = u, \label{part}
\end{equation}
where the partial derivatives of $u$ with respect to $\epsilon$ and $x$ have been interchanged. It should be clear now that (compare equations \eqref{uv2} and \eqref{part})
\begin{equation}
u_2^{p} = \frac{\partial u}{\partial\epsilon},
\end{equation}
is the particular solution of the inhomogeneous equation we were looking for. Finally, the general solution of equation \eqref{uv2} is given by
\begin{equation}
u_2= C u + D u^\perp + \frac{\partial u}{\partial\epsilon} .\label{u2a}
\end{equation}
From this equation we can easily calculate the Wronskian of the two solutions of the Jordan chain as
\begin{equation}
W(u_1,u_2)= D + W\left(u,\frac{\partial u}{\partial\epsilon}\right).
\end{equation}
Thus, the general Wronskian formula of equation \eqref{confformula} becomes now
\begin{equation}
g(x)=\left\{\ln\left[D + W\left(u,\frac{\partial u}{\partial\epsilon}\right)\right]\right\}',
\end{equation}
and the new potential $\widetilde{V}(x)$ is
\begin{equation}
\widetilde{V}(x)=V(x)-\left\{\ln\left[D + W\left(u,\frac{\partial u}{\partial\epsilon}\right)\right]\right\}'', \label{Dv}
\end{equation}
which represents an alternative way to calculate $\widetilde{V}(x)$ through the confluent second-order SUSY transformation.

Note that a special case of equation \eqref{Dv} has been addressed previously only for the free particle potential and with $D=0$ \citep{Mat92,Sta95}. In these works, the particular solution $\partial_\epsilon u$ was taken directly as the seed solution $u_2$ and thus the constant $D$, which arises from the non-trivial term involving the orthogonal function $u^\perp$ (see the second-term of the right-hand side of equation \eqref{u2a}), never appears in those treatments.

An additional point is worth to remark: without the constant $D$ the confluent second-order SUSY partner potential $\widetilde V$ will often have singularities. The freedom we have here for choosing this constant endows us with the possibility to generate families of non-singular potentials for a wide set of factorization energies.

We are going to use equation \eqref{Dv} now to implement a confluent second-order SUSY transformation for two simple systems. The first of them is the free particle, where both the differential and the integral versions of the confluent SUSY QM are easily applicable since the derivatives and the integrals involved are not difficult to calculate. The second one is the single-gap Lam\'e potential, for which the previously found integral equation \eqref{wconfluente} is not easy to apply, since the integrals of elliptic functions are complicated to evaluate. As far as we know, the confluent second-order SUSY transformation has been never applied before to this potential.

\subsection{Free particle}\label{secfree}
The free particle is not subject to any force so that the corresponding potential is constant; without loss of generality, let us take $V(x)=0$. In order to obtain non-singular confluent second-order SUSY partner potentials one has to use as transformation function, in general, a solution $u$ to the stationary Schr\"odinger equation \eqref{ucero} such that $W(u_1,u_2)\neq 0 \ \forall \ x\in{\mathbb R}$. This is achieved by demanding that $u$ vanishes at one of the boundaries of the $x$-domain (see \citep{FS03,FS05}). In particular, for the free particle these solutions are $\{e^{\kappa x},e^{-\kappa x}\}$ with the condition that $\kappa$ and $\epsilon$ satisfy the {\it dispersion relation} $2\epsilon = -\kappa^2, \ \kappa>0$.

We are going to use one of these solutions to perform the SUSY transformation, e.g., $u = e^{\kappa x}$; the other case can be obtained through a spatial reflection. Thus, the parametric derivative can be calculated using the chain rule as
\begin{equation}
\frac{\partial u}{\partial\epsilon} = \frac{\text{d}\kappa}{\text{d}\epsilon}\frac{\partial u}{\partial\kappa}
= -\frac{xu}{\kappa} = - \frac{xe^{\kappa x}}{\kappa}.
\end{equation}
We can easily evaluate the Wronskian of $u$ and $\partial_\epsilon u$ by using once again equation \eqref{wrons}:
\begin{equation}
W\left(u,\frac{\partial u}{\partial\epsilon}\right) = -\frac{u^2}{\kappa} = -\frac{e^{2\kappa x}}{\kappa}.\label{wfr}
\end{equation}
Now, inserting equation \eqref{wfr} into \eqref{Dv} to calculate the confluent second-order SUSY partner potential $\widetilde{V}$ of the free particle, we obtain
\begin{equation}
\widetilde{V}(x)=\frac{4D\kappa^{3}e^{2\kappa x}}{(D\kappa - e^{2\kappa x})^2}.\label{vfinal}
\end{equation}
Due to the {\it dispersion relation} ($2\epsilon = -\kappa^2, \ \kappa > 0$) there is a natural restriction on the factorization energy, namely, $\epsilon < 0$. Besides, in order to obtain non-singular transformations the parameter $D$ has to be restricted \citep{FS03,FS05}. Indeed, for $u = e^{\kappa x}$ we have that the non-singular domain is given by $D<0$, and reparametrizing as $D = -e^{2\kappa x_0}/\kappa$, with $x_0\in\mathbb{R}$, we can simplify \eqref{vfinal} to obtain
\begin{equation}
\widetilde{V}(x)=-\kappa^2\text{sech}^2 [\kappa(x-x_0)],\label{vfinal2}
\end{equation}
which is the P\"{o}schl-Teller potential with one bound state at the energy $E_0 = \epsilon = - \kappa^2/2$. It is worth to note that this result had also been obtained through first-order SUSY QM and by using the integral formulation for the confluent case \citep{FS03}. It is plausible that any non-singular SUSY transformation which departs from the free particle and creates just one bound state leads precisely to a P\"oschl-Teller potential (see also \citep{FS11}).

\begin{figure}
\begin{center}
\includegraphics[scale=0.5]{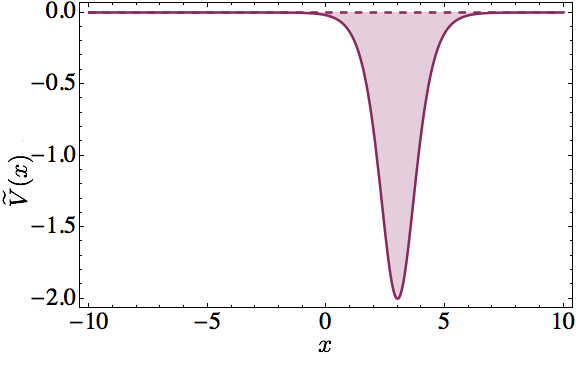}
\end{center}
\vspace{-5mm}
\caption{\small{Confluent second-order SUSY partner potential of the free particle obtained through equation~\eqref{vfinal2} for $\epsilon = -1$ and $x_0 = 3$.}} \label{figfree}
\end{figure}

An illustration of a confluent second-order SUSY partner potential $\widetilde V$, generated through this formalism from the free particle, is shown in figure~\ref{figfree}.

Note that in some previous works \citep{Mat92,Sta95}, the differential version of the confluent second-order SUSY transformation (Darboux transformation in those works) was implemented for the free particle with $D=0$ and using another transformation function, namely, $u = \sin [k(x+x_0)]$ with $2\epsilon = k^2 > 0$; however, by doing so, one deals only with singular transformations. Following the formalism of this work we have obtained a one-parameter family of non-singular potentials for each $\epsilon<0$.

For the free particle the integral and differential Wronskian formulas have been applied easily, since the involved integrals can be simply evaluated. Indeed, the reason to use this system was to check the effectiveness of our new formula. Nevertheless, there are some other potentials for which the calculation of the corresponding integrals is more complicated but the differential formalism can be applied straightforwardly. We will show next an example of this situation.

\subsection{Single-gap Lam\'e potential}
The Lam\'e periodic potentials are given by \citep{Ars81,FMRS02,FMRS02b}:
\begin{align}
V(x) &=\frac{1}{2}n(n+1) m\, \text{sn}^2(x|m)\nonumber\\
	&=\frac{1}{2}n(n+1)\left[\wp(x+iK(1-m))+\frac{1}{3}(m+1)\right],\label{lame}
\end{align}
where $\text{sn}(x|m)$ is a Jacobi elliptic function whose real period is $T=4K(m)$, $\wp(x)$ is the Weierstrass elliptic function, and
\begin{equation}
K(m)=\int_0^{\pi /2} \frac{\text{d}\theta}{(1-m\sin^2\theta)^{1/2}},
\end{equation}
is the real half-period of $V(x)$. The potentials \eqref{lame} have $2n+1$ band edges which define $n+1$ allowed and $n+1$ forbidden bands. They belong to a class of finite-gap periodic systems where the non-linear supersymmetry plays an important role. For example, Lam\'e potentials have been used to model a non-relativistic electron in periodic electric and magnetic field configurations which produce a 1D crystal \citep{CJNP08}. In addition, these potentials admit isospectral super-extensions \citep{CJP08} and they can be used to display hidden symmetries in quantum dynamical problems, specially in soliton dynamics \citep{AS09}. Note that Lam\'e potentials are particular cases of the associated Lam\'e potentials, which have been studied previously in the context of higher-order SUSY QM \citep{FG07}.

\begin{figure}
\begin{center}
\includegraphics[scale=0.3]{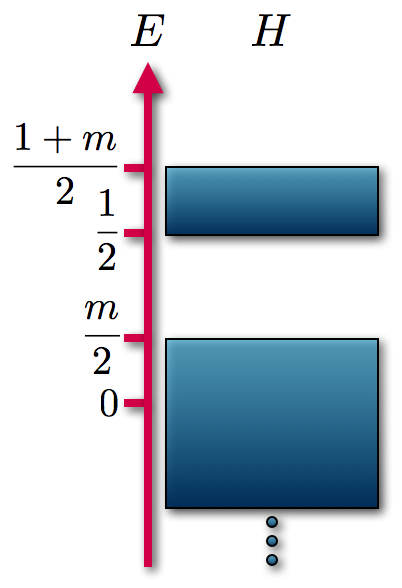}
\end{center}
\vspace{-5mm}
\caption{\small{Spectrum of the Lam\'e potential with $n=1$. The white bands correspond to the allowed energy region, a semi-infinite $[\frac{1+m}{2},\infty)$ and a finite one $[\frac{m}{2},\frac{1}{2}]$. The blue bands corresponds to the energy gaps, a semi-infinite $(-\infty,\frac{m}{2})$ and a finite one $(\frac{1}{2},\frac{1+m}{2})$.}} \label{figesplame}
\end{figure}

In this work we shall deal with the single-gap Lam\'e potential obtained fo $n=1$. The spectrum for the Hamiltonian associated with this specific potential is given by:
\begin{equation}
{\rm Sp}(H) = \left[\frac{m}{2},\frac{1}{2}\right] \cup \left[\frac{1+m}{2},\infty\right),
\end{equation}
i.e., it is composed by a finite energy band $[\frac{m}{2},\frac{1}{2}]$ plus a semi-infinite one $[\frac{1+m}{2},\infty)$ (see the white region in figure~\ref{figesplame}). The structure of the resolvent set of $H$ is similar, namely, there is a semi-infinite energy gap $(-\infty,\frac{m}{2})$ plus a finite one $(\frac{1}{2},\frac{1+m}{2})$ (observe the blue bands in figure~\ref{figesplame}).

As in the previous case, in order to implement the confluent second-order SUSY transformation we will use an appropriate seed solution $u$ associated with a factorization energy $\epsilon$ which is inside one of the energy gaps, i.e., in one of the blue bands in figure~\ref{figesplame} and such that $W(u_1,u_2) \neq 0$ $\forall \ x\in {\mathbb R}$. For our example this can be achieved by choosing $u$ as one of the two Bloch functions associated with $\epsilon$ \citep{FMRS02,FMRS02b}, i.e.,
\begin{subequations}
\begin{align}
u^{\beta}(x)&=\frac{\sigma(\omega ')}{\sigma(\delta +\omega ')}\frac{\sigma(x+\delta +\omega ')}{\sigma(x+\omega ')}e^{-x\zeta (\delta)},\\
u^{1/\beta}(x)&=\frac{\sigma(\omega ')}{\sigma(-\delta +\omega ')}\frac{\sigma(x-\delta +\omega ')}{\sigma(x+\omega ')}e^{x\zeta (\delta)},
\end{align}
\end{subequations}
where $\omega=K(m)$ and $\omega'=iK(1-m)$ are the real and imaginary half-periods of $\wp(x)$ \citep{AS72}, $\sigma$ and $\zeta$ are the non-elliptic Weierstrass functions \citep{Cha85}.

Note that $\beta$ is defined by the relation $u^{\beta}(x+T)=\beta u^{\beta}(x)$; then $\beta=\exp [ 2\delta\zeta(\omega)-2\omega\zeta(\delta) ]$. Besides, by expressing it as $\beta=\text{e}^{i\kappa}$, with $\kappa=2i[\omega\zeta(\delta)-\delta\zeta(\omega)]$ (up to an additive multiple of $2\pi i$) which is known as the quasi-momentum \citep{CJP08}. The displacement $\delta$ and the factorization energy $\epsilon$ are related by \citep{FMRS02b}:
\begin{equation}
\epsilon = \frac{1}{3}(m+1)-\frac{1}{2}\wp (\delta). \label{ed}
\end{equation}

In order to calculate the new potential from equation~\eqref{Dv}, let us choose the first Bloch function as transformation function, namely, $u=u^{\beta}$. It is worth pointing out that we are using one Bloch state to perform the SUSY transformation, even when these states are not normalized. Nevertheless, one of the advantages of the confluent algorithm is that it does not require normalized states to perform the transformation.

We are going to evaluate next its parametric derivative with respect to $\epsilon$, for which we will employ the following relations between $\sigma(x)$, $\zeta(x)$, and $\wp(x)$ \citep{Cha85}:
\begin{subequations}
\begin{align}
\sigma'(x)&=\sigma(x)\zeta(x),\label{d1}\\
\zeta'(x)&=-\wp(x),\\
\wp'(x)&=-\frac{\sigma(2x)}{\sigma^4(x)}.\label{d4}
\end{align}
\end{subequations}
Thus, using the chain rule and equation \eqref{ed}, we obtain
\begin{equation}
\frac{\partial u}{\partial\epsilon} = \frac{\text{d}\delta}{\text{d}\epsilon}\frac{\partial u}{\partial\delta} = -2\left(\frac{d\wp}{d\delta}\right)^{-1}\frac{\partial u}{\partial\delta},
\end{equation}
and an explicit calculation produces
\begin{equation}
\frac{\partial u}{\partial \delta}=[\zeta(x+\delta+\omega')-\zeta(\delta+\omega')+x\wp(\delta)]u.
\end{equation}
Thus, the Wronskian of equation \eqref{Dv} can be obtained by using once again equation \eqref{wrons}:
\begin{equation}
W\left(u,\frac{\partial u}{\partial\epsilon}\right) = 2\left(\frac{d\wp}{d\delta}\right)^{-1}[\wp(x+\delta+\omega')-\wp(\delta)]u^2\equiv f(x)u^{2},\label{wro}
\end{equation}
which defines the auxiliary function $f(x)$.

Finally, from equation \eqref{Dv} the new potential $\widetilde{V}$ can be calculated analytically as
\begin{equation}
\widetilde{V}(x)=V(x)+\frac{2[\zeta(x+\delta+\omega')-\zeta(x+\omega')-\zeta(\delta)]}{Du^{-2}+f}+\frac{1}{(Du^{-2}+f)^2}.\label{Vlame}
\end{equation}

\begin{figure}
\begin{center}
\includegraphics[scale=0.38]{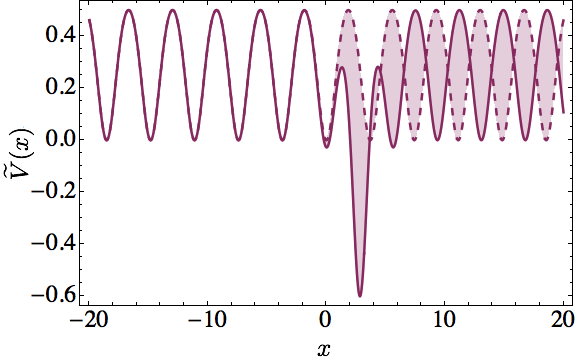}
\includegraphics[scale=0.38]{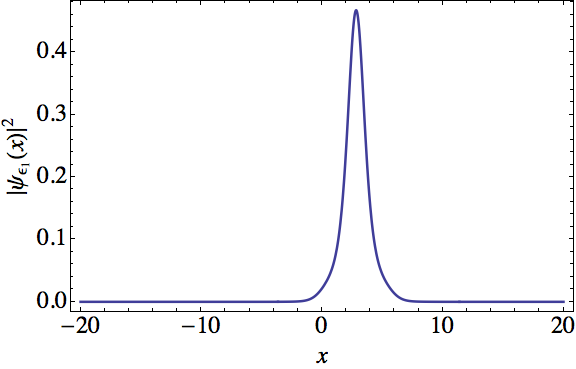}
\end{center}
\vspace{-5mm}
\caption{\small{SUSY partner potential $\widetilde V(x)$ (left) and the probability density of its new bound state (right), generated from the Lam\'e potential $V(x)$ for $n=1$. The parameters were taken as $m=1/2$, $\epsilon=5/100$, $x_0=0$, and $D=-45$.}} \label{figlame1}
\end{figure}

\begin{figure}
\begin{center}
\includegraphics[scale=0.38]{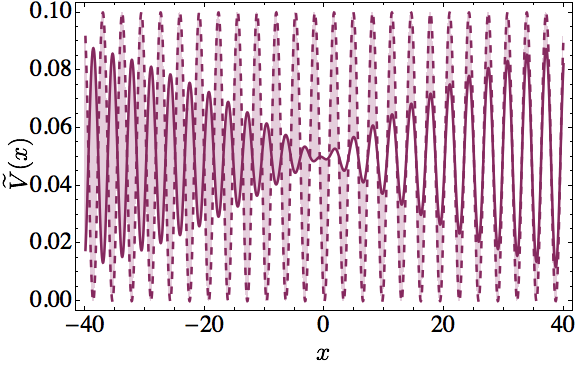}
\includegraphics[scale=0.38]{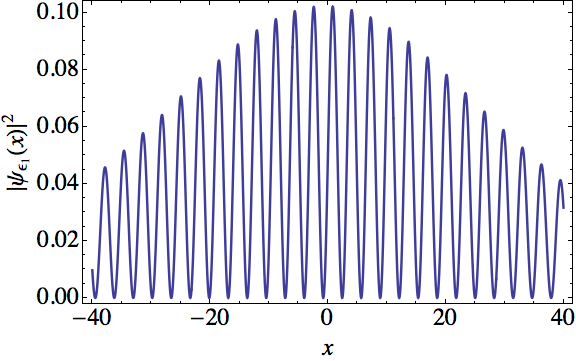}
\end{center}
\vspace{-5mm}
\caption{\small{SUSY partner potential $\widetilde V(x)$ (left) and the probability density of its new bound state (right), generated from the Lam\'e potential $V(x)$ for $n=1$. The parameters were taken as $m=1/10$, $\epsilon=21/40$, $x_0=0$, and $D=20$.}} \label{figlame2}
\end{figure}

Two potentials obtained through this method are shown in the left side of figures~\ref{figlame1} and \ref{figlame2}. They correspond to two different cases, for which either the factorization energy belongs to the infinite gap or to the finite one. Note that the shape of the new potentials (solid lines) are really different compared to the original one (dashed lines), and also between them. Indeed, it can be seen that the new potentials are in general non-periodic, although they become asymptotically periodic. Note that this periodicity defect of $\widetilde V(x)$ arises due to the creation of a bound state at an energy $\epsilon$ (inside an initial energy gap). The width and the position of this periodicity defect in general coincides with the $x$--domain where the new bound state
\begin{equation}
\psi_{\epsilon}^{(2)} (x) \propto \frac{u}{D + W\left(u,\partial_\epsilon u\right)},\label{state}
\end{equation}
has a non-trivial probability amplitude. For these two cases, the corresponding probability densities $\vert \psi_{\epsilon}^{(2)}(x)\vert^2$ are shown in the right side of figures~\ref{figlame1} and \ref{figlame2}.

Let us note that a similar physical situation, induced by a non-confluent second-order SUSY transformation, was found in \citet{FMRS02,FMRS02b}. The main advantage here is that we are using just one seed solution to create a bound state inside a given energy gap. Moreover, the explicit expressions obtained from our treatment become shorter than those derived by the non-confluent algorithm. Particularly interesting is the case in which the factorization energy $\epsilon$ is inside the finite gap, so that a bound state is created at this position. In such a situation, if the non-periodic potential $\widetilde{V}$ is perturbed by an additional interaction, the new bound state could be used as an intermediate state to perform transitions between the finite energy band and the infinite one. Note that the new bound state of equation~\eqref{state} is known as {\it localized impurity state} in solid state physics \citep[chapter 5]{Cal74}).

\subsection{Higher-order Wronskian differential formula}\label{kwro}
In this section we are going to calculate a generalization of the Wronskian differential formula that we just derived. Let us start with the Schr\"odinger equation
\begin{equation}
(H-\epsilon)u=0,\label{sch1}
\end{equation}
i.e., $u$ is an eigenfunction of the Hamiltonian $H$ and $\epsilon$ its eigenvalue. The function $u$ does not necessarily have physical interpretation, i.e., it could be a mathematical eigenfunction of $H$. In addition, we assume that $u=u(x,\epsilon)$ but $H=H(x)\neq H(\epsilon)$. Then, let us obtain the parametric derivative of equation~\eqref{sch1} with respect to $\epsilon$
\begin{equation}
(H-\epsilon)u_{\epsilon}=u,
\end{equation}
where $u_\epsilon$ represents the derivative of $u$ with respect to $\epsilon$ and we suppose that $u_{x\epsilon}=u_{\epsilon x}$. Deriving again
\begin{equation}
(H-\epsilon)\frac{u_{\epsilon\epsilon}}{2}=u_{\epsilon},
\end{equation}
and one more time
\begin{equation}
(H-\epsilon)\frac{u_{\epsilon\epsilon\epsilon}}{3}=u_{\epsilon\epsilon}.
\end{equation}
Now, we will prove by induction a formula for the $k$th derivative of $u$. Starting from the induction hypothesis given by
\begin{equation}
(H-\epsilon)\frac{\partial_\epsilon^k u}{k}=\partial_\epsilon^{k-1}u,
\end{equation}
we apply $\partial_\epsilon$ on both sides to obtain the general formula for the index $k+1$,
\begin{equation}
(H-\epsilon)\frac{\partial_\epsilon^{k+1} u}{k+1}=\partial_\epsilon^{k}u.\label{schu}
\end{equation}

Now, the Schr\"odinger equation is a second-order linear differential equation, therefore it has two linearly independent solutions for any given $\epsilon$. Let us call $v(x)$ the orthogonal function to $u$ defined by
\begin{equation}
W(u,v)=1,\label{defW}
\end{equation}
a Wronskian which can be made equal to any non-zero constant, but we choose 1 for simplicity. Solving equation~\eqref{defW} for $v$ we obtain
\begin{equation}
v(x)=u\int\frac{dx}{u^2}.
\end{equation}
We can easily prove that $v(x)$ also solves the Schr\"odinger equation~\eqref{sch1} by applying $\partial_x^2$ to the last equation, i.e.,
\begin{equation}
(H-\epsilon)v=0.\label{sch2}
\end{equation}
Following an analogous procedure to $u$ we can show that
\begin{equation}
(H-\epsilon)\frac{\partial_\epsilon^{k+1} v}{k+1}=\partial_\epsilon^{k}v.\label{schv}
\end{equation}

Now we are ready to study the general Jordan cycle or $k$th-order. To accomplish this, let us study first some low-order cycles and then address the $k$th-order one.

The second-order Jordan cycle is closed by two unknown functions $u_1,u_2$, i.e.,
\begin{subequations}
\begin{align}
(H-\epsilon)u_1 &=0,\label{jordan1}\\
(H-\epsilon)u_2 &=u_1.
\end{align}
\end{subequations}
Equation \eqref{jordan1} implies that $u_1$ can be a linear combination of $u$ and $v$, but without lost of generality let us choose $u_1=u$, and we should find now $u_2$ by using equation~\eqref{schu}. Initially, it looks like $u_2$ is equal to $\partial_\epsilon u$, but let us remember from section~\ref{wrodiff} that this is only a particular solution of the inhomogeneous equation. The general solution is given by
\begin{equation}
u_2=C_1 u +D_1 v +\partial_\epsilon u,
\end{equation}
where $C_1$ and $D_1$ are constants.

For the third-order Jordan cycle 
\begin{subequations}
\begin{align}
(H-\epsilon)u_1 &=0,\\
(H-\epsilon)u_2 &=u_1,\\
(H-\epsilon)u_3 &=u_2,
\end{align}
\end{subequations}
we do an analogous procedure. The solutions $u_1$ and $u_2$ are the same as before and the general solution of $u_3$ is given by
\begin{equation}
u_3=C_2u+D_2v+C_1\partial_\epsilon u +D_1\partial_\epsilon v+\frac{\partial_{\epsilon\epsilon}u}{2},
\end{equation}
where $C_2$ and $D_2$ are new constants. The apparent asymmetry between $u$ and $v$ due to the last term of this equation is produced by the initial choice of $u_1$ as $u$. Nevertheless, there is no loss of generality, it is just a parameter choice that simplifies the equations.

For the fourth-order Jordan cycle
\begin{subequations}
\begin{align}
(H-\epsilon)u_1 &=0,\\
(H-\epsilon)u_2 &=u_1,\\
(H-\epsilon)u_3 &=u_2,\\
(H-\epsilon)u_4 &=u_3,
\end{align}
\end{subequations}
the solutions $u_1$, $u_2$, and $u_3$ are the same as before, while $u_4$ is given by
\begin{equation}
u_4=C_3u+D_3v+C_2\partial_\epsilon u +D_2\partial_\epsilon v+\frac{C_1}{2}\partial_{\epsilon\epsilon}u+\frac{D_1}{2}\partial_{\epsilon\epsilon}v+\frac{\partial_{\epsilon\epsilon\epsilon}u}{2\cdot 3},
\end{equation}
where $C_3$ and $D_3$ are new constants. We can foresee from this equation the general structure  of the solutions for a higher-order Jordan cycle, and we use this as an induction hypothesis for the $k$th-order Jordan cycle in order to prove it for the $(k+1)$th-order case.

The induction hypothesis is that for the $k$th-order Jordan cycle
\begin{subequations}
\begin{align}
(H-\epsilon)u_1 &=0,\\
(H-\epsilon)u_2 &=u_1,\\
\ \ \ \vdots \nonumber\\
(H-\epsilon)u_k &=u_{k-1},
\end{align}
\end{subequations}
the system is closed by $u_1=u$ and the following $k-1$ functions $u_j$ ($j=2,\dots , k$):
\begin{equation}
u_j=\sum_{i=1}^{j-1}C_{j-i}\frac{\partial_{\epsilon}^{i-1}u}{(i-1)!}+\sum_{i=1}^{j-1}D_{j-i}\frac{\partial_{\epsilon}^{i-1}v}{(i-1)!}+
\frac{\partial_\epsilon^{j-1}u}{(j-1)!},\label{ind}
\end{equation}
where $\partial^0_\epsilon f=f$ and $0!=1$.

Now, let us assume that this equation is correct for the $k$ case and let us try to prove it for the $k+1$ case. The $(k+1)$th-order Jordan cycle is
\begin{subequations}
\begin{align}
(H-\epsilon)u_1 &=0,\\
(H-\epsilon)u_2 &=u_1,\\
\ \ \ \ \vdots \nonumber\\
(H-\epsilon)u_{k+1} &=u_{k},
\end{align}
\end{subequations}
and from equation~\eqref{ind} for $k+1$ we have that
\begin{equation}
u_{k+1}=\sum_{i=1}^{k}C_{k-i+1}\frac{\partial_{\epsilon}^{i-1}u}{(i-1)!}+\sum_{i=1}^{k}D_{k-i+1}\frac{\partial_{\epsilon}^{i-1}v}{(i-1)!}+\frac{\partial_\epsilon^{k}u}{k!}.\label{indk}
\end{equation}

By applying the operator $(H-\epsilon)$ to this function and using equations~\eqref{schu} and \eqref{schv} we obtain
\begin{equation}
(H-\epsilon)u_{k+1}=\sum_{i=2}^{k}C_{k-i+1}\frac{\partial_{\epsilon}^{i-2}u}{(i-2)!}+\sum_{i=2}^{k}D_{k-i+1}\frac{\partial_{\epsilon}^{i-2}v}{(i-2)!}+\frac{\partial_\epsilon^{k-1}u}{(k-1)!},\label{indk1}
\end{equation}
where we make use of the fact that for $i=1$ we have the Schr\"odinger equations~\eqref{sch1} and \eqref{sch2}. Then changing the labels $i\rightarrow i+1$ in equation~\eqref{indk1} we have
\begin{equation}
(H-\epsilon)u_{k+1}=\sum_{i=1}^{k-1}C_{k-i}\frac{\partial_{\epsilon}^{i-1}u}{(i-1)!}+\sum_{i=1}^{k-1}D_{k-i}\frac{\partial_{\epsilon}^{i-1}v}{(i-1)!}+\frac{\partial_\epsilon^{k-1}u}{(k-1)!},\label{indk2}
\end{equation}
i.e.,
\begin{equation}
(H-\epsilon)u_{k+1}=u_k,
\end{equation}
which finish the proof by induction.\hfill $\square$

Now, we can write down explicitly the formula for the new potential \citep{MNR00}
\begin{equation}
V_k(x)=V_0(x)-[\ln W(u_1,\dots , u_k)]'',
\end{equation}
which means that we need to calculate the Wronskian of the $k$ functions $u_j$, $j=1,\dots ,k$. For the second-order case we have seen that
\begin{subequations}
\begin{align}
u_1 &= u,\\
u_2 &= C_1u+D_1v+\partial_\epsilon u,
\end{align}\label{u1u2eqs}
\end{subequations}
\hspace{-1mm}then we arrive to
\begin{equation}
W(u_1,u_2)=D_1+W(u,\partial_\epsilon u).
\end{equation}

For the third-order case, in addition to equations \eqref{u1u2eqs} we have
\begin{equation}
u_3=C_2u+D_2v+C_1\partial_\epsilon u +D_1\partial_\epsilon v+\frac{\partial_{\epsilon\epsilon}u}{2},
\end{equation}
which leads to
\begin{align}
W(u_1,u_2,u_3)=&(C_1D_1-D_2)W(u,v,\partial_\epsilon u)+D_1^2W(u,v,\partial_\epsilon v)+D_1W(u,\partial_\epsilon u, \partial_\epsilon v)\nonumber\\
&+\frac{D_1}{2}W(u,v,\partial_\epsilon^2 u)+\frac{1}{2}W(u,\partial_\epsilon u,\partial_\epsilon^2 u).
\end{align}
Similar expressions for higher-order cases can be developed. We must remember that, although these formulas become increasingly complicated, we are performing a higher-order confluent transformation, also known as {\it hyperconfluent transformation} \citep{FS11}, in which $k$ states tend to have the same energy ($k\geq 2$), the corresponding integral equations are much more complicated to solve, and for several systems they cannot even be calculated.

\subsection{Conclusions}
In this section we have introduced a differential version of the confluent second-order SUSY transformation, as an alternative to generate new-exactly solvable potentials which avoids the need to evaluate some integrals arising in the formulation elaborated previously \citep{FS03,FS05,FS11}. Moreover, the differential formula we have found generalizes the one used in soliton theory \citep{Mat92,Sta95}. Its main advantage rests in the fact that families of non-singular potentials can be constructed by appropriately varying the new constant $D$ (see equation~\eqref{Dv}). We have successfully applied this technique to the free particle and to the single-gap Lam\'e potential. In the last case it was shown that, although the initial potentials are periodic, the SUSY generated ones become non-periodic, with a periodicity defect arising due to the creation of a bound state inside an initial energy gap. It was suggested that, under certain appropriate circumstances, this bound state could be used as an intermediate state to perform transitions from the lower energy band to the infinite one.

\section{Application. Inverted oscillator}\label{inverted}
In this section we will apply the first- and second-order supersymmetric quantum mechanics to obtain new exactly-solvable real potentials departing from the inverted oscillator. This system has some special properties; in particular, only very specific second-order transformations produce non-singular real potentials. It will be shown that these transformations turn out to be the so-called complex ones. Moreover, we will study the factorization method applied to the inverted oscillator and the algebraic structure of the new Hamiltonians.

\subsection{Introduction}
Let us consider now the following Hamiltonian
\begin{equation}
H=-\frac{\hbar^2}{2m}\frac{\text{d}^2}{\text{d} x^2}+\frac{1}{2}m\omega^2 x^2,
\end{equation}
where $m$ has units of mass and $\omega$ of frequency. In order to simplify it, we are going to use natural units, such that $\hbar=m=1$, to obtain
\begin{equation}
H=-\frac{1}{2}\frac{\text{d}^2}{\text{d} x^2}+\frac{1}{2}\omega^2 x^2.\label{Homega}
\end{equation}
Moreover, by choosing appropriately the value of $\omega$, three essentially different cases can be obtained
\begin{equation}
\omega=
\begin{cases}
1 & \text{harmonic oscillator},\\
0 & \text{free particle},\\
i & \text{inverted oscillator}.
\end{cases}
\end{equation}
These are three examples of exactly-solvable potentials in quantum mechanics. The first one, the harmonic oscillator, is a very well known system from which the technique of creation and annihilation operators and the whole formalism of the factorization method come from. The second, the free particle, has also been largely studied. This simple system allows us to work close to the limits of quantum theory, for example, with non-square-integrable wavefunctions with plenty of physical applications such as the plane waves. The third case is not so familiar: it is called either \textit{inverted oscillator}, \textit{repulsive oscillator}, \textit{inverse oscillator}, or \textit{parabolic potential barrier}. Although it started as an exercise from Landau's book \citep{LL58}, its physical applications have grown since the appearance of Barton's PhD thesis (published in \citep{Bar86}), v.g., as an instability model, as a mapping of the 2D string theory \citep{YKC06}, or as a toy model to study early time evolution in inflationary models \citep{GP91}.

It is interesting to observe that both oscillator potentials, harmonic and inverted, are produced simultaneously inside an ideal Penning trap, typically used to confine charged particles \citep{BG86,FV09}. In its standard setup, a quadrupolar electrostatic field creates a harmonic oscillator potential along the symmetry axis of the trap, inducing confinement along that direction. In addition, a two-dimensional inverted oscillator arises in the orthogonal plane, driving the particles towards the trap walls. In order to compensate for the last effect, a static homogeneous magnetic field along the symmetry axis of the trap is also applied, but for zero magnetic field the two kinds of oscillator potentials are created inside the cavity.

Mathematically, the harmonic and inverted oscillators are very much alike, and we will show that the solutions of one can be obtained almost directly from the other; nevertheless, we should remark that physically these two systems are very different. For example the harmonic oscillator has a discrete non-degenerate equidistant energy spectrum with square-integrable eigenfunctions, while the inverted oscillator has a continuous spectrum varying from $-\infty$ to $+\infty$, which is doubly degenerated, and whose eigenfunctions are not square-integrable.

On the other hand, we have seen that a standard technique for generating new exactly-solvable potentials departing from a given initial one is the supersymmetric quantum mechanics (SUSY QM) (for recent reviews see \citep{MR04,AC04,Suk05,FF05,Fer10,AI12}). Its simplest version, which makes use of first-order intertwining operators, has been employed for generating Hamiltonians whose spectra differ from the initial one in the ground state energy level. In addition, the higher-order variants, which involve differential intertwining operators of orders larger than one \citep{AIS93,AICD95,BS97,Fer97,AST01}, allow as well the modification of one or several excited state levels.

The SUSY techniques of first- and higher-order have been successfully applied to the harmonic oscillator \citep{FH99,BF11a} and the free particle \citep{MS91,BFF12} for generating plenty of exactly-solvable potentials. However, as far as we know, neither the first- nor the higher-order SUSY QM have been employed taking as a point of departure the inverted oscillator. In this section we aim to fill this gap by applying the supersymmetric transformations to the inverted oscillator. In order to do that, in section \ref{secinv2} we will get the general solution of the stationary Schr\"odinger equation (SSE) for the Hamiltonian \eqref{Homega} with an arbitrary energy $E$, which will remain valid even for $E\in{\mathbb C}$. In addition, the solutions which have a physical interpretation for the inverted oscillator will be identified. In section \ref{secinv3} we are going to explore the factorization method for both systems, obtaining the bound states for the harmonic oscillator and also several sets of mathematical polynomial solutions, a class of solutions which have been of interest along the time (see e.g. \citep{Que08,OS09,GKM10}). In section \ref{secinv4} we will work out the first-order SUSY QM for the inverted oscillator, while in section \ref{secinv5} we will apply the second-order one in its three different situations: real, confluent and complex cases. The last one will be the most important case of this section, as it is the only one that works out to obtain new real non-singular potentials. In section \ref{secinv6} we will explore the algebraic structure for the non-singular potentials generated through SUSY QM and their associated eigenfunctions. Finally, in section \ref{secinv7} we will present our conclusions. Appendix A contains the derivation of the orthogonality and completeness relations for the set of eigenfunctions of the inverted oscillator Hamiltonian (see also \citep{Wol79}). 

\subsection{General solution of the stationary Schr\"odinger equation}\label{secinv2}
First of all, let us solve the stationary Schr\"odinger equation for the Hamiltonian \eqref{Homega} with an arbitrary real energy $E$, although it is still valid for $E\in{\mathbb C}$:
\begin{align} \label{sseq}
H\psi(x) = \left(-\frac{1}{2}\frac{\text{d}^2}{\text{d} x^2}+\frac{\omega^2 x^2}{2}\right)\psi(x)  = E \psi(x).
\end{align}
The substitution $x=\omega^{-1/2}y$ with $\phi(y) \equiv \psi(\omega^{-1/2}y)$ leads to
\begin{equation}
\left(-\frac{1}{2}\frac{\text{d}^2}{\text{d} y^2}+\frac{y^2}{2}\right)\phi(y) = \frac{E}{\omega} \, \phi(y),
\end{equation}
which is the SSE for the harmonic oscillator potential in the variable $y$ associated with the {\it energy} $E/\omega$ \citep{JR98}. Here we assume that $\omega\neq 0$; otherwise we would immediately get the free particle problem. Thus, the general solution to the SSE \eqref{sseq} reads
\begin{equation}
\psi(x)= e^{-\omega x^2/2}
\left[C {}_1F_1\left(\frac{1}{4} - \frac{E}{2\omega},
\frac{1}{2}; \omega x^2 \right) + D \, x \, {}_1F_1\left(\frac{3}{4} - \frac{E}{2\omega},\frac{3}{2}; \omega x^2\right)\right],\label{roy1}
\end{equation}
where ${}_1F_1$ is the confluent hypergeometric function and $C,D\in\mathbb{R}$ are constants. Note that, in general $\psi(x)\notin \mathcal{L}^2(\mathbb{R})$, i.e., it does not belong to the space of square-integrable wavefunctions in one dimension for an arbitrary $E\in\mathbb{C}$.

The analysis of the solution for the harmonic oscillator (take $\omega = 1$ in equation~\eqref{roy1}) is widely known and can be found for example in \citet{JR98} or \citet{FF05}. On the other hand, since the inverted oscillator is not commonly studied, we will work it out in detail next.

The general solution of the inverted oscillator is obtained from equation~\eqref{roy1} for $\omega=i$. For simplicity, it will be expressed in terms of solutions with a definite parity, i.e., its even ($\psi_e$) and odd ($\psi_o$) solutions, which are given by
\begin{subequations}
\begin{align}
\psi_e(x) & = \text{e}^{-ix^2/2} {}_1F_1\left(\frac{1+2iE}{4},\frac{1}{2};ix^2\right) = 
\text{e}^{ix^2/2} {}_1F_1\left(\frac{1-2iE}{4},\frac{1}{2};-ix^2\right), \label{ue}\\
\psi_o(x) & = x \, \text{e}^{-ix^2/2} {}_1F_1\left(\frac{3+2iE}{4},\frac{3}{2};ix^2\right) = x \, \text{e}^{ix^2/2} {}_1F_1\left(\frac{3-2iE}{4},\frac{3}{2};-ix^2\right). \label{uo}
\end{align}\label{ueo}
\end{subequations}
\hspace{-1.8mm}This decomposition will simplify our mathematical work in sections \ref{secinv4} and \ref{secinv5}. Therefore the general solution is the following linear combination
\begin{equation}
\psi(x)=C \psi_e(x)+D \psi_o(x).\label{roy2}
\end{equation}
It is clear now that $\psi(x)$ is a real function, $\overline{\psi}(x)=\psi(x)$, for
any $E,C,D\in{\mathbb R}$.

Next, let us analyze the leading asymptotic behaviour of the solutions given in equations~\eqref{ue} and \eqref{uo}. To do this, we need the asymptotic behaviour of ${}_1F_1$ for $|z|\gg 1$ \citep{AS72}:
\begin{equation}
{}_1F_1(a,b,z)\simeq\frac{\Gamma(b)}{\Gamma(b-a)}\text{e}^{i \pi a}z^{-a}+\frac{\Gamma(b)}{\Gamma(a)}\text{e}^{z}z^{a-b}. \label{limit}
\end{equation}
Hence, the asymptotic behaviour for $\psi_e$ and $\psi_o$ can be straightforwardly obtained for $x\gg 1$:
\begin{subequations}
\begin{equation}
\psi_e(x) \simeq
\frac{\pi^{1/2} e^{-\pi E/4}}{x^{1/2}} \left[
\frac{e^{i(\pi/8 - x^2/2 )}x^{- i E}}{\Gamma(1/4-i E/2)} +
\frac{e^{-i(\pi/8 - x^2/2)}x^{iE}}{\Gamma(1/4 + i E /2)}
\right],\label{limue}
\end{equation}
\begin{equation}
\psi_o(x) \simeq
\frac{\pi^{1/2} e^{-\pi E/4}}{2x^{1/2}} \left[
\frac{e^{i(3\pi/8 - x^2/2)}x^{-iE}}{\Gamma(3/4-iE/2)} +
\frac{e^{-i(3\pi/8 - x^2/2)}x^{iE}}{\Gamma(3/4 + iE/2)}
\right].\label{limuo}
\end{equation}
\label{limus}
\end{subequations}
\hspace{-1.8mm}By taking into account these equations, a complex linear combination is found such that now the terms going as $x^{-iE}$ get cancelled
\begin{equation}
\psi_E^+(x) = N_E \left[\psi_e(x)- \frac{2\text{e}^{-i\pi /4} \, \Gamma(3/4 - i E /2)}{\Gamma(1/4 - i E/2)}  \psi_o(x)\right],
\label{uplus}
\end{equation}
where the {\it normalization} factor $N_E$ will be chosen in order to form an orthonormal set of eigenfunctions in the Dirac sense (see appendix A and \citep{Wol79}). On the other hand, another linearly independent solution can be found from $\psi_E^+(x)$ through the reflection $x\rightarrow -x$:
\begin{equation}
\psi_E^-(x) = N_E \left[\psi_e(x)+ \frac{2\text{e}^{-i\pi /4} \, \Gamma(3/4 - i E /2)}{\Gamma(1/4 - i E/2)}  \psi_o(x)\right].
\label{uminus}
\end{equation}

It is worth noting that $\{\psi_E^\sigma(x), \ \sigma = \pm, \ -\infty<E<\infty\}$ is a complete orthonormal set of eigenfunctions for the inverted oscillator Hamiltonian satisfying the following orthogonality and completeness relations:
\begin{subequations}
\begin{align}
(\psi_E^\sigma,\psi_{E'}^{\sigma'}) = \int_{-\infty}^{\infty} \overline{\psi}_E^{\, \sigma}(x)\psi_{E'}^{\sigma'}(x) dx 
& = \delta_{\sigma,\sigma'} \delta(E-E'), \\
\sum_{\sigma = \pm} \int_{-\infty}^{\infty} dE \psi_E^\sigma(x) \overline\psi_E^{\, \sigma}(x') & =  \delta(x-x'),
\end{align}
\end{subequations}
where $\delta_{\sigma,\sigma'}$ and $\delta(y-y')$ denote the Kronecker and Dirac delta functions respectively
(for a derivation of these equations we refer the reader to appendix A and to \citet*{Wol79}).

We can construct now a real linear combination with a specific physical interpretation \citep{MRW09,Wol10}, namely,
\begin{equation}
\psi_L(x) = \psi_e(x)- \left[ \frac{\text{e}^{i\pi /4} \, \Gamma(3/4 + i E /2)}{\Gamma(1/4 + i E/2)} +
\frac{\text{e}^{-i\pi /4} \, \Gamma(3/4 - i E /2)}{\Gamma(1/4 - i E/2)} \right] \psi_o(x). \label{ul}
\end{equation}
The subscript $L$ is employed because $\psi_L(x)$ represents a particle incident from the left since its probability amplitude for $x<0$ is substantially larger than the one for $x>0$ when $E<0$. On the other hand, another linearly independent real eigenfunction for the same $E$ with physical meaning can be obtained
from $\psi_L(x)$ through the reflection $x \rightarrow -x$:
\begin{equation}
\psi_R(x) = \psi_L(-x) = \psi_e(x)+ \left[ \frac{\text{e}^{i\pi /4} \, \Gamma(3/4 + i E /2)}{\Gamma(1/4 + i E/2)} + \frac{\text{e}^{-i\pi /4} \, \Gamma(3/4 - i E /2)}{\Gamma(1/4 - i E/2)} \right] \psi_o(x). \label{ur}
\end{equation}
The subscript $R$ denotes the fact that $\psi_R(x)$ represents now a particle incident from the right.

\begin{figure}
\begin{center}
\includegraphics[scale=0.6]{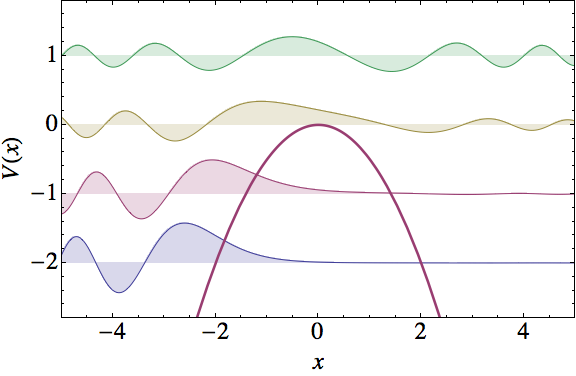}
\end{center}
\vspace{-5mm}
\caption{\small{Eigenfunctions $\psi_L(x)$ of the inverted oscillator Hamiltonian given by equation~\eqref{ul} for the energies $E \in \{ -2,-1,0,1 \}$.}} \label{inv1}
\end{figure}

The classical and quantum behaviours of a particle under the inverted oscillator potential is summarized in table~\ref{table}. In figure~\ref{inv1} we have plotted the eigenfunction $\psi_L(x)$ for different values of the energy $E$, i.e., the quantum behaviour of a particle incident from the left, which illustrates the results of table~\ref{table}. Since the complete set of orthonormal eigenfunctions of $H$ $\{\psi_E^\sigma(x), \ \sigma = \pm, \ -\infty<E<\infty\}$ has been found, one can conclude that the energy spectrum of the inverted oscillator Hamiltonian consists of the full real line, each eigenvalue $E\in{\mathbb R}$ being doubly degenerated.

\begin{table}
\begin{center}
\begin{tabular}{cll}
\hline
Value of $E$& Classical behaviour & Quantum behaviour\\
\hline
$E >0$& Goes over & Most probable goes over, some is reflected\\
$E =0$& Is trapped an infinite time & Is trapped a finite time (sojourn time)\\
$E <0$& Is reflected & Most probable is reflected, some goes over\\
\hline
\end{tabular}
\end{center}
\vspace{-3mm}
\caption{\small{The classical and quantum behaviours of a particle under the inverted oscillator potential.}} \label{table}
\end{table}

\subsection{Factorization method}\label{secinv3}
The Hamiltonian \eqref{Homega} is Hermitian for $\omega=1,0,i$ and it admits the well-known factorization method. Indeed, by using the analogues of the annihilation and creation operators for the harmonic oscillator,
\begin{equation}
a_{\omega}^{\pm}=\frac{1}{\sqrt{2}}\left(\mp \frac{\text{d}}{\text{d} x}+\omega x\right),\label{fac1}
\end{equation}
it can be shown that \citep{Shi00}
\begin{equation}\label{2wf}
H = a_{\omega}^{+}a_{\omega}^{-}+\frac{\omega}{2} = a_{\omega}^{-}a_{\omega}^{+}-\frac{\omega}{2}.
\end{equation}
Since for $\omega=1$ the operators $a_{1}^{+}$ and $a_{1}^{-}$ are mutually Hermitian conjugate, $(a_{1}^{-})^{\dag}=a_{1}^{+}$, then for the harmonic oscillator the two factorizations of equation~\eqref{2wf} are essentially different. On the other hand, for $\omega=i$ the operators $a_{i}^{\pm}$ are antihermitian,  $(a_{i}^{\pm})^{\dag}=-a_{i}^{\pm}$, which implies that for the inverted oscillator the two factorizations in equation~\eqref{2wf} are indeed the same (since $H$ is Hermitian).

It is clear now that the set of operators $\{H,a_{\omega}^{+},a_{\omega}^{-}\}$ satisfies the following algebra
\begin{subequations}
\begin{align}
\left[H,a_{\omega}^{\pm}\right]& =\pm\omega a_{\omega}^{\pm},\\
\left[a_{\omega}^{-},a_{\omega}^{+}\right]&=\omega,
\end{align}
\end{subequations}
which immediately leads to
\begin{equation}
Ha_{\omega}^{\pm}\psi(x)=(E\pm \omega)a_{\omega}^{\pm}\psi(x),
\end{equation}
where $\psi(x)$ satisfies equation~\eqref{sseq}, and applying $n$ times $a_{\omega}^{\pm}$ we get
\begin{equation}
H(a_{\omega}^{\pm})^n\psi(x)=(E\pm n\omega)(a_{\omega}^{\pm})^n\psi(x).\label{fac6}
\end{equation}
Note that for both cases, $\omega=1$ and $\omega=i$, the general solution of equation~\eqref{sseq} is given by equation~\eqref{roy1} in the extended domain $E\in\mathbb{C}$; however, these solutions are not always physically admissible. Next, we will examine in more detail each of the two algebras.

\subsubsection{Harmonic oscillator algebra}
For $\omega=1$, equations~(\ref{fac1}--\ref{fac6}) simplify to the Heisenberg-Weyl algebra of the harmonic oscillator Hamiltonian \citep{Per86}. The general solution of the SSE is given by equation~\eqref{roy1} for $\omega=1$, from which we can obtain its {\it bound states} or its {\it pure point spectrum}.


An alternative way is to take the first factorization of equation~\eqref{2wf}, $H=a_1^{+}a_1^{-}+1/2$, and look for the {\it extremal state} $\psi_0(x)$ which is annihilated by $a_1^{-}$,
\begin{equation}
a_1^{-}\psi_0(x)=0 \quad \Rightarrow \quad \psi_0(x)=\pi^{-1/4}\exp(-x^2/2).\label{base}
\end{equation}
In addition, $\psi_0(x)$ is an eigenfunction of $H$ with eigenvalue $E_0=1/2$, and if we apply iteratively the creation operator we will get the remaining bound states
\begin{equation}\label{excitados}
\psi_n(x)=\frac{(a_1^{+})^n}{(n!)^{1/2}}\psi_0(x)=\frac{1}{2^{n/2}\pi^{1/4}(n!)^{1/2}}\exp(-x^2/2)H_n(x),
\end{equation}
associated with the eigenvalues $E_n=n+1/2,\ n=0,1,\dots$, where $H_n(x)$ are the Hermite polynomials. This kind of algebra is known as {\it spectrum generating algebra}. Similarly, we can obtain another extremal state $\phi_0(x)$ from the second factorization, $H=a_1^{-}a_1^{+}-1/2$, as
\begin{equation}
a_1^{+}\phi_0(x)=0 \quad \Rightarrow\quad \phi_0(x)=\pi^{-1/4}\exp(x^2/2),
\end{equation}
which is a solution of the SSE associated with $E =-1/2 \equiv e_0$. We have added the constant factor $\pi^{-1/4}$ by symmetry with equation~\eqref{base}, even though the wavefunction $\phi_0(x)$ is not normalizable and therefore it is not a bound state. We have stressed this fact by choosing a different notation $\phi_0$ for this wavefunction.

Note that the extremal state $\phi_0(x)$ can also be obtained by acting $a_1^-$ on the {\it irregular wavefunction} $\varphi_0(x)$, defined as the second linearly independent solution of the SSE for $E_0=1/2$. Although this wavefunction is not normalizable, it does have some physical applications \citep{Leo10}.

\begin{figure}
\begin{center}
\includegraphics[scale=0.3]{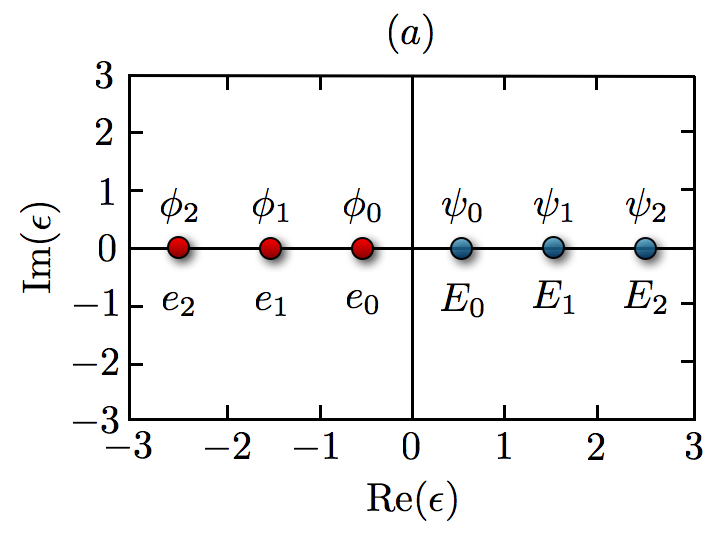}
\includegraphics[scale=0.3]{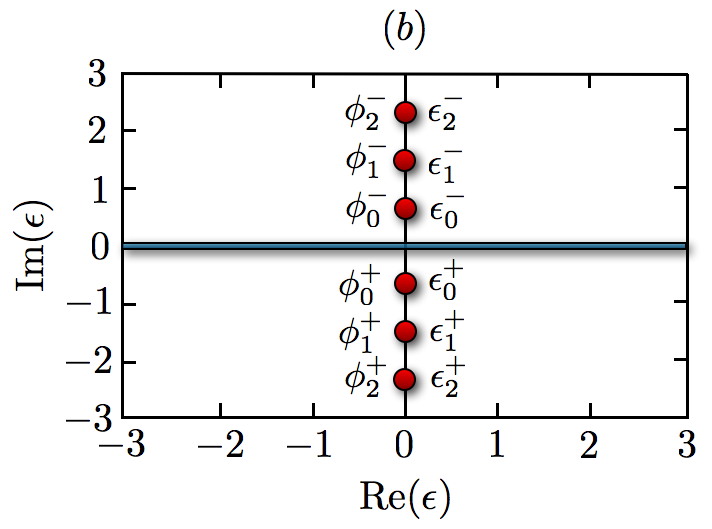}
\end{center}
\vspace{-5mm}
\caption{\small{The complex plane for which a general solution $\forall\, E\in\mathbb{C}$ can be found for both the harmonic and the inverted oscillators. In (a) we show the bound states (blue circles) and the polynomial solutions (red circles) for the harmonic oscillator. In (b) we show the scattering states on the real axis (blue line) and the polynomial solutions (red circles) for the inverted oscillator.}} \label{complexplane}
\end{figure}

Now, we will apply iteratively the annihilation operator in order to obtain a new ladder of wavefunctions expressed in terms of Hermite polynomials, although with a different argument as compared with the standard case. Hence, the ladder of nonphysical wavefunctions is given by
\begin{equation}\label{enesimosnofisicos}
\phi_n(x)=\frac{(a_1^{-})^n}{(n!)^{1/2}}\phi_0(x)=\frac{i^{-n}}{2^{n/2}\pi^{1/4}(n!)^{1/2}}\exp(x^2/2)H_n(ix),
\end{equation}
associated with the discrete nonphysical energies $e_n=-n-1/2,\ n=0,1,\dots$, which nevertheless provide an additional set of {\it polynomial solutions} for the harmonic oscillator potential. The factor $i^{-n}$ appears naturally when we factorize the Hermite polynomial with the correct argument. The position of the eigenvalues for the bound states \eqref{excitados} and the nonphysical energies associated with the polynomial solutions \eqref{enesimosnofisicos} on the complex $E$-plane can be seen in figure~\ref{complexplane}(a).

\subsubsection{Inverted oscillator}
For $\omega=i$ equations~(\ref{fac1}--\ref{fac6}) define a zeroth-order complex deformation of the Heisenberg-Weyl algebra. The general solution of the corresponding SSE is given by equation~\eqref{roy1} with $\omega=i$ in the full complex $E$-plane.

In this case there is no solution which fulfills the bound state condition, as it is expected because the inverted oscillator potential has neither local nor global wells which would support such bound states. Nevertheless, it is expected to have {\it dispersive states} that, under certain conditions, can be given a physical interpretation. Note that the most familiar system with these characteristic states is the free particle.

As it was shown in section \ref{secinv2}, the natural dispersive states of the inverted oscillator are those given by  
equations~(\ref{uplus}--\ref{uminus}) and (\ref{ul}--\ref{ur}) for any $E\in\mathbb{R}$. However, we can ask ourselves further: Are there any solutions related with the Hermite polynomials for the inverted oscillator? In order to answer this, let us find now the extremal states associated with the factorizations of equation~\eqref{2wf} for $\omega=i$.

Let us consider in the first place
\begin{equation}
a_i^{-}\phi_0^{-}(x)=0 \quad \Rightarrow \quad \phi_0^{-}(x)=\pi^{-1/4}\exp(-ix^2/2),
\end{equation}
for $\epsilon_0^{-}=i/2$. Furthermore, if we apply iteratively the creation operator $a_i^{+}$ we will get the following set of polynomial solutions of the SSE
\begin{equation}\label{enesimosnofisicosinvertedmenos}
\phi_n^{-}(x)=\frac{(a_i^{+})^n}{(n!)^{1/2}}
\phi_0^{-}(x)=\frac{i^{n/2}}{2^{n/2}\pi^{1/4}(n!)^{1/2}}\exp(-ix^2/2)H_n(i^{1/2}x),
\end{equation}
associated with $\epsilon_n^{-}=i(n+1/2),\ n=0,1,\dots$

Similarly, the use of the second factorization leads to
\begin{equation}
a_i^{+}\phi_0^{+}(x)=0 \quad \Rightarrow \quad \phi_0^{+}(x)=\pi^{-1/4}\exp(ix^2/2),
\end{equation}
associated with $\epsilon_0^{+}=-i/2$. In addition, if we apply $n$ times the annihilation operator $a_i^{-}$ we will get another set of polynomial solutions
\begin{equation}\label{enesimosnofisicosinvertedmas}
\phi_n^{+}(x)=\frac{(a_i^{-})^n}{(n!)^{1/2}}
\phi_0^{+}(x)=\frac{i^{-n/2}}{2^{n/2}\pi^{1/4}(n!)^{1/2}}\exp(ix^2/2)H_n(i^{3/2}x),
\end{equation}
associated with $\epsilon_n^{+}=-i(n+1/2),\ n=0,1,\dots$.

Let us stress once again that $\phi_n^{-}(x)$ and $\phi_n^{+}(x)$ are solutions of an SSE for complex $E$-values and thus, they do not have any physical  interpretation at all, neither as bound nor as dispersive states. Nevertheless, it turns out that they are related with polynomial solutions of the SSE. A diagram marking the positions in which this kind of solutions appear for the inverted oscillator on the complex $E$-plane is shown in figure~\ref{complexplane}(b).

Note that there is a direct extension of this factorization method, which is closely related to SUSY QM. In this generalization, instead of factorizing the Hamiltonian $H$ in two different ways in terms of a pair of first-order operators, as in equation~\eqref{2wf}, one looks for the most general first-order operators producing just a single factorization \citep{Mie84}. Thus, when the ordering of the operators factorizing $H$ is reversed, in general one arrives to a different Hamiltonian. This fact has been used to generate new exactly-solvable Hamiltonians departing from a given initial one \citep{Mie84,Fer84,Suk85a}.

Next, we are going to apply the first- and second-order SUSY QM to the inverted oscillator. This supplies us with the building bricks for implementing the higher-order transformations, since it is known nowadays that any non-singular transformation of order higher than two can always be factorized in terms of non-singular first- and second-order SUSY transformations. Note that, although this fact was conjectured for the first time by \citet{AICD95}and \citet{BS97}, however it was proven just recently in a rigorous way by \citet{AS07} (see also \citet{Sok08}). Then, the higher-order SUSY partners of the inverted oscillator can be obtained through iterations of the non-singular transformations which will be discussed here.

\subsection{First-order SUSY QM}\label{secinv4}
The first-order SUSY QM, which was studied in detail in section \ref{secsusy1}, is going to be applied now to the inverted oscillator potential
\begin{equation}
V_0(x)=-\frac{1}{2}x^2.
\end{equation}
As it was show, the new potential is given by
\begin{equation}
V_1(x) = V_0(x) - \alpha_1'(x)= V_0(x)-\left[\frac{u'(x)}{u(x)}\right]',
\end{equation}
where the transformation function $u(x)$ is the solution of the initial stationary Schr\"odin-ger equation associated with the factorization energy $E=\epsilon$ given in equations \eqref{ueo} and \eqref{roy2},  which is real for any $\epsilon, C, D\in{\mathbb R}$. It is clear now that if the {\it transformation function} $u(x)$ has zeroes, then the new potential $V_1(x)$ will have singularities at those points. Since the seed solution $u(x)$ given in equation~\eqref{roy2}, associated with an arbitrary {\it factorization energy} $\epsilon\in{\mathbb R}$, always has zeroes because it has oscillatory terms that cannot get cancelled (see figure~\ref{inv1}), it follows that it is impossible to perform real non-singular first-order SUSY transformations for the inverted oscillator. Note that the singular transformations are excluded because they change, in general, the domain of the initial potential and, consequently, the initial spectral problem (compare, e.g., \citep{MNN98}).

\subsection{Second-order SUSY QM}\label{secinv5}
The second-order SUSY QM was studied in detail in section \ref{secdir}. Here, we will make extensive use of the theory and equations developed in that section. What we will do next is to apply the three cases of the second-order SUSY QM to the inverted oscillator.  

\subsubsection{The real case for $c>0$}
As we studied in section~\ref{realcase}, In this case $\xi_1 = \sqrt{c}>0, \ \xi_2 = - \sqrt{c}$, $\epsilon_i = (d + \xi_i)/2 \in {\mathbb R}, \ i=1,2,$ $\epsilon_1\neq\epsilon_2$. The new potential $V_2(x)$ is given by
\begin{equation}
V_2(x)=V_0(x)-\{\ln[W(u_1,u_2)]\}'',\label{v2}
\end{equation}
i.e., to obtain a non-singular potential $V_2$, a $W(u_1,u_2)$ without zeroes is required (recall that in the first-order case it was directly the transformation function $u$ the one that should not have nodes). A $W(u_1,u_2)$ without zeroes could be achieved if $u_1$ and $u_2$ would have alternate nodes. For the inverted oscillator this requirement is true in some finite interval of the $x$-domain but not in the full real line. Therefore, we cannot produce real non-singular second-order SUSY transformations for the inverted oscillator. On the other hand, note that the zeroes of $u_1$, $u_2$ are closer to alternate in all ${\mathbb R}$ as $\epsilon_1$ and $\epsilon_2$ become closer. This fact hints us to use the confluent second-order SUSY QM, a well worked algorithm where the two factorization energies converge to a single one \citep{MNR00,FS03,BFF12}.

\subsubsection{The confluent case for $c=0$}
\label{confluent}
As we saw in section \ref{secconfluente}, in this case $\epsilon_1=\epsilon_2\equiv\epsilon\in\mathbb{R}$, the new potential is given by
\begin{equation}
V_2(x)=V_0(x)-\{\ln[w(x)]\}'',
\end{equation}
with
\begin{equation}
w(x)=w_0+\int_{x_0}^{x}u^2(z)\text{d}z,\label{omega}
\end{equation}
and the conditions \eqref{imas} or \eqref{imenos} must be satisfied, i.e., $u(x)$ should have a null asymptotic behaviour in one boundary and also be square-integrable over the corresponding semi-bounded interval.

Now, from the asymptotic behaviour of $\psi_e$ and $\psi_o$ given in equations~\eqref{limus} we can see that its leading terms fall off as $\sim |x|^{-1/2}$ (remarkably they do not depend on $\epsilon\in\mathbb{R}$); therefore they are at the frontier (but outside) of $\mathcal{L}^2(\mathbb{R})$, as any improper Dirac base. This means that they fulfill the first conditions of equations~\eqref{imas} and \eqref{imenos} but not the second ones, i.e., there is no linear combination of solutions which can be made square-integrable over a semi-bounded interval. This fact implies that the $w(x)$ of equation \eqref{omega} will always have one zero on the real axis, and consequently the confluent second-order SUSY transformation will always be singular. Thus, one can conclude that, although at first sight it seems possible to perform the non-singular SUSY transformation through the confluent algorithm, however the transformation functions $u(x)$ obtained from equation~(\ref{roy2}) by making $E=\epsilon$ do not fulfill neither equations~\eqref{imas} nor equations~\eqref{imenos}.

\subsubsection{The complex case for $c<0$}
For $c<0$ it turns out that  $\epsilon_1,\epsilon_2\in\mathbb{C}$ and $\epsilon_1=\overline{\epsilon}_2\equiv\epsilon$, as it was shown in section \ref{complexcase}. Now the new potential is given by
\begin{equation}
V_2(x)=V_0(x)-\{\ln [w(x)]\}''=V_0(x)-\left(\frac{u\overline{u}}{w}\right)', \label{gcom}
\end{equation}
with
\begin{equation}
w(x)=\frac{W(u,\overline{u})}{2(\epsilon-\overline{\epsilon})}.\label{funw}
\end{equation}
Furthermore, it is easy to show that $w(x)=\overline{w}(x)$, i.e., $w(x)\in \mathbb{R}$.

Once again, $w(x)$ must not have zeroes in ${\mathbb R}$ to avoid singularities in $V_2(x)$. Since $w'(x)=|u(x)|^{2}$, then $w(x)$ is a monotonically non-decreasing function. Thus, to assure that $w(x)\neq 0\ \forall \ x\in\mathbb{R}$, it is sufficient to fulfill either the conditions of equations~\eqref{imas} or those of equations~\eqref{imenos}, although both cannot be accomplished simultaneously since the transformation function $u(x)$ used to implement the SUSY algorithm cannot be physical, i.e., it is not square-integrable. The complex second-order transformations such that $u(x)$ fulfills one of these conditions can produce real potentials which are always strictly isospectral to the original one.

As it was discussed at the end of section \ref{confluent}, we cannot find a transformation function $u(x)$ that fulfills either equations~\eqref{imas} or equations~\eqref{imenos} for real factorization energies $\epsilon\in\mathbb{R}$. On the other hand, from the asymptotic behaviour for $\psi_e(x)$ and $\psi_o(x)$ given in equations~\eqref{limus}, we can see that both functions contain two terms, one going as $x^{-1/2-i\epsilon}$ and other as $x^{-1/2+i\epsilon}$. Then, for $\mbox{Im}(\epsilon)\neq 0$ we will have one term falling faster and other slower than $x^{-1/2}$. Moreover, if a linear combination of the two solutions which cancels the slower terms can be found, thus a solution that fulfills either the conditions of equations~\eqref{imas} or of equations~\eqref{imenos} is obtained.

\begin{figure}
\begin{center}
\includegraphics[scale=0.6]{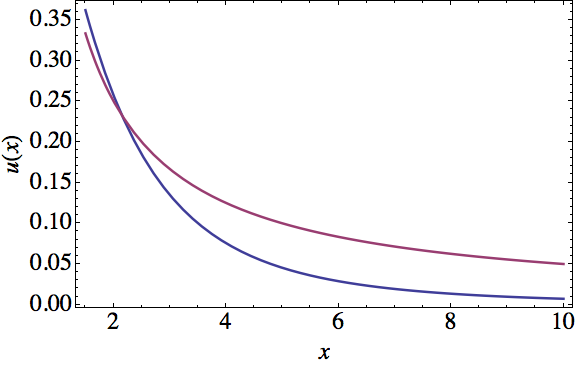}
\end{center}
\vspace{-5mm}
\caption{\small{Comparison between $|u_P(x,\epsilon)|^2$ for $\epsilon=5+i$ (blue line) and $\propto x^{-1}$ (magenta line). We can see here that the conditions of equation~\eqref{imas} are fulfilled.}} \label{figu}
\end{figure}

It is straightforward to obtain now the linear combinations that fulfill the condition of equations~\eqref{imas}, the resulting functions are given by
\begin{subequations}
\begin{align}
u_P(x,\epsilon)&=\psi_e(x) - \frac{2\, \text{e}^{-i\pi /4} \Gamma(3/4-i \epsilon/2)}{\Gamma(1/4 - i \epsilon /2)}\psi_o(x),\label{up}\\
u_N(x,\epsilon)&=\psi_e(x) - \frac{2\, \text{e}^{i\pi /4} \, \Gamma(3/4 + i \epsilon/2)}{\Gamma(1/4 + i \epsilon /2)}\psi_o(x),
\end{align}
\end{subequations}
where the labels $P$ and $N$ refer to {\it positive} and {\it negative}, according to the sign of $\mbox{Im}(\epsilon)$ which gives the right behaviour, and we have expressed explicitly the dependence of those functions from the complex factorization energy $\epsilon$. In figure~\ref{figu} it is shown a graph of $|u_P(x,\epsilon)|^2$ as compared with $\propto x^{-1}$. In this case the conditions of equation~\eqref{imas} are fulfilled and thus we can perform the non-singular complex SUSY transformation. Note that the transformation function $u_P(x,\epsilon)$ of equation~\eqref{up}, which is square-integrable in $\mathbb{R}^{+}$ for $\mbox{Im}(\epsilon)>0$, coincides with the solution found by \citet*[section 20.10]{Tit58}.

\begin{figure}[H]
\begin{center}
\includegraphics[scale=0.6]{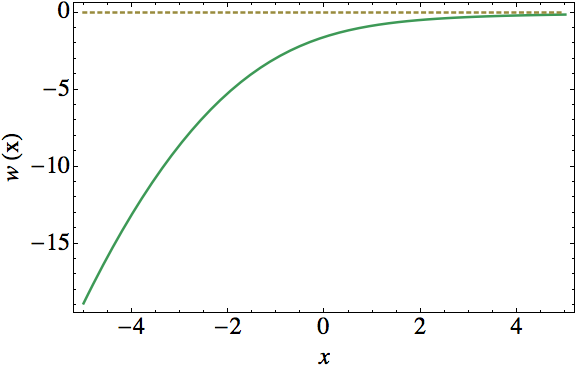}
\end{center}
\vspace{-5mm}
\caption{\small{The function $w(x)$ for the {\it complex energy} $\epsilon=5+i$ (green line). The zero of this function lies in $+\infty$.}} \label{figw}
\end{figure}

Next, we need to construct the function $w(x)$ defined by equation~\eqref{funw}, which is a real-defined function. In order to obtain a real non-singular SUSY partner potential \citep{FMR03}, this $w(x)$ should be built up by using only either $u_{P}(x,\epsilon)$ or $u_{N}(x,\epsilon)$ as transformation functions, which causes that the non-decreasing monotonic function $w(x)$ vanishes either at $x\rightarrow \infty$, if equations~(\ref{imas}) are obeyed, or at $x\rightarrow -\infty$, if equations~(\ref{imenos}) are satisfied. In figure~\ref{figw} we show the function $w(x)$ built from $u_{P}(x,\epsilon)$ for a specific factorization energy.

\begin{figure}[H]
\begin{center}
\includegraphics[scale=0.4]{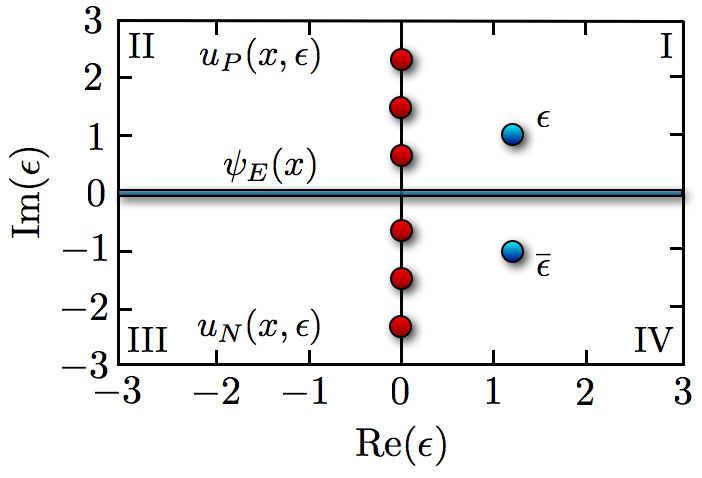}
\end{center}
\vspace{-5mm}
\caption{\small{Domain of the transformation functions $u_P(x,\epsilon)$ and $u_N(x,\overline{\epsilon})$ in the complex plane $\epsilon$. The real line ($\mbox{Im}(\epsilon)=0$) and the points $\epsilon = \pm i(m + 1/2), \ m=0,1,2,\dots$ are excluded. For quadrants I and II we should use $u_P(x,\epsilon)$ as transformation function, and $u_N(x,\epsilon)$ for quadrants III and IV.}} \label{inv4}
\end{figure}

This analysis shows that we can implement successfully the non-singular complex SUSY transformation through any of the two seed solutions $u_P(x,\epsilon)$ or $u_N(x,\epsilon)$. Nevertheless, we should remember that each of these functions works only for half of the complex plane $\epsilon$, i.e., the imaginary part of $\epsilon$ should be positive in order to use $u_P(x,\epsilon)$ or negative for $u_N(x,\epsilon)$. However, by noticing that
\begin{equation}
{\overline u}_P(x,\epsilon) = u_N(x,{\overline \epsilon}),
\end{equation}
and looking more carefully at the algorithm, it can be seen that for $\mbox{Im}(\epsilon)> 0$ both the complex factorization energy $\epsilon$ (with its corresponding transformation function $u_P(x,\epsilon)$) and its complex conjugate ${\overline \epsilon}$ (with ${\overline u}_P(x,\epsilon)= u_N(x,{\overline \epsilon})$) are both used for the transformation. This means that all possible non-singular transformations are already covered by using either $u_P(x,\epsilon)$ or $u_N(x,\epsilon)$ with $\epsilon$ lying on the right domain, as it is shown in figure~\ref{inv4}.

Note that the real line of the complex plane $\epsilon$ is excluded from the domain of non-singular SUSY transformations since our previous analysis is valid just for $\mbox{Im}(\epsilon)\neq 0$. Furthermore, the points $\epsilon = \pm i(m + 1/2), \ m=0,1,2,\dots$ are also excluded because at those points the solutions are reduced to the polynomials of Section 3.2 and the asymptotic behaviour either of $\psi_e$ or $\psi_o$, given by equations~\eqref{limus}, is no longer valid.

The analytic expression for the new potential $V_2$ can be found by substituting equations \eqref{gcom} and \eqref{funw} into \eqref{gammaa} in order to obtain
\begin{equation}
V_2(x)=-\frac{x^2}{2}-\left(\frac{u\overline{u}' + u'\overline{u}}{w}-\frac{(u\overline{u})^2}{w^2}\right).
\end{equation}

\begin{figure}
\begin{center}
\includegraphics[scale=0.6]{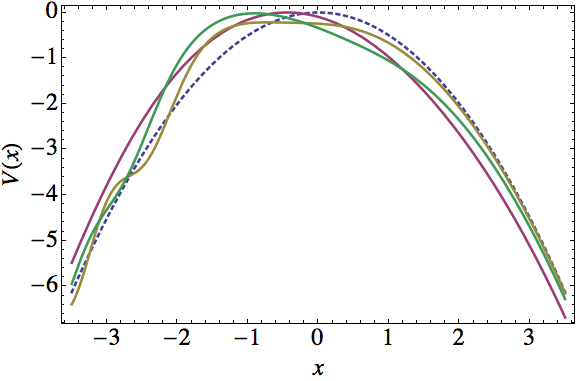}
\end{center}
\vspace{-5mm}
\caption{\small{Inverted oscillator potential (blue dashed line) and its second-order SUSY partners generated by using $u_P(x,\epsilon)$ as transformation function for different complex values of $\epsilon$, $\{ 10^{-5}+5i,(1+i)/5,10^{-2}+i \}$, which correspond to the green, magenta, and yellow lines; respectively.}} \label{inv5}
\end{figure}

In figure~\ref{inv5} we show several supersymmetric partners of the inverted oscillator potential, built through this complex algorithm. We want to remark that, in these cases, there are no new bound states at all being created by the transformations.

\subsection{Algebra of the new Hamiltonians}\label{secinv6}
In the last section we have finally obtained new potentials through SUSY QM departing from the inverted oscillator. In fact, they constitute a two-parametric family of potentials, whose parameters are $\mbox{Re}(\epsilon)$, $\mbox{Im}(\epsilon)$. We have also shown that $\epsilon$ and $\overline{\epsilon}$ induce the same transformation and, consequently, the same potential $V_2(x)$.

Remember that $H_2$ is isospectral to $H_0$ for the only second-order SUSY transformation that works, namely, for the complex case. Note that $B^{-}\equiv (B^{+})^{\dag}$, then it turns out that
\begin{subequations}
\begin{align}
B^{+}B^{-}=(H_2-\epsilon)(H_2-\overline{\epsilon}),\\
B^{-}B^{+}=(H_0-\epsilon)(H_0-\overline{\epsilon}).
\end{align}
\end{subequations}

Let us define now a pair of new operators $L_i^{\pm}$ as
\begin{equation}
L_i^{\pm}\equiv B^{+}a_i^{\pm}B^{-},
\end{equation}
which act onto the eigenfunctions of the new Hamiltonian $H_2$ as ladder operators because they satisfy the following algebra
\begin{figure}
\begin{center}
\includegraphics[scale=0.3]{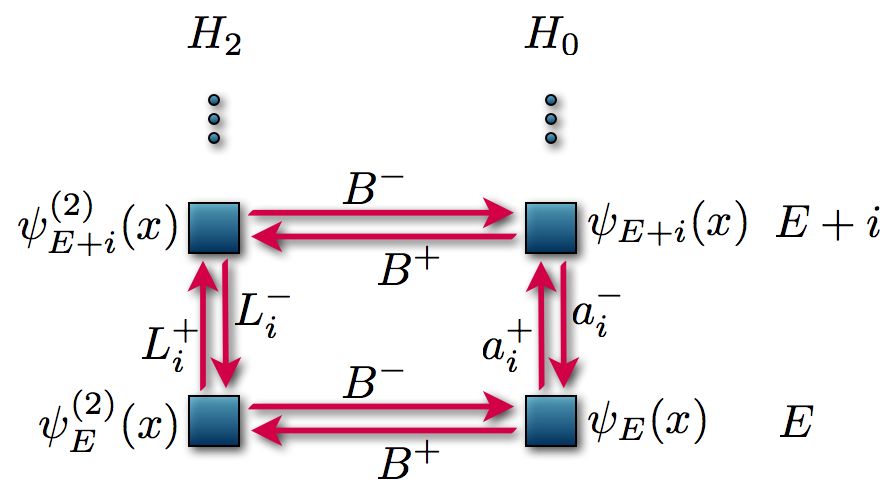}
\end{center}
\vspace{-5mm}
\caption{\small{Diagram of the complex second-order SUSY transformation. $H_0$ has ladder operators $a_i^{\pm}$, and $H_2$ has also ladder operators denoted by $L_i^{\pm}$, which are built from $B^{\pm}$ and $a_i^{\pm}$.}} \label{figsusy2}
\end{figure}
\begin{subequations}
\begin{align}
[H_2,L_i^{\pm}]&=\pm i L_i^{\pm},\\
[L_i^{-},L_i^{+}]&= N_5(H_2+i)-N_5(H_2)=P_4(H_2),
\end{align}
\end{subequations}
where $P_4,N_5$ are fourth- and fifth-order complex polynomials, respectively. They are given by the analogue of the number operator for $H_2$
\begin{equation}
L_i^{+}L_i^{-} = (H_2-\epsilon)(H_2-\overline{\epsilon})(H_2-\epsilon-i)(H_2-\overline{\epsilon}-i)(H_2-i/2) \equiv P_5(H_2).
\end{equation}
Note that the ladder operators $L_i^{\pm}$ of $H_2$ are also antihermitian, $(L_i^{\pm})^\dagger = - L_i^{\pm}$, as it happens for the inverted oscillator.

\begin{figure}
\begin{center}
\includegraphics[scale=0.6]{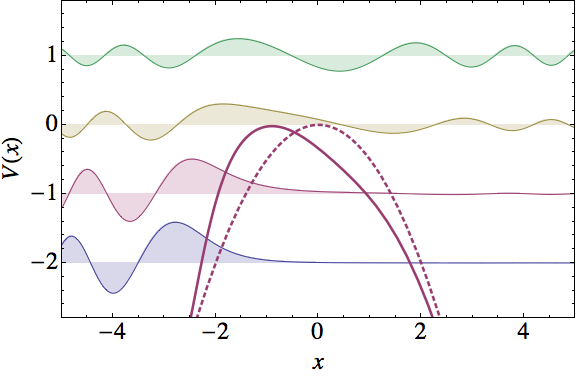}
\end{center}
\vspace{-5mm}
\caption{\small{SUSY partner potential (solid curve) of the inverted oscillator (dashed line) and four of its eigenfunctions associated with the energies $E\in\{-2,-1,0,1\}$. The factorization energy involved in this transformation is $\epsilon = 10^{-5}+5i$.}} \label{fignew}
\end{figure}

Next we can obtain the analytic expression for the eigenfunctions of the new Hamiltonian $H_2$. Using equation~\eqref{H2B2} we get \citep{FF05}
\begin{subequations}
\begin{align}
B^{+}\psi_E & =\sqrt{(E-\epsilon)(E-\overline{\epsilon})}\psi_E^{(2)}, \label{bmastimesb}\\
B^{-}\psi_E^{(2)} & =\sqrt{(E-\epsilon)(E-\overline{\epsilon})}\psi_E,
\end{align}
\end{subequations}
where $\psi_E(x)$ denotes an eigenfunction of $H_0$ associated with an arbitrary real energy $E$ (it can be the $\psi_E^\pm(x)$ of equations~\eqref{uplus} and \eqref{uminus}, the $\psi_L(x)$ of \eqref{ul}, the $\psi_R(x)$ of \eqref{ur}, or a linear combination of both) and $\psi^{(2)}_E(x)$ denotes the corresponding eigenfunction of $H_2$ (see diagram in figure~\ref{figsusy2}). Indeed, by substituting in the definition of $B^{+}$ (see equation~(\ref{2S_Bmas})) the expressions for $g(x)$ and $h(x)$ of equations \eqref{gammai} and \eqref{gcomp2}, after several simplifications we get
\begin{equation}
\psi_E^{(2)}(x) \propto \frac{w'(x)}{w(x)} \left[ - \psi_E'(x) + \frac{u'(x)}{u(x)}\psi_E(x) \right] + 2(\epsilon - E) \psi_E(x).
\end{equation}
We should recall that $u(x)$ is a complex transformation function, $\epsilon\in\mathbb{C}$ is the corresponding factorization energy, and $E\in\mathbb{R}$ is the energy associated with the initial and transformed eigenfunctions $\psi_E(x),\psi_E^{(2)}(x)$. Note that the asymptotic behaviour for $\psi_E^{(2)}(x)$ is the same as for the initial eigenfunction. In figure~\ref{fignew} a new potential $V_2(x)$ and four of its associated eigenfunctions are shown.

\subsection{Conclusions}\label{secinv7}
In this part of the thesis, SUSY QM has been successfully employed to generate new exactly-solvable potentials departing from the inverted oscillator. We have shown that the first-order, as well as the real and confluent second-order SUSY transformations always produce singular potentials. On the other hand, through the complex second-order SUSY QM it is possible to generate new real non-singular exactly-solvable potentials.

It has been shown that this transformation can be achieved through two specific complex seed solutions which are the only ones fulfilling the appropriate conditions in order to produce new non-singular real potentials isospectral to the inverted oscillator. Furthermore, we have obtained a simple analytic expression for the eigenfunctions associated with the new Hamiltonian $H_2$. Let us note that the potentials generated in this section can be used as models in every physical situation where the inverted oscillator has been employed before \citep{Bar86,GP91,Shi00,YKC06}. This is because the new Hamiltonians are isospectral to the original one and the form of their potentials is quite similar. These facts, in particular, could be important to foresee alternative models for describing the small imperfections appearing when a real Penning trap is built up \citep{BG86,CF11,CFV11}, specially if they do not change the spectrum of the corresponding ideal arrangement.

Furthermore, we have studied the general algebraic structure of the original system with arbitrary $\omega$, which is reduced to the harmonic oscillator case for $\omega=1$. We have also analyzed in detail the case with $\omega=i$, which is related to the inverted oscillator and turns out to have a deformed complex Heisenberg-Weyl algebraic structure. For the new Hamiltonians obtained through the complex second-order SUSY QM applied to the inverted oscillator, we have examined as well the related algebra in some detail.

In the future, we would like to analyze further the algebras associated with a more general system with an arbitrary $\omega$, different from $1$ and $i$, since the symmetry in the wavefunctions $\psi_n,\phi_n,\phi_n^{-},\phi_n^{+}$ of equations~\eqref{excitados}, \eqref{enesimosnofisicos}, \eqref{enesimosnofisicosinvertedmenos}, and \eqref{enesimosnofisicosinvertedmas} suggest the existence of a common structure that, perhaps, would allow us to understand deeper this family of oscillators.
\chapter{Polynomial Heisenberg algebras}\label{pha}

Lie algebras and their deformations play an important role in various problems of physics, for example, Higgs algebra \citep{Hig79} is applied to several Hamiltonians with analytic solution \citep{BDK94}. In the Lie algebras, the commutators are linear combinations of the generators; on the other hand, in their deformations some commutators are non-linear functions of the generators \citep{DGRS99}.

In this chapter we will study the polynomial Heisenberg algebras (PHA), i.e., systems for which the commutators of the Hamiltonian $H$ with the ladder operators $\mathcal{L}^{\pm}$ (also known as creation and annihilation operators) are the same as for the harmonic oscillator, but the commutator $[\mathcal{L}^{-},\mathcal{L}^{+}]$ is a polynomial $P(H)$ of $H$. Some of these algebras are constructed by taking $\mathcal{L}^{\pm}$ as $m$th-order differential operators \citep{Fer84D,DEK92,SRK97,FH99,ACIN00}.

Furthermore, it is important to study not only these specific algebras, but also the characterization of the general systems ruled by them. We will see in this chapter that the difficulties in the study of this problem grow with the order $m$ of the polynomial: for zeroth- and first-order PHA, the systems become the harmonic and the radial oscillators, respectively \citep{Fer84D,DEK92,Adl93,SRK97}. On the other hand, for second- and third-order PHA, the determination of the potentials is reduced to find solutions of Painlev\'e IV and V equations, denoted as $P_{IV}$ and $P_{V}$, respectively \citep{Adl93,WH03}.

This means that, in order to have a system described by these PHA, we need solutions of $P_{IV}$ and $P_V$. Nevertheless, in this thesis we will use this connection but for the inverse problem, i.e., first we look for systems which are certainly described by PHA and then we develop a method to find solutions of the Painlev\'e equations.

The structure of this chapter is the following: first, in section \ref{secpha1} we will introduce the definition of the Heisenberg-Weyl algebra. Then, in section \ref{secpha2} we will study their polynomial deformations or PHA. Finally, in section \ref{secpha3} we will study the general systems described with PHA from zeroth- up to third-order.

\section{Heisenberg-Weyl algebras}
\label{secpha1}
Let $R$ be a commutative ring and $M$ a module over $R$ generated freely by two sets ${Q_i}$, ${P_i}$ and an element $c$, with $i \in I$, where $I$ is the set of indices. We define the product $[\cdot,\cdot]:M\times M \rightarrow M$ as a bilinear extension such that
\begin{subequations}
\begin{align}
[c,c]=[Q_i,c]=[P_i,c]&=0,\\
[Q_i,Q_j]=[P_i,P_j]&=0,\quad \forall \  i,j \in I,\\
[Q_i,P_j]&=0, \quad \forall \ i\neq j \in I,\\
[Q_i,P_i]&=c, \quad \forall \ i \in I.
\end{align}
\end{subequations}
A Heisenberg-Weyl algebra usually takes a standard definition of the $[\cdot,\cdot]$ operation and therefore it is usually defined only by the operators $Q_i,P_j$ and sometimes by the number $c$, which is the scale. Nevertheless, there are some special cases where $c$ is not an element of the domain, but rather a polynomial of the operators. In that case we talk about polynomial deformation of the Heisenberg-Weyl algebra. Additional information can be found in \citet*{Per86}.

\section{Polynomial Heisenberg algebras}
\label{secpha2}
A polynomial Heisenberg algebra can be defined by two typical commutation relations
\begin{equation}
[H,\mathcal{L}_m^{\pm}]=\pm \mathcal{L}_m^{\pm}, \label{pha1}
\end{equation}
and one atypical relation that characterizes the deformation
\begin{equation}
[\mathcal{L}_m^{-},\mathcal{L}_m^{+}]\equiv N_m(H+1) - N_m(H)= P_{m-1} (H),\label{pha2}
\end{equation}
where $\mathcal{L}_m^{\pm}$ are $m$th-order differential ladder operators, $P_{m-1}(H)$ is a $(m-1)$th-order polynomial of $H$ and $N_m(H)\equiv \mathcal{L}_m^{+}\mathcal{L}_m^{-}$ is a $m$th-order polynomial in $H$ which is analogous to the number operator of the harmonic oscillator and is factorized as
\begin{equation}
N_m(H)=\prod_{i=1}^{m}(H-\mathcal{E}_{i}),\label{facN}
\end{equation}
where $\mathcal{E}_i$ are the energies associated with the {\it extremal states}. These algebras have $m$th-order differential ladder operators, nevertheless, the polynomial $P_{m-1}(H)$ in the commutator that characterizes the deformation (see equation \eqref{pha2}) is of $(m-1)$th-order. Then we will say that this is a polynomial Heisenberg algebra (PHA) of $(m-1)$th-order.

These deformed algebras were developed in recent years. We can trace them back to studies of the SUSY partner potentials of the harmonic oscillator in order to obtain a generalized definition of coherent states by \citet*{FHN94} and \citet*{FNR95}. Soon after, \citet{AS97} obtained explicitly the first-order PHA as a deformation of the Abraham-Moses-Mielnik potentials \citep{Mie84}. Nevertheless, in that work they are considered as Lie algebras of infinite dimension. Then, \citet{FH99} finally proposed a general definition but the PHA were not studied in detail until the seminal work by \citet*{CFNN04}, where the explicit study of the general systems described by these algebras was obtained.

The corresponding systems are described by the Hamiltonian
\begin{equation}
H=-\frac{1}{2}\frac{\text{d}^2}{\text{d}x^2} + V(x),\label{pha3}
\end{equation}
i.e., we are using natural units as in chapter \ref{capsusyqm}. The algebraic structure generated by $\{H,\mathcal{L}_m^{-},\mathcal{L}_m^{+}\}$ provides information about the spectrum of $H$, $\text{Sp}(H)$ \citep{DEK92,FH99,ACIN00}. In fact, let us consider the $m$th-dimensional solution space of the $m$th-order differential equation $\mathcal{L}_m^{-}\psi=0$, called the {\it kernel} of $\mathcal{L}_m^{-}$ and denoted as $\mathcal{K}_{\mathcal{L}_m^{-}}$. Then
\begin{equation}
\mathcal{L}_m^{+}\mathcal{L}_m^{-}\psi = \prod_{i=1}^{m}(H-\mathcal{E}_i )\psi=0.
\end{equation}
Since $\mathcal{K}_{\mathcal{L}_m^{-}}$ is invariant with respect to $H$, then it is natural to select the eigenfunctions of $H$ as basis for the solution space, i.e.,
\begin{equation}
H\psi_{\mathcal{E}_i}=\mathcal{E}_i \psi_{\mathcal{E}_i}.
\end{equation}

\begin{figure}\centering
\includegraphics[scale=0.3]{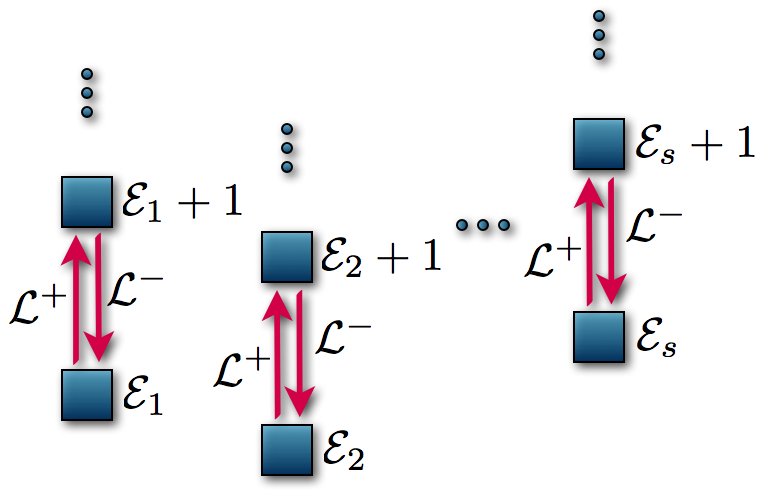}
\caption{\small{Spectrum for a Hamiltonian with $s$ physical extremal states. In general, each one of them has associated one infinite ladder.}}\label{fig.pha1a}
\end{figure}

Therefore, $\psi_{\mathcal{E}_i}$ are the extremal states of $m$ mathematical ladders with spacing $\Delta E=1$ that start from $\mathcal{E}_i$. Let $s$ be the number of those states with physical significance, $\{ \psi_{\mathcal{E}_i} ;i=1,\dots ,s \}$; then, operating iteratively with $\mathcal{L}_m^{+}$ we can construct $s$ physical energy ladders, as it is shown in figure \ref{fig.pha1a}.

\begin{figure}\centering
\includegraphics[scale=0.3]{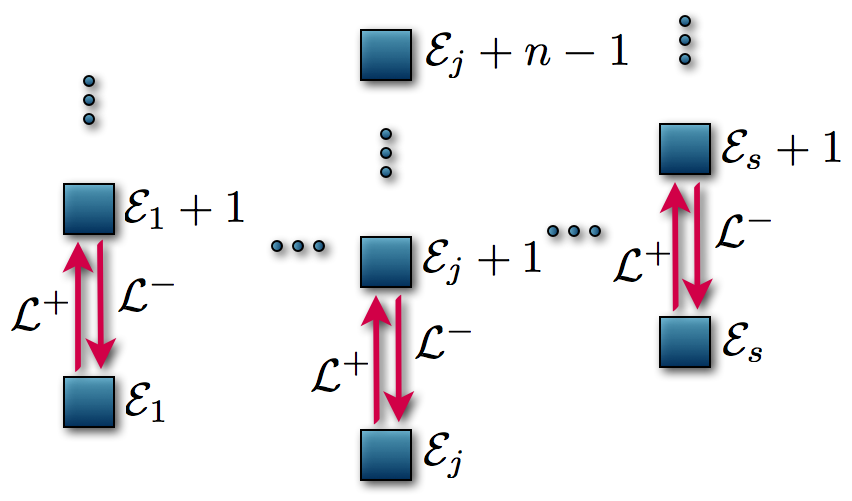}
\caption{\small{Spectrum of a Hamiltonian with $s$ physical extremal states, where $\psi_{\mathcal{E}_j}$ fulfill condition \eqref{finita} and therefore the system has $s-1$ infinite and one finite (the $j$-th) ladders.}}\label{fig.pha1b}
\end{figure}

It is possible that for a ladder starting from $\mathcal{E}_j$ there exists an integer $n\in N$ such that
\begin{subequations}
\begin{align}
(\mathcal{L}_m^{+})^{n-1}\psi_{\mathcal{E}_j} & \neq 0,\\
(\mathcal{L}_m^{+})^{n}\psi_{\mathcal{E}_j} & = 0.
\end{align}\label{finita}
\end{subequations}
Then, if we analize $\mathcal{L}_m^{-}(\mathcal{L}_m^{+})^{n}\psi_{\mathcal{E}_j} = 0$ we can see that other roots of equation~\eqref{facN} must fulfill $\mathcal{E}_k = \mathcal{E}_j +n$, where $k\in \{s+1,\dots ,m\}$ and $\ j\in \{1,\dots,s \}$. Therefore, Sp$(H)$ contains $s-1$ infinite ladders and a finite one with length $n$, that starts in $\mathcal{E}_j$ and finish in $\mathcal{E}_j +n-1$, as it is shown in figure \ref{fig.pha1b}.

We conclude that the spectrum of systems described by an $(m-1)$th-order PHA can have at most $m$ infinite ladders. Note that the ladder operators of the harmonic oscillator $a^\pm$ together with the Hamiltonian satisfy equations (\ref{pha1}--\ref{facN}). Moreover, a higher-order algebra with odd $m$ can be constructed simply by taking $\mathcal{L}_m^{-}=a^{-}P(H)$, $\mathcal{L}_m^{+}=P(H)a^{+}$, where $a^{+}$ and $a^{-}$ are the standard creation and annihilation operators, while $P(H)$ is a real polynomial of $H$ \citep{DGRS99}. These deformations are called \emph{reducible}, and in this context they are artificial because for our system we have already operators $a^{+}$ and $a^{-}$ that fulfill an algebra of lower-order.

We will show in chapters \ref{cappain} and \ref{5painleve} that some ladder operators of order larger than three (four) can also be factorized as $L^{+}=P(H)l^{+}$ ($L^{+}=P(H)\ell^{+})$, where $l^{+}$ ($\ell^{+}$) is a differential ladder operator of third- (fourth-) order which ultimately leads to solutions of Painlev\'e IV (Painlev\'e V) equation.

\section{General systems with PHA}
\label{secpha3}

Let us determine the general systems described by the PHA studied in section \ref{secpha2}. Since for $m\geq 4$ the calculations are quite involved, we will only analyze the cases for $m = 0,1,2,3$. The results of this section are summarized in table~\ref{tablepha}.

\begin{table}
\begin{center}
\begin{tabular}{ccl}
\hline
PHA ($m$) & Ladder operators & System\\
\hline
0th-order&1st-order&Harmonic oscillator (HO)\\
1st-order&2nd-order&Radial oscillator (RO)\\
2nd-order&3rd-order&Connected with $P_{IV}$\\
3rd-order&4th-order&Connected with $P_{V}$\\
\hline
\end{tabular}
\end{center}
\vspace{-3mm}
\caption{\small{The first four PHA and their systems.}} \label{tablepha}
\end{table}

\subsection{Zeroth-order PHA. First-order ladder operators.}
Let us look for the general Hamiltonian $H$ and the first-order ladder operators $\mathcal{L}_1^{\pm}$, defined by
\begin{equation}
\mathcal{L}_1^+=\frac{1}{2^{1/2}}\left[-\frac{\text{d}}{\text{d}x}+f(x)\right],\quad \mathcal{L}_1^-=(\mathcal{L}_1^+)^\dag,
\end{equation}
that satisfy equation \eqref{pha1}. Thus, a system involving $V$, $f$, and their derivatives is obtained
\begin{equation}
f' - 1 = 0,\quad V' - f = 0.
\end{equation}

Up to coordinate and energy displacements, it turns out that $f(x) = x$ and $V(x) = x^2/2$. This potential has one equidistant infinite ladder starting from the extremal state $\psi_{\mathcal{E}_1}=\pi^{-1/4} \exp(-x^2/2)$, which is a normalized eigenfunction of $H$ with eigenvalue $\mathcal{E}_1 = 1/2$ annihilated by $\mathcal{L}_1^-$. Here, the number operator is linear in $H$, $N_1(H) = H - \mathcal{E}_1$, i.e., the general system obeying the zeroth-order PHA of section \ref{secpha2} is the harmonic oscillator. The natural ladder operators of the system are the first-order creation and annihilation operators, i.e., $\mathcal{L}_1^\pm=a^\pm$. This algebra is the Heisenberg-Weyl algebra which has been widely studied.

\subsection{First-order PHA. Second-order ladder operators.}
Let us suppose now that
\begin{equation}
\mathcal{L}_2^+ = \frac{1}{2}\left[\frac{\text{d}^2}{\text{d}x^2} + g(x)\frac{\text{d}}{\text{d}x} + h(x)\right],\quad \mathcal{L}_2^- = (\mathcal{L}_2^+)^\dag.
\end{equation}
Then, equation \eqref{pha1} leads to a system of equations for $V$, $g$, $h$, and their derivatives
\begin{subequations}
\begin{align}
g' + 1 &= 0,\\
h' + 2V' + g &= 0,\\
h'' +2V'' +2gV' +2h&=0.
\end{align}
\end{subequations}
The general solution (up to coordinate and energy displacements) is given by
\begin{subequations}
\begin{align}
g(x)&=-x,\\
h(x)&= \frac{x^2}{4} - \frac{\gamma}{x^2} - \frac{1}{2},\\
V(x)&= \frac{x^2}{8} +\frac{\gamma}{2x^2},\label{m1sols3}
\end{align}\label{m1sols}
\end{subequations}
\hspace{-2mm}where $\gamma$ is a constant of integration. The potential from equation~\eqref{m1sols3} have two equidistant energy ladders (not necessarily physical) generated by acting with powers of $\mathcal{L}_2^+$ on the two extremal states
\begin{equation}
\psi_{\mathcal{E}_1}\propto x^{1/2+\sqrt{\gamma+1/4}} \exp\left(-\frac{x^2}{4}\right),\quad
\psi_{\mathcal{E}_2}\propto x^{1/2-\sqrt{\gamma+1/4}} \exp\left(-\frac{x^2}{4}\right).
\end{equation}

Let us recall that $\mathcal{L}_2^- \psi_{\mathcal{E}_j}=0=(H-\mathcal{E}_j)\psi_{\mathcal{E}_j}$, where
\begin{equation}
\mathcal{E}_1=\frac{1}{2}+\frac{1}{2}\sqrt{\gamma+\frac{1}{4}},\quad
\mathcal{E}_2=\frac{1}{2}-\frac{1}{2}\sqrt{\gamma+\frac{1}{4}}.
\end{equation}
Now $N_2(H)$ is quadratic in $H$, i.e., $N_2(H)=(H-\mathcal{E}_1)(H-\mathcal{E}_2)$. The potentials can be expressed as
\begin{equation}
V(x)= \frac{x^2}{8} + \frac{\ell(\ell+1)}{2x^2},\quad x>0, \quad \ell\geq 0,
\end{equation}
that are obtained by making $\gamma = \ell(\ell+1),$ $\ell \geq 0$. Thus, the general systems having second-order ladder operators are described by the radial oscillator potentials. The natural ladder operators of the first-order PHA are the second-order ones of the radial oscillator, i.e., $\mathcal{L}_2^\pm=b_\ell^\pm \equiv b^\pm$. This is the $\mathcal{SO}(2,1)$ algebra.

\subsection{Second-order PHA. Third-order ladder operators.}\label{secondPHA}
\begin{figure}
\begin{center}
\includegraphics[scale=0.32]{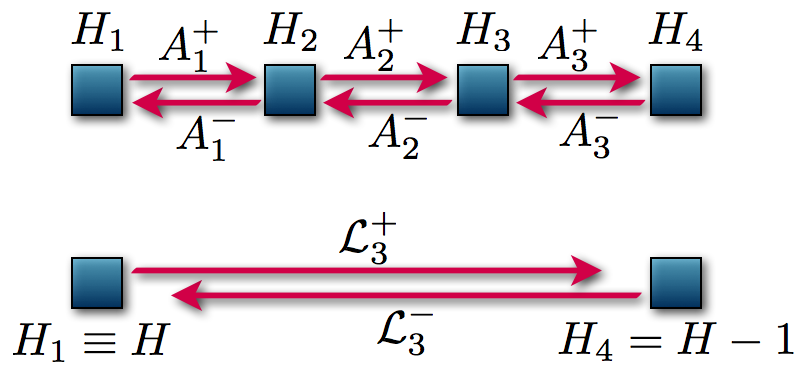}
\end{center}
\vspace{-5mm}
\caption{\small{Diagram of the two equivalent SUSY transformations. Above: the three-step first-order operators $A_1^{\pm}$, $A_2^{\pm}$, and $A_3^{\pm}$ allow to accomplish the transformation. Below: the direct transformation achieved through the third-order operators $\mathcal{L}_3^{\pm}$.}}
\label{diasusy}
\end{figure}

In this case, both $\mathcal{L}_3^{\pm}$ will be third-order differential ladder operators. Now, we propose a closed-chain of three SUSY transformations \citep{VS93,Adl94,DEK94,ACIN00,Gra04,Mar09b} so that $\mathcal{L}_3^{\pm}$ are expressed as
\begin{subequations}
\begin{align}
\mathcal{L}_3^{+}&=A_{3}^{+}A_{2}^{+}A_{1}^{+}=
\frac{1}{2^{3/2}}\left(\frac{\text{d}}{\text{d}x}-f_3\right)\left(\frac{\text{d}}{\text{d}x}-f_2\right)\left(\frac{\text{d}}{\text{d}x}-f_1\right),\\
\mathcal{L}_3^{-}&=A_{1}^{-}A_{2}^{-}A_{3}^{-}=
\frac{1}{2^{3/2}}\left(-\frac{\text{d}}{\text{d}x}-f_1\right)\left(-\frac{\text{d}}{\text{d}x}-f_2\right)\left(-\frac{\text{d}}{\text{d}x}-f_3\right)\label{lmenos}.
\end{align}\label{eles}
\end{subequations}
\hspace{-2mm}In general, $(\mathcal{L}_3^{-})^{\dag}\neq \mathcal{L}_3^{+}$, except in the case where all $f_j \in \mathbb{R}$. The pair $A_j^\pm$ fulfill two intertwining relations of kind
\begin{equation}
H_{j+1}A^{+}_j =A^{+}_{j}H_{j},\quad
H_{j}A^{-}_j =A^{-}_{j}H_{j+1},\label{fac2}
\end{equation}
where $j=1,2,3$. In figure~\ref{diasusy} we present a diagram of the intertwining relations.

If we equate the two different factorizations associated with each $H_i$ in equation \eqref{fac2} which lead to the same Hamiltonians, we get
\begin{subequations}
\begin{align}
H_{1}&=A_{1}^{-}A_{1}^{+}+\epsilon_{1},\\
H_{2}&=A_{1}^{+}A_1^{-}+\epsilon_1=A_{2}^{-}A_{2}^{+}+\epsilon_{2},\\
H_{3}&=A_{2}^{+}A_2^{-}+\epsilon_2=A_3^{-}A_{3}^{+}+\epsilon_{3},\\
H_{4}&=A_{3}^{+}A_3^{-}+\epsilon_3.
\end{align}\label{facs2pha}
\end{subequations}
In addition, the closure condition is given by
\begin{equation}
H_4=H_1-1\equiv H-1.\label{clocon}
\end{equation}
By making the corresponding operator products we get the following system of equations \citep{VS93,Adl94,MN08}
\begin{subequations}
\begin{align}
f_1'+f_2'&=f_1^2-f_2^2+2(\epsilon_1-\epsilon_2),\label{f1}\\
f_2'+f_3'&=f_2^2-f_3^2+2(\epsilon_2-\epsilon_3),\label{f2}\\
f_3'+f_1'&=f_3^2-f_1^2+2(\epsilon_3-\epsilon_1+1).\label{f3}
\end{align}
\end{subequations}

Eliminating $f_2^2$ from equations \eqref{f1} and \eqref{f2} we get
\begin{equation}
f_1'+2f_2'+f_3'=f_1^2-f_3^2+2(\epsilon_1-\epsilon_3),
\end{equation}
and from here we substitute $f_3^2$ from equation \eqref{f3} to obtain
\begin{equation}
f_1'+f_2'+f_3'=1,
\end{equation}
which, after integration becomes
\begin{equation}
f_1+f_2+f_3=x.\label{f4}
\end{equation}
Now, substituting equation~\eqref{f4} into \eqref{f1} to eliminate $f_2$
\begin{equation}
f_1=\frac{x-f_3}{2}+\frac{1-f_3'}{2(x-f_3)}-\frac{\epsilon_1-\epsilon_2}{x-f_3}.
\end{equation}
Let us define now a useful new function as $g \equiv f_3-x$, from which we get
\begin{equation}
f_1=-\frac{g}{2}+\frac{g'}{2g}+\frac{\epsilon_1-\epsilon_2}{g}.\label{f1b}
\end{equation}
Similarly, by plugging equation~\eqref{f4} into \eqref{f1} to eliminate $f_1$ and using $g$ it turns out that
\begin{equation}
f_2=-\frac{g}{2}-\frac{g'}{2g}-\frac{\epsilon_1-\epsilon_2}{g}.\label{f2b}
\end{equation}
Now that we have $f_1,f_2,f_3$ in terms of $g$, we replace them in equation~\eqref{f3} in order to obtain
\begin{equation}
gg'' = \frac{1}{2}(g')^2 + \frac{3}{2}g^4 + 4g^3x+ 2g^2\left(x^2-\epsilon_1-\epsilon_2+2\epsilon_3+1\right)-2(\epsilon_1-\epsilon_2)^2,\label{PIV}
\end{equation}
which is the Painlev\'e IV equation ($P_{IV}$) \citep{IKSY91,VS93,Adl94,BCH95,ACIN00,WH03} (compare with \eqref{PIVlib}). The standard notation is
\begin{equation}
gg'' = \frac{1}{2}(g')^2 + \frac{3}{2}g^4 + 4g^3x+ 2g^2\left(x^2-a\right)+b,
\end{equation}
with parameters
\begin{equation}
a=\epsilon_1+\epsilon_2-2\epsilon_3 -1,\quad b=-2(\epsilon_1-\epsilon_2)^2.\label{abe}
\end{equation}
Since, in general $f\in\mathbb{C}$ then $g\in\mathbb{C}$. In addition, $\epsilon_i\in\mathbb{C}$ which implies that $a,\, b\in\mathbb{C}$ and therefore $g$ is a complex solution to $P_{IV}$ associated with the complex parameters $a,\, b$.

With the solution of $g(x)$ one can find the new potential $V(x)$ as
\begin{equation}
V(x)=\frac{x^2}{2}-\frac{g'}{2}+\frac{g^2}{2}+xg+\epsilon_3+\frac{1}{2}.\label{VPIV}
\end{equation}
Until now we have used some auxiliary energies $\epsilon_j$ to prove that the closure relation of the second-order PHA leads to systems ruled by $P_{IV}$. Now, the energies of the extremal states are defined as the roots of the generalized number operator, which is cubic in this case
\begin{equation}
N_3(H)=(H-\mathcal{E}_1)(H-\mathcal{E}_2)(H-\mathcal{E}_3).\label{q3}
\end{equation}
Using the definitions from equations~(\ref{eles}--\ref{clocon}) it turns out that
\begin{equation}
\mathcal{E}_1=\epsilon_1+1,\quad\mathcal{E}_2=\epsilon_2+1,\quad \mathcal{E}_3=\epsilon_3+1.
\end{equation}

As can be seen, if one solution $g(x)$ of $P_{IV}$ is obtained for certain values of ${\cal E}_1, \ {\cal E}_2, \ {\cal E}_3$, then the potential $V(x)$ as well as the corresponding ladder operators $\mathcal{L}_3^\pm$ are completely determined, see equations \eqref{f1b}, \eqref{f2b} and \eqref{VPIV}. Moreover, the three extremal states, some of which could have physical interpretation, are obtained from
\begin{equation}
\mathcal{L}_3^{-} \psi_{\mathcal{E}_j}=(H-\mathcal{E}_j)\psi_{\mathcal{E}_j}=0, \quad j=1,2,3,
\end{equation}
which leads to the following expressions
\begin{subequations}
\begin{align}
 \psi_{{\cal E}_1} &\propto \left( \frac{g'}{2g} - \frac{g}{2} - \frac{1}{g}\sqrt{-\frac{b}{2}} - x\right)
\exp\left[\int\left( \frac{g'}{2g} + \frac{g}{2} - \frac{1}{g}\sqrt{-\frac{b}{2}} \right) \text{d}x \right], \label{exes1} \\
\psi_{{\cal E}_2} &\propto \left( \frac{g'}{2g} - \frac{g}{2} + \frac{1}{g}\sqrt{-\frac{b}{2}} - x\right)
\exp\left[\int\left( \frac{g'}{2g} + \frac{g}{2} + \frac{1}{g}\sqrt{-\frac{b}{2}} \right) \text{d}x \right], \label{exes2} \\
\psi_{{\cal E}_3} &\propto \exp\left( - \frac{x^2}{2} - \int g\, \text{d}x\right). \label{exes3}
\end{align}\label{exes}
\end{subequations}
\hspace{-1.8mm}The corresponding physical ladders of our system are obtained departing from the extremal states with physical meaning. In this way we can determine the spectrum of the Hamiltonian $H$.

On the other hand, if we have identified a system with third-order differential ladder operators, it is possible to design a mechanism for obtaining solutions of the Painlev\'e IV equation. The key point of this procedure is to identify the extremal states of our system; then, from equation~\eqref{exes3} it is straightforward to see that
\begin{equation}
g(x) = - x - \{\ln[\psi_{{\cal E}_3}(x)]\}'.\label{exes3des}
\end{equation}
Notice that, by making cyclic permutations of the indices of the initially assigned extremal states $\psi_{{\cal E}_1},\, \psi_{{\cal E}_2},\, \psi_{{\cal E}_3}$, we will obtain three solutions of $P_{IV}$ with different parameters $a,b$.

Hence, we have found a recipe for building systems ruled by second-order PHA defined by equations~(\ref{pha1}--\ref{facN}): first find a function $g(x)$ that solves $P_{IV}$ from equation~\eqref{PIV}; then calculate the potential using equation~\eqref{VPIV}, and its three ladders from the extremal states given by equations~\eqref{exes}. In order to test the effectiveness of this recipe, let us analyze some systems associated with particular solutions $g(x)$ of $P_{IV}$.

\subsubsection{Harmonic oscillator}
Let us consider the following $P_{IV}$ solution of equation~\eqref{PIV}:
\begin{equation}
g(x)=-x-\alpha(x),
\end{equation}
where $\mathcal{E}_1=\mathcal{E}_3$, $\alpha(x)=u'/u$ satisfies the Riccati equation
\begin{equation}
\alpha'(x)+\alpha^2(x)=x^2-2\epsilon,\label{ricc}
\end{equation}
with $\epsilon=\mathcal{E}_3-\mathcal{E}_2+1/2$, and $u(x)$ is the Schr\"odinger solution given by
\begin{equation}
u(x,\epsilon)=\text{e}^{-x^2/2}\left[_1F_1\left(\frac{1-2\epsilon}{4},\frac{1}{2};x^2\right)
+2x\nu\frac{\Gamma\left(\frac{3-2\epsilon}{4}\right)}{\Gamma\left(\frac{1-2\epsilon}{4}\right)}
\ _1F_1\left(\frac{3-2\epsilon}{4},\frac{3}{2};x^2\right)\right],\label{usch}
\end{equation}
where $\nu\in \mathbb{R}$, $|\nu |<1$. This function $g(x)$ substituted in \eqref{VPIV} gives the new potential
\begin{equation}
V(x)=\frac{x^2}{2}+\mathcal{E}_2-\frac{1}{2},
\end{equation}
which is the harmonic oscillator potential displaced in energy. The three extremal states from equations~\eqref{exes} become now
\begin{equation}
\psi_{\mathcal{E}_1}=0,\quad \psi_{\mathcal{E}_2}\propto \exp\left(-\frac{x^2}{2}\right), \quad  \psi_{\mathcal{E}_3}\propto u(x).
\end{equation}
We see that the only physical ladder is the one generated from $\psi_{\mathcal{E}_2}$. Here, we have a case where the deformed algebra is reducible in the sense explained at the end of section \ref{secpha2}. In fact, it is easy to show that $\mathcal{L}_3^{-}=a^{-}(H-\mathcal{E}_1)$.

\subsubsection{First-order SUSY partners of the harmonic oscillator}\label{2pha1}
They arise for $g(x)$ taking the form
\begin{equation}
g(x)=-x+\alpha(x),
\end{equation}
where $\alpha=u'/u$ satisfies equation~\eqref{ricc}, but now $\mathcal{E}_1=\mathcal{E}_3+1$, $\epsilon=\mathcal{E}_3-\mathcal{E}_2+1/2$, and $u(x)$ is again the Schr\"odinger solution given in equation~\eqref{usch}. This $g(x)$ leads to the exactly solvable potentials
\begin{equation}
V(x)=\frac{x^2}{2}-\alpha'(x)+\mathcal{E}_2-\frac{1}{2},
\end{equation}
which are the 1-SUSY partners of the harmonic oscillator. The extremal states become
\begin{equation}
\psi_{\mathcal{E}_1}\propto A^+ a^+ u(x),\quad \psi_{\mathcal{E}_2}\propto A^+\exp\left(-\frac{x^2}{2}\right),\quad \psi_{\mathcal{E}_3}\propto \frac{1}{u(x)},
\end{equation}
where $A^+$ is the first-order intertwining operator defined in equation~\eqref{A1}.

\subsubsection{$k$-SUSY partners of the harmonic oscillator}
Some years ago it was conjectured a method to obtain second-order PHA from the $k$th-order SUSY partners of the harmonic oscillator, which usually have $2k$th-order PHA with $(2k+1)$th-order ladder operators \citep{FNN04}. Nevertheless, this conjecture was proven just recently \citep{BF11a}, as a matter of fact, as part of the work of this thesis. Moreover, the corresponding systems will be connected with $P_{IV}$ and they will give us new solutions of this equation. This process will be explained in detail in chapter \ref{cappain}, but in any case, we will describe the systems here for completeness.

The process consists in taking $k$ transformation functions $u_j$ of a non-physical ladder, i.e.,
\begin{equation}
Hu_j=\epsilon_ju_j,
\end{equation}
but with the condition that they are connected through the operator $a^-$, as
\begin{equation}
u_{j+1}=a^{-}u_j,\quad \epsilon_{j+1}=\epsilon_1-j<\frac{1}{2},\quad j=1,\dots , k-1,
\end{equation}
with $u_1$ being a solution of equation \eqref{usch}. With this choice, the $k-1$ factorization energies $\epsilon_1,\dots , \epsilon_{k-1}$ appear twice in the analogue of the number operator for the $k$-SUSY partners
\begin{equation}
N_{2k+1}(H)=\left(H-\frac{1}{2}\right)\prod_{j=1}^{k}(H-\epsilon_j-1)(H-\epsilon_j),
\end{equation}
which implies that the natural $(2k+1)$th-order ladder operator $L_k^{+}$ of the $k$-SUSY partner can be written as
\begin{equation}
L_k^{+}=\left(\prod_{j=1}^{k-1}(H-\epsilon_j)\right)l_k^{+},
\end{equation}
where $l_k^{+}$ is a third-order differential ladder operator \citep{FNN04,BF11a}. We say that the algebra is {\it reducible} from $2k$th- to second-order. This fact will be proven in chapter~\ref{cappain} as a theorem, where we will write down specifically the necessary conditions to perform this reduction. The operator $l_k^{+}$ turns out to be essential to obtain the solutions of $P_{IV}$ in chapter \ref{cappain} and also to derive several sets of coherent states for related systems in chapter \ref{p4cs}. Indeed, the extremal states are now
\begin{equation}
\psi_{\mathcal{E}_1}\propto B^+ a^+ u_1(x),\quad
\psi_{\mathcal{E}_2}\propto B^+\exp\left(-\frac{x^2}{2}\right),\quad
\psi_{\mathcal{E}_3}\propto \frac{W(u_1,\dots ,u_{k-1})}{W(u_1,\dots ,u_{k})},\label{psisksusy}
\end{equation}
where $B^{+}$ is the $k$th-order intertwining operator of equation~\eqref{Bdag}, $\alpha(x)=u_k'/u_k$ satisfies equation~\eqref{ricc} but with energies $\mathcal{E}_1=\mathcal{E}_3+k$, $\epsilon=\mathcal{E}_3-\mathcal{E}_2+1/2$. The expression for $\psi_{\mathcal{E}_3}$ will be justified in appendix B. By comparing equations \eqref{exes3des} and \eqref{psisksusy}, we can see that the solution $g(x)$ of $P_{IV}$ is
\begin{equation}
g(x)=-x-[ \ln W(u_1,\dots ,u_{k-1})]'+[\ln W(u_1,\dots ,u_k)]',
\end{equation}
and the corresponding potential becomes
\begin{equation}
V(x)=\frac{x^2}{2}-[\ln W(u_1,\dots ,u_k)]''+\mathcal{E}_2-\frac{1}{2}.
\end{equation}

\subsection{Third-order PHA. Fourth-order ladder operators.}
\begin{figure}
\begin{center}
\includegraphics[scale=0.3]{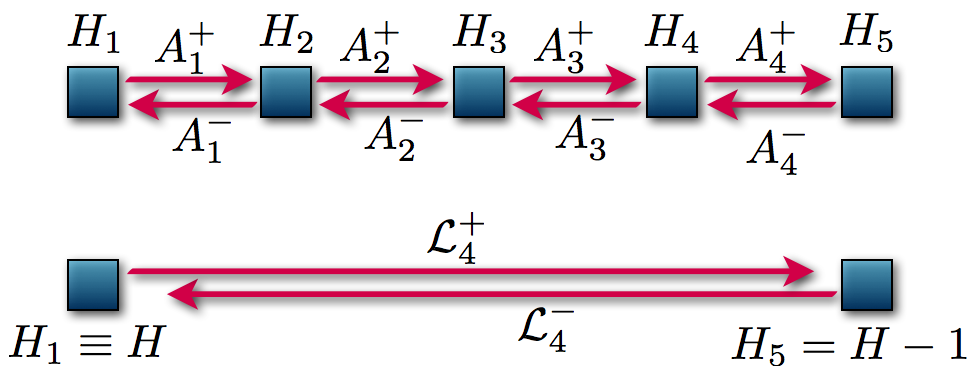}
\end{center}
\vspace{-5mm}
\caption{\small{Diagram representing the two equivalent SUSY transformations. Above: the four-step first-order operators $A_1^{\pm}$, $A_2^{\pm}$, $A_3^{\pm}$ and $A_4^\pm$. Below: the direct transformation achieved through the fourth-order operators $\mathcal{L}_4^{\pm}$.}}
\label{diasusyRO}
\end{figure}

In this case $\mathcal{L}_4^{\pm}$ will be fourth-order ladder operators. We propose a closed-chain as follows \citep{Adl94}
\begin{subequations}
\begin{align}
\mathcal{L}_4^{+}&=A_4^+A_{3}^{+}A_{2}^{+}A_{1}^{+}=
\frac{1}{2^{2}}\left(\frac{\text{d}}{\text{d}x}-f_4\right)\left(\frac{\text{d}}{\text{d}x}-f_3\right)\left(\frac{\text{d}}{\text{d}x}-f_2\right)\left(\frac{\text{d}}{\text{d}x}-f_1\right),\\
\mathcal{L}_4^{-}&=A_{1}^{-}A_{2}^{-}A_{3}^{-}A_4^-=
\frac{1}{2^{2}}\left(-\frac{\text{d}}{\text{d}x}-f_1\right)\left(-\frac{\text{d}}{\text{d}x}-f_2\right)\left(-\frac{\text{d}}{\text{d}x}-f_3\right)\left(-\frac{\text{d}}{\text{d}x}-f_4\right)\label{lmenosRO}.
\end{align}\label{elesRO}
\end{subequations}
\hspace{-1mm}Each pair of operators $A_j^-$, $A_j^+$ intertwines two Hamiltonians $H_j$ and $H_{j+1}$ in the way
\begin{equation}
H_{j+1}A^{+}_j =A^{+}_{j}H_{j},\quad
H_{j}A^{-}_j =A^{-}_{j}H_{j+1},
\end{equation}
where $j=1,2,3,4$. This leads to the following factorizations of the Hamiltonians
\begin{subequations}
\begin{align}
H_{1}&=A_{1}^{-}A_{1}^{+}+\epsilon_{1},\\
H_{2}&=A_{1}^{+}A_1^{-}+\epsilon_1=A_{2}^{-}A_{2}^{+}+\epsilon_{2},\\
H_{3}&=A_{2}^{+}A_2^{-}+\epsilon_2=A_3^{-}A_{3}^{+}+\epsilon_{3},\\
H_{4}&=A_{3}^{+}A_3^{-}+\epsilon_3=A_4^{-}A_{4}^{+}+\epsilon_{4},\\
H_{5}&=A_{4}^{+}A_4^{-}+\epsilon_4.
\end{align}
\end{subequations}
To accomplish the closed-chain we need the closure condition given by
\begin{equation}
H_5=H_1-1\equiv H-1.\label{cloconRO}
\end{equation}
In figure~\ref{diasusyRO} we show a diagram representing the transformation and the closure relation.

By making the corresponding operator products we obtain the following systems of equations
\begin{subequations}
\begin{align}
f_1'+f_2'&=f_1^2-f_2^2+2(\epsilon_1-\epsilon_2),\label{f1RO}\\
f_2'+f_3'&=f_2^2-f_3^2+2(\epsilon_2-\epsilon_3),\label{f2RO}\\
f_3'+f_4'&=f_3^2-f_4^2+2(\epsilon_3-\epsilon_4),\label{f3RO}\\
f_4'+f_1'&=f_4^2-f_1^2+2(\epsilon_4-\epsilon_1+1).\label{f4RO}
\end{align}\label{efesRO}
\end{subequations}

Up to now, the method employed for this case is very similar to the one for the second-order PHA, and indeed can be taken as its generalization. Nevertheless, the similarity ends now, because the technique used to solve the system of equations~\eqref{efesRO} is really different. In fact, this system which leads to $P_V$ is much more complicated to solve than the one leading to $P_{IV}$.

Let us simplify the notation making $\alpha_1=\epsilon_1-\epsilon_2$, $\alpha_2=\epsilon_2-\epsilon_3$, $\alpha_3=\epsilon_3-\epsilon_4$, and $\alpha_4=\epsilon_4-\epsilon_1+1$. If we sum all equations~\eqref{efesRO} we obtain
\begin{equation}
f_1+f_2+f_3+f_4=x.\label{sumas}
\end{equation}
Since the system is over-determined, we can use a constrain $A$ as
\begin{equation}
f_1^2-f_2^2+f_3^2-f_4^2=\alpha_4-\alpha_3+\alpha_2-\alpha_1\equiv A.\label{restr1}
\end{equation}
With the two equations~\eqref{sumas} and \eqref{restr1} we can reduce the system of equations in \eqref{efesRO} to a second-order one. Let us denote $g\equiv f_3+f_4$, $q\equiv f_2+f_3$, $p\equiv f_3-f_4$. Then equations~\eqref{f2RO} and \eqref{f3RO} are written as
\begin{subequations}
\begin{align}
g' & =gp+2\alpha_3,\label{gqp1}\\
q' & =q(q-g-p)+2\alpha_2,
\end{align}\label{gqp}
\end{subequations}
\hspace{-1mm}and the restriction $A$ is expressed as
\begin{equation}
x p + (g-x)(2q-x)=A.\label{restr2}
\end{equation}
Now we have the system of the three equations \eqref{gqp} and \eqref{restr2}. We define $t\equiv 2q-x$ and then we clear $p$ from equation~\eqref{restr2}
\begin{equation}
p=\frac{1}{x}[A+(x-g)t],
\end{equation}
Then we substitute the last equation into both equations~\eqref{gqp} to obtain a two-equation system
\begin{subequations}
\begin{align}
g' & =\frac{g}{x}[A+(x-g)t]+2\alpha_3,\\
t' & =(t+ x)\left(\frac{gt-A}{x}+\frac{x-t}{2}-g\right)+4\alpha_2-1.
\end{align}\label{gh}
\end{subequations}

Now, let us define two new functions $w$ and $v$ as
\begin{subequations}
\begin{align}
xt(x) & = v(x^2),\\
g(x) & = \frac{x}{1-w(x^2)},\label{conecgw}
\end{align}\label{vw}
\end{subequations}
now we change $x^2\rightarrow z$, which takes the system to
\begin{subequations}
\begin{align}
v' & = \left(\frac{z}{4}-\frac{v^2}{4z}\right)\frac{w+1}{w-1}+(\alpha_1+\alpha_3)\frac{v}{z}+\alpha_2-\alpha_4,\label{vw1}\\
w' & = \frac{\alpha_3}{z}(w-1)^2+\frac{\alpha_1+\alpha_3}{z}(w-1)-\frac{vw}{2z},\label{vw2}
\end{align}
\end{subequations}
where the derivatives are now with respect to $z$. Then, we clear $v$ from equation~\eqref{vw2}, derive the resulting equation and substitute $v$ and $v'$ from equation~\eqref{vw1}. After some long calculations we finally obtain one equation for $w$ given by
\begin{equation}
w''=\left(\frac{1}{2w}+\frac{1}{w-1}\right)(w')^2-\frac{w'}{z}+\frac{(w-1)^2}{z^2}\left(aw+\frac{b}{w}\right)+c\frac{w}{z}+d\frac{w(w+1)}{w-1},\label{PV}
\end{equation}
with the parameters
\begin{equation}
a=\frac{\alpha_1^2}{2},\quad b=-\frac{\alpha_3^2}{2},\quad c=\frac{\alpha_2-\alpha_4}{2},\quad d=-\frac{1}{8}.\label{paraPV}
\end{equation}
This is Painlev\'e V equation ($P_V$). In general $w\in\mathbb{C}$ and also the parameters $a,b,c,d\in\mathbb{C}$.

The corresponding spectrum contains four independent equidistant ladders starting from the extremal states \citep{CFNN04}:
\begin{subequations}
\begin{align}
\psi_{\mathcal{E}_1} \propto &\left[\frac{h}{2}\left(\frac{g'}{2g}-\frac{h'}{2h}+\frac{g+h}{2}-\frac{\alpha_1}{g}\right)-\alpha_1-\alpha_2-\frac{\alpha_3}{2}\right]\nonumber\\
&\exp\left[\int\left(\frac{g'}{2g}+\frac{g}{2}-\frac{\alpha_1}{g}\right)\text{d}x\right],\\
\psi_{\mathcal{E}_2} \propto & \left[\frac{h}{2}\left(\frac{g'}{2g}-\frac{h'}{2h}+\frac{g+h}{2}+\frac{\alpha_1}{g}\right)-\alpha_2-\frac{\alpha_3}{2}\right]\nonumber\\
& \exp\left[\int\left(\frac{g'}{2g}+\frac{g}{2}+\frac{\alpha_1}{g}\right)\text{d}x\right],
\end{align}
\begin{align}
\psi_{\mathcal{E}_3} \propto & \exp\left[\int\left(\frac{h'}{2h}+\frac{h}{2}-\frac{\alpha_3}{h}\right)\text{d}x\right],\label{extRO3}\\
\psi_{\mathcal{E}_4} \propto & \exp\left[\int\left(\frac{h'}{2h}+\frac{h}{2}+\frac{\alpha_3}{h}\right)\text{d}x\right],\label{extRO4}
\end{align}\label{extremalRO}
\end{subequations}
\hspace{-1mm}where
\begin{equation}
h(x)=-x-g(x).\label{hdeg}
\end{equation}

The number operator $N_4(H)$ for this system will be of fourth-order
\begin{equation}
N_4(H)=(H-\mathcal{E}_1)(H-\mathcal{E}_2)(H-\mathcal{E}_3)(H-\mathcal{E}_4).
\end{equation}
From the definitions in equations (\ref{elesRO}--\ref{cloconRO}) we can obtain the energies of the extremal states in terms of the factorization energies as
\begin{equation}
\mathcal{E}_1=\epsilon_1+1,\quad \mathcal{E}_2=\epsilon_2+1,\quad \mathcal{E}_3=\epsilon_3+1,\quad \mathcal{E}_4=\epsilon_4+1.\label{eeps}
\end{equation}

Therefore, if we have a solution $w$ of $P_V$, equation~\eqref{PV}, we obtain a system characterized by a third-order PHA. Let us discuss some explicit examples.

\subsubsection{Radial oscillator}
Let us consider the following function \citep{CFNN04}
\begin{equation}
g(x)=-\frac{x}{2}-\frac{\ell}{x}-\alpha(x),\label{gRO}
\end{equation}
which is connected with the solution to $P_V$ through equation \eqref{conecgw}. In this case we have $\mathcal{E}_1=\mathcal{E}_3$ and $\alpha(x)$ is a solution of the Riccati equation
\begin{equation}
\alpha'(x)+\alpha^2(x)=2\left[\frac{x^2}{8}+\frac{\ell(\ell+1)}{2x^2}-\epsilon\right],
\end{equation}
where $\epsilon=\mathcal{E}_3-(\mathcal{E}_2+\mathcal{E}_4)/2$. As we have done before, we can convert this Riccati equation into a Schr\"odinger one through the substitution $\alpha=u'/u$:
\begin{equation}
-\frac{u''}{2}+\left(\frac{x^2}{8}+\frac{\ell(\ell+1)}{2x^2}\right)u=\epsilon u,
\end{equation}
whose general solution is given by equation \eqref{solRO}. The expression \eqref{gRO} for $g$ leads now to the potential we are studying:
\begin{equation}
V(x)=\frac{x^2}{8}+\frac{\ell(\ell+1)}{2x^2}+\frac{\mathcal{E}_4-\mathcal{E}_2-1}{2},
\end{equation}
which is the radial oscillator potential. The extremal states from equation~\eqref{extremalRO} are
\begin{subequations}
\begin{align}
\psi_{\mathcal{E}_1}& = 0,\\
\psi_{\mathcal{E}_2}& \propto x^{-\ell+1}\exp\left(-\frac{x^2}{4}\right),\\
\psi_{\mathcal{E}_3}& \propto \left[u'-\left(\frac{x}{2}-\frac{\ell}{x}\right)u\right],\\
\psi_{\mathcal{E}_4}& \propto x^\ell \exp\left(-\frac{x^2}{4}\right).
\end{align}\label{extremalRO1}
\end{subequations}
\hspace{-1mm}We must remember that the radial oscillator was also described by a first-order PHA. Now we are studying a third-order case, then this system must be reduced to the original first-order PHA.

\subsubsection{First-order SUSY partners of the radial oscillator}
We can take now
\begin{equation}
g(x)=-\frac{x}{2}-\frac{\ell+1}{x}+\alpha(x).
\end{equation}
Again $\alpha(x)$ fulfills a Riccati equation and $u(x)$ a Schr\"odinger one, but now we have $\mathcal{E}_1=\mathcal{E}_3+1$, $\ell=\mathcal{E}_4-\mathcal{E}_2-1/2$, $\epsilon=\mathcal{E}_3-(\mathcal{E}_2+\mathcal{E}_4-1)/2$, and $u(x)$ is given by the general solution for the radial oscillator of equation~\eqref{solRO}. Then the associated potential is now
\begin{equation}
V(x)=\frac{x^2}{8}+\frac{\ell(\ell+1)}{2x^2}+\frac{\mathcal{E}_2+\mathcal{E}_4-1}{2}-\alpha'(x),
\end{equation}
which are the first-order SUSY partners of the radial oscillator. The four extremal states are
\begin{subequations}
\begin{align}
\psi_{\mathcal{E}_1}& \propto A^+b^{+}u,\\
\psi_{\mathcal{E}_2}& \propto A^+\left[x^{-\ell}\exp\left(-\frac{x^2}{4}\right)\right],\\
\psi_{\mathcal{E}_3}& \propto \frac{1}{u},\\
\psi_{\mathcal{E}_4}& \propto A^+\left[x^{\ell+1} \exp\left(-\frac{x^2}{4}\right)\right],
\end{align}\label{extremalRO2}
\end{subequations}
\hspace{-1.5mm}where $A^+$ is the first-order intertwining operator of SUSY QM and $b^+$ is the second-order ladder operator from section~\ref{SUSYRO}.

\subsubsection{$k$th-order SUSY partners of the radial oscillator}
As in the case of the second-order PHA and the higher-order SUSY partners of the harmonic oscillator, it was also conjectured a generalization of a method to reduce the usual $(2k+2)$th-order PHA of the radial oscillator to a fourth-order one \citep{FNN04}. Nevertheless, we will prove this generalization as part of the work of this thesis in form of a theorem in chapter~\ref{5painleve}. Moreover, the corresponding systems will be connected with $P_{V}$ and they will supply us with new solutions of this equation. Again, we will work out the results here for completeness.

Let us take $k$ transformation functions $u_j$, solutions of the Schr\"odinger equation for the radial oscillator with energy $\epsilon_j$ as
\begin{equation}
u_{j+1}\propto (b^-)^j u_1, \quad \epsilon_{j+1}=\epsilon_1-j<\frac{\ell}{2}+\frac{3}{4},
\end{equation}
for $j=1,\dots , k-1$. From equation~\eqref{extremalRO2} it is easy to see that
\begin{subequations}
\begin{align}
\psi_{\mathcal{E}_1}& \propto B^+b^{+}u_1,\\
\psi_{\mathcal{E}_2}& \propto B^+\left[x^{-\ell}\exp\left(-\frac{x^2}{4}\right)\right],\\
\psi_{\mathcal{E}_3}& \propto \frac{W(u_1,\dots,u_{k-1})}{W(u_1,\dots,u_k)},\\
\psi_{\mathcal{E}_4}& \propto B^+\left[x^{\ell+1} \exp\left(-x^2/4\right)\right]
\propto \frac{W\left(u_1,\dots,u_k,x^{\ell+1}\exp\left(-\frac{x^2}{4}\right)\right)}{W(u_1,\dots,u_k)},\label{RO34}
\end{align}\label{extremalRO3}
\end{subequations}
\hspace{-1mm}where now $B^+$ is the $k$th-order intertwining operator of $k$-SUSY. The parameters are now $\mathcal{E}_1=\mathcal{E}_3+k$, $\ell=\mathcal{E}_4-\mathcal{E}_2-1/2$, $\epsilon=\mathcal{E}_3- (\mathcal{E}_2+\mathcal{E}_4-1)/2$, and $u$ is solution of equation~\eqref{solRO}. The formula for equation~\eqref{RO34} is discussed in appendix B.

If we compare the expressions for the extremal states from equation~\eqref{extremalRO} with the ones from \eqref{extremalRO3} we can immediately obtain
\begin{align}
h(x)&=\frac{2\alpha_3}{[\ln(\psi_{\mathcal{E}_4})-\ln(\psi_{\mathcal{E}_3})]'}=\left\{\ln\left[W(\psi_{\mathcal{E}_3},\psi_{\mathcal{E}_4})\right]\right\}'\nonumber \\
 &= \frac{2\alpha_3 W(u_1,\dots,u_{k-1})W(u_1,\dots, u_k,x^{\ell+1}\exp(-x^2/4))}
{W\left( W(u_1,\dots,u_{k-1}), W(u_1,\dots, u_k,x^{\ell+1}\exp(-x^2/4))\right)}.\label{ache}
\end{align}
Therefore, we obtain an expression for $g$ as
\begin{align}
g(x)&=-x-\frac{2\alpha_3}{[\ln(\psi_{\mathcal{E}_4})-\ln(\psi_{\mathcal{E}_3})]'}=-x-\left\{\ln\left[W(\psi_{\mathcal{E}_3},\psi_{\mathcal{E}_4})\right]\right\}'\nonumber\\
 &= -x-\frac{2\alpha_3 W(u_1,\dots,u_{k-1})W(u_1,\dots, u_k,x^{\ell+1}\exp(-x^2/4))}
{W\left( W(u_1,\dots,u_{k-1}), W(u_1,\dots, u_k,x^{\ell+1}\exp(-x^2/4))\right)}.\label{solPV}
\end{align}
Remember that $g(x)$ is directly related with $w(z)$, the solution of $P_V$ through
\begin{equation}
w(z)=1+\frac{z^{1/2}}{g(z^{1/2})}.
\end{equation}
Finally, the corresponding potentials are given by
\begin{equation}
V(x)=\frac{x^2}{8}+\frac{\ell(\ell+1)}{2x^2}-\{\ln[W(u_1,\dots,u_k)]''\}+\frac{\mathcal{E}_2+\mathcal{E}_4-1}{2}.
\end{equation}
\chapter{Painlev\'e IV equation}
\label{cappain}

Nowadays there is a growing interest in the study of nonlinear phenomena and their corresponding description. This motivates us to look for the different relations which can be established between physical subjects and nonlinear differential equations \citep{Sac91}. For example, the standard treatment for supersymmetric quantum mechanics (SUSY QM) leads to the Riccati equation \citep{AIS93,FF05}, which is the simplest non-linear first-order differential equation naturally associated with the general eigenvalue problem for the Schr\"odinger equation. Moreover, there are other specific links for particular potentials, e.g., the SUSY partners of the free particle are connected with solutions of the KdV equation \citep{Lam80}. Is there something similar for other potentials different from the free particle?

In this thesis we will work out the connection with Painlev\'e equations \citep{VS93,Adl94}. Although these equations were discovered from strictly mathematical considerations, nowadays they are widely used to describe several physical phenomena \citep{AC92}. In particular, the Painlev\'e IV equation ($P_{IV}$) is relevant in fluid mechanics, non-linear optics, and quantum gravity \citep{Win92}, while the Painlev\'e V equation ($P_V$) appears in condense matter \citep{Kan02}, electrodynamics \citep{Win92}, and solid state \citep{Lev92}.

Since its birth, the SUSY QM catalyzed the study of exactly-solvable Hamiltonians and gave a new insight into the algebraic structure characterizing these systems. Historically, the essence of SUSY QM was developed first as Darboux transformation in mathematical physics \citep{MS91} and later as factorization method in quantum mechanics \citep{IH51,Mie84}.

We are going to explore further the relation established between the SUSY partners of the harmonic and radial oscillators and some analytic solutions of $P_{IV}$ and $P_V$. This has been widely studied both in the context of dressing chains \citep{VS93,DEK94,Adl94} and in the framework of SUSY QM \citep{ACIN00,FNN04,CFNN04,MN08}. The key point is the following: the determination of general Schr\"odinger Hamiltonians having third-order differential ladder operators requires to find solutions of $P_{IV}$. At algebraic level, this means that the corresponding systems are characterized by second-order polynomial deformations of the Heisenberg-Weyl algebra, i.e., by the polynomial Heisenberg algebras (PHA) studied in chapter~\ref{pha}. Furthermore, those systems having fourth-order differential ladder operators require solutions of $P_V$ and they are described by third-order PHA.

On the other hand, if one wishes to obtain solutions of Painlev\'e equations, the mechanism works in the opposite way: first one looks for systems having third- or fourth-order differential ladder operators; then the corresponding solutions of Painlev\'e equations can be identified. It is worth to note that the first-order SUSY partners of the harmonic and the radial oscillators have associated natural differential ladder operators of third- and fourth-order, so that families of solutions to Painlev\'e equations can be easily obtained through this approach. Up to our knowledge, \citet*{ARS80} and \citet*{Fla80} were the first people who realize the connection between PHA (called commutator representation in these works) and Painlev\'e equations. Later, \citet*{VS93}; \citet*{DEK94}; and \citet*{Adl94} connected both subjects with first-order SUSY QM. This relation has been further explored in the higher-order case by \citet*{ACIN00,FNN04,CFNN04,MN08,SHC05,FC08,Mar09a,Mar09b}; among others.

In this chapter we will focus on the connection between higher-order SUSY partners of the harmonic oscillator and $P_{IV}$. In the chapter \ref{5painleve} we will work out the connection between the radial oscillator and $P_V$.

Note that the need to avoid singularities in the new potential $V_k(x)$ and the requirement for the Hamiltonian $H_k$ to be Hermitian lead to some restrictions \citep{BF11a} which are usually not taken into account: (i) first of all, the relevant transformation function has to be real, which implies that the associated factorization energy is real; (ii) as a consequence, the spectrum of $H_k$ consists of two independent physical ladders, an infinite one departing from the ground state energy $E_0 = 1/2$ of $H_0$, plus a finite one with $k$ equidistant levels, all of which have to be placed below $E_0$. Regarding $P_{IV}$, these two restrictions imply that non-singular real solutions $g(x;a,b)$ can be obtained only for certain real parameters $a,b$.

However, from the point of view of spectral design, it would be important to overcome restriction (ii) so that some (or all) steps of the finite ladder could be placed above $E_0$. In this way we would be able to manipulate not just the lowest part of the spectrum \citep{FNN04,CFNN04,BF11a}, but also the excited state levels, which would endow us with improved tools for the spectral design. In this chapter we are also going to show that this can be achieved if one relaxes restriction (i) as well, which will lead us to use complex transformation functions \citep{AICD99} and will permit us to generate complex solutions to $P_{IV}$ associated with real parameters $a,b$ \citep{BF11b,BF12,BF13a,Ber12}.

We will also use the scheme of SUSY QM to obtain complex partner potentials. As far as we know, they were first studied by \citet{AICD99}, where complex potentials with real energy spectra were obtained through complex transformation functions associated with real factorization energies, and later by \citet{FMR03} for complex factorization {\it energies}. On the other hand, in this work we will study complex potentials with \textit{complex energy spectra} as well.

Furthermore, we will introduce the method we developed to obtain new solutions to $P_{IV}$ equation \citep{Ber10,Ber12,BF11a,BF12,BF13a}. In order to accomplish that, we will show that under certain conditions on the positions of the $k$ new levels and on the used Schr\"odinger solutions, the combined results of SUSY QM (chapter~\ref{capsusyqm}) and PHA (chapter~\ref{pha}) lead to new solutions to $P_{IV}$. After that, we will formulate and prove a {\it reduction theorem}, which imposes some restrictions on the SUSY transformations to reduce the natural ladder operators associated with $H_k$ from higher- to third-order. Then, we will study the properties of the different ladder operators, both the natural and the reduced ones, in order to analyze the consequences of the theorem. In particular, we will study the different type of $P_{IV}$ solutions that can be obtained with this method.

Finally, after we have obtained very general formulas for these new solutions of $P_{IV}$, we classify them into several hierarchies. We do this to be able to compare them with other solutions which are spread in the literature \citep{WH03,CM08,FC08}. This classification is based upon the special functions used to explicitly write down the solutions.

\section{Painlev\'e equations}\label{peqs}
The special functions play an important role in the study of linear differential equations, that are also of great importance in mathematical physics. Examples of this functions are
\begin{itemize}
\item Airy functions $Ai(z)$.
\item Bessel functions $J_{\nu}(z)$.
\item Parabolic cylindrical functions $D_{\nu}(z)$.
\item Whittaker functions $M_{\kappa,\mu}(z)$.
\item Confluent hypergeometric functions $_1F_1(a,b;z)$.
\item Hypergeometric functions $_2F_1(a,b,c;z)$.
\end{itemize}

Some of these functions are solutions of linear ordinary differential equations with rational coefficients which receive the same name as the functions. For example, the Bessel functions are solutions of Bessel equation, which is the simplest second-order linear differential equation with one irregular singularity. Furthermore, Bessel functions are used to describe the motion of planets through Kepler equation.

Painlev\'e equations play an analogous role for the non-linear differential equations. In fact some specialists \citep{IKSY91,CM08} consider that during the 21st century, Painlev\'e functions will be new members of the special functions. Nowadays, physicists and mathematicians already use these functions. The corresponding equations are non-linear second-order ordinary differential equations and they have been found by Painlev\'e and others at the beginning of the 20th century by purely mathematical reasons. Since then, physicists have used them to describe a growing variety of systems. Next, we point out some of them
\begin{itemize}
\item Asymptotic behaviour of non-linear evolution equations \citep{SA81}.
\item Statistical mechanics \citep{ASM95}.
\item Correlation functions of the XY model \citep{STN01}.
\item Bidimensional Ising model \citep{MPS02}.
\item Superconductivity \citep{KSSLW09}.
\item Bose-Einstein condensation \citep{KSSLW09}.
\item Stimulated Raman dispersion \citep{MK10}.
\item Quantum gravity and quantum field theory \citep{FIK91}.
\item Aleatory matrix models \citep{ASM95}.
\item Topologic field theory (WDVV equations) \citep{Dub98}.
\item General relativity \citep{GMS08}.
\item Solutions of Einstein axialsymmetric equations \citep{GMS08}.
\item Negative curvature surfaces \citep{CX10}.
\item Plasma physics \citep{KSSLW09}.
\item Hele-Shaw problems \citep{LTW09}.
\item Non-linear optics \citep{FG90}.
\end{itemize}

During the last years, more and more researchers are interested in these equations and they have found interesting analytic, geometric, and algebraic properties.

Next, we will present some properties of the six Painlev\'e equations, giving an special emphasis to the Painlev\'e IV equation ($P_{IV}$), in the next chapter we will rather focus on the Painlev\'e V equation ($P_V$). We will use then the theory developed in the last chapter, regarding systems described by PHA with third-order ladder operators. They include the harmonic oscillator and special families of its SUSY partner potentials. We will show that there is a connection between general systems described by third-order differential ladder operators and solutions of $P_{IV}$. Finally, we will apply this technique to appropriate SUSY partner potentials of the harmonic oscillator to obtain explicit solutions of the $P_{IV}$.

\subsection{The six Painlev\'e equations}
\label{secant}
The ideas of Paul Painlev\'e allowed to distinguish six families of non-linear second-order differential equations, traditionally represented by \citep{IKSY91,Sac91}
\begin{subequations}
\begin{align}
P_I\ :\ & w'' = 6 w^2 + z, \\
P_{II} \ :\ & w'' = 2 w^3 + zw + a , \\
P_{III} \ :\ & w'' = \frac{1}{w}{w'}^2  - \frac{1}{z}w' + \frac{1}{z}(a w^2 + b) + c w^3 + \frac{d}{w}, \\
{\color[rgb]{0.85,0.0,0.3} P_{IV} \ :}\ & {\color[rgb]{0.85,0.0,0.3} w'' = \frac{1}{2w}{w'}^2  + \frac{3}{2}w^3 + 4 z w^2 + 2(z^2 - a)w + \frac{b}{w},}\label{PIVlib}\\
 {\color[rgb]{0.85,0.0,0.3} P_{V} \ :}\ &  {\color[rgb]{0.85,0.0,0.3}w'' = \left[\frac{1}{2w} + \frac{1}{w - 1}\right]{w'}^2 - \frac{1}{z}w'  + \frac{(w - 1)^2}{z^2}\left[a w + \frac{b}{w}\right]} \nonumber\\
\phantom{.} &\hspace{10mm} {\color[rgb]{0.85,0.0,0.3}+ \frac{c w}{z} + \frac{d w(w + 1)}{w - 1},}\\
P_{VI} \ :\ & w'' =  \frac12 \left[\frac{1}{w} + \frac{1}{w -1} + \frac{1}{w - z}\right]{w'}^2 - \left[\frac{1}{z} + \frac{1}{z - 1} + \frac{1}{w - z}\right]w'  \nonumber\\
\phantom{.} & \hspace{9mm} + \frac{w(w-1)(w-z)}{z^2(z-1)^2}\left[a + \frac{b z}{w^2} + \frac{c(z -1)}{(w-1)^2} + \frac{d z(z -1)}{(w - z)^2}\right] ,
\end{align}\label{painleves}
\end{subequations}
\hspace{-1mm}where $a,b,c,d\in\mathbb{C}$ are constants.

It is curious that only equations $P_I$, $P_{II}$, and $P_{III}$ were actually discovered by \citet{Pai00,Pai02}, the next ones were developed by \citet{Fuc07} and \citet{Gam10} following Painlev\'e ideas and actually correcting one error in Painlev\'e calculations. Moreover, from  $P_{VI}$ one can obtain the rest of equations by combining them or through a limit process.

The general form of non-linear second-order differential equation is
\begin{equation}
w''=f(z,w,w'),
\end{equation}
where $f$ is a rational function in $w$ and $w'$ and it is {\it locally analytic} in $z$, i.e., analytic except for isolated singularities in $\mathbb{C}$. The essential point here is that, in general, the singularities in the solution $w(z)$ are {\it movable}, i.e., their location depends on the constants of integration associated with the boundary conditions. An equation is said to have the {\it Painlev\'e property} if all its solutions are free from movable branch points; however, the solutions may have movable poles or movable isolated essential singularities.

A clear example of movable singularity is given by the equation
\begin{equation}
w''=w^2,
\end{equation}
whose general solution is
\begin{equation}
w(z)=\frac{6}{(z-z_0)^2}.
\end{equation}

There are fifty equations with the Painlev\'e property, most of them can be either reduced to linear equations or solved in terms of elliptic functions. After that, one still has six equations left, which are the six Painlev\'e equations from \eqref{painleves}.

For arbitrary values of the parameters $a,b,c,d$, the general solutions of the Painlev\'e equations are transcendental, i.e., they cannot be expressed in closed form in terms of elementary functions. However, for specific values of the parameters they have special solutions in terms of elementary or special functions \citep{BCH95}.

We know that second-order differential equations  play a fundamental role in the description of physical phenomena, in contrast to higher-order equations which are not found so commonly at the moment. Nevertheless, until recently, scientists were not aware of any non-linear ordinary differential equation without essential singularities that have physical significance.

One could ask immediately: how is it possible that these functions, discovered purely by mathematical reasons can appear so predominantly in physics? The answer is that these functions are defined by ordinary differential equations, just as the exponential function is defined by $u'-u=0$ and, as soon as the ordinary differential equation governing the physical system possesses some single-valuedness, the elliptic and Painlev\'e functions are likely to contribute to the corresponding single-valued expression \citep{CM08}. Therefore, they can emerge whenever a physical system can be described by such type of equations, besides of fulfilling some other conditions.

\subsection{Painlev\'e IV equation}

$P_{IV}$ appears in several areas of physics, for example, it is used to describe processes in fluid mechanics \citep{Win92},  non-linear optics \citep{FG90}, and quantum gravity \citep{FIK91}. It also arises in mathematics, in the symmetry reduction of several partial differential equations, including Boussinesq equations \citep{CK89}, modified Boussinesq equations \citep{Cla89},  dispersive wave equations \citep{PW89}, cubic Schr\"odinger equation \citep{BP80}, fifth-order Schr\"odinger equation \citep{GW89}, N-wave interaction equations \citep{QC83}, and auto-dual Yang-Mills field equations \citep{AC92}.

We know a variety of solutions to $P_{IV}$ which depend on the two parameters $a,b$ of the equation, i.e., $w=w(a,b;z)$, see \eqref{PIVlib}. Indeed, most of the solutions found in the literature have an explicit expression as infinite series related with {\it complementary error function} ($\text{erfc}(z)$) or with {\it parabolic cylinder function} ($D_\nu(z)$). In this chapter we will prove  that all of them can be included in a general solution family related with the {\it confluent hypergeometric function} \citep{BF11a,BF12,BF13a}. Some of them can also be expressed as a ratio of polynomials, this type of solutions are obtained through different B\"acklund transformations. A wide classification of these solutions have been obtained by \citet{BCH95} and \citet{ BF11a,BF13a}.

Also, in section~\ref{realhie} we will show that the existent solutions in the literature \citep{BCH95,MN08,CM08} can be obtained with our method, by applying SUSY QM to the harmonic oscillator for specific values of the parameters $\epsilon$ and $\nu$.

\section{Reduction theorem for third-order ladder operators}\label{sectma}
As we showed in chapter \ref{pha}, systems ruled by second-order PHA are connected with $P_{IV}$. What we do here is to invert the problem, i.e., instead of having to solve $P_{IV}$ to obtain Hamiltonians ruled by this kind of algebra, we use systems which already have third-order ladder operators and then we obtain solutions to $P_{IV}$. In section \ref{2pha1} we showed that the 1-SUSY partners of the harmonic oscillator are ruled by second-order PHA. Are there any other systems with this algebra? In this section we will prove a {\it reduction theorem} in which it is shown that special cases of $k$th-order SUSY partners of the harmonic oscillator, which are normally ruled by $2k$th-order algebras, can be reduced to second-order ones.\\

\noindent {\bf Theorem.} Suppose that the $k$th-order SUSY partner $H_k$ of the harmonic oscillator Hamiltonian $H_0$ is generated by $k$ Schr\"odinger seed solutions, which are connected by the standard annihilation operator in the way:
\begin{equation}
u_j = (a^{-})^{j-1} u_1, \quad \epsilon_j = \epsilon_1 - (j-1), \quad j=1,\dots,k, \label{restr}
\end{equation}
where $u_1(x)$ is a nodeless Schr\"odinger seed solution given by equation~\eqref{hyper} for $\epsilon_1 < 1/2$ and $\vert \nu_1 \vert < 1$.
\smallskip
Therefore, the natural ladder operator $L_k^+ = B_k^{+} a^{+} B_k^{-}$ of $H_k$, which is of $(2k+1)$th-order, is factorized in the form
\begin{equation}
L_k^+ = P_{k-1}(H_k) l_k^+,\label{hipo}
\end{equation}
where $P_{k-1}(H_k) = (H_k - \epsilon_1)\dots(H_k - \epsilon_{k-1})$ is a polynomial of $(k-1)$th-order in $H_k$, $l_k^+$ is a third-order differential ladder operator such that
\begin{equation}
[H_k,l_k^+] = l_k^+, \label{conmHl}
\end{equation}
and
\begin{equation} \label{annumk3}
 l_k^+ l_k^- = (H_k - \epsilon_k)\left(H_k - \frac{1}{2} \right)(H_k - \epsilon_1 - 1).
\end{equation}
\medskip
\noindent {\bf Proof (by induction).} For $k=1$ the result is obvious since
\begin{equation}
 L_1^+ = P_0(H_1)l_1^+ , \quad P_0(H_1) = 1.
\end{equation}

Let us suppose now that the theorem is valid for a given $k$; then, we are going to show that it is valid as well for $k+1$.
\begin{align}
\mbox{Hypothesis} 				\quad \ \ \quad & \quad \quad \ \ \mbox{To be shown} \nonumber\\
L_{k}^{+}=P_{k-1}(H_k)l_{k}^{+}\quad & \quad L_{k+1}^{+}=P_{k}(H_{k+1})l_{k+1}^{+} 
\end{align}

From the intertwining technique it is clear that we can go from $H_k$ to $H_{k+1}$ and vice versa through a first-order SUSY transformation
\begin{equation}
H_{k+1} A_{k+1}^+ = A_{k+1}^+ H_k, \quad H_kA_{k+1}^{-}=A_{k+1}^{-}H_k.
\end{equation}
\begin{figure}[H]\centering
\includegraphics[scale=0.3]{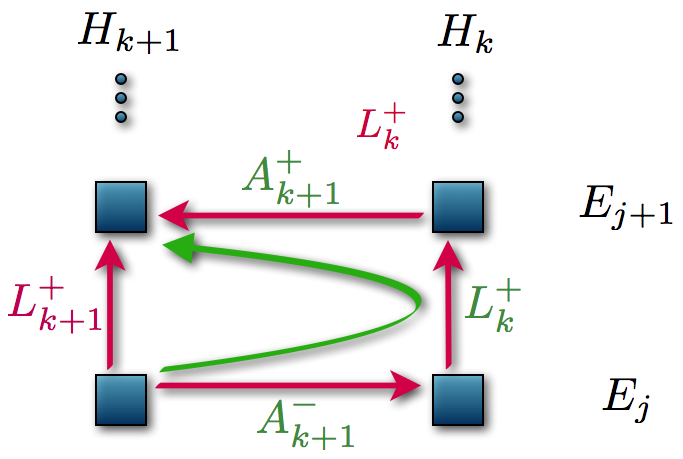}
\caption{\small{Diagram of the action of the operators in equation \eqref{Lentre}. The effective action of ${\color[rgb]{0.85,0.0,0.3}L_{k+1}^{+}}$ is the same as the joint action of ${\color[rgb]{0,0.6,0} A_{k+1}^{+}L_{k}^{+}A_{k+1}^{-}}$. }}\label{fig.sigL}
\end{figure}
\begin{figure}[H]\centering
\includegraphics[scale=0.35]{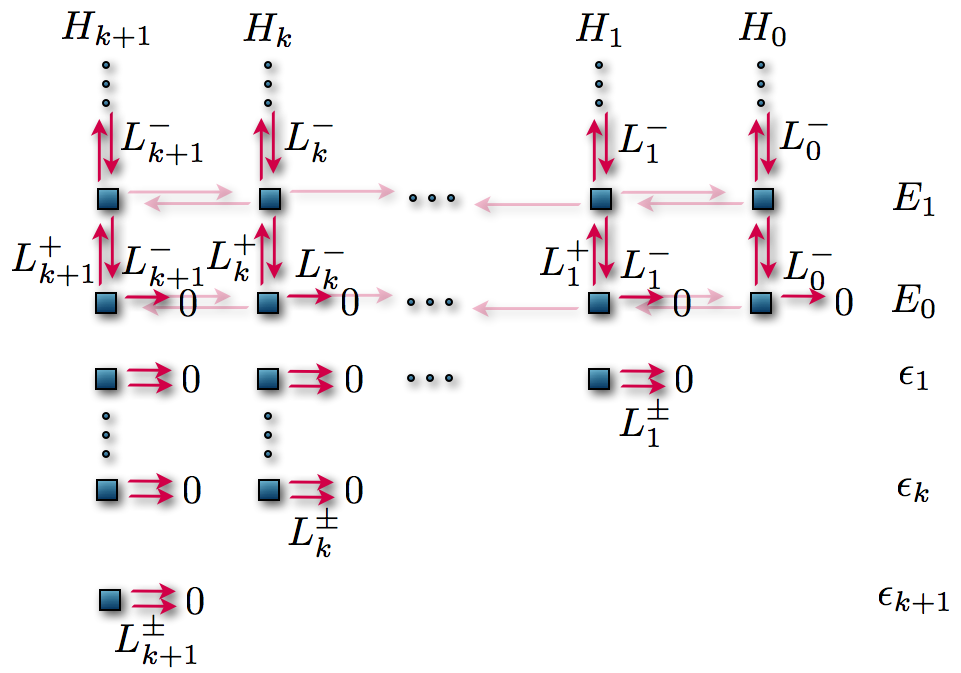}
\caption{\small{Action of the operators $L_{j}^{\pm}$ on the eigenstates of the SUSY Hamiltonians $H_j$, where $L_0^{\pm}=a^{\pm}$. We can see that the operators $L_{j}^{\pm}$ annihilate all new states while $L_{j}^{-}$ also does it with the transformed ground state of $H_0$.}}\label{fig.tma1}
\end{figure}
Moreover, it is straightforward to show that
\begin{equation}
L_{k+1}^+ = A_{k+1}^+ L_{k}^+ A_{k+1}^- .\label{Lentre}
\end{equation}

From the induction hypothesis it is obtained
\begin{equation}
L_{k+1}^{+} = A_{k+1}^{+}P_{k-1}(H_k)l_{k}^{+}A_{k+1}^{-} 
= P_{k-1}(H_{k+1})\underbrace{A_{k+1}^{+}l_{k}^{+}A_{k+1}^{-}}_{\widetilde{l}_{k+1}^{+}},
\label{pl}
\end{equation}
where
\begin{equation}
\widetilde{l}_{k+1}^{+}\equiv A_{k+1}^{+}l_{k}^{+}A_{k+1}^{-},\label{l52}
\end{equation}
is a fifth-order differential ladder operator for $H_{k+1}$. Now, it can be shown that
\begin{equation}
\widetilde{l}_{k+1}^+ \widetilde{l}_{k+1}^- = (H_{k+1} - \epsilon_k)^2 (H_{k+1} - \epsilon_{k+1})\left(H_{k+1} - \frac{1}{2}\right) (H_{k+1} - \epsilon_1 - 1).
\end{equation}
Note that the term $(H_{k+1} - \epsilon_{k+1})\left(H_{k+1} - \frac{1}{2}\right)(H_{k+1} - \epsilon_1 - 1)$ in this equation is precisely the result that would be obtained from the product $l_{k+1}^+ l_{k+1}^-$ for the third-order ladder operators of $H_{k+1}$. Thus, it is concluded that
\begin{equation}
\widetilde{l}_{k+1}^+ = q(H_{k+1}) l_{k+1}^+,
\end{equation}
where $q(H_{k+1})$ is a polynomial of $H_{k+1}$. By remembering that $\widetilde{l}_{k+1}^+, l_{k+1}^+$, and $H_{k+1}$ are differential operators of fifth-, third-, and second-order respectively, one can conclude that $q(H_{k+1})$ is linear in $H_{k+1}$ and therefore
\begin{equation}
\widetilde{l}_{k+1}^+ = (H_{k+1} - \epsilon_k) l_{k+1}^+ .\label{N52}
\end{equation}
By substituting this result in equation~\eqref{pl} we finally obtain
\begin{equation}
L_{k+1}^+ = P_{k-1}(H_{k+1})(H_{k+1} - \epsilon_k) l_{k+1}^+ = P_{k}(H_{k+1})l_{k+1}^+.
\end{equation}
With this we conclude our proof. \hfill $\square$

\section{Operator $l_k^{+}$}\label{secope}
In this section we will prove some properties of the newly defined operators $l_k^{\pm}$. In the upper part of the spectrum, i.e., the part that is isospectral to the harmonic oscillator $\{E_0,E_1,\dots \}$, they have a similar action as $L_k^{\pm}$. In particular, in this infinite ladder both operators $L_k^-$ and $l_k^-$ only annihilate the eigenstate associated with $E_0$ (the ground state energy of $H_0$). Nevertheless, in the lower spectral part, operators $L_k^{\pm}$ annihilate all eigenstates with energies $\{ \epsilon_1,\epsilon_2,\dots ,\epsilon_k \}$, while $l_k^{\pm}$ do not.

\begin{figure}[H]\centering
\includegraphics[scale=0.35]{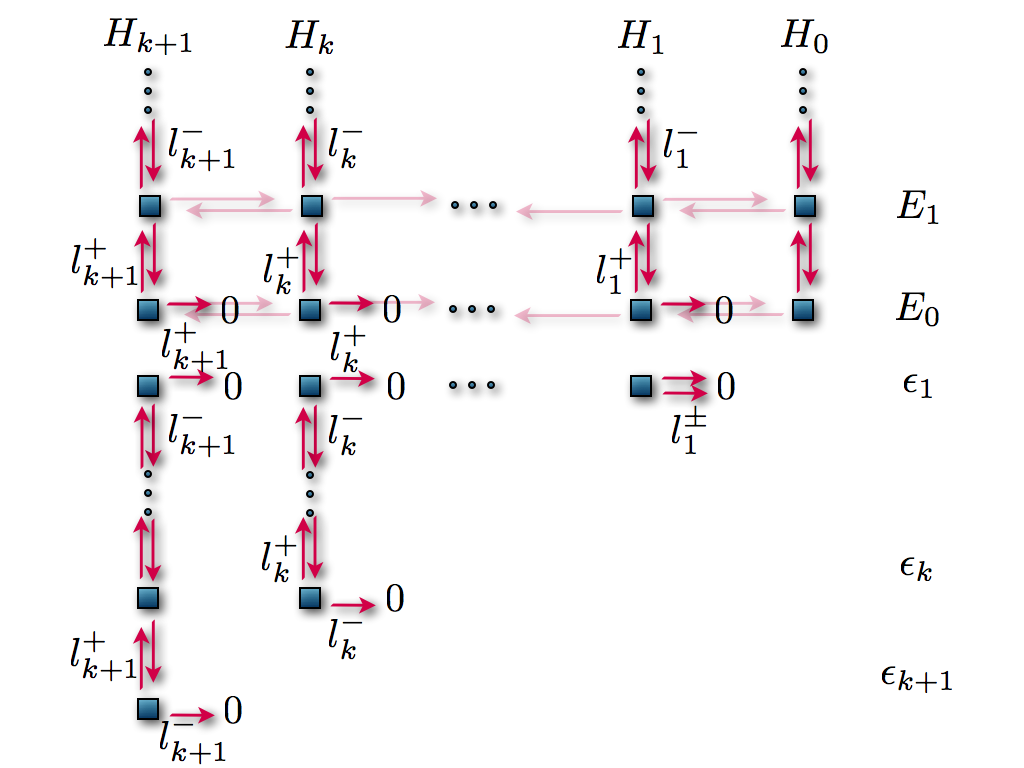}
\caption{\small{The action of the operators $l_{j}^{\pm}$ over the eigenstates of the SUSY Hamiltonians $H_j$. Note that $l_1^{\pm}=L_1^{\pm}$ and it is seen that $l_{j}^{-}$ always annihilates the eigenstates associated with $E_0$ and $\epsilon_j$. Also, $l_{j}^{+}$ annihilates the one associated with $\epsilon_1$.}}\label{fig.tma2}
\end{figure}

Indeed, the new operators $l_k^{\pm}$ do actually allow the displacement between the new eigenstates of the finite ladder. In that ladder, $l_{k}^{-}$ only annihilates the eigenstate associated with $\epsilon_k$ (the new ground state energy). On the other hand, $l_{k}^{+}$ only annihilates the new eigenstate with the highest-energy $\epsilon_1$ in that ladder. A diagram representing the action of the new operators $l_j^{\pm}$ on all eigenstates of the SUSY Hamiltonians $H_j$ is shown in figure~\ref{fig.tma2}.

\subsection{Relation with $A_{k+1}^{+}$}\label{secAl}
In analogy to the relation in equation~\eqref{Lentre} for the operators $L_k^{+}$, we will try to obtain something similar for operators $l_k^{\pm}$, even though the corresponding expression cannot be the same because equation~\eqref{Lentre} raises the differential order, from $2k+1$ of $L_{k}^{+}$ to $2k+3$ of $L_{k+1}^{+}$. On the other hand, operators $l_k^{\pm}$ and $l_{k+1}^{\pm}$ are always of third-order. Nevertheless, this does not exclude that a useful relation can be calculated using operators $l_{k+1}^{+}$, $l_k^{+}$ and $A_{k+1}^{+}$. This equation is obtained now.
\begin{figure}[H]\centering
\includegraphics[scale=0.35]{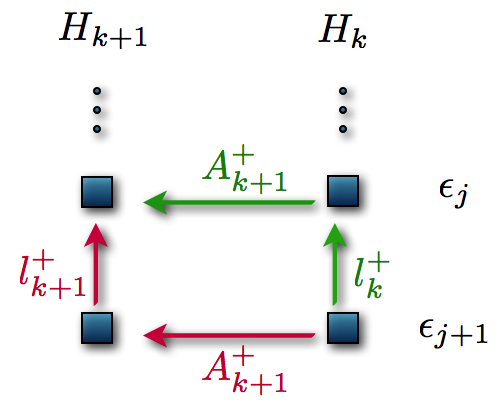}
\caption{\small{Action of the operators ${\color[rgb]{0,0.6,0}A_{k+1}^{+}l_{k}^{+}} $ and ${\color[rgb]{0.85,0.0,0.3} l_{k+1}^{+}A_{k+1}^{+}}$ over the eigenstates of the Hamiltonians $H_k$.}}\label{fig.tma3}
\end{figure}

From equations~\eqref{N52} and \eqref{l52} we get
\begin{equation}
A_{k+1}^{+}l_{k}^{+}A_{k+1}^{-} = (H_{k+1} - \epsilon_{k})l_{k+1}^{+}.
\end{equation}
Next, we multiply by the operator $A_{k+1}^{+}$ on the right and make use of equation~\eqref{HAA}, $A_{k+1}^{-}A_{k+1}^{+}=H_{k}-\epsilon_{k+1}$, to obtain
\begin{equation}
(H_{k+1} - \epsilon_{k})l_{k+1}^{+}A_{k+1}^{+} = A_{k+1}^{+}l_{k}^{+}(A_{k+1}^{-}A_{k+1}^{+})
=A_{k+1}^{+}l_{k}^{+}(H_{k}-\epsilon_{k+1}).
\end{equation}
Now we use the commutation relations given by equations~\eqref{conmHl} and \eqref{HAAH}, i.e., $l_k^{+}H_k=(H_k-1)l_k^{+}$ and $H_{k+1}A_{k+1}^{+}=A_{k+1}^{+}H_k$ leading to
\begin{align}
(H_{k+1} - \epsilon_{k})l_{k+1}^{+}A_{k+1}^{+}&=A_{k+1}^{+}(H_{k}-\epsilon_{k+1}-1)l_{k}^{+}\nonumber\\
&=A_{k+1}^{+}(H_{k}-\epsilon_{k})l_{k}^{+}\nonumber\\
&=(H_{k+1}-\epsilon_{k})A_{k+1}^{+}l_{k}^{+}.
\end{align}
Therefore
\begin{equation}
A_{k+1}^{+}l_{k}^{+} = l_{k+1}^{+}A_{k+1}^{+}.\label{AllAmas}
\end{equation}
The consecutive action of these operators can be seen in the diagram of figure~\ref{fig.tma3}.

In a similar way it can be found that
\begin{equation}
A_{k+1}^{+}l_{k}^{-} = l_{k+1}^{-}A_{k+1}^{+}.\label{AllAmenos}
\end{equation}
On the other hand, using equations~\eqref{AllAmas} and \eqref{AllAmenos}, we obtain two more relations which represent the inverse action; thus, we obtain four equations that allow the displacement of the complete spectrum of the SUSY Hamiltonians $H_k$ and $H_{k+1}$.

\begin{figure}\centering
\includegraphics[scale=0.35]{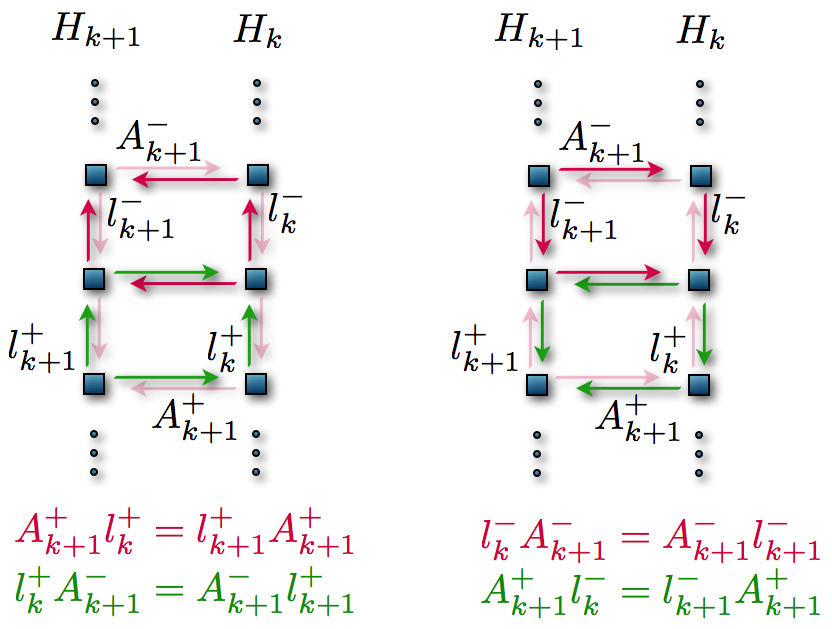}
\caption{\small{Action of the operators from equations~\eqref{finalAl} over the eigenstates of the SUSY Hamiltonians $H_k$ and $H_{k+1}$. We can see that this relations allow the displacement to any level of the spectrum, including the new eigenstates.}}\label{fig.tma4}
\end{figure}

Note that these four relations are general, i.e., they can be applied on any eigenstate, including those that are annihilated by the first operator. A full diagram can be seen in figure~\ref{fig.tma4}. The four relations can be summarized as
\begin{subequations}
\begin{align}
A_{k+1}^{+}l_{k}^{\pm} &= l_{k+1}^{\pm}A_{k+1}^{+},\\
l_{k}^{\pm}A_{k+1}^{-} &= A_{k+1}^{-}l_{k+1}^{\pm}.
\end{align}\label{finalAl}
\end{subequations}

\subsection{Operator $l_{k}^{+}l_k^{-}$}
If we apply the standard number operator $a^{+}a^{-}$ on the eigenstates of the initial Hamiltonian $H_0$, we effectively multiply it by its eigenvalue, i.e.,
\begin{equation}
a^{+}a^{-}|\psi_{n}^{(0)}\rangle = n |\psi_{n}^{(0)}\rangle.
\end{equation}

On the other hand, once we have the ladder operators $L_k^{\pm}$ for the SUSY Hamiltonians $H_k$, we have defined in analogous way the new {\it number operator} as $L_k^{+}L_k^{-}$, whose action on the mapped eigenstates $|\psi_{n}^{(k)}\rangle$ of the original energy levels and on the eigenstates $|\psi_{\epsilon_j}^{(k)}\rangle$ of the new ones is
\begin{subequations}
\begin{align}
L_k^{+}L_k^{-}|\psi_{n}^{(k)}\rangle &=\left[ n \prod_{j=1}^{k}\left(n-\epsilon_j -\frac{1}{2} \right)\left(n-\epsilon_j+\frac{1}{2}\right)\right] |\psi_{n}^{(k)}\rangle ,\\
L_k^{+}L_k^{-}|\psi_{\epsilon_j}^{(k)}\rangle &= 0.
\end{align}
\end{subequations}
Recall that Sp$(H_k)=\{ \epsilon_k,  \epsilon_{k-1},\dots,  \epsilon_1,E_0,E_1,\dots \}$; then we can see that when the operator $L_k^{+}L_k^{-}$ acts on the states $|\psi_{n}^{(k)}\rangle$ associated with the original eigenvalues, it multiplies them by a $(2k+1)$th-order polynomial of $n$ and annihilates the eigenstates $|\psi_{\epsilon_j}^{(k)}\rangle$ associated with the new energy levels.

Regarding the ladder operators $l_k^{\pm}$, which are always of third-order, we can define in an analogous way another {\it number operator} through $l_k^{+}l_k^{-}$. Its action on the eigenstates of $H_k$ is now
\begin{subequations}
\begin{align}
l_k^{+}l_k^{-} | \psi_{n}^{(k)}\rangle &=n \left(n-\epsilon_1 -\frac{1}{2}\right)\left(n-\epsilon_k+\frac{1}{2}\right) | \psi_{n}^{(k)}\rangle , \\
l_k^{+}l_k^{-} | \psi_{\epsilon_j}^{(k)}\rangle &= \left(\epsilon_j-\frac{1}{2}\right)(\epsilon_j -\epsilon_1 -1)(\epsilon_j-\epsilon_k) | \psi_{\epsilon_j}^{(k)}\rangle .
\end{align}
\end{subequations}
We should note that this new operator only annihilates the eigenstates associated with the old ground state eigenvalue $E_0$ and the new lower energy level $\epsilon_k$.

\subsection{Consequences of the theorem}\label{secnuevasg}
In section~\ref{sectma} we have proven a theorem that shows the conditions under which the following factorization is fulfilled
\begin{equation}
L_{k}^{+}=P_{k-1}(H_k)l_{k}^{+}.\label{facttma}
\end{equation}

Let us recall that in section~\ref{secpha2} we saw that there are some PHA that are \emph{reducible}, i.e., they fulfill  $\mathcal{L}_m^{+}=P(H)a^{+}$, such that the ladder operator is factorized as a product of a polynomial in the Hamiltonian times a first-order ladder operator $a^{+}$. In the case addressed in the theorem we have a similar situation.

When the SUSY transformation fulfills the requirements given in the theorem, the algebra generators are factorized as shown in equation~\eqref{facttma}, i.e., as a product of a $(k-1)$th-order polynomial in the SUSY Hamiltonians $H_k$ times a third-order ladder operator $l_k^{+}$. This means that the $2k$th-order PHA, obtained through a SUSY transformation of $k$th-order as specified in the theorem with $\epsilon_j=\epsilon_1-(j-1),\ j=1,\dots ,k$, can be \emph{reduced} to a second-order PHA with third-order ladder operators.

Let us remind that these algebras are closely related to $P_{IV}$. This implies that when we reduce the higher-order algebras we open the possibility of generating new solutions of $P_{IV}$. In fact, this is the main driving force of this thesis since we were able to find solutions of $P_{IV}$ through this method. First, we obtained solution families given in the literature \citep{BF11a}, then we worked to expand the solution space in the parameters $a,b$. We generated first real solutions associated with real parameters \citep{BF11a}, then complex solutions associated with real parameters \citep{BF12,BF13a}, and finally, complex solutions associated with complex parameters \citep{Ber12}. In the next three sections we will show the method used to obtain these solutions and then, in sections \ref{realhie} and \ref{comphie}, we classify them into several solution hierarchies \citep{BF11a,BF13a}.

\section{Real solutions of $P_{IV}$ with real parameters}\label{realsols}
\subsection{First-order SUSY QM}
\begin{figure}
\begin{center}
\includegraphics[scale=0.37]{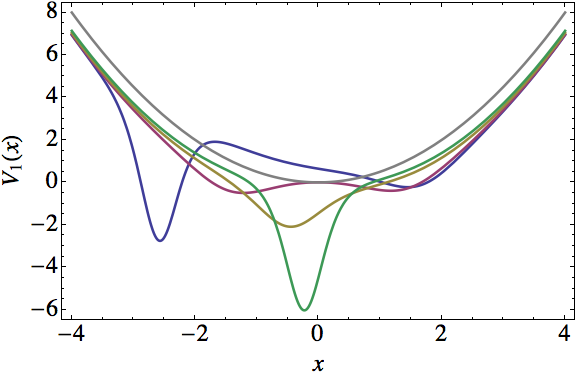} \hskip0.4cm
\includegraphics[scale=0.37]{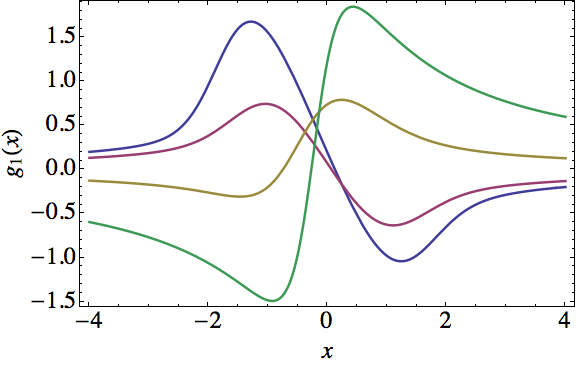}
\end{center}
\vspace{-5mm}
\caption{\small{First-order SUSY partner potentials $V_1(x)$ (left) of the harmonic oscillator and the $P_{IV}$ solutions $g_1(x,\epsilon_1)$ (right) for $\epsilon_1 = 0.25$, $\nu_1=0.99$ (blue); $\epsilon_1=0$, $\nu_1=0.1$ (magenta); $\epsilon_1=-1$, $\nu_1=0.5$ (yellow); and $\epsilon_1=-4$, $\nu_1=0.5$ (green).}}\label{PIV1}
\end{figure}

For $k=1$ we saw that the ladder operators $L_1^\pm$ are of third-order. This means that the first-order SUSY transformation applied to the harmonic oscillator could provide solutions to $P_{IV}$. To find them, we need to identify first the extremal states, which are annihilated by $L_1^-$ and at the same time are eigenstates of $H_1$. From the corresponding spectrum, one realizes that the transformed ground state of $H_0$ and the eigenstate of $H_1$ associated with $\epsilon_1$ are two physical extremal states associated with our system. Since the other root of $N_3(H_1)$ is $\epsilon_1 + 1 \not\in{\rm Sp}(H_1)$, then  the corresponding extremal state will be non-physical, which can be simply constructed from the non-physical seed solution used to implement the transformation as $A_1^+ a^{+} u_1$. Due to this, the three extremal states for our system and their corresponding factorization energies (see equation~\eqref{q3}) become
\begin{subequations}
\begin{alignat}{3}
\psi_{{\cal E}_1} & \propto A_1^+ e^{-x^2/2}, &\quad  {\cal E}_1 &= \frac{1}{2},\\
\psi_{{\cal E}_2} & \propto A_1^+ a^{+} u_1, &\quad  {\cal E}_2 &= \epsilon_1 + 1,\\
\psi_{{\cal E}_3} & \propto \frac{1}{u_1}, &\quad {\cal E}_3 &= \epsilon_1.
\end{alignat}
\end{subequations}
The first-order SUSY partner potential $V_1(x)$ of the harmonic oscillator and the corresponding non-singular solution of $P_{IV}$ are
\begin{subequations}
\begin{align}
V_1(x) &= \frac{x^2}2 - \{\ln [u_1(x)]\}', \\
 g_1(x,\epsilon_1) &=  - x - \{\ln [\psi_{{\cal E}_3}(x)]\}' = - x  + \{\ln[u_1(x)]\}',\label{solg1a}
\end{align}
\end{subequations}
where we label the $P_{IV}$ solution with an index characterizing the order of the transformation employed and we explicitly indicate the dependence on the factorization energy.  Notice that two additional solutions of the $P_{IV}$ can be obtained by cyclic permutations of the indices $(1,2,3)$. However, they will have singularities at some points and thus we drop them in this approach. An illustration of the first-order SUSY partner potentials $V_1(x)$ of the harmonic oscillator as well as its corresponding solutions $g_1(x,\epsilon_1)$ of $P_{IV}$ are shown in figure~\ref{PIV1}.

\subsection{$k$th-order SUSY QM}
With the reduction theorem of section \ref{sectma} we have proven that through higher-order SUSY QM we can obtain more systems ruled by second-order PHA, which are also connected with $P_{IV}$. Nevertheless, it is still not clear how to generate new solutions of $P_{IV}$. In this subsection it will be shown the way this is accomplished, by imposing certain restrictions on the seed solutions used to implement the SUSY transformation.

We have determined the restrictions on the Schr\"odinger seed solutions $u_j$ to reduce the order of the natural algebraic structure of the Hamiltonian $H_k$ from $2k$ to $2$. Now, suppose we stick to these constraints for generating $H_k$. As the reduced ladder operator $l_k^{+}$ is of third-order, it turns out that we can obtain solutions of $P_{IV}$ once again. To get them, we just need to identify the extremal states of our system. Since the roots of the polynomial of equation~\eqref{annumk3} are now $E_0,\, \epsilon_1 + 1,$ and $\epsilon_k = \epsilon_1 - (k-1)$; the spectrum of $H_k$ consists of two physical ladders: an infinite one departing from $E_0$ and a finite one starting from $\epsilon_k$ and ending at $\epsilon_1$. Thus, the two physical extremal states correspond to the mapped eigenstate of $H_0$ with eigenvalue $E_0$ and the eigenstate of $H_k$ associated with $\epsilon_k $. The other extremal state (which corresponds to $\epsilon_1 +1 \not\in{\rm Sp}(H_k)$) is non-physical, proportional to $B_k^{+} a^{+} u_1$. Thus, the three extremal states are
\begin{figure}
\begin{center}
\includegraphics[scale=0.37]{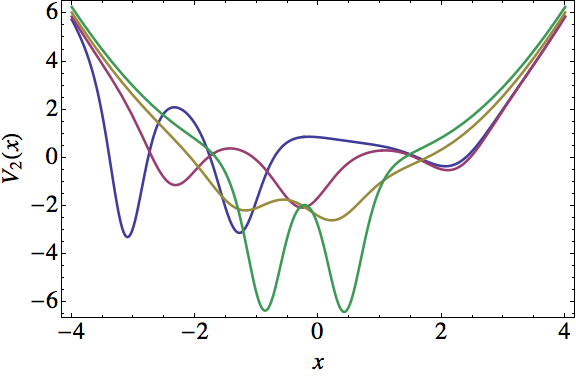} \hskip0.4cm
\includegraphics[scale=0.37]{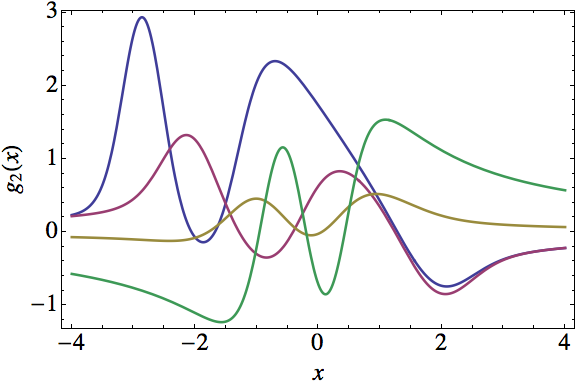}
\end{center}
\vspace{-5mm}
\caption{\small{Second-order SUSY partner potentials $V_2(x)$ (left) of the harmonic oscillator and the corresponding $P_{IV}$ solutions $g_2(x,\epsilon_1)$ (right) for $\epsilon_1 = 0.25,\nu_1=0.99$ (blue), and $\epsilon_1=\{0.25 \text{ (magenta)}$,$-0.75\text{ (yellow)}$,$-2.75\text{ (green)}\}$ with $\nu_1 = 0.5$.}}\label{PIV2}
\end{figure}
\begin{subequations}
\begin{alignat}{3}
\psi_{{\cal E}_1} & \propto B_k^+ e^{-x^2/2}, \quad &  \quad {\cal E}_1& = \frac{1}{2}, \label{edo1}\\
\psi_{{\cal E}_2} & \propto B_k^+ a^{+} u_1, \quad &  \quad {\cal E}_2 & = \epsilon_1 + 1,\label{edo2}\\
\psi_{{\cal E}_3} & \propto \frac{W(u_1,\dots,u_{k-1})}{W(u_1,\dots,u_k)}, \quad & {\cal E}_3 & = \epsilon_k = \epsilon_1 - (k - 1).\label{edo3}
\end{alignat}\label{edo123}
\end{subequations}
\hspace{-1mm}The $k$th-order SUSY partner potential of the harmonic oscillator and the corresponding non-singular solution of the $P_{IV}$ become
\begin{subequations}
\begin{align} \label{parafig3}
V_k(x) &= \frac{x^2}2 - \{\ln [W(u_1,\dots,u_k)]\}'' , \quad k\geq 2,\\
g_k(x,\epsilon_1) &= - x - \{\ln[\psi_{{\cal E}_3}(x)]\}' =  - x - \left\{\ln \left[\frac{W(u_1,\dots,u_{k-1})}{W(u_1,\dots,u_{k})}\right]\right\}'.\label{solg}
\end{align}\label{parafig3ysolg}
\end{subequations}

Let us recall that the $k$ Schr\"odinger seed solutions in the previous expressions are no longer arbitrary, rather, they have to obey the restrictions imposed by our theorem (see equation~\eqref{restr}).

\begin{figure}
\begin{center}
\includegraphics[scale=0.37]{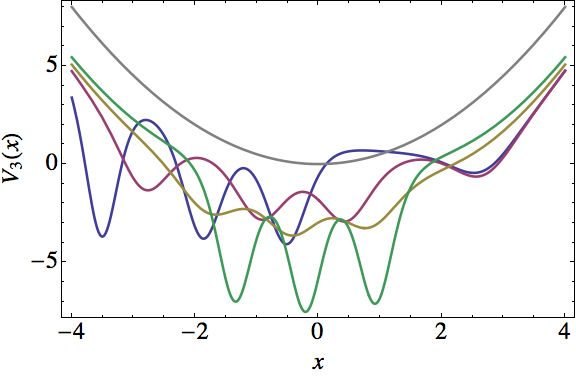} \hskip0.4cm
\includegraphics[scale=0.37]{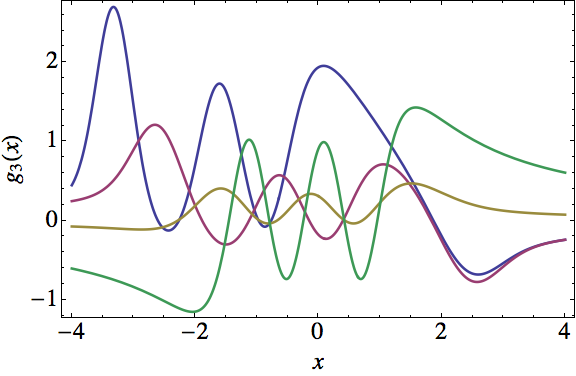}
\end{center}
\vspace{-5mm}
\caption{\small{Third-order SUSY partner potentials $V_3(x)$ (left) of the harmonic oscillator and the corresponding $P_{IV}$ solutions $g_3(x,\epsilon_1)$ (right) for $\epsilon_1 = 0.25,\nu_1=0.99$ (blue), and $\epsilon_1=\{0.25 \text{ (magenta)}$,$-0.75\text{ (yellow)}$,$-2.75\text{ (green)}\}$ with $\nu_1 = 0.5$.}}\label{PIV3}
\end{figure}

We have illustrated the $k$th-order SUSY partner potentials $V_k(x)$ of the harmonic oscillator and the corresponding $P_{IV}$ solutions $g_k(x,\epsilon_1)$ in figure \ref{PIV2} for $k=2$ and in figure~\ref{PIV3} for $k=3$.

\begin{figure}
\begin{center}
\includegraphics[scale=0.45]{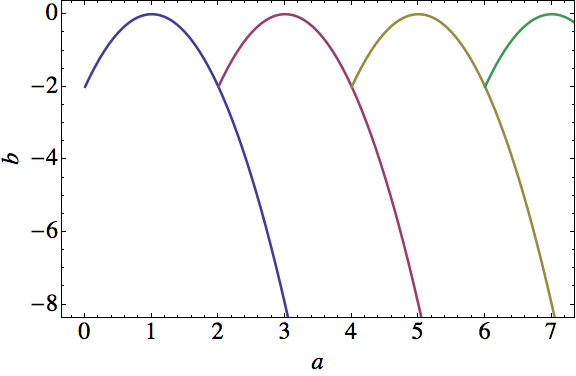}
\end{center}
\vspace{-5mm}
\caption{\small{Parameter space $a,b$ where real non-singular solutions to $P_{IV}$ with real parameters can be found. We plotted the first four families for $k=\{1,2,3,4\}$ as blue, magenta, yellow, and green; respectively.}}\label{paraRR}
\end{figure}

Using this algorithm we are able to find solutions to $P_{IV}$ with specific parameters $a,b$, i.e., not for any combination of them. Actually, we can write $a,b$ in terms of the two parameters of the transformation $\epsilon_1,k$. However, as $a,b,\epsilon_1\in\mathbb{R}$ but $k\in\mathbb{Z}^{+}$, we cannot expect to cover all the parameter space $a,b$. Indeed, we have
\begin{equation}
a=-\epsilon_1+2k-\frac{3}{2},\quad b=-2\left(\epsilon_1+\frac{1}{2}\right)^2.
\end{equation}
We have plotted this parametric relations in figure~\ref{paraRR} for $k=\{1,2,3,4\}$.

\section{Complex solutions of $P_{IV}$ with real parameters}\label{complexsols1}
We have shown in chapter \ref{pha} that the first-order SUSY partner Hamiltonians of the harmonic oscillator are naturally described by second-order PHA, which are connected with $P_{IV}$. Furthermore, we have proven a theorem stating the conditions for the Hermitian higher-order SUSY partners Hamiltonians of the harmonic oscillator to have this kind of algebras \citep{BF11a}. The main requirement is that the $k$ Schr\"odinger seed solutions have to be connected in the way
\begin{subequations}
\begin{align}
u_j=(a^{-})^{j-1}u_1,& \quad \label{us}\\
\epsilon_j=\epsilon_1-(j-1), & \quad j=1,\dots , k,
\end{align}
\end{subequations}
where $a^{-}$ is the standard annihilation operator of $H_0$ so that the only free seed $u_1$ has to be a real solution of equation~\eqref{usch} without zeros, associated with a real factorization energy $\epsilon_1$ such that $\epsilon_1<E_0=1/2$.

In this section we intend to overcome this restriction, although if we use the formalism as in \citet{BF11a} with $\epsilon_1 > E_0$, we would only obtain singular SUSY transformations. In order to avoid this, we will instead employ complex SUSY transformations. The simplest way to implement them is to use a complex linear combination of the two standard linearly independent real solutions which, up to an unessential factor, leads to the following complex solutions depending on a complex constant $\lambda + i \kappa$ ($\lambda, \kappa \in \mathbb{R}$) \citep{AICD99}:
\begin{equation}
u(x;\epsilon ) = e^{-x^2/2}\left[ {}_1F_1\left(\frac{1-2\epsilon}{4},\frac12;x^2\right)
 + x(\lambda + i\kappa)\, {}_1F_1\left(\frac{3-2\epsilon}{4},\frac32;x^2\right)\right]. \label{u1}
\end{equation}
The known results for the real case \citep{JR98} are obtained by making $\kappa=0$ and expressing $\lambda$ as
\begin{equation}
\lambda= 2 \nu\frac{\Gamma(\frac{3 - 2\epsilon}{4})}{\Gamma(\frac{1-2\epsilon}{4})}, \label{nu}
\end{equation}
with $\nu \in \mathbb{R}$.

Hence, through this formalism we will obtain the $k$th-order SUSY partner potential $V_k(x)$ of the harmonic oscillator and the corresponding $P_{IV}$ solution $g(x;\epsilon_1)$, both of which will be complex, using once again equations \eqref{parafig3ysolg}. In addition, the extremal states of $H_{k}$ and their corresponding energies are given by equations \eqref{edo123}. Recall that all $u_j$ satisfy equation~\eqref{us} and $u_1$ corresponds to the general solution given in equation~\eqref{u1}.

For $k=1$, the first-order SUSY transformation  and equation~\eqref{solg} lead to what is known as \emph{one-parameter solutions} to $P_{IV}$, due to the restrictions imposed by equation~\eqref{abe} onto the two parameters $a,b$ of $P_{IV}$ which makes them both depend on $\epsilon_1$ \citep{BCH95}. For this reason, this family of solutions cannot be found in any point of the parameter space $(a,b)$, but only in the subspace defined by the curve $\{\left( a(\epsilon_1), b(\epsilon_1)\right),\ \epsilon_1 \in \mathbb{R}\}$ consistent with equations~\eqref{abe}. 

\begin{figure}
\begin{center}
\includegraphics[scale=0.45]{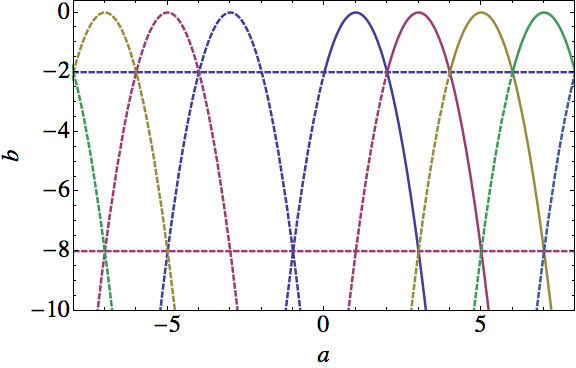}
\end{center}
\vspace{-5mm}
\caption{\small{Parameter space $(a,b)$ for non-singular $P_{IV}$ solutions. The curves represent the solution subspace for real or complex (solid lines) and only complex (dashed lines) solutions. The colors for $k=\{1,2,3,4\}$ are blue, magenta, yellow, and green; respectively.}} \label{parameterspace}
\end{figure}

Then, by increasing the order of the SUSY transformation to an arbitrary integer $k$, we will expand this subspace to obtain $k$ different families of one-parameter solutions. This procedure is analogous to iterated auto-B\"acklund transformations \citep{RS82}. Also note that by making cyclic permutations of the indices of the three energies $\mathcal{E}_j$ and the corresponding extremal states of equations~\eqref{edo123}, we expand the solution families to three different sets, defined by
\begin{subequations}
\begin{alignat}{3}
a_{i}&=-\epsilon_1 + 2k -\frac{3}{2}, & \quad  b_{i} & =-2\left(\epsilon_1+\frac{1}{2}\right)^{2}, \label{ab1}\\
a_{ii}&= 2\epsilon_1 -k, & \quad b_{ii} & =-2k^2, \\
a_{iii}&=-\epsilon_1-k-\frac{3}{2}, & \quad b_{iii} & =-2\left(\epsilon_1 - k +\frac{1}{2}\right)^2,
\end{alignat}
\end{subequations}
where we have added an index to distinguish them. The first pair ($a_i,b_i$) can provide non-singular real or complex solutions, while the second and third ones can only give non-singular complex solutions, due to singularities in the real case. A part of the non-singular solution subspace for both real and complex cases is shown in figure~\ref{parameterspace}. One can check that those points which belong to two different sets lead to the same $P_{IV}$ solutions.

\begin{figure}
\begin{center}
\includegraphics[scale=0.36]{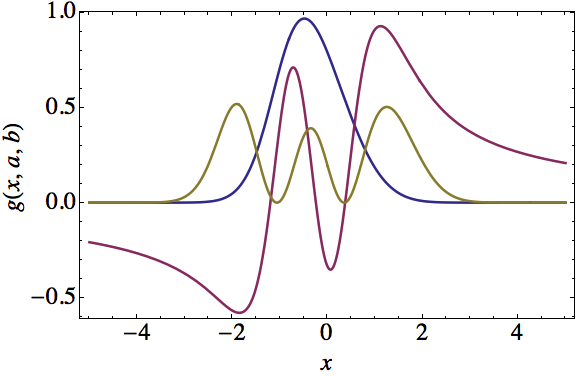}
\end{center}
\vspace{-5mm}
\caption{\small{Some real solutions to $P_{IV}$, corresponding to $a_i=1$, $b_i=0$ ($k=1$, $\epsilon_1=-1/2$, $\nu=0.7$) (blue), $a_i=4$, $b_i=-2$ ($k=2$, $\epsilon_1=-3/2$, $\nu=0.5$) (magenta), and  $a_i=7$, $b_i=-8$ ($k=3$, $\epsilon_1=-1/2$, $\nu=0.3$) (yellow).}} \label{greal}
\end{figure}

In turn, let us briefly describe some of the $P_{IV}$ solutions obtained by this method (a deeper analysis is given in sections \ref{realhie} and \ref{comphie} further ahead). The real solutions arise by taking  $\kappa=0$, and expressing $\lambda$ as in equation~\eqref{nu} with $\epsilon_1<E_0$. They can be classified into three relevant solution hierarchies, namely, confluent hypergeometric, complementary error and rational hierarchies. Let us note that the same set of real solutions to $P_{IV}$ can be obtained through inverse scattering techniques \citep{AC92} (compare the solutions of \citet{BCH95} with those of \citet{BF11a}). In figure~\ref{greal}, three real solutions to $P_{IV}$ are presented, which belong to the complementary error hierarchy.

\begin{figure}
\begin{center}
\includegraphics[scale=0.36]{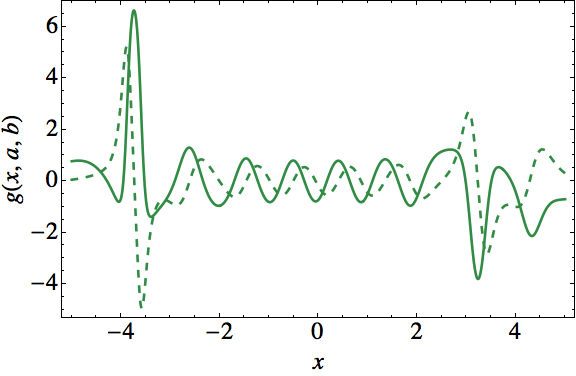}\hspace{5mm}
\includegraphics[scale=0.36]{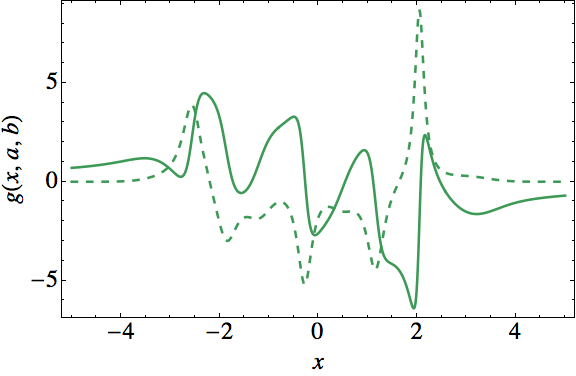}
\end{center}
\vspace{-5mm}
\caption{\small{Real (solid line) and imaginary (dashed line) parts of some complex solutions to $P_{IV}$. The left plot corresponds to $a_{ii}=12$, $b_{ii}=-8$ ($k=2$, $\epsilon_1=7$, $\lambda=\kappa=1$) and the right one to $a_{iii}=-5$, $b_{iii}=-8$ ($k=1$, $\epsilon_1=5/2$, $\lambda=\kappa=1$).}} \label{gcomplex1}
\end{figure}

Next, we study the complex solutions subspace, i.e., we allow now that $\epsilon_1 \geq E_0$. The real and imaginary parts of the complex solutions $g(x;a,b)$ for two particular choices of real parameters $a,b$, which belong to different solution sets, are plotted in figure~\ref{gcomplex1}.

\begin{figure}
\begin{center}
\includegraphics[scale=0.4]{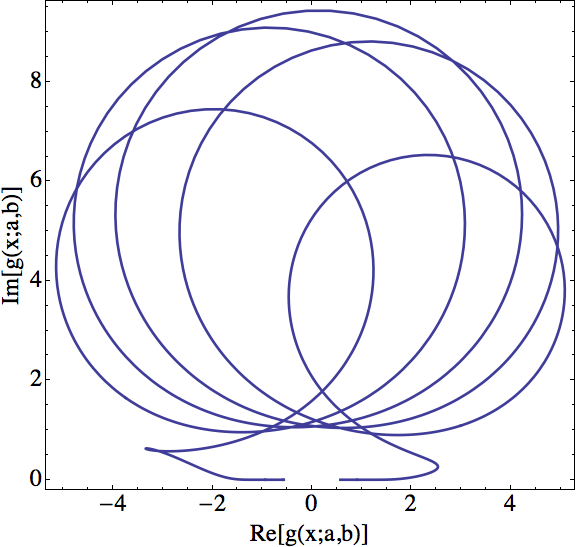}
\end{center}
\vspace{-5mm}
\caption{\small{Parametric plot of the real and imaginary parts of $g(x;a,b)$ for  $a_i=-9/2$, $b_i=-121/2$ ($k=1$, $\epsilon_1=5$, $\lambda=\kappa=2$) and $|x| \leq 10$. For larger values of $x$, the curve slowly approaches the origin in both sides.}} \label{complexpara1}
\end{figure}

Note that, in general, $\psi_{\mathcal{E}_j}\neq 0\ \forall\ x \in \mathbb{R}$ , i.e., the solutions $g(x;a,b)$ are not singular. Moreover, both real and imaginary parts have an null asymptotic behaviour ($g\rightarrow 0$ as $|x|\rightarrow \infty$). This property becomes evident in figure~\ref{gcomplex1}, as well as in the parametric plot of the real and imaginary parts of $g(x;a,b)$ of figure~\ref{complexpara1}.

\begin{figure}
\includegraphics[scale=0.37]{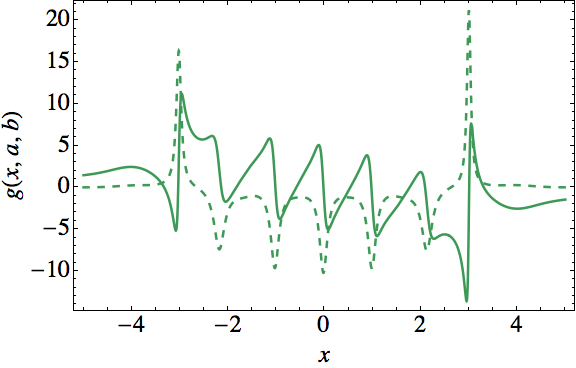}\hspace{5mm}
\includegraphics[scale=0.37]{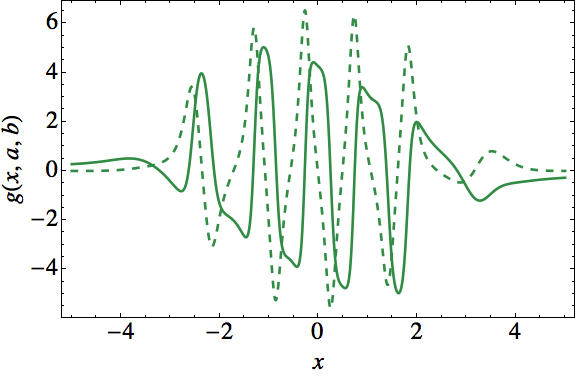}
\vspace{-5mm}
\caption{\small{Real (solid curve) and imaginary (dashed curve) parts of some complex solutions to $P_{IV}$. The left plot corresponds to $a_{iii}=-5/2$, $b_{iii}=-121/2$ ($k=2$, $\epsilon_1=5$, $\lambda=1$, $\kappa=5$) and the right one to $a_{iii}=9$, $b_{iii}=-2$ ($k=1$, $\epsilon_1=5$, $\lambda=\kappa=1$).}} \label{gcomplex2}
\end{figure}

The real and imaginary parts of the complex solutions $g(x;a,b)$ for two particular choices of real parameters $a,b$ are plotted in figure~\ref{gcomplex2}.

\section{Complex solutions of $P_{IV}$ with complex parameters}\label{complexsols2}
In this section we will use complex first-order SUSY transformations to obtain complex solutions to $P_{IV}$ with complex parameters $a,b$.

First of all, let us note that the ladder operators associated with $H_1$ are given by
\begin{equation}
L_1^{\pm}=A_1^{+}a^{\pm}A_1^{-},\label{lmas2}
\end{equation}
or explicitly,
\begin{subequations}
\begin{align}
L_1^{+}&=\frac{1}{2^{3/2}}\left(-\frac{\text{d}}{\text{d}x} +\beta\right)\left(-\frac{\text{d}}{\text{d}x}+x\right)\left(\frac{\text{d}}{\text{d}x}+\beta\right)\nonumber\\
&=\frac{1}{2^{3/2}}\left(\frac{\text{d}}{\text{d}x} -\beta\right)\left(\frac{\text{d}}{\text{d}x}-x\right)\left(\frac{\text{d}}{\text{d}x}+\beta\right),\\
L_1^{-}&=\frac{1}{2^{3/2}}\left(-\frac{\text{d}}{\text{d}x} +\beta\right)\left(\frac{\text{d}}{\text{d}x}+x\right)\left(\frac{\text{d}}{\text{d}x}+\beta\right)\nonumber\\
&=\frac{1}{2^{3/2}}\left(-\frac{\text{d}}{\text{d}x} +\beta\right)\left(-\frac{\text{d}}{\text{d}x}-x\right)\left(-\frac{\text{d}}{\text{d}x}-\beta\right).\label{eles2}
\end{align}
\end{subequations}
This system has third-order differential ladder operators, therefore, it is ruled by a second-order PHA and so we can apply our analysis of section~\ref{secondPHA} by identifying the ladder operators $L_1^{\pm}$ given in equations \eqref{eles2} with those of equations \eqref{eles}. This identification leads to
\begin{equation}
f_1= -\beta,\quad
f_2= x,\quad
f_3=\beta. \label{efes}
\end{equation}
Recall that the function $g=f_3-x$ fulfills the Painlev\'e IV equation, then one solution to $P_{IV}$ is
\begin{equation}
g=\beta-x.
\end{equation}

Let $\psi_{\mathcal{E}_j}$, $j=1,2,3$, be the states annihilated by $L_1^{-}$, where $\mathcal{E}_j$ represents the factorization energy for the corresponding extremal state. In particular, let $\psi_{\mathcal{E}_3}$ be annihilated by $A_1^{-}$, and from equation~\eqref{lmenos} we can see that it is also annihilated by $L_1^{-}$. By solving $A_1^{-}\psi_{\mathcal{E}_3}=0$ we get
\begin{equation}
\psi_{\mathcal{E}_3}\propto \exp\left[-\int f_3(y)\text{d}y\right],
\end{equation}
from which it can be shown that the corresponding solution to $P_{IV}$ reads
\begin{equation}
g(x,\epsilon)=-x-{\ln[\psi_{\mathcal{E}_3}(x)]}'.\label{sols2}
\end{equation}

For our complex first-order SUSY partner potential $V_1(x)$, generated by using the complex seed solution $u(x)$ of equation~\eqref{solu}, the three extremal states (up to a numerical factor), and their corresponding energies consistent with equations~\eqref{efes} are given by
\begin{subequations}
\begin{alignat}{3}
\psi_{\mathcal{E}_1}&= A^{+}\exp(-x^2/2), &\quad \mathcal{E}_1&= 1/2,\\
\psi_{\mathcal{E}_2}&= A^{+}a^{+}u  &\quad \mathcal{E}_2&= \epsilon+1,\\
\psi_{\mathcal{E}_3}&= u^{-1},  & \mathcal{E}_3&=\epsilon.
\end{alignat}\label{psiscom}
\end{subequations}

\begin{figure}
\begin{center}
\includegraphics[scale=0.37]{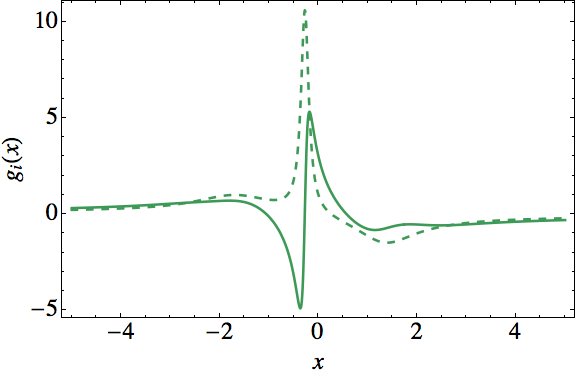}
\includegraphics[scale=0.37]{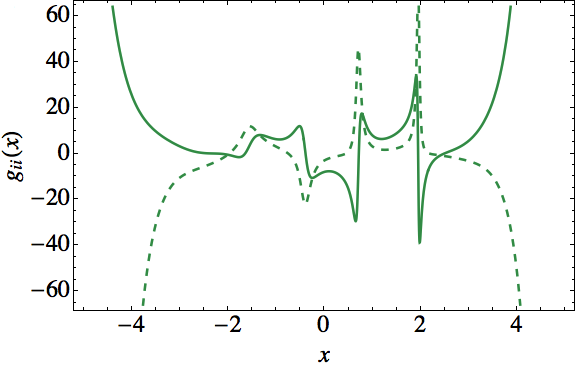}
\includegraphics[scale=0.37]{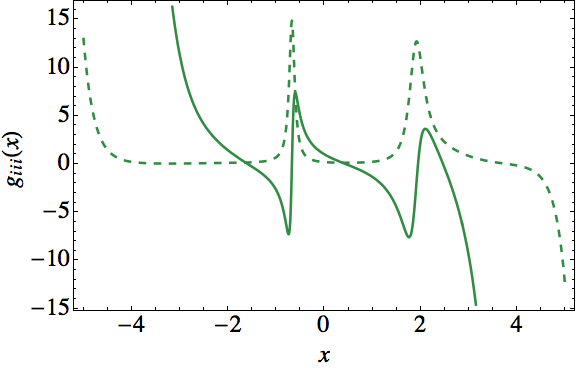}
\end{center}
\vspace{-5mm}
\caption{\small{Complex solutions to $P_{IV}$, the solid line corresponds to the real part and the dashed to the complex one: $g_i(x)$ for $\epsilon=-1+i10^{-2}$ and $\lambda=\kappa=1$, $g_{ii}(x)$ for $\epsilon=4+i 2^{-1}$ and $\lambda=\kappa=1$, and $g_{iii}(x)$ for $\epsilon=1+i$ and $\lambda=3$, $\kappa=1$.}}
\label{sols}
\end{figure}

Nevertheless, the labeling given in equations~\eqref{psiscom} is not essential, so by making cyclic permutations of the indices associated with the three extremal states of equations~\eqref{psiscom} we get three solutions to $P_{IV}$
\begin{subequations}
\begin{align}
g_i(x,\epsilon)&=-x-{\ln[\psi_{\mathcal{E}_1}(x)]}',\\
g_{ii}(x,\epsilon)&=-x-{\ln[\psi_{\mathcal{E}_2}(x)]}',\\
g_{iii}(x,\epsilon)&=-x-{\ln[\psi_{\mathcal{E}_3}(x)]}'.
\end{align}
\end{subequations}
The corresponding parameters $a,b$ of $P_{IV}$ are given by
\begin{subequations}
\begin{alignat}{3}
a_i=&-\epsilon+\frac{1}{2}, &\quad b_i=& -2\left(\epsilon+\frac{1}{2}\right)^2,\\
a_{ii}=& 2\epsilon-1, &\quad  b_{ii}=& -2,\\
a_{iii}=& -\epsilon-\frac{5}{2}, &\quad b_{iii}=& -2\left(\epsilon-\frac{1}{2}\right)^2,
\end{alignat}\label{abs}
\end{subequations}
\hspace{-2mm}where we have added the subscript corresponding to the extremal state begin used. In figure~\ref{sols} we have presented one example for each of the three families of solutions.

\begin{figure}
\begin{center}
\includegraphics[scale=0.32]{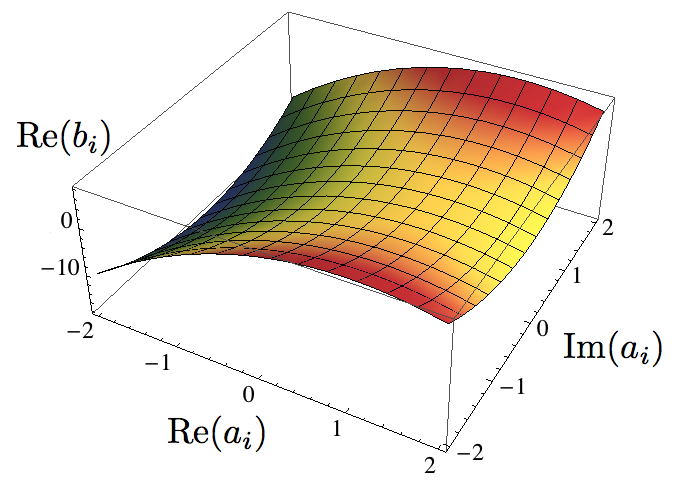}
\includegraphics[scale=0.32]{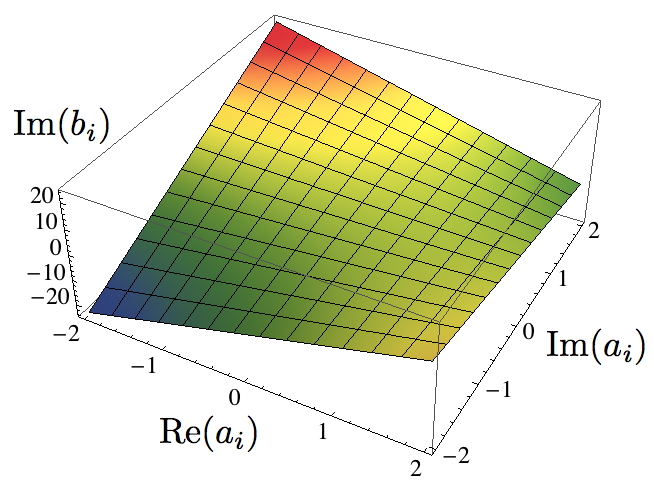}
\end{center}
\vspace{-5mm}
\caption{\small{Parameter space where we show $\text{Re}(b_i)$ (left) and $\text{Im}(b_i)$ (right) as functions of $\text{Re}(a_i)$ and $\text{Im}(a_i)$. The plots for $b_{ii}(a_{ii})$ and $b_{iii}(a_{iii})$ are similar.}}
\label{figpara}
\end{figure}

From equations~\eqref{abs} we can see that $a_j$ is linear in $\epsilon$ for all three cases. So, instead of studying the parametric relation of $a,b$ in terms of $\epsilon$, let us analyze $b_j=b_j(a_j)$, namely,
\begin{equation}
b_i= -2(a_i-1)^2,\quad
b_{ii}= -2,\quad
b_{iii}= -2(a_{iii}+3)^2.\label{b1a1}
\end{equation}
Then, we can choose $a\in\mathbb{C}$ but $b$ will be fixed by its corresponding relation with $a$. In figure~\ref{figpara} we show the domain for $b_i$ from equations~\eqref{b1a1}. There are similar plots for $b_{iii}$, but for $b_{ii}$ we have $b_{ii}=-2$ $\forall\, a_{ii}\in\mathbb{C}$, which is a plane in $\mathbb{C}$.

\section{Real solution hierarchies}\label{realhie}
The solutions $g_k(x,\epsilon_1)$ of the Painlev\'e IV equation can be classified according to the explicit special functions of which they depend on \citep{BCH95,BF11a,Ber12,BF13a}. Our general formulas, given by equations~\eqref{solg} and \eqref{sols2}, are expressed in terms of the confluent hypergeometric function $_1F_1$, although for specific values of the parameter $\epsilon_1$ they can be simplified to different special functions, including the error function $\text{erf}(z)$ and rational functions.

Let us remark that, in this thesis, we are interested in non-singular SUSY partner potentials and the corresponding non-singular solutions of $P_{IV}$. In this section we will deal with real solutions, then we restrict the parameters to $\epsilon_1 <1/2$ and $\vert \nu_1 \vert < 1$. The complex case will be treated in section~\ref{comphie}.

\subsection{Confluent hypergeometric function hierarchy}
In general, the solutions of $P_{IV}$ given in equation~\eqref{solg} are expressed in terms of two confluent hypergeometric functions. Next we write down the explicit formula for $g_1(x,\epsilon_1)$ in terms of the parameters $\epsilon_1,\, \nu_1$ (with $\epsilon_1 < 1/2$ and $\vert \nu_1 \vert < 1$ to avoid singularities):
\begin{align}\displaystyle
g_1(x,\epsilon_1)=&  \frac{2\nu_1\Gamma\left(\frac{3-2\epsilon_1}{4}\right) \left[(3-6x^2)\,{}_1F_1\left(\frac{3-2\epsilon_1}{4},\frac{3}{2};x^2\right)+x^2(3-2\epsilon_1)\,{}_1F_1\left(\frac{7-2\epsilon_1}{4},\frac{5}{2};x^2\right) \right]}
{3 \Gamma\left(\frac{1-2\epsilon_1}{4}\right)\,{}_1F_1\left(\frac{1-2\epsilon_1}{4},\frac{1}{2};x^2\right) + 6\nu_1 x \Gamma\left(\frac{3-2\epsilon_1}{4}\right)\,{}_1F_1(\frac{3-2\epsilon_1}{4},\frac{3}{2};x^2) }\nonumber\\
&+\frac{x\Gamma\left(\frac{1-2\epsilon_1}{4}\right)\left[ -2\,{}_1F_1\left(\frac{1-2\epsilon_1}{4},\frac{1}{2};x^2\right)+(1-2\epsilon_1)\,{}_1F_1\left(\frac{5-2\epsilon_1}{4},\frac{3}{2};x^2\right)	 \right]}
{\Gamma\left(\frac{1-2\epsilon_1}{4}\right)\,{}_1F_1\left(\frac{1-2\epsilon_1}{4},\frac{1}{2};x^2\right) + 2\nu_1 x \Gamma\left(\frac{3-2\epsilon_1}{4}\right)\,{}_1F_1(\frac{3-2\epsilon_1}{4},\frac{3}{2};x^2) }. \label{g1}
\end{align}

Notice that the solutions $g_1(x,\epsilon_1)$ plotted in figure~\ref{PIV1} for specific values of $\epsilon_1,\, \nu_1$ correspond to particular cases of this hierarchy. The explicit analytic formulas for higher-order solutions $g_k(x,\epsilon_1)$ can be obtained through equation~\eqref{solg}, and they have a similar form as in equation~\eqref{g1}, although with more terms.

\subsection{Error function hierarchy}
It would be interesting to analyze the possibility of reducing the explicit form of the $P_{IV}$ solution to the error function. To do that, let us fix the factorization energy in such a way that any of the two hypergeometric series of equation~\eqref{hyper} reduce to that function. This can be achieved for
\begin{equation}
\epsilon\in\left\{-\frac12,-\frac32,-\frac52,\dots ,-\frac{(2m+1)}{2},\dots \right\}.\label{ener}
\end{equation}
If we define the auxiliary function $\varphi_{\nu_1}(x)\equiv \sqrt{\pi}e^{x^2}[1+\nu_1\, \text{erf}(x)]$ to simplify the formulas, we can get simple expressions for $g_k(x,\epsilon_1)$ with some specific parameters $k$ and $\epsilon_1$:

\begin{figure}
\begin{center}
\includegraphics[scale=0.37]{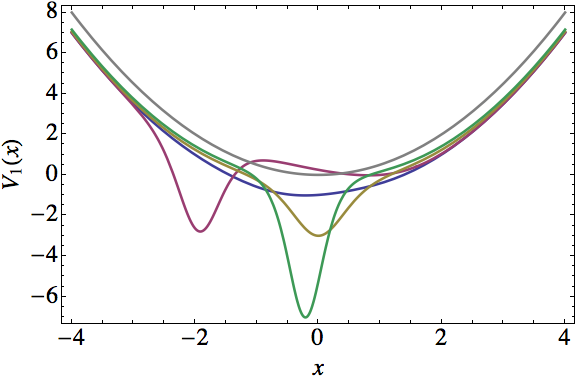}
\includegraphics[scale=0.37]{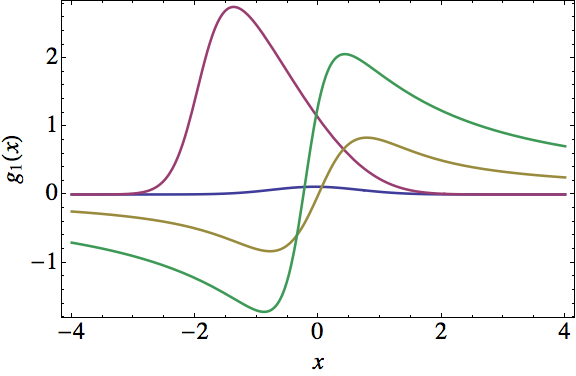}
\end{center}
\vspace{-5mm}
\caption{\small{First-order SUSY partner potentials $V_1(x)$ (left) of the harmonic oscillator and the $P_{IV}$ solutions $g_1(x,\epsilon_1)$ (right) for $\epsilon_1 =-0.5,\nu_1=0.1$ (blue); $\epsilon_1 =-0.5,\nu_1=0.99$ (magenta); $\epsilon_1 =-1.5,\nu_1=0.001$ (yellow); and $\epsilon_1 =-3.5,\nu_1=0.5$ (green), which belong to the error function hierarchy of solutions.}}\label{PIVerf}
\end{figure}

\begin{subequations}
\begin{align}
g_1(x,-1/2)&=\frac{2\nu_1}{\varphi_{\nu_1}(x)}, \label{erf1}\\
g_1(x,-3/2)&=\frac{\varphi_{\nu_1}(x)}{1+x\varphi_{\nu_1}(x)},\\
g_1(x,-5/2)&=\frac{4[\nu_1 + \varphi_{\nu_1}(x)]}{2\nu_1 x +(1+2x^2)\varphi_{\nu_1}(x)},\label{erf2}\\
g_2(x,-1/2)&=\frac{4\nu_1[\nu_1 + 6\varphi_{\nu_1}(x)]}{\varphi_{\nu_1}(x)[\varphi_{\nu_1}^2(x) -2\nu_1 x \varphi_{\nu_1}(x) -2\nu_1^2]}.
\end{align}\label{erfs}
\end{subequations}

An illustration of the first-order SUSY partner potentials $V_1(x)$ of the harmonic oscillator and the corresponding $P_{IV}$ solutions $g_1(x,\epsilon_1)$ of equations~\eqref{erfs} is given in figure~\ref{PIVerf}.

\subsection{Rational hierarchy}
\begin{figure}
\begin{center}
\includegraphics[scale=0.37]{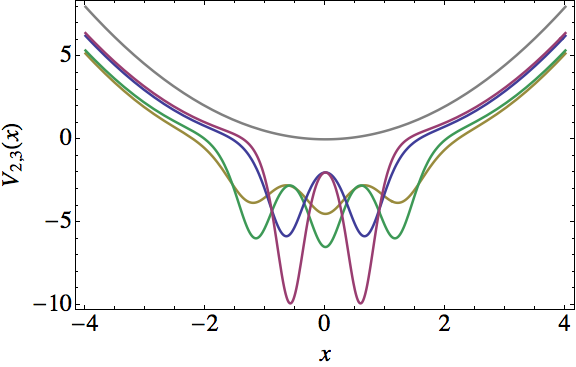}
\includegraphics[scale=0.37]{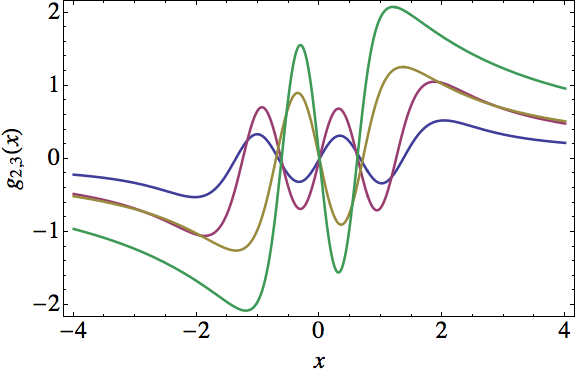}
\end{center}
\vspace{-5mm}
\caption{\small{Two second-order $V_2(x)$ and two third-order $V_3(x)$ partner potentials (left) of the harmonic oscillator and the corresponding $P_{IV}$ solutions (right) given by equations~\eqref{pol1} in blue, \eqref{pol2} in magenta, \eqref{pol3} in green, and \eqref{pol4} in yellow.}}\label{PIVpol}
\end{figure}

Our previous formalism allows us to generate solutions of $P_{IV}$ involving in general the confluent hypergeometric series, which has an infinite sum of terms. Let us look for the restrictions needed to reduce the explicit form of equation~\eqref{solg} to non-singular rational solutions. To achieve this, once again the factorization energy $\epsilon_1$ has to take a value in the set given by equation~\eqref{ener}, but depending on the $\epsilon_1$ taken, just one of the two hypergeometric functions is reduced to a polynomial. Thus, we need to choose additionally the parameter $\nu_1=0$ or $\nu_1\rightarrow\infty$ to keep the appropriate hypergeometric function.
However, for the values $-(4m-1)/2, m=1,2,\dots$ and $\nu_1\rightarrow \infty$, it turns out that $u_1$ will have always a zero at $x=0$, which will produce one singularity for the corresponding $P_{IV}$ solution. In conclusion, the rational non-singular solutions $g_k(x,\epsilon_1)$ of the $P_{IV}$ arise by making in equation~\eqref{hyper} $\nu_1 = 0$ and
\begin{equation}
\epsilon_1 \in \left\{-\frac12,-\frac52,\dots, -\frac{(4m+1)}2, \dots \right\}.
\end{equation}
Taking as the point of departure the Schr\"odinger solutions with these $\nu_1$ and $\epsilon_1$ and using our previous expressions \eqref{solg} for a given order of the transformation we get the following explicit expressions for $g_k(x,\epsilon_1)$
\begin{subequations}
\begin{align}
g_1(x,-5/2)&=\frac{4 x}{1 + 2 x^2},\label{rg1}\\
g_1(x,-9/2)&= \frac{8 (3 x + 2 x^3)}{3 + 12 x^2 + 4 x^4},\\
g_1(x,-13/2)&= \frac{12 (15 x + 20 x^3 + 4 x^5)}{15 + 90 x^2 + 60 x^4 + 8 x^6},\\
g_2(x,-5/2)&= -\frac{4 x}{1 + 2 x^2} + \frac{16 x^3}{3 + 4 x^4},\label{pol1}\\
g_2(x,-9/2)&= -\frac{8 (3 x + 2 x^3)}{3 + 12 x^2 + 4 x^4} + \frac{32 (15 x^3 + 12 x^5 + 4 x^7)}{45 + 120 x^4 + 64 x^6 + 16 x^8},\label{pol2}
\end{align}
\begin{align}
g_2(x,-13/2)=& -\frac{12 (15 x + 20 x^3 + 4 x^5)}{15 + 90 x^2 + 60 x^4 + 8 x^6} \nonumber\\
	&+ \frac{48 (525 x^3 + 840 x^5 + 600 x^7 + 160 x^9 + 16 x^{11})}{1575 + 6300 x^4 + 6720 x^6 + 3600 x^8 + 768 x^{10} + 64 x^{12}},\\
g_3(x,-5/2)=& \frac{4 x (27 - 72 x^2 + 16 x^8)}{27 + 54 x^2 + 96 x^6 - 48 x^8 + 32 x^{10}},\label{pol3}\\
g_3(x,-9/2)=& -\frac{32 (15 x^3 + 12 x^5 + 4 x^7)}{45 + 120 x^4 + 64 x^6 + 16 x^8} \nonumber\\
	& +\frac{24 (225 x - 150 x^3 + 120 x^5 + 240 x^7 + 80 x^9 + 32 x^{11})}{675 + 2700 x^2 - 900 x^4 + 480 x^6 + 720 x^8 + 192 x^{10} + 64 x^{12}},\label{pol4}
\end{align}
\end{subequations}
some of which are illustrated in figure~\ref{PIVpol}.

\begin{figure}
\begin{center}
\includegraphics[scale=0.4]{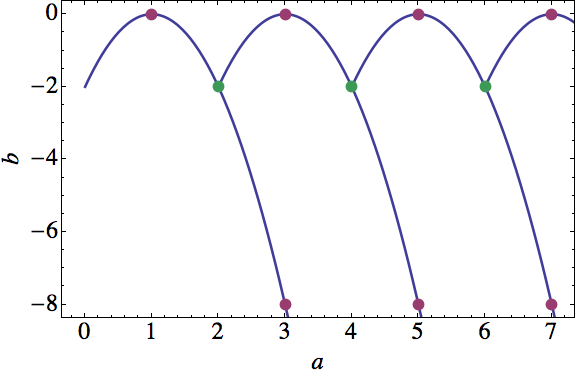}
\end{center}
\vspace{-5mm}
\caption{\small{Parameter space for the real $P_{IV}$ solution hierarchies. The lines represent solutions for the confluent hypergeometric function hierarchy, in the magenta dots appear the error function hierarchy, and in the green dots arise both, the rational and error function hierarchies.}} \label{realpara}
\end{figure}

In figure~\ref{realpara} we show where we can find the specific hierarchies in the solution parameter space $a,b$. It is clear that all of them are also inside the more general hierarchy of the confluent hypergeometric function.

\section{Complex solution hierarchies}\label{comphie}
Let us study now the complex solutions subspace associated with real parameters $a,b$ of $P_{IV}$, i.e., we use the complex linear combination of equation~\eqref{u1} and the associated $P_{IV}$ solution of equation~\eqref{solg}. This allows us to use seeds $u_1$ with $\epsilon_1 \geq E_0$ but without producing singularities. Moreover, the complex case is richer than the real one, since all three extremal states of equations~\eqref{edo123} lead to non-singular complex $P_{IV}$ solution families.

\subsection{Confluent hypergeometric hierarchy}
As in the real case, in general the solutions of $P_{IV}$ are expressed in terms of two confluent hypergeometric functions. In particular,
the explicit formula for the first family $g_i(x;\epsilon_1)$ in terms of the parameters $\epsilon_1,\, \Lambda=\lambda+i\kappa$ is given by
\begin{align}\displaystyle
g_i(x,\epsilon_1)=&  \frac{\Lambda\left[(1-2x^2)\,{}_1F_1\left(\frac{3-2\epsilon_1}{4},\frac{3}{2};x^2\right)+(1-\frac{2}{3}\epsilon_1)x^2 {}_1F_1\left(\frac{7-2\epsilon_1}{4},\frac{5}{2};x^2\right)\right]}
{\,{}_1F_1\left(\frac{1-2\epsilon_1}{4},\frac{1}{2},x^2\right)+\Lambda \,x\,{}_1F_1\left(\frac{3-2\epsilon_1}{4},\frac{3}{2},x^2\right)}\nonumber\\
&+\frac{x\left[(2\epsilon_1 - 1)\,{}_1F_1\left(\frac{5-2\epsilon_1}{4},\frac{3}{2};x^2\right)-2\,{}_1F_1\left(\frac{1-2\epsilon_1}{4},\frac{1}{2};x^2\right)\right]}
{\,{}_1F_1\left(\frac{1-2\epsilon_1}{4},\frac{1}{2},x^2\right)+\Lambda\,x\,{}_1F_1\left(\frac{3-2\epsilon_1}{4},\frac{3}{2},x^2\right)}. \label{g1complex}
\end{align}
Once again, for all families the explicit analytic formulas for the higher-order solutions $g_k(x;\epsilon_1)$ can be obtained through equation \eqref{solg}.

\subsection{Error function hierarchy}
If we choose the parameter $\epsilon_1=-(2m+1)/2$ with $m\in\mathbb{N}$, as in the real case, we obtain the error function hierarchy. In terms of the auxiliary function $\phi_{\Lambda}=\text{e}^{x^2}[4 + \Lambda \pi^{1/2}\text{erf}(x)]$, a solution from the third family is written as
\begin{equation}
g_{iii}(x;-5/2)=\frac{4\Lambda + 4x\phi_{\Lambda}(x)}{2\Lambda x+(1+2x^2)\phi_{\Lambda}(x)}.
\end{equation}
\subsection{Imaginary error function hierarchy}
\begin{figure}
\begin{center}
\includegraphics[scale=0.4]{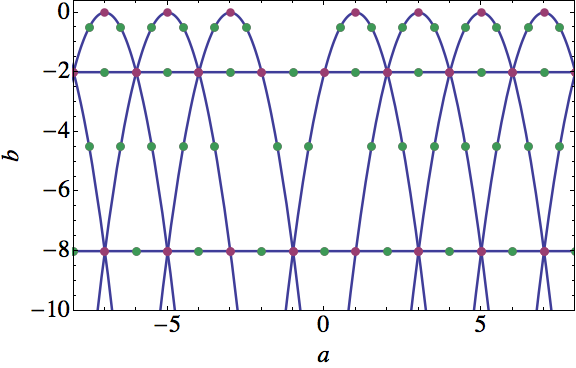}
\end{center}
\vspace{-5mm}
\caption{\small{Parameter space for complex solution hierarchies. The lines correspond to the confluent hypergeometric function, the magenta dots to the error or imaginary error functions, and the green dots to the first kind modified Bessel function.}} \label{complexpara2}
\end{figure}

On the contrary of the real case, now we can use $\epsilon_1\geq E_0$, giving place to more solution families. This is clear by comparing the real and complex parameter spaces of solutions from figures~\ref{realpara} and \ref{complexpara2}. By defining a new auxiliary function $\phi^i_{\Lambda}=\text{e}^{-x^2}[4 + \Lambda \pi^{1/2}\text{erfi}(x)]$, where $\text{erfi}(x)$ is the {\it imaginary error function}, we can write down an explicit solution from the third family and for $k=1$ as
\begin{equation}
g_{iii}(x;5/2)=\frac{4\Lambda(1-x^2)+2x(-3+2x^2)\phi^i_{\Lambda}(x)}
{2\Lambda x+(1-2x^2)\phi^i_{\Lambda}(x)}.
\end{equation}

\subsection{First kind modified Bessel function hierarchy}
\begin{figure}
\begin{center}
\includegraphics[scale=0.4]{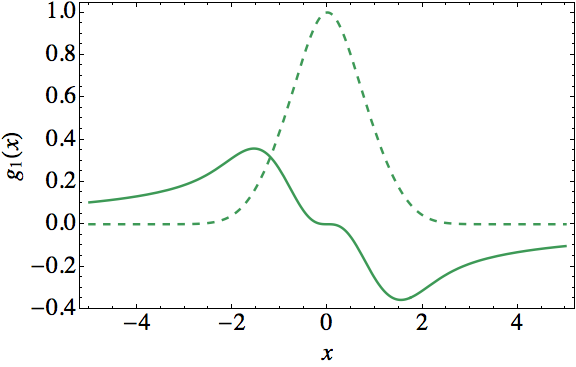}
\end{center}
\vspace{-5mm}
\caption{\small{Real (solid curve) and imaginary (dashed curve) parts of a complex solution to $P_{IV}$ that belong to the first-kind modified Bessel function hierarchy. The plot corresponds to $k=1$, $\epsilon_1=0$, $\lambda=0$, and $\kappa=1$.}} \label{complexsol}
\end{figure}

Let us write down an example of one solution of this hierarchy for $\lambda=0$, $\kappa=1$, $\Lambda = i$, i.e., $u_1$ is a purely imaginary linear combination of the two standard real solutions associated with $\epsilon_1 = 0$ and for $k=1$:
\begin{equation}
g_i(x;0)=\frac{\Gamma\left(\frac{3}{4}\right)\left[I_{-\frac{5}{4}}\left(\frac{x^2}{2}\right)+(1-x^2)I_{-\frac{1}{4}}\left(\frac{x^2}{2}\right)\right]+2 i x^2 \Gamma\left(\frac{5}{4}\right)\left[I_{-\frac{3}{4}}\left(\frac{x^2}{2}\right)-I_{\frac{1}{4}}\left(\frac{x^2}{2}\right)\right]}
{x\Gamma\left(\frac{3}{4}\right)I_{-\frac{1}{4}}\left(\frac{x^2}{2}\right)+2 i x \Gamma\left(\frac{5}{4}\right)I_{\frac{1}{4}}\left(\frac{x^2}{2}\right)}.
\end{equation}
Its real and imaginary parts are plotted in figure~\ref{complexsol}.

\section{Non-Hermitian Hamiltonian}
Let us analize now the Hamiltonian $H_k$ obtained by the complex SUSY transformation. Note that the real case, which leads to Hermitian Hamiltonians, has been studied previously \citep{Mie84,AIS93}, allowing to obtain information related to the structure of the energy spectrum, $\text{Sp}(H_k)$, the number of zeroes of the eigenfunctions of $H_k$, and the way in which they are connected by the third-order ladder operators $l_k^{\pm}$. This action is in agreement with the fact that $\text{Sp}(H_k)$ consists of an infinite ladder plus a finite one: there are two extremal states (both annihilated by $l_k^{-}$) from which the two ladders start, one associated with $\epsilon_k$ and the other one to $E_0=1/2$; since the ladder starting from $\epsilon_k$ ends at $\epsilon_1$, the eigenfunction associated with $\epsilon_1$ is annihilated by $l_k^{+}$. The actions of $l_k^{\pm}$ onto any other physical eigenstate of $H_k$ are non-null and only connect the eigenstates belonging to the same ladder.

As far as we know, complex SUSY transformations with real factorization energies were used for the first time by \citet{AICD99} to obtain non-Hermitian Hamiltonians with real spectra. These topics have been of great interest in the context of both parity-time (PT) symmetric Hamiltonians developed by \citet{BB98} and pseudo-Hermitian Hamiltonians, studied by \citet{MB04}. Next, we will examine the structure of the non-Hermitian SUSY generated Hamiltonians $H_k$.

\begin{figure}
\begin{center}
\includegraphics[scale=0.3]{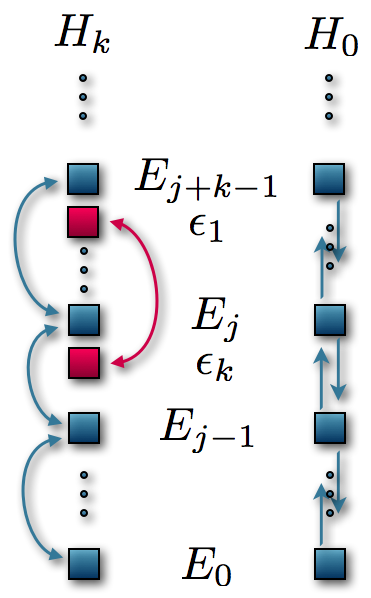}
\end{center}
\vspace{-5mm}
\caption{\small{Spectrum of the SUSY partner Hamiltonians $H_0$ (right) and $H_k$ (left) for $\epsilon_1>E_0$, $\epsilon_1\neq E_j$; Sp($H_k$) contains one finite and one infinite ladder. The blue squares represent the original and mapped eigenstates of $H_0$ and $H_k$, while the red ones the $k$ new levels of $H_k$ created by the transformation. All of them have associated square-integrable eigenfunctions.}} \label{espectros}
\end{figure}

First of all, the new Hamiltonians necessarily have complex eigenfunctions, although the associated eigenvalues are still real. In previous works, the factorization energy associated with the real transformation function $u_1$ was bounded by $\epsilon_1<E_0=1/2$. In this section we are using complex transformation functions to be able to overcome this restriction and yet obtain non-singular solutions. This naturally leads to complex solutions to $P_{IV}$ generated through factorization energies which could be placed now above $E_0$. The resulting spectra for the non-Hermitian Hamiltonians $H_k$ obey the same criteria as the real case, namely, they are composed of an infinite ladder plus a finite one, which now could be placed, either fully or partially, above $E_0$. The eigenfunctions associated with the energy levels of the original harmonic oscillator are given by equation~\eqref{psin} and the ones associated with the new energy levels by equation~\eqref{psie}, all of them are square-integrable. A diagram of the described spectrum is shown in figure~\ref{espectros}.

\begin{figure}
\begin{center}
\includegraphics[scale=0.37]{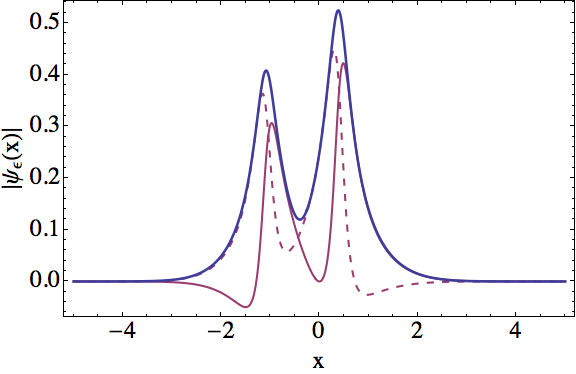}
\includegraphics[scale=0.37]{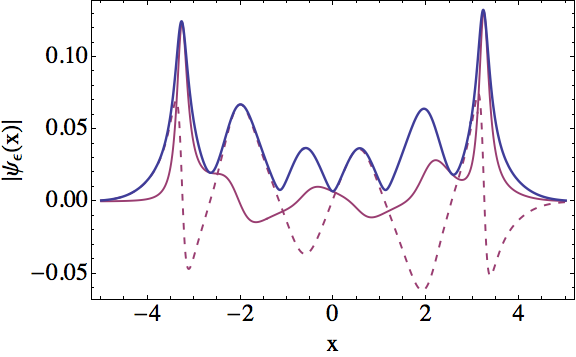}
\end{center}
\vspace{-5mm}
\caption{\small{Plot of the absolute value (blue line), the real, and the imaginary parts (solid and dashed magenta lines) of the eigenfunction $\psi_{\mathcal{E}_1}(x)$ given by equation~\eqref{edo1} for the values $k=2$, $\epsilon_1=-1$, $\lambda=1$, $\kappa=1/2$ (left) and $k=2$, $\epsilon_1=4$, $\lambda=1$, $\kappa=6$ (right).}} \label{waves}
\end{figure}

The extremal states of the SUSY generated Hamiltonian $H_k$ are given by equations~\eqref{edo123}. These are non-singular complex solutions of the stationary Schr\"odinger equation for $H_k$ and, from their asymptotic behaviour, we conclude that those given by equations~\eqref{edo1} and \eqref{edo3} are square-integrable. Note that in this case the oscillation theorem does not hold anymore, neither for the real nor for the imaginary parts, although a related node structure emerges. The absolute value and the real and imaginary parts of $\psi_{\mathcal{E}_1}(x)$ for two particular cases are shown in figure~\ref{waves}.

\begin{figure}
\begin{center}
\includegraphics[scale=0.3]{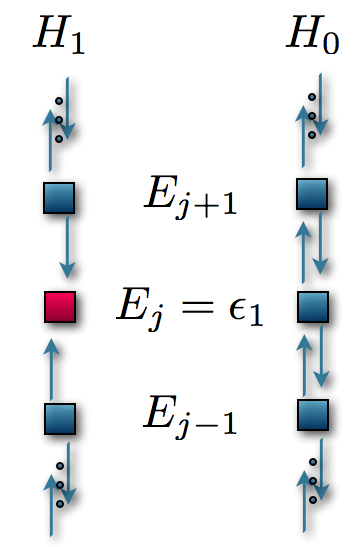}
\end{center}
\vspace{-5mm}
\caption{\small{Spectra of the harmonic oscillator Hamiltonian $H_0$ and its complex first-order SUSY partner $H_1$ when we use the factorization energy $\epsilon_1=E_j \in \text{Sp}(H_0)$. The level $E_j$ of $H_1$ is connected with its adjacent ones just in one way.}} \label{1susy}
\end{figure}

On the other hand, the complex transformations for $\epsilon_1=E_j$ are worth of a detailed study, namely, when the factorization energy $\epsilon_1$ belongs to the spectrum of the original harmonic oscillator Hamiltonian. For instance, let us consider a first-order SUSY transformation with $\epsilon_1=E_j$ and $u_1$ given by equation~\eqref{u1}, i.e., $u_1$ is a complex linear combination of the eigenfunction $\psi_j$ of $H_0$ and the other linearly independent solution of the Schr\"odinger equation. It is straightforward to see that the action of the ladder operator $L_1^{-}=l_1^-=A_1^{+}a^{-}A_1^{-}$ is given by
\begin{subequations}
\begin{align}
E_{s},&\quad L_1^{-}(A_1^{+}\psi_{s})  \propto A_1^{+}\psi_{s-1},\\
E_j, &\quad L_1^{-}(A_1^{+}\psi_{j})  = 0,\\
E_{0}, &\quad L_1^{-}(A_1^{+}\psi_{0})  = 0,
\end{align}
\end{subequations}
where $s \neq j$, $s \neq 0$, the energies shown correspond to the departure state, and we have used that $A_1^{+}\psi_j \propto u_1^{-1}$. For $L_1^{+}=A_1^{+}a^{+}A_1^{-}$ we have
\begin{subequations}
\begin{align}
E_{s},&\quad L_1^{+}(A_1^{+}\psi_{s})  \propto \psi_{s+1},\\
E_j, &\quad L_1^{+}(A_1^{+}\psi_{j})  =0,
\end{align}
\end{subequations}
which does not match with the established criteria for the non-singular real and complex cases with $\epsilon_1 \neq E_j$, since now it turns out that
\begin{subequations}
\begin{align}
E_{j+1},&\quad L_1^{-}(A_1^{+}\psi_{j+1})  \propto A_1^{+}\psi_{j} \propto u_1^{-1} \neq 0,\\
E_{j-1},&\quad L_1^{+}(A_1^{+}\psi_{j-1})   \propto A_1^{+}\psi_{j} \propto u_1^{-1} \neq 0.
\end{align}
\end{subequations}
The resulting Hamiltonian is isospectral to the harmonic oscillator but with a special algebraic structure because now one state (the one associated with $E_j$) is connected just in one way with the adjacent ones (associated with $E_j \pm 1$). A diagram representing this structure is shown in figure~\ref{1susy}. We are currently studying the $k$th-order case and expect to find the new criteria which will be valid for these special transformations.
\chapter{Painlev\'e IV coherent states}
\label{p4cs}

\section{Introduction}
The study of coherent states (CS) has taken an important place in quantum physics since they can be used to establish an essential connection between quantum and semi-classical analysis of a system \citep{Kla63a,Kla63b,Per86,GK99,BHK12}. The term {\it coherent} itself comes from the current language of quantum optics, and it was introduced by \citet{Gla63a,Gla63b}. Moreover, they have also become important for the mathematical formulation of quantum theory. In fact, \citet{GK99} proved that a set of CS can be constructed for any formal Hilbert space with discrete or continuous basis, which includes as special cases those Hilbert spaces related with physical quantum systems.

There are several ways to address this problem; nevertheless, most of them can be studied as part of the formalism of Gazeau-Klauder (GK) \citep{GK99}. In that work, the authors presented an axiomatic approach to CS, developing a set of axioms to be fulfilled for a set of states to be called {\it coherent}. The four axioms introduced by GK are: normalization condition, continuity in the labels, resolution of the identity, and temporal stability.

On the other hand, according to Glauber the CS can be constructed following three different definitions: as eigenstates of the annihilation operator, as the action of the displacement operator on a preferred state (in this work they will be the {\it extremal} states), and finally as those which satisfy a given uncertainty principle.

In this work we will develop sets of CS for what we called {\it Painlev\'e IV Hamiltonian systems}, i.e., special cases of the $k$-SUSY partners of the harmonic oscillator which have always associated third-order differential ladder operators $l_k^\pm$, as we studied in chapter~\ref{cappain}, and that are closely related with $P_{IV}$. We will call these new sets as {\it Painlev\'e IV coherent states} (PIVCS).

Let us note that this is not a simple task because, as we saw in chapter~\ref{cappain}, these systems have a Hilbert space $\mathcal{H}$ formed by the sum of two subspaces: one of finite dimension, related with the semi-infinite ladder for the original spectrum of the harmonic oscillator and other one of finite dimension, associated with the new levels created by the $k$-SUSY transformation.

We will try to construct these new CS using the previous definitions but now with the ladder operators $l_k^\pm$ and we will see that neither of them works perfectly well. Nevertheless, under a modification of $l_k^\pm$ called {\it linearization}, which was inspired by \citet{FH99} (see also \citep{FHN94,FNR95}), we can obtain a suitable set of CS for the Hilbert space of the Painlev\'e IV Hamiltonian systems.

\section{Coherent states}\label{cs}
At the beginning of the study of quantum mechanics, \citet{Sch26} was interested in making a connection between the new science and classical mechanics. With this interest in mind he developed some quantum states, nowadays called coherent states, that restore the classical behaviour for the position operator of a quantum system \citep{AAG00}. Let the Hamiltonian of the system be given by
\begin{equation}
H=\frac{P^2}{2m}+V(Q),
\end{equation}
where $Q$ is the position and $P$ the momentum operators in the Schr\"odinger's picture. Then, the position operator $Q(t)$ in the Heisenberg picture evolves in the following way
\begin{equation}
Q(t)=\exp\left(\frac{i}{\hbar}Ht\right)Q\exp\left(-\frac{i}{\hbar}Ht\right).
\end{equation}
For Schr\"odinger, {\it classical behaviour} meant that the expectation value, or average, $\overline{q}(t)$ of $Q(t)$ would have to obey the equations of motion of classical mechanics
\begin{equation}
m\ddot{\overline{q}}(t)+\frac{\overline{\partial V}}{\partial q}=0.
\end{equation}

The first example of CS discovered by Schr\"odinger is given for the harmonic oscillator potential
\begin{equation}
V(q)=\frac{1}{2}m\omega^2q^2,
\end{equation}
i.e., one of the easier problems to solve in quantum mechanics. The CS will be labeled as $\ket{z}$, $z\in\mathbb{C}$, and are defined in such a way that we recover the classical sinusoidal solution given by
\begin{equation}
\bra{z}Q(t)\ket{z}=Q_0|z|\cos(\omega t - \varphi),
\end{equation}
where $z=|z|\exp(i\varphi)$ and $Q_0=(2\hbar/m\omega)^{1/2}$ is a fundamental quantum length built from the universal constant $\hbar$ and the constants $m,\omega$ that characterize the harmonic oscillator.

This is how Schr\"odinger defined the CS to make a smooth transition between classical and quantum mechanics. Nevertheless, one should always remember that the CS are truly quantum states which, however, allow us to make a classical reading of a quantum situation.

\subsection{Properties of CS}\label{defsCS}

The CS for the harmonic oscillator turned out to be really special when it was realized that they have properties that no other quantum system fulfill \citep{Sch26,Kla63a,Kla63b,Gla63a,Gla63b}, that is why they are called {\it canonical CS}. Its most important properties are the following \citep{AAG00}:

\begin{enumerate}
\item The states $\ket{z}$ saturate the Heisenberg inequality
\begin{equation}
\esp{\Delta Q}_z\esp{\Delta P}_z=\frac{1}{2}\hbar,
\end{equation}
where $\esp{\Delta Q_z} \equiv [\bra{z}Q^2\ket{z}-\bra{z}Q\ket{z}^2]^{1/2}$.

\item They are eigenstates of the annihilation operator $a^-$, with eigenvalue $z$:
\begin{equation}
a^-\ket{z}=z\ket{z},
\end{equation}
where $a^-=(2m\hbar\omega)^{-1/2}(m\omega Q+iP)$.

\item They are obtained from a  unitary action of the Heisenberg-Weyl group onto the ground state $\ket{0}$ of the harmonic oscillator. This is a key Lie group in quantum mechanics, whose Lie algebra is generated by $\{Q,P,\mathbb{I}\}$, with $[Q,P]=i\hbar\mathbb{I}$, so that
\begin{equation}
\ket{z}=\exp(za^+-\overline{z}a^-)\ket{0}.
\end{equation}

\item The set of CS $\{\ket{z}\}$ constitute an over-complete family of vectors in the Hilbert space of states for the harmonic oscillator. This property is encoded in the following resolution of the identity operator
\begin{equation}
\mathbb{I}=\frac{1}{\pi}\int_C \text{d}\,\text{Re}(z)\,\text{d}\,\text{Im}(z)\ket{z}\bra{z}.\label{p4reside}
\end{equation}
\end{enumerate}

These four properties are, to some extent, the basis of the many generalizations of the canonical CS. It is important to note that the four properties are all satisfied by the canonical CS, which set them apart from any other quantum systems and even among the wave packets of the harmonic oscillator.

\subsection{Gazeau-Klauder axioms}
In this section we will describe an axiomatic approach for families of states to be called {\it coherents}. This CS definition, which is based on the resolution of the identity in equation \eqref{p4reside}, will be given by the four Gazeau-Klauder (GK) axioms \citep{BHK12}

\begin{enumerate}
\item \emph{Normalization condition,}
\begin{equation}
\inner{z}{z}=1.
\end{equation}
\item \emph{Continuity in labels}, which means that as $\norm{z'-z}\rightarrow 0$, we have
\begin{equation}
\norm{\ket{z'}-\ket{z}}\rightarrow 0.
\end{equation}

\item \emph{Resolution of identity}, 
\begin{equation}
\int \ket{z}\bra{z}\text{d}\mu (z) = \mathbb{I}.
\end{equation}

\item \emph{Temporal stability}. The condition is
\begin{equation}
U(t)\ket{z}= \exp(-i\theta)\ket{z(t)},
\end{equation}
where $\exp(-i\theta)$ is an unimportant phase factor.
\end{enumerate}

\subsection{Generalized CS}
In this work, we have already discussed the usual ways of approaching the standard CS, i.e., Glauber's original three definitions, one to three of section \ref{cs} \citep{Gla63a,Gla63b}. Although the harmonic oscillator is important and useful, one often finds that other systems cannot be described in such a way. Therefore, it is natural to ask for some generalized CS that could be used to describe and analyze other quantum systems. Nevertheless, a natural question arises: how do we generalize the concept of CS for other quantum systems? It turns out that this question has been answered in different ways by different people as follows.

\begin{itemize}
\item {\it Generalized CS as minimum uncertainty states}. This is the original motivation of Schr\"odinger in his construction of 1926. The generalization along this direction was carried out by \citet{ACST74,ACS76} and \citet{NS78,NS79} and named {\it intelligent states}. This case has several limitations, as it can only be constructed for classically integrable systems described by the standard Lie algebras, which means that it must employ the usual creation and annihilation operators. Furthermore, the states obtained are not unique, since the so called {\it squeezed states} \citep{Per86} are also included.

\item {\it Generalized CS as eigenstates of the annihilation operator}. In this case a generalized definition of the annihilation operator is used. This approach was adopted by \citet{BG71} in their discussion about {\it new coherent states}. Nevertheless, this definition has some drawbacks: first, these CS cannot be defined in Hilbert spaces with finite dimension and second, they usually do not correspond to physically realizable states.

\item {\it Generalization based on the displacement operator}. In this case we generalize the displacement operator and also the states from which the CS are built of, i.e., we do not necessarily depart from the ground state but rather from an {\it extremal state}. This point of view was pursued by \citet{Gil72} and \citet{Per72,Per75} in terms of the so called {\it generalized CS}. Here, the choice of the displacement operator and the extremal states is fundamental.
\end{itemize}

\section{Painlev\'e IV coherent states} \label{secp4cs}
The CS for systems different from the harmonic oscillator are generated through several definitions, as we just reviewed in last section. Let us recall that the CS can be built up as eigenstates of the annihilation operator and also as the result of a certain displacement operator acting onto an extremal state.

In this section we will use the third-order ladder operators $l_k^\pm$ obtained in chapter~\ref{cappain} to generate families of CS. Recall that these operators appear for very specific systems, which are ruled by second-order PHA and are directly connected with solutions of $P_{IV}$. To accomplish this, we will divide the Hilbert space in two subspaces: one generated by the transformed eigenfunctions $\psi_n^{(k)}$ (or the equivalent eigenstates $\ket{n^k}$) associated with the original spectrum of the harmonic oscillator, whose subspace will be denoted as $\mathcal{H}_\text{iso}$; the other is generated by the eigenfunctions $\psi_{\epsilon_j}^{(k)}$ (or the equivalent eigenstates $\ket{\epsilon_j^k}$) associated with the new energy levels, which will be denoted as $\mathcal{H}_\text{new}$. 

In this chapter we will reverse the order of index $j$ in the factorization energies $\epsilon_j$, i.e., $\epsilon_j \rightarrow \epsilon_{k-j}$, in such a way that now we have $\epsilon_0 < \epsilon_1 < \cdots <\epsilon_{k-1}$. We will see that this notation is more appropriate for this chapter.

The action of $l_k^{\pm}$ on the eigenstates $\ket{n^k} \in \mathcal{H}_{\text{iso}}$ of the Hamiltonian $H_k$ is given by
\begin{subequations}
\begin{align}
l_k^{-}\ket{n^k}&=\sqrt{(E_n-E_0)(E_n-\epsilon_0)(E_{n}-\epsilon_0-k)}\ket{n-1^{k}}, \label{cslkiso}\\
l_k^{+}\ket{n^k}&=\sqrt{(E_{n+1}-E_0)(E_{n+1}-\epsilon_0)(E_{n+1}-\epsilon_0-k)}\ket{n+1^{k}}. \label{cslka}
\end{align}
\end{subequations}
Meanwhile, on $\ket{\epsilon_j^{k}} \in \mathcal{H}_{\text{new}}$ we have
\begin{subequations}
\begin{align}
l_k^{-}\ket{\epsilon_j^k}&=\sqrt{(\epsilon_j-\epsilon_0)(\epsilon_{j}-E_0)(\epsilon_j-\epsilon_k)}\ket{\epsilon_{j-1}^k}, \label{cslknew}\\
l_k^{+}\ket{\epsilon_j^k}&=\sqrt{(\epsilon_{j+1}-\epsilon_0)(\epsilon_{j+1}-E_0)(\epsilon_{j+1}-\epsilon_{k})}\ket{\epsilon_{j+1}^k},\label{cslk}
\end{align}\label{cslk2}
\end{subequations}
\hspace{-1.8mm}where $E_0$ and $\epsilon_0$ are the lowest energy levels of $H_k$ in each of the two subspaces $\mathcal{H}_\text{iso}$ and $\mathcal{H}_\text{new}$, respectively. Recall that we have $E_n=E_0+n$ and, with the new ordering, $\epsilon_j= \epsilon_0+j$. We should remark that we also get the correct results for the two extremal states $j=0$ and $j=k$ in $\mathcal{H}_\text{new}$. The index $k$ in a generic vector $\ket{a^k}$ indicates the same label as the Hamiltonian $H_k$. Thus, for the eigenvectors the label $a$ refers to the energy level, meanwhile for the CS we have $a=z\in\mathbb{C}$ and we still will write the index $k$, in order that we can distinguish the new CS from those of the harmonic oscillator $\ket{z}$. 

It is worth to note that with this convention we can have a confusion when $z=\epsilon_j$ or when $z=n$; nevertheless, we believe that through the context it will be clear the situation we are dealing with. From equations \eqref{cslkiso} and \eqref{cslknew} we can see that operator $l_k^-$ annihilates the eigenstates $\ket{0^k}$ and $\ket{\epsilon_0^k}$, while from equations~\eqref{cslk} and \eqref{cslka} $l_k^+$ only annihilates $\ket{\epsilon_{k-1}^k}$ (see figure~\ref{fig.tma2}).

\subsection{Painlev\'e IV coherent states from the annihilation operator}
Now we will generate the CS as eigenstates of the annihilation operator, which are usually defined as $a^-\ket{z}=z\ket{z}$. In this case the definition with the annihilation operators $l_k^-$ for the Painlev\'e IV Hamiltonian systems takes the form
\begin{equation}
l_k^-\ket{z^k}=z\ket{z^k}. \label{csAOCSP4}
\end{equation}
In principle, we could generate independent CS in the two subspaces $\mathcal{H}_\text{iso}$ and $\mathcal{H}_\text{new}$.

\subsubsection{PIVCS in the subspace $\mathcal{H}_{\text{iso}}$}
In order to find the PIVCS $\ket{z^k_\text{iso}}$ in this subspace, we need to express this state as a linear combination of the eigenvectors of $H_k$, i.e., $\left\{\ket{n^k},\, n=0,1,2, \dots \right\}$, which form a complete orthogonal set in $\mathcal{H}_\text{iso}$. Therefore
\begin{equation}
\ket{z^k_\text{iso}}=\sum_{n=0}^\infty c_n\ket{n^k},
\end{equation}
where the constants $c_n$ are still to be determined.

We have to check if they satisfy the four axioms of the GK approach to CS.

\begin{itemize}
\item \emph{Normalization condition.}
Applying $l_k^-$ on this expression and requiring that equation~\eqref{csAOCSP4} is fulfilled we finally obtain
\begin{equation}
\ket{z^k_\text{iso}}=c_0 \sum_{n=0}^{\infty}\frac{z^n}{\sqrt{n!}} \sqrt{\frac{\Gamma(E_0-\epsilon_0+1)\Gamma(E_0-\epsilon _0-k+1)}{\Gamma(E_0-\epsilon_0+1+n)\Gamma(E_0-\epsilon _0-k+1+n)}}\ket{n^k},\label{p4iso1}
\end{equation}
which depends only on the constant $c_0$. Without lost of generality we can choose it as real positive, and by normalization of $\ket{z_\text{iso}^k}$ we get that is given by
\begin{equation}
c_0= [{}_0F_2(E_0-\epsilon_0+1,E_0-\epsilon _0-k+1;|z|^2)]^{-1/2}, 
\end{equation} 
where ${}_pF_q$ is a generalized hypergeometric function, defined as
\begin{equation}
_pF_q (a_1,\dots, a_p; b_1, \dots, b_q,x)\equiv \sum_{n=0}^\infty  \frac{(a_1)_n \dots (a_p)_n}{(b_1)_n \dots (b_q)_n} \frac{x^n}{n!}.\label{p4iso3}
\end{equation}
We can also define an {\it auxiliary function} $c_0(a,b)$, which will be useful later on, as
\begin{equation}
c_0(a,b)=[{}_0F_2(E_0-\epsilon_0+1,E_0-\epsilon _0-k+1;a^{*}b)]^{-1/2}.
\end{equation}

\item \emph{Continuity of the labels.}
It is easy to check this axiom if we see that
\begin{equation}
\norm{\ket{z'^k_\text{iso}}-\ket{z^k_\text{iso}}}^2=\inner{z'^k_\text{iso}-z^k_\text{iso}}{z'^k_\text{iso}-z^k_\text{iso}}=2\left[1-\text{Re}(\inner{z'^k_\text{iso}}{z^k_\text{iso}}) \right].  
\end{equation}
We can write down the projection of two CS in the subspace, the so called {\it reproducing kernel}, using the auxiliary function $c_0(a,b)$, as
\begin{equation}
\inner{z'^k_\text{iso}}{z^k_\text{iso}}=\frac{c_0(z',z')c_0(z,z)}{c_0^2(z',z)}.
\end{equation}
This implies that in the limit $z' \rightarrow z$ its found that $\ket{z'^k_\text{iso}} \rightarrow \ket{z^k_\text{iso}}$. In figure~\ref{figproj2} we show a projection $|\inner{z'^k_\text{iso}}{z_\text{iso}^k}|$ as function of $z$ for a fixed $z'$. For the harmonic oscillator, this plot would be a Gaussian function, but in this case we find a deformation.
\begin{figure}\centering
\includegraphics[scale=0.28]{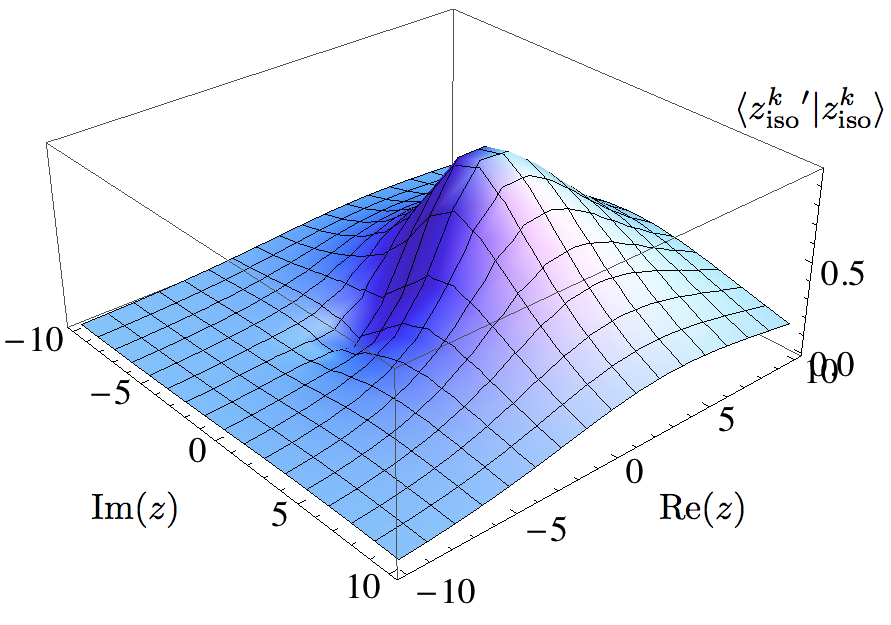}
\caption{\small{We show the absolute value of the projection of the CS $\ket{z_\text{iso}^k}$ which is an eigenstate of the annihilation operator $l_k^-$ to another CS $\ket{z'^k_\text{iso}}$ for $z'=5+i$, $\epsilon_0=-2$, and $k=2$.}}\label{figproj2}
\end{figure}

\item \emph{Resolution of the identity.}
We have to look for a function $\text{d}\mu(z)$ such that the following equation is fulfilled
\begin{equation}
\int\ket{z^k_\text{iso}}\bra{z^k_\text{iso}}\text{d}\mu (z)=\mathbb{I}_{\text{iso}}.\label{csidentity}
\end{equation}
To accomplish this we propose
\begin{equation}
\text{d}\mu (z)={}_0F_2(E_0-\epsilon_0+1,E_0-\epsilon_0-k+1;r)h(r)r\text{d}r\text{d}\theta,
\end{equation}
and if we insert this equation into \eqref{csidentity} and change $x=|z|^2=r^2$ we arrive to
\begin{equation}
\int_0^\infty x^n h(x)\text{d}x=\frac{\Gamma(E_0-\epsilon_0+1+n)\Gamma(E_0-\epsilon_0-k+1+n)\Gamma(n+1)}{\pi c_0^2\Gamma(E_0-\epsilon_0+1)\Gamma(E_0-\epsilon_0-k+1)}.
\end{equation}
This means that $h(r)$ is the {\it inverse Mellin transform} of the right-hand side of the last equation. It turns out that $h(x)$ is given in terms of the {\it Meijer} $G$ {\it function}, defined as
\begin{equation}
\MeijerG{m}{n}{p}{q}{a_1,\ldots,a_p}{b_1,\ldots,b_q}{x}\equiv \mathcal{M}^{-1}\left[\frac{\prod_{j=1}^{m}\Gamma(b_j+s)\prod_{j=1}^n\Gamma(1-a_j-s)}{\prod_{j=m+1}^{q}\Gamma(1-b_j-s)\prod_{j=n+1}^p\Gamma(a_j+s)};x\right],  
\end{equation}
with $m=3$, $n=0$, $p=0$, $q=3$, $b_1=0$, $b_2=E_0-\epsilon_0$, $b_3=E_0-\epsilon_0-k$, i.e.,
\begin{equation}
h(r)=\frac{\MeijerG{3}{0}{0}{3}{0,E_0-\epsilon_0,E_0-\epsilon_0-k}{}{r^2}}{\pi c_0^2\Gamma(E_0-\epsilon_0+1)\Gamma(E_0-\epsilon_0-k+1)}.  
\end{equation}
Notice that for $k=1$ we obtain the CS calculated by \citet{FH99}. This is because in that work, they calculated the CS using the $(2k+1)$th-order differential ladder operators $L_k^\pm$ from chapter~\ref{pha},f while now we are using the third-order ladder operators $l_k^\pm$ and they coincide for $k=1$. However, for $k>1$ these CS are different and completely new for the subspace $\mathcal{H}_{\text{iso}}$.

\item \emph{Temporal stability.}
We need to apply now the evolution operator to the CS in the subspace $\ket{z^k_{\text{iso}}}$. If we do this, we obtain
\begin{align}
\hspace{-2mm}U(t)\ket{z^k_{\text{iso}}}=&c_0\exp(-iH_kt)\sum_{n=0}^\infty\frac{z^n}{\sqrt{n!}}\sqrt{\frac{\Gamma(E_0-\epsilon_0+1)\Gamma(E_0-\epsilon _0-k+1)}{\Gamma(E_0-\epsilon_0+1+n)\Gamma(E_0-\epsilon _0-k+1+n)}}\ket{n^k}   \nonumber\\
=& \exp(-iE_0t)\ket{z\exp(-it)^k_\text{iso}}\equiv \exp(-iE_0t)\ket{z^k_\text{iso}(t)}.
\end{align}
This means that a CS evolves into another CS, up to a global phase factor.

\item \emph{State probability.}
It is also useful to calculate the probability $p_n(z)$ for a given CS $\ket{z^k_{\text{iso}}}$ that an energy measurement results in the value $E_n$. This probability $p_n(z)$ is given by
\begin{equation}
p_n(z) =|\inner{n^k}{z^k_\text{iso}}|^2=c_0^2\frac{|z|^{2n}}{n!}\frac{\Gamma(E_0-\epsilon_0+1)\Gamma(E_0-\epsilon _0-k+1)}{\Gamma(E_0-\epsilon_0+1+n)\Gamma(E_0-\epsilon _0-k+1+n)}.
\end{equation}\label{prob1}
\end{itemize}

This means that the states $\ket{z_\text{iso}^k}$ are a proper set of CS in the subspace $\mathcal{H}_{\text{iso}}$.

\subsubsection{PIVCS in the subspace $\mathcal{H}_{\text{new}}$}
The subspace $\mathcal{H}_{\text{new}}$ is $k$-dimensional, therefore, the operator $l_k^-$ can be represented as a $k \times k$ matrix with elements given by
\begin{equation}
(l_k^-)_{mn}=\inner{\epsilon_m^k}{l_k^-|\epsilon_n^k}.
\end{equation}
Then, from equation~\eqref{cslknew} we find that the only non-zero elements are in the so called {\it superdiagonal}, i.e., directly above of the main diagonal. Furthermore, it is straightforward to check that this matrix is {\it nilpotent}, in this case its $k$th-power is the {\it zero matrix}.

Now, multiplying the eigenvalue equation $(l_k^-) {\bf x} = z {\bf x}$, by $(l_k^-)^{k-1}$ we obtain
\begin{equation}
(l_k^-)^k {\bf x} = z^k {\bf x} = {\bf 0} \qquad \Rightarrow \qquad z=0,  
\end{equation}
which means that the only possible eigenvalue for the matrix $(l_k^-)$ is $z=0$. The same can be proven for $l_k^-$ and then, its only eigenvector is $\ket{\epsilon_0^k}$. Therefore, through this definition we cannot generate a family of CS in the subspace $\mathcal{H}_{\text{new}}$ that solves the identity operator. This is due to the finite dimension of this subspace.

\subsection{Painlev\'e IV coherent states from the displacement operator}
The CS defined as displaced versions of the ground state are not simple to generate for the $k$-SUSY partner potentials of the harmonic oscillator. The reason is that the commutator of $l_k^-$ and $l_k^+$ is no longer the identity operator. Therefore, if we make the changes $a^- \rightarrow l_k^-$ and $a^+ \rightarrow l_k^+$ in the displacement operator for the harmonic oscillator, it turns out that
\begin{equation}
\widetilde{D}(z) = \exp\left(z l_k^+ - z^*l_k^- \right) \neq \exp\left(-\frac{1}{2}|z|^2 \right) \exp \left(z l_k^+ \right) \exp  \left( -z^*l_k^- \right),
\end{equation}
since now $[l_k^-,l_k^+]=P_2(H_k)$ and therefore we cannot simply apply the Baker-Campbell-Hausdorff formula to separate the exponentials. For that reason, we decide to propose instead the operator already separated from the very beginning, i.e., this last expression is going to be taken as the displacement operator for the new systems,
\begin{equation}
D(z)=\exp\left(-\frac{1}{2}|z|^2 \right) \exp \left(z l_k^+ \right) \exp  \left( -z^*l_k^- \right).
\end{equation}

Let us recall that for the harmonic oscillator, the ground state is annihilated by $a^-$. As we mentioned in section~\ref{cs}, a generalization of this procedure consists in using not the ground state but an extremal state, i.e., a non-trivial eigenstate of $H_k$ belonging as well to the kernel of the annihilation operator $l_k^-$. In this case, the extremal states are $\ket{0^k}$ for the subspace $\mathcal{H}_\text{iso}$ and $\ket{\epsilon^k_0}$ for $\mathcal{H}_\text{new}$.

\subsubsection{PIVCS in the subspace $\mathcal{H}_{\text{iso}}$}
Let us apply the newly defined displacement operator $D(z)$ on the extremal state $\ket{0^k}\in\mathcal{H}_{\text{iso}}$, also adding a normalization constant $C_z$ for convenience,
\begin{equation}
\ket{z^k_{\text{iso}}}= C_z D(z) \ket{0^k}= C_z \exp\left(-\frac{1}{2}|z|^2 \right)\sum_{n=0}^\infty \frac{(zl_k^+)^n}{n!}\ket{0^k}. \end{equation}
After several simplifications we have
\begin{equation}
\ket{z^k_{\text{iso}}}=C_z \exp\left(-\frac{1}{2}|z|^2 \right) \sum_{n=0}^\infty \frac{z^n}{\sqrt{n!}}\left[\prod_{m=1}^n\sqrt{(m+E_0-\epsilon_0)(m+E_0-\epsilon_0-k)}\right]\ket{n^k}.  
\end{equation}
At first sight one could think that this is a right set of CS in this subspace. Nevertheless, if we analyze its normalization it is found that
\begin{align}
\inner{z_\text{iso}^k}{z_\text{iso}^k}=&|C_z|^2 \exp\left(-|z|^2 \right)\sum_{n=0}^\infty\frac{|z|^{2n}}{n!}\prod_{m=1}^n (m+E_0-\epsilon_0)(m+E_0-\epsilon_0-k) \nonumber \\
=&|C_z|^2 \exp\left(-|z|^2 \right)\, {}_2F_0(E_0+1-\epsilon_0,E_0+1-\epsilon_0-k;|z|^2).
\end{align}
The fact that it is written in terms of the generalized hypergeometric function ${}_2F_0(E_0+1-\epsilon_0,E_0+1-\epsilon_0-k;|z|^2)$ shows that the norm can be equal to $1$ only when $z=0$, but it diverges for all $z\neq 0\in\mathbb{C}$ \citep{Erd53}. Therefore, the only normalized CS that we obtain when we apply this displacement operator on the extremal state in the subspace $\mathcal{H}_{\text{iso}}$ is precisely the extremal state $\ket{0^k}$. For $z\neq 0$ we obtain an expression that does not correspond to any vector in the Hilbert space of the system.

\subsubsection{PIVCS in the subspace $\mathcal{H}_{\text{new}}$}
In the subspace $\mathcal{H}_{\text{new}}$, we apply now the displacement operator $D(z)$ on the extremal state $\ket{\epsilon_0^k}$, which is also annihilated by the operator $l_k^-$. This will lead us to
\begin{equation}
\ket{z_\text{new}^k}= N_z' \exp\left(-\frac{|z|^2}{2}\right) \left[\sum_{j=0}^{k-1}\left( \prod_{i=1}^j \sqrt{(E_0-\epsilon_0-i)(k-i)}\right) \frac{z^j}{\sqrt{j!}} \ket{\epsilon_j^k}  \right]. \label{csdisp1}
\end{equation}
We can absorb factor $\exp(-|z|^2/2)$ in the constant $N_z'$ to redefine it as $N_z$, and also use the definition of Pochhammer symbols
\begin{equation}
(x)_n\equiv x(x+1)(x+2)\dots(x+n-1)=\frac{\Gamma(x+n)}{\Gamma(x)},
\end{equation}
to rewrite
\begin{subequations}
\begin{align}
\prod_{i=1}^j(k-i)&=\frac{\Gamma(k)}{\Gamma(k-j)} = (k-j)_j,\\
\prod_{i=1}^j(E_0-\epsilon_0-i)&=\frac{\Gamma(E_0-\epsilon_0)}{\Gamma(E_0-\epsilon_0-j)}=(E_0-\epsilon_0-j)_j.
\end{align}
\end{subequations}
Then we have
\begin{equation}
\ket{z_\text{new}^k}= N_z \left[\sum_{j=0}^{k-1}\sqrt{(E_0-\epsilon_0-j)_j (k-j)_j} \frac{z^j}{\sqrt{j!}} \ket{\epsilon_j^k}  \right].\label{csdisp3}
\end{equation}

Once we have obtained this explicit expression for the states $\ket{z_\text{new}^k}\in\mathcal{H}_\text{new}$, we can check now if they satisfy the GK axioms for being called coherent states.

\begin{itemize}
\item \emph{Normalization condition.}
In this case we do not have any problem with the normalization, because the involved sum is finite. Without lost of generality we can choose $N_z$ to be real positive such that
\begin{equation}
N_z=\left[\sum_{j=0}^{k-1}\frac{|z|^{2j}}{j!}(E_0-\epsilon_0-j)_j (k-j)_j\right]^{-1/2}.\label{csdisp2}
\end{equation}

\item \emph{Continuity of the labels.}
We prove it similarly as for the annihilation operator CS
\begin{equation}
\norm{\ket{z'^k_\text{new}}-\ket{z^k_\text{new}}}^2=\inner{z'^k_\text{new}-z^k_\text{new}}{z'^k_\text{new}-z^k_\text{new}}=2\left[1-\text{Re}(\inner{z'^k_\text{new}}{z^k_\text{new}}) \right].  
\end{equation}
We redefine now the normalization factor $N_z$ to be the more general function
\begin{equation}
N(a,b)=\left[\sum_{j=0}^{k-1}\frac{(a^*b)^{j}}{j!}(E_0-\epsilon_0-j)_j (k-j)_j \right]^{-1/2}.
\end{equation}
Substituting the definition of $N(a,b)$ in equations \eqref{csdisp1} and \eqref{csdisp2}, we obtain for the inner product
\begin{equation}
\inner{z'^k_\text{new}}{z^k_\text{new}}=\frac{N(z',z')N(z,z)}{N^2(z',z)}.
\end{equation}
This means that in the limit $z' \rightarrow z$ it is found that $\ket{z'^k_\text{new}} \rightarrow \ket{z^k_\text{new}}$. 

\item \emph{Resolution of the identity.}
In this case we should show that
\begin{equation}
\int\ket{z_\text{new}^k}\bra{z_\text{new}^k}\text{d}\mu (z)=\mathbb{I}_{new}.
\end{equation}
By plugging the expression of equation \eqref{csdisp3} for the CS, it turns out that
\begin{equation}
\mathbb{I}_{\text{new}}=2\pi\sum_{j=0}^{k-1}\frac{\ket{\epsilon_j^k}\bra{\epsilon_j^k}}{j!}\frac{\Gamma(E_0-\epsilon_0)}{\Gamma(E_0-\epsilon_0-j)}\frac{\Gamma(k)}{\Gamma(k-j)}\int_0^\infty N_z^2 r^{2j+1}\mu (r)\text{d}r,
\end{equation}
where we have used polar coordinates and integrate the angle variable. Then we propose that 
\begin{equation}
\mu(r)=\frac{g(r)}{\pi N_z^2\Gamma(E_0-\epsilon_0)\Gamma(k)}.
\end{equation}
Now we can change $x=r^2$ and $s=j+1$ to obtain the condition on $g(r)$ as
\begin{equation}
\int_0^\infty x^{s-1}g(x)\text{d}x=\Gamma(s)\Gamma(1+k-s)\Gamma(1+E_0-\epsilon_0-s).
\end{equation}
Then we get $g(r)$ as a Meijer $G$ function,
\begin{equation}
g(r)=\MeijerG{1}{2}{2}{1}{-k,\epsilon_0-E_0}{0}{r^2}.
\end{equation}

\item \emph{Temporal stability.}
If we apply the evolution operator to a CS in the subspace $\mathcal{H}_{\text{new}}$ we obtain
\begin{align}
U(t)\ket{z^k_{\text{new}}}&=\exp(-iH_kt)N_z \sum_{j=0}^{k-1}\sqrt{(E_0-\epsilon_0-j)_j(k-j)_j}\frac{z^j}{\sqrt{j!}}\ket{\epsilon_j^k}  \nonumber\\
&=\exp(-i\epsilon_0 t)\ket{z\exp(-it)^k_{\text{new}}}\equiv \exp(-i\epsilon_0t)\ket{z^k_{\text{new}}(t)}.
\end{align}
We can see that one of these CS always evolves into another CS in this subspace, up to a global phase factor.
\end{itemize}

\subsection{Painlev\'e IV linearized coherent states}
Until now, we have seen that the definition of CS as eigenstates of the annihilation operator $l_k^-$ works appropriately only for $\mathcal{H}_{\text{iso}}$ and the one associated to the displacement operator $D(z)$ only for $\mathcal{H}_{\text{new}}$, i.e., no definition allows us to obtain sets of CS in the two subspaces $\mathcal{H}_{\text{iso}}$ and $\mathcal{H}_{\text{new}}$ simultaneously when we use the third-order ladder operators $l_k^\pm$. Nevertheless, we still have the alternative to {\it linearize} these operators $l_k^\pm$, defining some new ones as
\begin{subequations}\begin{align}
\ell_k^{+}&\equiv \sigma(H_k)l_k^{+},   \\
\ell_k^{-}&\equiv \sigma(H_k+1)l_k^{-},\label{csellk}
\end{align}\label{p4linell}\end{subequations}
\hspace{-1mm}with
\begin{equation}
\sigma(H_k)=[(H_k-\epsilon_0)(H_k-\epsilon_0-k)]^{-1/2},  
\end{equation}
where, by convention we take the positive square root. The infinite-order differential ladder operators $\ell_k^\pm$ are defined through their action onto the basis of $\mathcal{H}_{\text{iso}}$ and $\mathcal{H}_\text{new}$. We must remark that, although we use the notation $\ell_k^\pm$, they are not related with the fourth-order ladder operators that will appear in chapter \ref{5painleve} related with the $k$-th order SUSY partners of the radial oscillator.

It would seem more natural to define $\ell_k^{-}$ as $(\ell_k^{+})^\dag$, but in such a case we would not have the right action when applied to $\ket{\epsilon_0^k}$, i.e., on this eigenstate the action of the two alternative definitions is different. However, with the new ladder operators of equations \eqref{p4linell} the action on the eigenvectors of the Hamiltonian $H_k$ is strongly simplified. This linearization process has been previously applied to the general SUSY partners of the harmonic oscillator in order to obtain some CS using different annihilation operators \citep{FHN94,FNR95,FH99, FHR07}.

\subsubsection{PIVCS in the subspace $\mathcal{H}_{\text{iso}}$}
The action of the new ladder operators on the eigenvectors of $H_\text{iso}$ is given by
\begin{subequations}
\begin{align}
\ell_k^{-}\ket{n^k} & =\sqrt{n}\ket{n-1^k},\\
\ell_k^{+}\ket{n^k} & =\sqrt{n+1}\ket{n+1^k}.  
\end{align}
\end{subequations}
Now we can define the analogous of the number operator in $\mathcal{H}_{\text{iso}}$ as $N_{\text{iso}}\equiv \ell_k^{+}\ell_k^{-}$, given that $N_{\text{iso}}\ket{n^k}=n\ket{n^k}$. Furthermore, we can easily show that the operators $\{\ell_k^{+},\ell_k^{-},H_k\}$ obey the following commutation rules
\begin{equation}
[\ell_k^{-},\ell_k^{+}]=\mathbb{I}_{\text{iso}},\quad [H_k,\ell_k^{\pm}]=\pm\ell_k^{\pm},\label{csalgebra}
\end{equation}
where $\mathbb{I}_{\text{iso}}$ is the identity operator in the subspace $\mathcal{H}_{\text{iso}}$. Equations~\eqref{csalgebra} mean that the linearized ladder operators satisfy a Heisenberg-Weyl algebra on $\mathcal{H}_\text{iso}$.

Now, to generate a set of linearized CS using the displacement operator, we define an analogous operator as
\begin{equation}
\mathcal{D}(z)=\exp\left(z \ell_k^+ - z^*\ell_k^- \right) = \exp\left(-\frac{1}{2}|z|^2 \right) \exp \left(z \ell_k^+ \right) \exp  \left( -z^*\ell_k^- \right),\label{lineD}
\end{equation}
i.e., with the linearized ladder operators $\ell_k^\pm$ instead of the $l_k^\pm$. It is worth to remark that the equality given by this equation is valid as long as it is applied in the subspace $\mathcal{H}_\text{{iso}}$. Then, the CS are given by
\begin{equation}
\ket{z^k_{\text{iso}}}= \mathcal{D}(z)\ket{0^k}=\exp\left(-\frac{|z|^2}{2}\right)\sum_{n=0}^{\infty}\frac{z^n}{\sqrt{n!}}\ket{n^k},\label{linstates}
\end{equation} 
which means that when we use the linearized ladder operator we obtain an expression similar to the CS of the harmonic oscillator. The difference rely in the states that are added. For the harmonic oscillator the states are the eigenstates of $H_0$, while in this case the added states are the eigenstates of $H_k$ in $\mathcal{H}_{\text{iso}}$. Next, let us analyze some properties of this new set of CS.

\begin{figure}\centering
\includegraphics[scale=0.4]{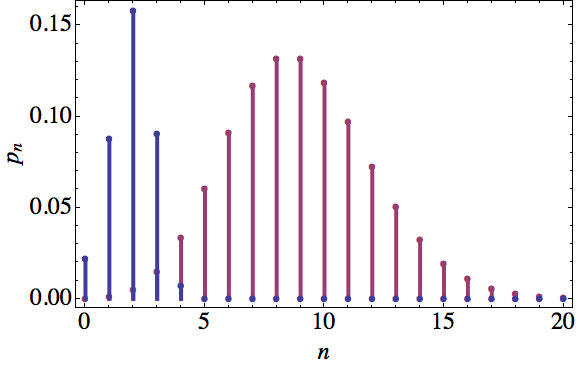}
\caption{\small{Probability distribution for an energy mesurement to obtain $n+1/2$ in a system described by a CS $\ket{z^k_{\text{iso}}}$ constructed through the annihilation operator (blue), equation~\eqref{prob1} for $z=3$ and $\epsilon_0=-3$, and by means of the linearized displacement operator, equation~\eqref{prob2} (magenta) for $z=3$.}}\label{figcs1}
\end{figure}

\begin{itemize}
\item \emph{Normalization condition.} The states of equation \eqref{linstates} are already normalized.

\item \emph{Continuity of the labels.}
The proof that $\ket{z'^k_{\text{iso}}} \rightarrow \ket{z^k_{\text{iso}}}$ when $z' \rightarrow z$ is the same as for the CS of the annihilation operator in $\mathcal{H}_{\text{iso}}$.

\item \emph{Resolution of the identity.}
To prove the identity resolution we follow the same procedure as for the harmonic oscillator to obtain
\begin{equation}
\frac{1}{\pi}\iint \ket{z^k_{\text{iso}}} \bra{z^k_{\text{iso}}}\, \text{d}\, \text{Re}(z)\, \text{d}\, \text{Im}(z)= \mathbb{I}_{\text{iso}}.
\end{equation}
In this way we assure that all vector that belong to $\mathcal{H}_{iso}$ can be expressed in terms of these CS.

\item \emph{Temporal stability.}
When the evolution operator is applied to the CS $\ket{z^k_{\text{iso}}}$ we obtain
\begin{align}
U(t)\ket{z^k_{\text{iso}}}&=\exp(-iH_kt)\exp\left(-\frac{|z|^2}{2}\right)\sum_{n=0}^\infty\frac{z^n}{\sqrt{n!}}\ket{n^k} \nonumber\\
&=\exp(-iE_0t)\exp\left(-\frac{|z|^2}{2}\right)\sum_{n=0}^\infty\frac{[z\exp(-it)]^n}{\sqrt{n!}}\ket{n^k} \nonumber\\
&=\exp(-iE_0t)\ket{z\exp(-it)^k_\text{iso}}\equiv \exp(-iE_0t)\ket{z^k_\text{iso}(t)}.
\end{align}
This means that one of these CS evolves in another CS, up to a global phase factor.


\item \emph{States probability}
If we have a particle in a CS $\ket{z_{\text{iso}}^k}$ and we perform an energy measurement, the system has a probability $p_n(z)$ of getting the value $n+1/2$. This probability is given by
\begin{equation}
p_n(z)=|\inner{n^k}{z_{\text{iso}}^k}|^2=e^{-|z|^2} \frac{|z|^{2n}}{n!},\label{prob2}
\end{equation}
which is a Poisson distribution with mean value given by $|z|^2$. This transition probability is plotted in figure~\ref{figcs1}, where we show the transition probability for these states and also for the states that were developed in a previous section as eigenstates of the annihilation operator $l_k^-$ in $\mathcal{H}_\text{iso}$.
\end{itemize}

\subsubsection{PIVCS in the subspace $\mathcal{H}_{\text{new}}$}
The action of the linearized ladder operators $\ell_k^\pm$ onto the basis of $\mathcal{H}_{\text{new}}$, $\left\{ \ket{\epsilon_j^k} , j=0,\dots,\right.$ $\left.k-1 \right\}$, is given by
\begin{subequations}\begin{align}
\ell_k^{+}\ket{\epsilon_j^k}&=(1-\delta_{j,k-1})\sqrt{\epsilon_{j+1}-E_0}\ket{\epsilon_{j+1}^k} , \\ \ell_k^{-}\ket{\epsilon_j^k}&=(1-\delta_{j,0})\sqrt{\epsilon_{j}-E_0}\ket{\epsilon_{j-1}^k},
\end{align}\end{subequations}
which come from the definition of $\ell_k^\pm$ in equations \eqref{p4linell}. It turns out that $\ell_k^-$ annihilates the eigenstate $\ket{\epsilon_0^k}$, while $\ell_k^+$ annihilates $\ket{\epsilon_{k-1}^k}$. If we operate now the commutator $\conm{\ell_k^-}{\ell_k^+}$ on the same basis we obtain
\begin{subequations}\begin{align}
[\ell_k^{-},\ell_k^{+}]\ket{\epsilon_0^k}&=(\epsilon_0+1-E_0)\ket{\epsilon_0^k},   \\
[\ell_k^{-},\ell_k^{+}]\ket{\epsilon_j^k}&=\ket{\epsilon_j^k},\quad \quad j=1,2,\dots ,k-2,   \\
[\ell_k^{-},\ell_k^{+}]\ket{\epsilon_{k-1}^k}&=(E_0+1-\epsilon_0-k)\ket{\epsilon_{k-1}^k},   
\end{align}\end{subequations}
i.e., in the subspace $\mathcal{H}_{\text{new}}$ the operator $\conm{\ell_k^-}{\ell_k^+} \neq \mathbb{I}_{\text{new}}$, due to its action on the states $\ket{\epsilon_0^k}$ and $\ket{\epsilon_{k-1}^k}$. Nevertheless, if we use the same analogous displacement operator as for the subspace $\mathcal{H}_\text{iso}$, given by equation~\eqref{lineD}, and we apply it to the ground state $\ket{\epsilon_0^k}$, we obtain
\begin{equation}
\ket{z_\text{new}^k}=C_z' D(z)\ket{\epsilon_0^k}=C_z' \sqrt{\Gamma(E_0-\epsilon_0)}\exp\left(-\frac{|z|^2}{2}\right)\sum_{j=0}^{k-1}\frac{(iz)^j}{j!}\frac{1}{\sqrt{\Gamma(E_0-\epsilon_0-j)}}\ket{\epsilon_j^k},  \label{csketk2}
\end{equation}
where $C_z'$ is a normalization constant, which we can choose to include the factor $\sqrt{\Gamma(E_0-\epsilon_0)}\exp(-|z|^2/2)$ so that
\begin{equation}
\ket{z_\text{new}^k}=C_z \sum_{j=0}^{k-1}\frac{(iz)^j}{j!}\sqrt{\frac{1}{\Gamma(E_0-\epsilon_0-j)}}\ket{\epsilon_j^k}. \label{csketk}
\end{equation}
These CS satisfy the following properties:

\begin{itemize}
\item \emph{Normalization condition.} The normalization constant $C_z$ can be chosen as real and positive, without lost of generality. Then
\begin{equation}
C_z=\left[ \sum_{j=0}^{k-1}\frac{|z|^{2j}}{(j!)^2}\frac{1}{\Gamma(E_0-\epsilon_0-j)}   \right]^{-1/2}.  
\end{equation}

\item \emph{Continuity of the labels.}
In order to check this property, we can see that
\begin{equation}
\norm{\ket{z'^k_\text{new}}-\ket{z^k_\text{new}}}^2=\inner{z'^k_\text{new}-z^k_\text{new}}{z'^k_\text{new}-z^k_\text{new}}=2\left[1-\text{Re}(\inner{z'^k_\text{new}}{z^k_\text{new}}) \right].  
\end{equation}
To simplify notation, we can define a complex function $C(a,b)$ as
\begin{equation}
C(a,b)= \left[\sum_{j=0}^{k-1} \frac{(a^*b)^j}{(j!)^2} \frac{1}{\Gamma(E_0-\epsilon_0-j)}\right]^{-1/2}.  
\end{equation}
Therefore
\begin{equation}
\inner{z'^k_\text{new}}{z^k_\text{new}}=\frac{C(z',z')C(z,z)}{C^2(z',z)},  
\end{equation}
which implies that in the limit $z' \rightarrow z$ it is found that $\ket{z'^k_\text{new}} \rightarrow \ket{z^k_\text{new}}$. 

\item \emph{Resolution of the identity.}
Recall that this property refers to the fulfillment of the following expression
\begin{equation}
\int_\mathbb{C} \ket{z^k_\text{new}} \bra{z^k_\text{new}}\mu(z)\text{d}z=\mathbb{I}_\text{new},
\end{equation}
where $\mu(z)$ is a positive definite function to be found. If we substitute the expression $\ket{z^k_\text{new}}$ given by equation \eqref{csketk}, write $z$ in polar coordinates, and integrate the angular one we obtain
\begin{equation}
\mathbb{I}_{\text{new}}= 2\pi\sum_{j=0}^{k-1}\frac{\ket{\epsilon_j^k} \bra{\epsilon_j^k}}{(j!)^2\Gamma(E_0-\epsilon_0-j)}\int_0^\infty C_z^2 r^{2j+1}\mu(r)\text{d}r.  
\end{equation}
In order to simplify the last equation, we introduce the function $f(r)$ as
\begin{equation}
\mu(r)=\frac{f(r)}{\pi C_z^2}, \label{csmu}
\end{equation}
in such a way that the following equation must be fulfilled
\begin{equation}
2\int_0^\infty r^{2j+1}f(r)\text{d}r=\Gamma^2(j+1)\Gamma(E_0-\epsilon_0-j).  
\end{equation}
With the change of variable $r^2=x$ and of index $j=s-1$ we obtain
\begin{equation}
\int_0^\infty x^{s-1}f(x)\text{d}x  =\Gamma^2(s)\Gamma(E_0+1-\epsilon_0-s)\equiv \mathcal{M}[f(x);s],  
\end{equation}
where $\mathcal{M}$ is the Mellin transform that we saw earlier in this chapter. Now we need to find the {\it inverse Mellin transform}
\begin{equation}
f(x)=\mathcal{M}^{-1}[\Gamma^2(s)\Gamma(E_0+1-\epsilon_0-s);x]. \label{cseqf}
\end{equation}
It is possible to find several inverse Mellin transforms in tables, for example in \citet{Erd54}. In this case, the function $f(x)$ of equation~\eqref{cseqf} turns out to be a Meijer $G$ function with $m=2,\, n=1,\, p=1,\, q=2,\, a_1=\epsilon_0-E_0,\, b_1=b_2=0$, i.e.,
\begin{equation}
f(r)=\MeijerG{2}{1}{1}{2}{\epsilon_0-E_0}{0,\, 0}{r^2}.  
\end{equation}
Even more, in \citet{Erd53} one can find some expressions for the Meijer $G$ function in terms of other special functions, in particular of the Whittaker function $W_{\kappa,\mu}(z)$, which in turn can be written in terms of the logarithmic solution of the confluent hypergeometric function $U(a,c;z)$ \citep{AS72}. Then we have
\begin{equation}
f(r)=\Gamma^2(E_0+1-\epsilon_0) U(E_0+1-\epsilon_0,1;r^2).
\end{equation}
 
We still need to prove the positiveness of $\mu(z)$. To accomplish this, we follow the work by \citet{SP00}, where a similar problem is solved using the {\it convolution property} for the inverse Mellin transform, also called {\it generalized Parseval formula}, which is given by
\begin{align}
\mathcal{M}^{-1}[g^*(s)h^*(s);x] &=\frac{1}{2\pi i}\int_{-i\infty}^{i\infty}g^*(s)h^*(s)x^{-s}\text{d}x \nonumber\\ &=\int_0^{\infty}g(xt^{-1})h(t)t^{-1}\text{d}t.\label{parseval2}
\end{align}
In this way, if we choose $g^*(s)=\Gamma(s)\Gamma(E_0+1-\epsilon_0-s)$ and using the following equation \citep{Erd54} 
\begin{equation}
\mathcal{M}^{-1}[\Gamma(\alpha+s)\Gamma(\beta-s)]=\Gamma(\alpha+\beta)x^\alpha(1+x)^{\alpha-\beta},  
\end{equation}
it is obtained
\begin{equation}
g(x)=\Gamma(E_0+1-\epsilon_0)(1+x)^{\epsilon_0-E_0-1}.  
\end{equation}
Now, if we employ $h^*(s)=\Gamma(s)$, from the definition of the Gamma function we have
\begin{equation}
h(x)=\exp(-x),
\end{equation}
and using the generalized Parseval formula from equation \eqref{parseval2} it turns out that
\begin{align}
f(x)&=\int_0^{\infty}\Gamma(E_0+1-\epsilon_0)(1+xt^{-1})^{\epsilon_0-E_0-1}\exp(-t)t^{-1}\text{d}t   \nonumber\\
&=\Gamma(E_0+1-\epsilon_0)\int_0^\infty t^{E_0-\epsilon_0}(t+x)^{\epsilon_0-E_0-1}\exp(-t)\text{d}t.  
\end{align}
If we replace this last result in equation~\eqref{csmu} we obtain
\begin{equation}
\mu(r)=\frac{\Gamma(E_0+1-\epsilon_0)}{\pi C_z^2} \int_0^\infty t^{E_0-\epsilon_0}(t+r^2)^{\epsilon_0-E_0-1}\exp(-t)\text{d}t.
\end{equation}
Besides, taking into account that $E_0>\epsilon_0$, and that the domain of $r$ and $t$ is $[0,\infty)$ we can conclude that we have found, at least, a positive definite measure, i.e, this CS do resolve the identity in the subspace $\mathcal{H}_{\text{new}}$. 

\item \emph{Temporal stability.}
If we apply the evolution operator to a CS in the subspace $\mathcal{H}_{\text{new}}$ we obtain
\begin{align}
U(t)\ket{z^k_{\text{new}}}&=C_z \sum_{j=0}^{k-1}\frac{(iz)^j}{j!}\sqrt{\frac{1}{\Gamma(E_0-\epsilon_0-j)}}\exp(-iH_kt)\ket{\epsilon_j^k} \nonumber\\
&=\exp(-i\epsilon_0t)C_z \sum_{j=0}^{k-1}\frac{[iz\exp(-it)]^j}{j!}\sqrt{\frac{1}{\Gamma(E_0-\epsilon_0-j)}}\ket{\epsilon_j^k} \nonumber\\
&=\exp(-i\epsilon_0t)\ket{z\exp(-it)^k_{\text{new}}}\equiv \exp(-i\epsilon_0t)\ket{z^k_{\text{new}}(t)}.
\end{align}
We can see that one of these CS always evolves into another CS in this subspace, up to a global phase factor.

\item \emph{State probability.}
Now we can easily calculate the probability $p_j(z)$ that, if the system is in a CS $\ket{z^k_{\text{new}}}$, an energy measurement will give as a result the eigenvalue $\epsilon_j$, namely,
\begin{equation}
p_j(z)=|\inner{\epsilon_j^k}{z^k_\text{new}}|^2 =\frac{|z|^{2n}}{(n!)^2} \frac{1}{\Gamma(E_0-\epsilon_0-n)} \left[ \sum_{s=0}^{k-1} \frac{|z|^{2s}}{(s!)^2} \frac{1}{\Gamma(E_0-\epsilon_0-s)} \right]^{-1}.  
\end{equation}
\end{itemize}

\section{Conclusions}
In this chapter we have studied the CS for the special kind of Hamiltonian systems that are connected with $P_{IV}$ through second-order PHA. To do that, first we did a brief introduction to the CS. Then, using the third-order ladder operators, characteristic for these systems we employed two definitions to obtain the corresponding CS. We have built up the CS as eigenstates of the annihilation operator, as arising from the displacement operator action, and as linearized CS from the corresponding displacement operator. We also showed that they fulfill the four Gazeau-Klauder axioms.

We must remember that the Painlev\'e IV Hamiltonian systems have two energy ladders, one semi-infinite starting from $E_0=1/2$, and one finite that starts from $\epsilon_0$ and has $k$ levels, where $k$ is the order of the SUSY transformation used to generate the potential. Thus it is natural that the system is described by two subspaces: one generated by the eigenstates that are isospectral to the harmonic oscillator, denoted as $\mathcal{H}_{\text{iso}}$, and another one generated by the eigenstates associated with the new levels, denoted as $\mathcal{H}_{\text{new}}$.

For the PIVCS which are eigenstates of the annihilation operator $l_k^-$, we were able to obtain a suitable set of CS only in subspace $\mathcal{H}_{\text{iso}}$. For the PIVCS arising from the displacement operator action, with $D(z)$ built from $l_k^\pm$, we find a suitable set only in the complementary subspace $\mathcal{H}_{\text{new}}$. Finally, for the linearized PIVCS arising from the corresponding displacement operator action, we have found good sets of CS in both subspaces $\mathcal{H}_{\text{iso}}$ and $\mathcal{H}_{\text{new}}$, i.e., through the same definition we were able to find CS for the whole space $\mathcal{H}=\mathcal{H}_{\text{iso}}\oplus\mathcal{H}_{\text{new}}$.

We must remark that the sets of CS that were found for the separated subspaces with different definitions are also good ones. Indeed, we can even define a new set of CS for the complete $\mathcal{H}$ if we use both definitions simultaneously, i.e., use the PIVCS which are eigenstates of the annihilation operator $l_k^-$ for $\mathcal{H}_{\text{iso}}$ and the PIVCS arising from the displacement operator associated with $l_k^\pm$ for $\mathcal{H}_{\text{new}}$ and they will generate any state in $\mathcal{H}$.
\chapter{Painlev\'e V equation}
\label{5painleve}

In this chapter we will study the Painlev\'e V equation, given by
\begin{equation}
w''=\left(\frac{1}{2w}+\frac{1}{w-1}\right)(w')^2-\frac{w'}{z}+\frac{(w-1)^2}{z^2}\left(aw+\frac{b}{w}\right)+c\frac{w}{z}+d\frac{w(w+1)}{w-1}.\label{PVa}
\end{equation}
As we can see, the solutions $w(z)$ of $P_V$ will depend on the four parameters of the equation, i.e., $w=w(a,b,c,d;z)$. The functions $w(z)$ that solve the equation are called {\it Painlev\'e V trascendents}. These will be new {\it special functions} in the sense that, in general, they cannot be written in terms of other special functions. Nevertheless, for special values of the parameters $a,b,c,d$, they can actually be expressed in terms of other special functions. Solutions of this type which are found in the literature can be given in terms of rational functions or of special functions as Laguerre polynomials, Hermite polynomials, Weber functions, Bessel functions, among others. In this chapter we will obtain explicit solutions to $P_V$ initially expressed in terms of the more general special function ${}_1F_1$, the confluent hypergeometric function. Then we will obtain some of the solution families related with other special functions to compare them with the ones existent in the literature. Throughout this chapter we will call these families {\it solution hierarchies}.

Let us note that $P_{V}$ appears in several areas of physics and mathematics, e.g., it is used in the study of correlation functions in condense matter \citep{Kan02}, in the analysis of Maxwell-Bloch systems in electrodynamics \citep{Win92}, and in the symmetry reduction for the stimulated Raman scattering in solid state \citep{Lev92}.

\section{Reduction theorem for fourth-order ladder operators}\label{sectmaRO}
As we study in chapter \ref{pha}, systems ruled by second-order PHA are connected with $P_{IV}$. After that, in chapter \ref{cappain} we prove that there is a reduction theorem for some even-order PHA, which are associated with a subset of SUSY partners of the harmonic oscillator. We show that under certain conditions, those algebras are reduced to second-order PHA, generated by third-order ladder operators. Now we will do the same for the odd-order PHA associated with the SUSY partners of the radial oscillator. These algebras will be reduced here to third-order PHA, which are generated by fourth-order ladder operators.

To accomplish that, in this section we will prove another {\it reduction theorem} through which we will identify the special higher-order SUSY partners of the radial oscillator, normally ruled by a $(2k+1)$th-order algebra, which are also ruled by a third-order one. In this chapter we will use $j$ as the angular momentum index in order to free $\ell$, which is going to be used as the reduced ladder operator and it is not related with the linearized ladder operators for the Painlev\'e IV Hamiltonian systems developed in chapter \ref{p4cs}. We also stop writing explicitly the dependence of the radial oscillator Hamiltonian $H_0$, its eigenvalues $E_n$, and its ladder operators $b^\pm$ on the angular momentum index.\\

\noindent {\bf Theorem.} Let $H_k$ be the $k$th-order SUSY partner of the radial oscillator Hamiltonian $H_0$ generated by $k$ Schr\"odinger seed solutions. These solutions $u_i$ are connected by the annihilation operator of the radial oscillator $b^-$ as
\begin{equation}
u_i = (b^{-})^{i-1} u_1, \quad \epsilon_i = \epsilon_1 - (i-1), \quad i=1,\dots,k, \label{restrRO}
\end{equation}
where $u_1(x)$ is a Schr\"odinger solution without zeroes, given by equation~\eqref{solRO} for $\epsilon_1 < E_0=j/2+3/4$ and
\begin{equation}
\nu_1\geq -\frac{\Gamma\left(\frac{1-2j}{2}\right)}{\Gamma\left(\frac{1-2j-4\epsilon_1}{4}\right)}.
\end{equation}
Therefore, the natural $(2k+2)$th-order ladder operator $L_k^+ = B_k^{+} b^{+} B_k^{-}$ of $H_k$ turn out to be factorized in the form
\begin{equation}
L_k^+ = P_{k-1}(H_k) \ell_k^+,\label{hipoRO}
\end{equation}
where $P_{k-1}(H_k) = (H_k - \epsilon_1)\dots(H_k - \epsilon_{k-1})$ is a polynomial of $(k-1)$th-order in $H_k$ and $\ell_k^+$ is a fourth-order differential ladder operator,
\begin{equation}
[H_k,\ell_k^+] = \ell_k^+, \label{conmHlRO}
\end{equation}
such that
\begin{equation} \label{annumk3RO}
\ell_k^+ \ell_k^- =\left(H_k -E_0 \right)\left(H_k + E_0 -1 \right)(H_k - \epsilon_k)(H_k - \epsilon_1-1).
\end{equation}
\medskip

\noindent {\bf Proof (by induction).} The proof is similar as for the third-order ladder operators of section \ref{sectma}, the only changes are the order of the operators and the energies of the extremal states that appear in the analogous of the number operator.

For $k=1$ the result is direct
\begin{equation}
 L_1^+ = P_0(H_1)\ell_1^+ , \quad P_0(H_1) = 1.
\end{equation}

Let us suppose now that the theorem is valid for a given $k$ (induction hypothesis) and then we are going to show that it is also valid for $k+1$.
\begin{align}
\mbox{Hypothesis} 				\quad \ \ \quad & \quad \quad \ \ \mbox{To be shown} \nonumber\\
L_{k}^{+}=P_{k-1}(H_k)\ell_{k}^{+}\quad & \quad L_{k+1}^{+}=P_{k}(H_{k+1})\ell_{k+1}^{+} 
\end{align}

From the intertwining technique it is clear that we can go from $H_k$ to $H_{k+1}$ and vice versa through a first-order SUSY transformation
\begin{equation}
H_{k+1} A_{k+1}^+ = A_{k+1}^+ H_k, \quad H_kA_{k+1}^{-}=A_{k+1}^{-}H_k.
\end{equation}
Moreover, it is straightforward to show that
\begin{equation}
L_{k+1}^+ = A_{k+1}^+ L_{k}^+ A_{k+1}^- ,\label{LentreRO}
\end{equation}
whose action can be seen in figure~\ref{fig.sigL} from chapter \ref{cappain}.

From the induction hypothesis one obtains
\begin{equation}
L_{k+1}^{+}  = A_{k+1}^{+}P_{k-1}(H_k)\ell_{k}^{+}A_{k+1}^{-}
				 	 = P_{k-1}(H_{k+1})\underbrace{A_{k+1}^{+}\ell_{k}^{+}A_{k+1}^{-}}_{\widetilde{\ell}_{k+1}^{+}},\label{plRO}
\end{equation}
where
\begin{equation}
\widetilde{\ell}_{k+1}^{+}\equiv A_{k+1}^{+}\ell_{k}^{+}A_{k+1}^{-},\label{l52RO}
\end{equation}
is a sixth-order differential ladder operator for $H_{k+1}$. A direct calculation leads to
\begin{equation}
\widetilde{\ell}_{k+1}^+ \widetilde{\ell}_{k+1}^- = (H_{k+1} - \epsilon_k)^2 \left(H_{k+1} -E_0 \right)\left(H_{k+1} + E_0 -1 \right)(H_{k+1} - \epsilon_{k+1})(H_{k+1} - \epsilon_1-1).
\end{equation}
Note that the last four terms of this equation are precisely the result that would be obtained from the product $\ell_{k+1}^+ \ell_{k+1}^-$ of the fourth-order ladder operators of $H_{k+1}$. Thus, it is concluded that
\begin{equation}
\widetilde{\ell}_{k+1}^+ = q(H_{k+1}) \ell_{k+1}^+,
\end{equation}
where $q(H_{k+1})$ is a polynomial of $H_{k+1}$. By remembering that $\widetilde{\ell}_{k+1}^+, \ell_{k+1}^+$, and $H_{k+1}$ are differential operators of sixth-, fourth-, and second-order respectively, one can conclude that $q(H_{k+1})$ is linear in $H_{k+1}$, and we already know that $\epsilon_k$ is a root of $q(H_{k+1})$, therefore
\begin{equation}
\widetilde{\ell}_{k+1}^+ = (H_{k+1} - \epsilon_k) \ell_{k+1}^+ .\label{N52RO}
\end{equation}
By substituting this result in equation~\eqref{plRO} we finally obtain
\begin{equation}
L_{k+1}^+ = P_{k-1}(H_{k+1})(H_{k+1} - \epsilon_k) \ell_{k+1}^+ = P_{k}(H_{k+1})\ell_{k+1}^+.
\end{equation}
With this we conclude our proof. \hfill $\square$

\section{Operators $\ell_k^{+}$}\label{secope5}
In this section we will prove some properties of the newly defined operators $\ell_k^{\pm}$. First of all, operators $L_k^\pm$ act exactly as the ones defined in chapter~\ref{cappain} for the SUSY partner potentials of the harmonic oscillator, which were shown in figure~\ref{fig.tma1}. They connect as usual ladder operators the eigenstates belonging to the original part of the spectrum $E_n$, but annihilate all the eigenstates for the newly created levels at $\epsilon_i$.

\begin{figure}\centering
\includegraphics[scale=0.35]{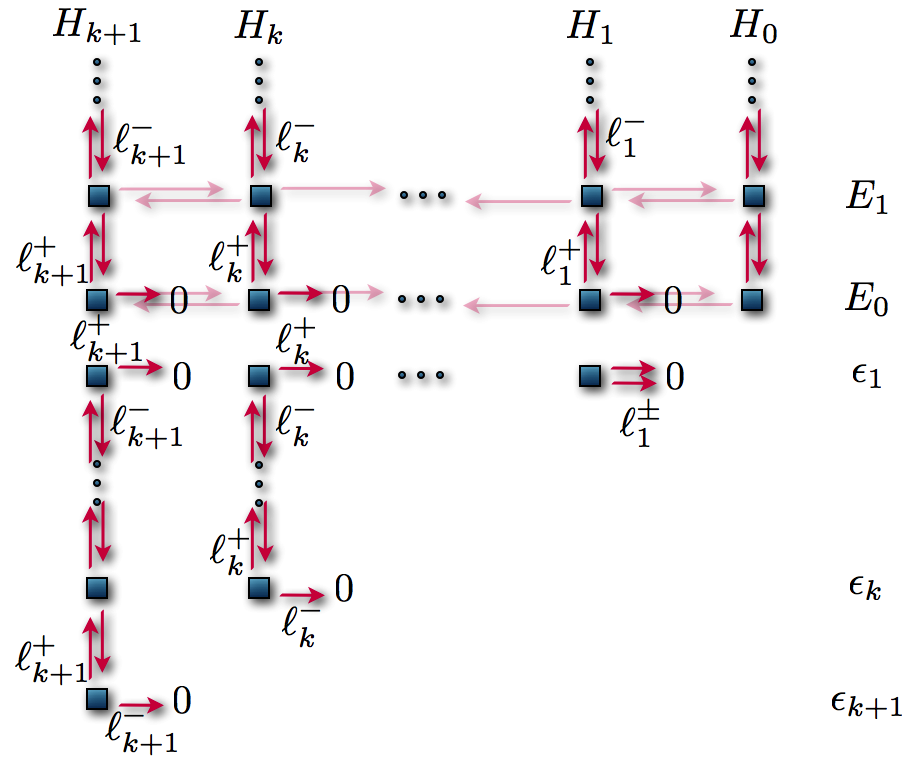}
\caption{\small{Action of the operators $\ell_{k}^{\pm}$ over the eigenstates of the SUSY Hamiltonians $H_k$. Note that $\ell_1^{\pm}=L_1^{\pm}$, and one can see here that the operator $\ell_{k}^{-}$ always annihilate the eigenstates associated with $E_0$ and $\epsilon_k$, while $\ell_{k}^{+}$ annihilates the one associated with $\epsilon_1$.}}\label{fig.tma2RO}
\end{figure}

On the other hand, the new operators $\ell_k^\pm$ do actually allow the displacement between the eigenstates of the finite ladder. In the physical ladders, $\ell_{k}^{-}$ only annihilates the eigenstates associated with $E_0$ (the initial ground energy level) and $\epsilon_k$ (the new ground energy level), and $\ell_{k}^{+}$ annihilates only the new eigenstate with the highest-energy $\epsilon_1$. A diagram representing the action of the fourth-order ladder operators $\ell_k^{\pm}$ on all the eigenstates of the new SUSY Hamiltonians $H_k$ is shown in figure~\ref{fig.tma2RO}.

\subsection{Relation with $A_{k+1}^{+}$}
\begin{figure}\centering
\includegraphics[scale=0.35]{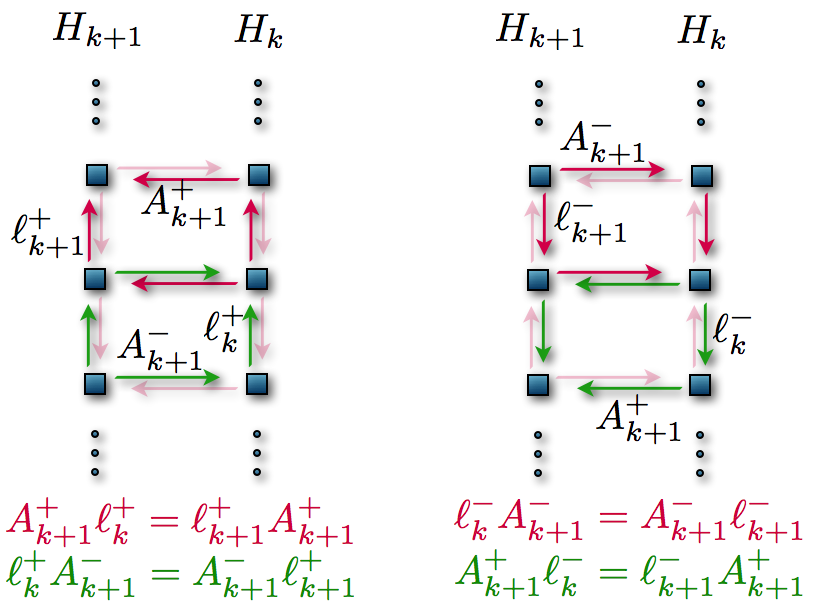}
\caption{\small{Action of the operators from equations~\eqref{finalAlRO} over the eigenstates of the SUSY Hamiltonians $H_k$ and $H_{k+1}$. We can see that this relations allow the displacement to any level of the spectrum, including the new eigenstates.}}\label{fig.tma4RO}
\end{figure}

In analogy to the relations in equations~\eqref{finalAl} for the operators $l_k^\pm$, we can obtain similar equations for the operators $\ell_k^{\pm}$. In the same way as in section~\ref{secAl} we obtain
\begin{subequations}
\begin{align}
A_{k+1}^{+}\ell_{k}^{\pm} &= \ell_{k+1}^{\pm}A_{k+1}^{+},\\
\ell_{k}^{\pm}A_{k+1}^{-} &= A_{k+1}^{-}\ell_{k+1}^{\pm}.
\end{align}\label{finalAlRO}
\end{subequations}
\hspace{-1.5mm}Note that these four relations are general, i.e., they can be applied to any eigenstate in the physical ladders, including those that are annihilated by any operator. A full diagram can be seen in figure~\ref{fig.tma4RO}.

\subsection{Operator $\ell_{k}^{+}\ell_k^{-}$}
If we apply the analogue of the number operator $b^{+}b^{-}$ for the radial oscillator Hamiltonian $H_0$ on its eigenstates $|\psi_{n}^{(0)}\rangle$, we effectively multiply them by a second-order polynomial in $n$, i.e.,
\begin{equation}
b^{+}b^{-}\ket{\psi_{n}^{(0)}} = n(n+2E_0-1) \ket{\psi_{n}^{(0)}}.
\end{equation}

Also, with the natural $(2k+2)$th-order ladder operators $L_k^{\pm}$ for the SUSY Hamiltonians $H_k$ we can define a new analogue of the number operator $L_k^{+}L_k^{-}$, whose action on the eigenstates $\ket{\psi_{n}^{(k)}}$ associated with $E_n$ and the eigenstates $\ket{\psi_{\epsilon_j}^{(k)}}$ for the new levels is
\begin{subequations}
\begin{align}
L_k^{+}L_k^{-}\ket{\psi_{n}^{(k)}} &=\left[ n(n+2E_0-1) \prod_{i=1}^{k}\left(n+E_0-\epsilon_i \right)\left(n+E_0-\epsilon_i-1\right)\right] \ket{\psi_{n}^{(k)}} ,\\
L_k^{+}L_k^{-}\ket{\psi_{\epsilon_j}^{(k)}} &= 0.
\end{align}
\end{subequations}
Recall that Sp$(H_k)=\{ \epsilon_k,  \epsilon_{k-1},\dots,  \epsilon_1,E_0,E_1,\dots \}$, then we can see that for the eigenstates $\ket{\psi_{n}^{(k)}}$ associated with the original eigenvalues, the operator $L_k^{+}L_k^{-}$ multiplies them by a $(2k+2)$th-order polynomial of $n$ and annihilates the eigenstates for the new energy levels $\ket{\psi_{\epsilon_j}^{(k)}}$.

Now, departing from the new fourth-order ladder operators $\ell_k^{\pm}$ we can define yet another analogue of the number operator as $\ell_k^{+}\ell_k^{-}$. Its action on the eigenstates of $H_k$ is now
\begin{subequations}
\begin{align}
\ell_k^{+}\ell_k^{-} | \psi_{n}^{(k)}\rangle &=n(n+2E_0-1)\left(n+E_0-\epsilon_1 -1\right)\left(n+E_0-\epsilon_k\right) | \psi_{n}^{(k)}\rangle , \\
\ell_k^{+}\ell_k^{-} | \psi_{\epsilon_i}^{(k)}\rangle &= (\epsilon_i-E_0)(\epsilon_i+E_0-1)(\epsilon_i -\epsilon_1 -1)(\epsilon_i-\epsilon_k) | \psi_{\epsilon_i}^{(k)}\rangle .
\end{align}
\end{subequations}
We should note that the only physical eigenstates that $\ell_k^{+}\ell_k^{-}$ annihilates are the ones associated with the old ground state energy $E_0$ and the new lower level $\epsilon_k$.

\subsection{Consequences of the theorem}\label{secnuevasgRO}
In section~\ref{sectmaRO} we have proven a theorem that establishes the conditions under which the following factorization is fulfilled
\begin{equation}
L_{k}^{+}=P_{k-1}(H_k)\ell_{k}^{+}.\label{facttmaRO}
\end{equation}

With this factorization, the natural ladder operators $L_k^\pm$ can be written as a product of a $(k-1)$th-order polynomial of the SUSY Hamiltonians $H_k$ and the fourth-order ladder operator $\ell_k^\pm$. This means that the $(2k+1)$th-order PHA, obtained through a SUSY transformation as specified in the theorem with $\epsilon_i=\epsilon_1-(i-1),\ i=1,\dots ,k$, can be \emph{reduced} to a third-order PHA with fourth-order ladder operators.

Recall from chapter~\ref{pha} that these algebras are closely related to $P_{V}$. This means that when we reduce the higher-order algebras, we open the possibility of obtaining new solutions of $P_{V}$, similar to the case of second-order PHA and $P_{IV}$. In the following sections we will explain the method we developed to obtain solutions of $P_V$ and then classify them into solution hierarchies.

\section{Real solutions to $P_V$ with real parameters}

\subsection{First-order SUSY QM}
If we calculate the first-order SUSY partners of the radial oscillator, we get a system naturally ruled by a third-order PHA. We obtain now the solutions of $P_V$ following \citet{CFNN04}. To do that, we need to identify the extremal states of $H_1$ and its energies. From the spectrum of the radial oscillator we have already two extremal states, one physical associated with $E_0=j/2+3/4$ and one non-physical with $-E_0+1$. The other two roots are added by the SUSY transformation, i.e., one is the new level at $\epsilon_1$ and the other is $\epsilon_1+1$ to have a finite ladder. Then

\begin{subequations}
\begin{alignat}{3}
\psi_{{\cal E}_1} & \propto A_1^+b^+u, &\quad {\cal E}_1 & = \epsilon_1+1,\\
\psi_{{\cal E}_2} & \propto A_1^+ \left[x^{-j} \exp(-x^2/4)\right], & \quad  {\cal E}_2 & = -E_0+1,\\
\psi_{{\cal E}_3} & \propto u^{-1}, & \quad  {\cal E}_3 & = \epsilon_1,\\
\psi_{{\cal E}_4} & \propto A_1^+ \left[x^{j+1} \exp(-x^2/4)\right], & \quad  {\cal E}_4 & = E_0,
\end{alignat}\label{extremalp5}
\end{subequations}
\hspace{-1mm}where $A_1$ is the first-order intertwining operator and $b^+$ is the ladder operator for the radial oscillator.

\begin{figure}\centering
\includegraphics[scale=0.37]{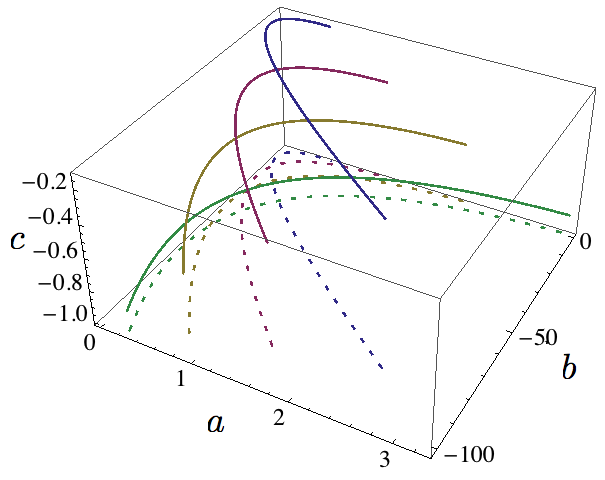}
\caption{\small{Parameters space for the solution of $P_V$. We vary the parameters $\epsilon_1\in (5/4,-10)$ and $j=1,2,3,4$; with colors blue, magenta, yellow, and green; respectively. The dashed lines are projections of the solid lines in the bottom of the box for visual help.}}\label{paraRO}
\end{figure}

For this system we are able to connect with $P_V$ and specific parameters $a,b,c,d\in\mathbb{C}$. From equations \eqref{paraPV}, \eqref{eeps} and \eqref{extremalp5} we obtain $a,b,c,d$ in terms of one parameter of the original system $E_0$ and one of the SUSY transformation $\epsilon_1$ as
\begin{equation}
a=\frac{(E_0+\epsilon_1)^2}{2},\quad b=-\frac{(E_0-\epsilon_1)^2}{2},\quad c=\frac{1-2E_0}{2},\quad d=-\frac{1}{8}.
\end{equation}
Actually, $E_0=E_0(j)=j/2+3/4$. Then we can write the parametrization as
\begin{equation}
a=\frac{(2j+4\epsilon_1+3)^2}{32},\quad b=-\frac{(2j-4\epsilon_1+3)^2}{32},\quad c=-\frac{2j+1}{4},\quad d=-\frac{1}{8}.
\end{equation}
In general, the four parameters $a,b,c,d$ are written in terms of $j\in\mathbb{R}^+$ and $\epsilon_1\in\mathbb{C}$, although in this section we study the case where both, the $P_V$ parameters and the factorization energy, are real, i.e., $a,b,c,d,\epsilon_1\in\mathbb{R}$. In the next sections we will study the complex case. We must remark that usually in the physical studies of the radial oscillator systems $j\in\mathbb{Z}^+$, as it is the angular momentum index, but in this case we use this system as an auxiliary system in order to obtain solutions to $P_V$, that is why we rather use the generalized radial oscillator with $j\in\mathbb{R}^+$. In figure~\ref{paraRO}, we show a parametric plot of the three parameters $a,b,c$ (remember that $d$ is constant) as function of $j$ and $\epsilon_1$. Along these curves in the parameter space we can find solutions of $P_V$, indeed a one-parametric family of solutions for each one of those points.

If we restrict ourselves to non-singular real solutions of $P_V$ with real parameters we also have the restriction $\epsilon_1\leq E_0=j/2+3/4$. Moreover, for each one of those points we have a one-parameter family of solutions, labelled by the parameter $\nu_1$ from equation~\eqref{solRO} under the restriction from \eqref{condRO}.

\begin{figure}\centering
\includegraphics[scale=0.37]{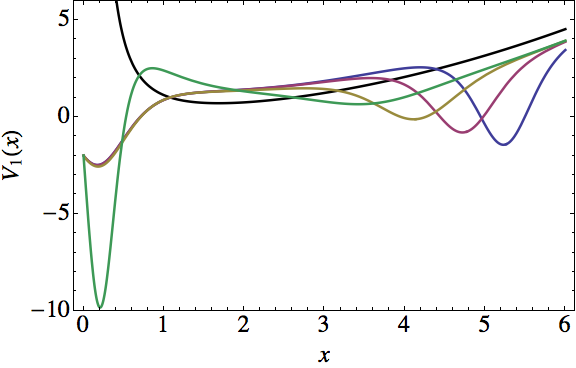}
\includegraphics[scale=0.37]{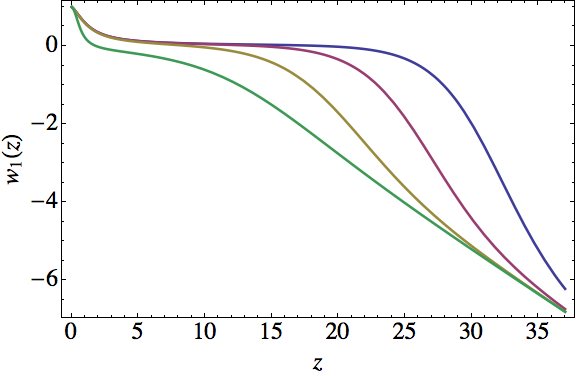}
\caption{\small{SUSY partner potential $V_1(x)$ of the radial oscillator (black) (left) and the solutions $w_1(z)$ to $P_V$ for $j=1$, $\epsilon_1=1$, and $\nu_1=\{$0.905 (blue), 0.913 (magenta), 1 (yellow), 10 (green)$\}$.}}\label{pv1}
\end{figure}

Then from 1-SUSY we can obtain the following partner potential and the function $g(x)$ related with the solution $w(z)$ of $P_V$
\begin{subequations}
\begin{align}
V_1(x)&=\frac{x^2}{8}+\frac{j(j+1)}{2x^2}-[\ln u(x)]'',\\
g_1(x)&=-\frac{x}{2}-\frac{j+1}{x}+[\ln u(x)]',
\end{align}
\end{subequations}
where we have added an index to indicate the order of the SUSY transformation. Since $g_1(x)$ is connected with the solution $w_1(z)$ of $P_V$ through
\begin{equation}
w_1(z)=1+\frac{z^{1/2}}{g_1(z^{1/2})},
\end{equation}
then
\begin{equation}
w_1(z)=1+\frac{2zu(z^{1/2})}{2z^{1/2}u'(z^{1/2})-(z+2j+1)u(z^{1/2})}.
\end{equation}
An illustration of the first-order SUSY partner potentials of the radial oscillator $V_1(x)$ and the corresponding solutions $w_1(z)$ of $P_V$ are shown in figure~\ref{pv1}.

\subsection{$k$th-order SUSY QM}
With the reduction theorem for the fourth-order ladder operators of section~\ref{sectmaRO} we are able to reduce the ($2k+1$)th-order PHA induced by the natural ladder operators for the SUSY partners of the radial oscillator to third-order PHA. Basically, the $k$ transformation functions have to be connected through the annihilation operator $b^-$ and therefore their energies will be given by $\epsilon_i=\epsilon_1-(i-1)$. This means that we will add one equidistant ladder with $k$ steps, one for each first-order SUSY transformation. There is also a restriction on the factorization energy, i.e., $\epsilon_1<E_0$. 

Once again, we need to identify the extremal states of our system. The roots of the polynomial in equation~\eqref{annumk3RO} are
$E_0, -E_0 +1, \epsilon_k, \epsilon_1+1$, two of them are physical extremal states, the ones associated with $E_0$ and $\epsilon_k$, a non-physical one coming from the radial oscillator at $-E_0+1$, and another non-physical that will make the new ladder finite, $\epsilon_1+1$. The four extremal states are thus
\begin{subequations}
\begin{alignat}{3}
\psi_{{\cal E}_1} & \propto B_k^+b^+u_1, & \quad {\cal E}_1 & = \epsilon_1+1,\\
\psi_{{\cal E}_2} & \propto B_k^+ \left[x^{-j} \exp(-x^2/4)\right], & \quad  {\cal E}_2 & = -E_0+1,\\
\psi_{{\cal E}_3} & \propto \frac{W(u_1,\dots ,u_{k-1})}{W(u_1,\dots ,u_k)}, & \quad  {\cal E}_3 & = \epsilon_k,\\
\psi_{{\cal E}_4} & \propto B_k^+ \left[x^{j+1} \exp(-x^2/4)\right], & \quad  {\cal E}_4 & = E_0.
\end{alignat}\label{extremalp5k}
\end{subequations}
\hspace{-1mm}From appendix B, we can write $\psi_{{\cal E}_4} $ as
\begin{equation}
\psi_{{\cal E}_4} \propto B_k^+ \left[x^{j+1} \exp(-x^2/4)\right]\propto \frac{W(u_1,\dots , u_k,x^{j+1}\exp(-x^2/4))}{W(u_1,\dots , u_k)}.
\end{equation}
From equation \eqref{ache} we can obtain the auxiliary function $h(x)$ as
\begin{equation}
h(x)=\left\{\ln\left[W(\psi_{\mathcal{E}_3},\psi_{\mathcal{E}_4})\right]\right\}',
\end{equation}
and then from \eqref{solPV} one arrives at
\begin{equation}
g(x)=-x-h(x)=-x-\left\{\ln\left[W(\psi_{\mathcal{E}_3},\psi_{\mathcal{E}_4})\right]\right\}'.
\end{equation}

Therefore, the $k$th-order SUSY partner potential $V_k(x)$ of the radial oscillator and its corresponding $g_k(x)$ function are
\begin{subequations}
\begin{align}
V_k(x)&=\frac{x^2}{8}+\frac{j(j+1)}{2x^2}-[\ln W(u_1,\dots , u_k)]'',\\
g_k(x)&=-x+\frac{2(E_0+\epsilon_1-k)W(u_1,\dots ,u_{k-1})W(u_1,\dots ,u_{k},x^{j+1}\exp(-x^2/4))}{W\left(W(u_1,\dots ,u_{k-1}), W(u_1,\dots ,u_{k},x^{j+1}\exp(-x^2/4))\right)}.\label{VgkROb}
\end{align}\label{VgkRO}
\end{subequations}
\hspace{-1mm}Recall that $g_k(x)$ is directly related with the function $w_k(z)$ through
\begin{equation}
w_k(z)=1+\frac{z^{1/2}}{g_k(z^{1/2})},\label{p5k}
\end{equation}
which is a solution to $P_V$ with parameters given by
\begin{equation}
a=\frac{(E_0+\epsilon_1)^2}{2},\quad b=-\frac{(E_0-\epsilon_1+k-1)^2}{2},\quad c=\frac{k-2E_0}{2},\quad d=-\frac{1}{8}.
\end{equation}

\begin{figure}\centering
\includegraphics[scale=0.37]{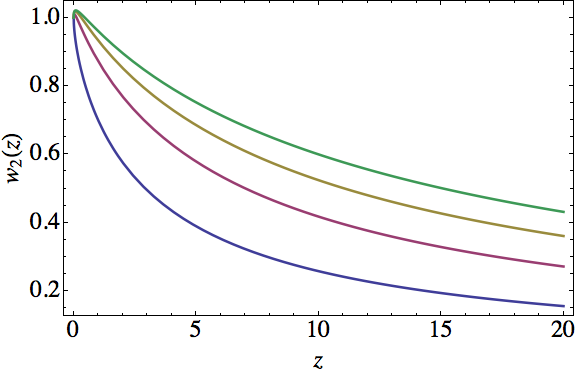}
\includegraphics[scale=0.37]{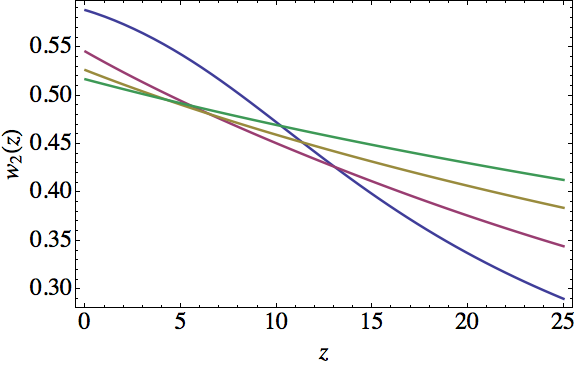}
\caption{\small{$P_V$ solutions $w_2$ generated by the second-order SUSY QM. The first plot is for the parameters $j=0$, $\nu_1=0$, and $\epsilon_1=\{1/4$ (blue), $-3/4$ (magenta), $-7/4$ (yellow), $-11/4$ (green)$\}$. The second plot is for $\epsilon_1=0$, $\nu_1=0$, and $j=\{$1 (blue), 3 (magenta), 6 (yellow), 10 (green)$\}$.}}\label{solskp5}
\end{figure}

\begin{figure}\centering
\includegraphics[scale=0.37]{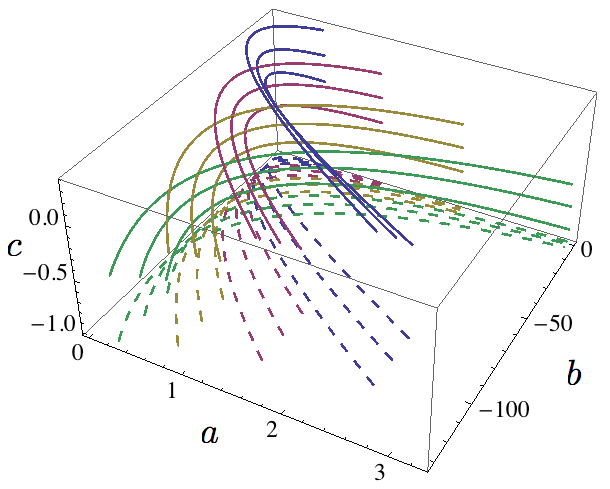}
\caption{\small{Solution parameters space for $P_V$ through $k$ SUSY QM. We vary $\epsilon_1\in (5/4,-10)$ and $j=1,2,3,4$ with colors blue, magenta, yellow, green; respectively. We show the first three solution families for $k=1,2,3$. The dashed lines are projections of the solid lines in the bottom of the box for visual help.}}\label{paraRO2}
\end{figure}

In figure~\ref{solskp5} we show some $P_V$ solutions $w_2(z)$ obtained with the second-order SUSY transformation. Furthermore, with the $k$th-order SUSY QM we are able to expand the solution space $(a,b,c)$ by the inclusion of $k$. In figure~\ref{paraRO2} we show a plot of the solution space for $k=1,2,3$.

\section{Complex solutions to $P_V$ with real parameters}
We can use the theorem proven in section~\ref{sectmaRO} even with complex transformation functions. The simplest way to implement them is to use a complex linear combination of two standard linearly independent real solutions with a complex constant $\lambda+i\kappa$, as
\begin{align}
u(x,\epsilon)=\, & x^{-j}\text{e}^{-x^2/4}\left[{}_1F_1\left(\frac{1-2j-4\epsilon}{4},\frac{1-2j}{2};\frac{x^2}{2}\right) \right.\nonumber\\
&+\left.(\lambda+i\kappa)\left(\frac{x^2}{2}\right)^{j+1/2}{}_1F_1\left(\frac{3+2j-4\epsilon}{4},\frac{3+2j}{2};\frac{x^2}{2}\right)\right].\label{solRO2}
\end{align}
The result for the real solution given in equation~\eqref{solRO} is accomplished with the conditions
\begin{equation}
\lambda=\nu \frac{\Gamma\left(\frac{3+2j-4\epsilon}{4}\right)}{\Gamma\left(\frac{3+2j}{2}\right)},\quad \kappa=0.
\end{equation}

Compared with the case when we were only looking for real solutions, we have two restrictions that can now be surpassed. The first of them is the restriction $\epsilon_1<E_0$, and the second one is that we had to choose our extremal states in the order of equation \eqref{extremalp5k}. Now we can perform permutations on the indices of the extremal states and we still do not obtain singularities, because in general $\psi_{\mathcal{E}_i}\neq 0$ in the complex plane. Therefore, the solution space is even bigger for the complex solutions. 

In equation \eqref{paraPV} we have the four parameters of $P_V$ in terms of the four extremal states but we also have symmetry in the exchanges $\mathcal{E}_1 \leftrightarrow \mathcal{E}_2$ and $\mathcal{E}_3 \leftrightarrow \mathcal{E}_4$. Thus from the $4!=24$ possible permutations of the four indexes we have six different solutions to $P_V$. Next we show the six solution families in terms of $\epsilon_1$, $j$, and $k$. We have added an index to distinguish them
\begin{subequations}
\begin{alignat}{5}
a_1 & =\frac{(2j+4\epsilon_1+3)^2}{32}, & \ b_1 & =-\frac{(2j-4\epsilon_1+4k-1)^2}{32}, & \  c_1 & =\frac{-2j+2k-3}{4},\\
a_2 & =\frac{(2j+4\epsilon_1-4k+3)^2}{32}, & \ b_2 & =-\frac{(2j-4\epsilon_1-1)^2}{32}, & \  c_2 & =-\frac{2j+2k+1}{4},\\
a_3 & =\frac{(2j-4\epsilon_1+4k-1)^2}{32}, & \ b_3 & =-\frac{(2j+4\epsilon_1+3)^2}{32}, & \  c_3 & =\frac{2j-2k-1}{4},\\
a_4 & =\frac{(2j-4\epsilon_1-1)^2}{32}, & \ b_4 & =-\frac{(2j+4\epsilon_1- 4k+3)^2}{32}, & \  c_4 & =\frac{2j+2k+1}{4},\\
a_5 & =\frac{k^2}{2}, & \ b_5 & =-\frac{(2j+1)^2}{8}, & \  c_5 & =\frac{2\epsilon_1-k}{2},\\
a_6 & =\frac{(2j+1)^2}{8}, & \ b_6 & =-\frac{k^2}{2}, & \  c_5 & =-\frac{2\epsilon_1+k-1}{2}.
\end{alignat}\label{4paraRO} 
\end{subequations}

\begin{figure}\centering
\includegraphics[scale=0.37]{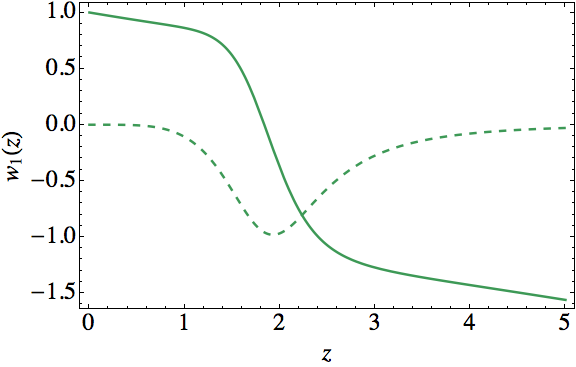}
\includegraphics[scale=0.37]{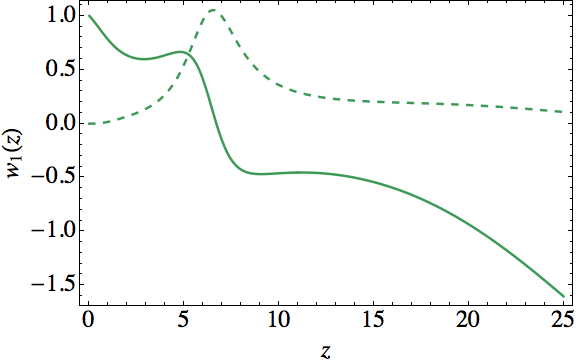}
\caption{\small{Real (solid) and imaginary (dashed) parts of the solution $w_1(z)$ to $P_V$ for $j=3$, $\epsilon_1=0$, and $\nu_1=100i$ (left) and $j=2$, $\epsilon_1=2$, and $\nu_1=i$ (right).}}\label{comp5}
\end{figure}

Then, the same solutions from equations~\eqref{VgkRO} hold. In figure~\ref{comp5} we show two complex solutions to $P_V$ with real parameters $a,b,c,d$.

\section{Complex solutions to $P_V$ with complex parameters}
We can also obtain complex solutions to $P_V$ simply by allowing the factorization {\it energy} in equation~\eqref{solRO2} to be complex. Then, the solutions will also be complex but now, as the parameters $a,b,c$ of $P_V$ will also be complex, as they depend on $\epsilon_1$.

\begin{figure}\centering
\includegraphics[scale=0.37]{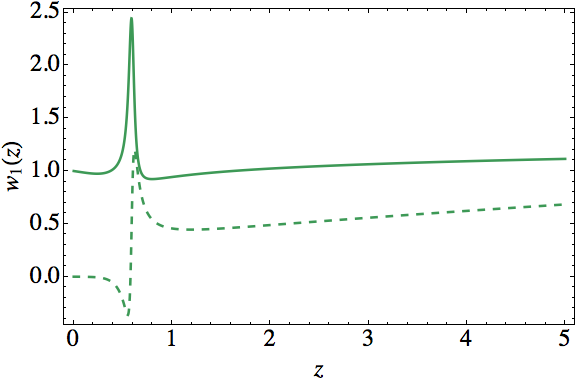}
\includegraphics[scale=0.37]{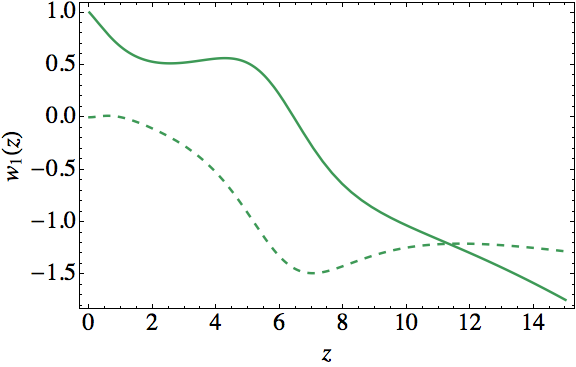}
\caption{\small{Real (solid) and imaginary (dashed) parts of the solution $w_1(z)$ to $P_V$ for $j=3$, $\epsilon_1=1+11i$, and $\nu_1=100i$; and $j=1$, $\epsilon_1=1-0.6i$, and $\nu_1=1-i$.}}\label{comp52}
\end{figure}

For example, in figure~\ref{comp52} we show two complex solutions to $P_V$ but now associated with the complex parameters of $P_V$ given by
\begin{subequations}
\begin{alignat}{5}
a&=-\frac{115}{4}+i\frac{429}{16}, & \quad b& = \frac{1911}{32}+i\frac{55}{4}, & \quad c &= \frac{49}{4},\\
a&=-\frac{1881}{800}-i\frac{27}{20}, & \quad b& = \frac{119}{800}-i\frac{3}{20}, & \quad c &= -\frac{3}{4}.
\end{alignat}
\end{subequations}

\section{Solution hierarchies for $P_V$}
The solutions $w(z)$ that we have found for $P_V$ are expressed in terms of $g(x)$ in equation~\eqref{p5k}, and at the same time $g(x)$ is expressed in terms of the functions $u_i$ of equation~\eqref{VgkROb}. Recall that $u_i$ are eigenfunctions of the radial oscillator Hamiltonian with fixed eigenvalues, which are determined only by the two parameters $\epsilon_1$ and $\nu_1$. Also remember that all of them are written in terms of the confluent hypergeometric function ${}_1F_1$ in equation~\eqref{solRO}. Therefore our solutions to $P_V$ will be written as functions that depend on ${}_1F_1$.

Recall that the Painlev\'e equations themselves define new {\it special functions}, the Painlev\'e trascendents, which are the functions that solve the corresponding equation. Nevertheless, for some special values of the parameters, they can be written in terms of known special functions. This is useful to define {\it solution hierarchies}, as we saw in sections~\ref{realhie} and \ref{comphie} for $P_{IV}$. We will do the same classification for the solutions $w(z)$ of $P_V$.

\subsection{Laguerre polynomials hierarchy}
When the following two conditions are fulfilled
\begin{subequations}
\begin{alignat}{3}
 \epsilon_1= & \ n-\frac{j}{2}+\frac{1}{4}, & \quad \nu_1 & =0, \\
\epsilon_1= & \ n+\frac{j}{2}+\frac{3}{4}, & \quad \nu_1 & \rightarrow\infty,
\end{alignat}
\end{subequations}
then the confluent hypergeometric function reduces to a Laguerre polynomial due to the following identity
\begin{equation}
L_n^{(\alpha)}(x)=\frac{(\alpha+1)_n}{n!}{}_1F_1(-n,\alpha+1,x).
\end{equation}
Two examples of solutions to $P_V$ that belong to this hierarchy are
\begin{subequations}
\begin{align}
w_1(z)=& \ 1-z^{-1/2},\\
w_1(z)=& \  1-\frac{z^{3/2}L_1^{(\alpha)}(z^2/2)}{2L_1^{(\alpha)}(z^2/2)-2\alpha-1},
\end{align}
\end{subequations}
where $\alpha=-(2j+1)/2$.

\subsection{Hermite polynomials hierarchy}
\begin{figure}\centering
\includegraphics[scale=0.37]{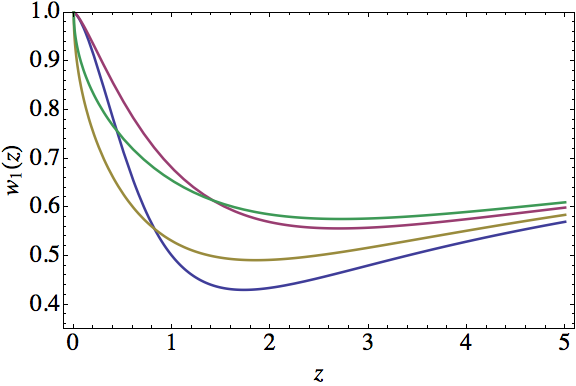}
\caption{\small{Solutions $w_1(z)$ that belong to the Hermite polynomial hierarchy given by equation \eqref{w1hermitea} for $n=0,-1$ (blue, magenta) and by \eqref{w1hermiteb} for $n=-1,-2$ (yellow, green).}}\label{w1her}
\end{figure}

Take now
\begin{subequations}
\begin{alignat}{5}
j & =0, & \quad \epsilon_1 & =n+\frac{1}{4}, & \quad \nu_1 & =0,\\
j & =0, & \quad \epsilon_1 & =n+\frac{3}{4}, & \quad \nu_1 & \rightarrow\infty.
\end{alignat}
\end{subequations}
We obtain then the Hermite polynomials $H_n(x)$. Two examples belonging to this hierarchy $P_V$ solutions are
\begin{subequations}\begin{align}
w_1(z)&=1-\frac{z^{3/2}H_{2n}(z)}{(z^2+1)H_{2n}(z)-4nzH_{2n-1}(z)},\label{w1hermitea}\\
w_1(z)&=1+\frac{z^{1/2}H_{2n}(z)}{4nH_{2n-1}(z)-zH_{2n}(z)}\label{w1hermiteb}.
\end{align}\label{w1hermite}\end{subequations}
\hspace{-1mm}In figure~\ref{w1her} we have plotted two members for each of these two solution families with $n=1,2$. In the plots, it looks like the solutions present a singularity but they do not, as can be proven analytically using equations~\eqref{w1hermite}.

\subsection{Weber or Parabolic cylinder hierarchy}
In order to reduce the confluent hypergeometric function ${}_1F_1$ into a Weber or parabolic cylinder function $E_\nu (x)$; the conditions to be fulfilled are
\begin{equation}
j=0,  \quad \epsilon_1  =\frac{2\nu+1}{4}, \quad \nu_1 =0.
\end{equation}
Two examples of solutions $w_1(z)$ that are obtained through 1-SUSY are
\begin{subequations}
\begin{align}
w_1(z)&=1-\frac{2z^{3/2}E_\nu(z)}{2(z^2+1)E_\nu(z)-zE_{\nu-1}(z)+zE_{\nu+1}(z)},\\
w_1(z)&=1-\frac{2z^{1/2}E_\nu(z)}{2(z^2+1)E_\nu(z)-zE_{\nu-1}(z)+zE_{\nu+1}(z)}.
\end{align}
\end{subequations}

\subsection{Modified Bessel hierarchy}
\begin{figure}\centering
\includegraphics[scale=0.37]{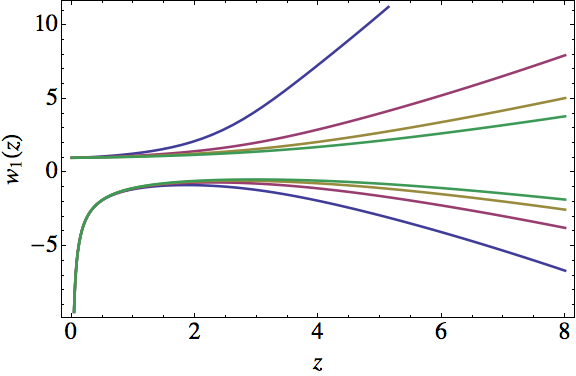}
\caption{\small{Solutions $w_1(z)$ of $P_V$ given by equations~\eqref{w1bessel} that belong to the modified Bessel hierarchy. The positive solutions belong the the first family and negative solutions to the second one, for $\nu=1$ (blue), $\nu=2$ (magenta), $\nu=3$ (yellow), and $\nu=4$ (green).}}\label{w1bes}
\end{figure}

Under the conditions
\begin{subequations}
\begin{alignat}{5}
j & =-\frac{4\nu+1}{2}, & \quad \epsilon_1 & =0, & \quad \nu_1 & =0,\\
j & =-\frac{4\nu +3}{2}, & \quad \epsilon_1 & =0, & \quad \nu_1 & =0,\\
j & =\frac{4\nu-1}{2}, & \quad \epsilon_1 & =0, & \quad \nu_1 & \rightarrow\infty,\\
j & =\frac{4\nu+1}{2}, & \quad \epsilon_1 & =0, & \quad \nu_1 & \rightarrow\infty,
\end{alignat}
\end{subequations}
the function ${}_1F_1$ reduces to the modified Bessel function $I_\nu(z)$. Then, two examples of the corresponding  $P_V$ solutions are
\begin{subequations}
\begin{align}
w_1(z)&=1-\frac{2z^{3/2}I_\nu (z^2/4)}{(z^2-8\nu)I_\nu (z^2/4)-z^2I_{\nu+1}(z^2/4)},\\
w_1(z)&=1+\frac{2I_\nu (z^2/4)}{z^{1/2}[I_{\nu+1}(z^2/4)-I_{\nu}(z^2/4)]}.
\end{align}\label{w1bessel}
\end{subequations}
\hspace{-1.5mm}We present four examples for each of the two solution families of equation \eqref{w1bessel} in figure \ref{w1bes}.

\subsection{Exponential hierarchy}
\begin{figure}\centering
\includegraphics[scale=0.37]{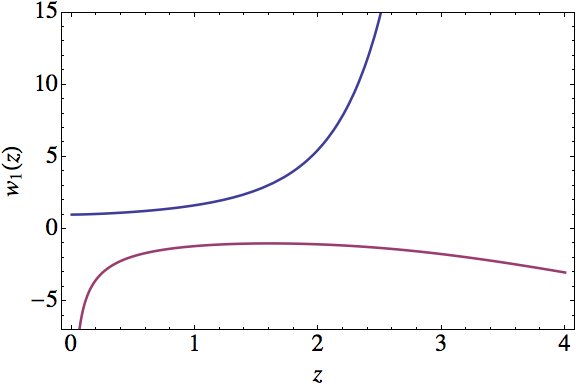}
\caption{\small{Solutions $w_1(z)$ of $P_V$ given by equations~\eqref{w1expsol} that belong to the exponential hierarchy.}}\label{w1exp}
\end{figure}

For special values of the parameters of the transformation, the confluent hypergeometric functions reduce to a polynomial, but there is still the exponential function in the general solution $u(x)$. The conditions
\begin{equation}
j =\frac{1}{2}, \quad \epsilon_1 =0,  \quad \nu_1 \rightarrow\infty,
\end{equation}
illustrate this situation. The two solutions that are obtained through $1$-SUSY are
\begin{subequations}
\begin{align}
w_1(z)&=1+\frac{\exp(z^2/2)-1}{z^{1/2}},\\
w_1(z)&=1-\frac{z^{3/2}}{2}+\frac{z^{7/2}}{2z^{2}+4-4\exp(z^2/2)}.
\end{align}\label{w1expsol}
\end{subequations}
\hspace{-1mm}In figure \ref{w1exp} we show these two solutions that belong to the exponential hierarchy.

\subsection{Other polynomial hierarchy}
Usually, there are some conditions for the ${}_1F_1$ to reduce to an exponential. In this case, this exponential cancels with the exponential of the general solution $u(x)$, which causes the $P_V$ solution to be a polynomial, but different from any other hierarchy. Two examples are obtained for
\begin{subequations}
\begin{alignat}{3}
\epsilon_1 & =\frac{j}{2}-\frac{1}{4}, & \quad \nu_1 & =0,\\ 
\epsilon_1 & =-\frac{j}{2}-\frac{3}{4}, & \quad \nu_1 & \rightarrow\infty.
\end{alignat}
\end{subequations}
An explicit solution of $P_V$ is given by
\begin{equation}
w_1(z)=1-\frac{z^{3/2}}{2j+1}.
\end{equation}
\chapter{Conclusions and perspectives}

\section{Conclusions}
We begin the study of SUSY QM, PHA, and Painlev\'e equations with several questions in mind and throughout the work of this thesis we have answered some of those questions, although now we have some new ones. First, we proved that higher-order SUSY QM also leads to solutions of Painlev\'e IV and V equations, then we find the specific conditions to accomplish it. After that, we studied those solutions and classify them into solution hierarchies, we also study the parameter space of possible solutions. In order to prove this thesis statement, we have followed the method resumed next. 

In chapter \ref{capsusyqm}, we studied the general framework of SUSY QM, from first- to $k$th-order transformations. Specifically, we analyzed in more depth the cases of the harmonic and radial oscillators, which were widely used in the subsequent chapters. Then, in the same chapter we presented a couple of original contributions of this thesis in this area. First, a new formula to calculate the confluent SUSY transformation given in terms of Wronskian of parametric derivatives of the transformation function \citep*{BFF12}. Second, the application of SUSY QM to the {\it inverted oscillator} potential, which was not addressed before \citep{BF13b}.

Then, in chapter \ref{pha} we presented the definition of the Heisenberg-Weyl algebra and their polynomial deformations, i.e., PHA. After this, we studied the general systems described by PHA. For zeroth- and first-order we obtain the harmonic and radial oscillator, respectively. For second- and third-order PHA, the determination of the potential for these systems is reduced to find solutions of $P_{IV}$ and $P_{V}$, respectively.

We begun the analysis of these equations in chapter \ref{cappain}. First we gave a general overview of the six Painlev\'e equations. Then we studied specifically the case of $P_{IV}$. First we prove a {\it reduction theorem} \citep{Ber10,BF11a} for $(2k)$th-order algebras to be reduced to second-order PHA and therefore connected with $P_{IV}$. Through this theorem we were able to find solutions to $P_{IV}$, given in terms of confluent hypergeometric functions ${}_1F_1$ \citep{BF11b}. For particular cases, these solutions can be classified in {\it solution hierarchies} expressed in terms of several special functions, e.g., rational functions, error function, complementary error function, Bessel function, among others \citep{BF11a,BF13a}. Even more, we expanded the solution families using complex transformation functions associated to either real or complex factorization {\it energies}; thus, we were able to find real solutions associated with real parameters $a,b$ of $P_{IV}$ and complex solutions for real and complex parameters \citep{Ber12,BF12}.

To complement the study of $P_{IV}$, in chapter \ref{p4cs} we studied the {\it coherent states} associated with the specific SUSY partners of the harmonic oscillator that are connected with $P_{IV}$ through the reduction theorem, which we called {\it Painlev\'e IV coherent states}. These systems have always third-order ladder operators $l_k^\pm$. In order to do this, we studied first the usual definitions of CS and their generalizations, then we wrote down the four Gazeau-Klauder axioms for defining generalized CS. Later, we used the third-order ladder operators $l_k^\pm$ to obtain the CS, first, as eigenstates of the annihilation operator $l_k^-$, then through the action of an analogue of the displacement operator $D(z)$ onto the extremal states. Here we got partial results, i.e., CS that work appropriately only on a subspace of the Hilbert space $\mathcal{H}$ of the $P_{IV}$ Hamiltonian systems. Then, we applied a process called {\it linearization} to the ladder operators to define a new displacement operator $\mathcal{D}(z)$ that do works out in the entire $\mathcal{H}$.

After that, in chapter \ref{5painleve} we studied the analogous scheme for $P_V$. First we wrote down and prove the {\it reduction theorem} for $(2k+1)$th-order PHA to be reduced to third-order ones, which are connected with $P_V$. Then,  we addressed the inverse problem, i.e., to use systems described by third-order algebras in order to find solutions to $P_V$ \citep{CFNN04}. We were able to identify real solutions associated with real parameters $a,b,c,d$ of $P_V$ and also complex solutions, associated with real and complex parameters. Furthermore, these solutions are also given in terms of confluent hypergeometric functions ${}_1F_1$, and thus we also classified them in terms of the special functions they depend on to generate several {\it solution hierarchies}. We obtained Laguerre, Hermite, modified Bessel, Weber, and exponential function hierarchies. 

\section{Perspectives}
As we have mentioned throughout this thesis, some of its results are already published in the scientific literature. Nevertheless, there are still some parts of the thesis that have not been published yet. The immediate future work plan is to publish this results. We are talking specifically about the Painlev\'e IV coherent states, presented in chapter \ref{p4cs}, and about the solutions to Painlev\'e V equation, including real and complex ones and their classification into solution hierarchies.

We have also presented in section~\ref{kwro} a higher-order generalization of our Wronskian formula for the confluent SUSY transformation. With this generalization it would be possible to obtain the so called {\it hyperconfluent} SUSY transformation. Not only that, but we have recently found a work where the parametric derivative of the confluent hypergeometric function ${}_1F_1(a,c;z)$ with respect to $a,b$ is already reported \citep{AG08}. This can expand the application range of the Wronskian formula we proved in section \ref{difconfluente}. It would be possible to apply this formula to the harmonic oscillator, radial oscillator, among others.

About the main subject of the thesis, the generation of solutions to $P_{IV}$ and $P_V$, we can generate more of them if we use different systems ruled by PHA and we prove the analogous {\it reduction theorems} for potentials different from the harmonic and radial oscillators. For example, we are currently working with the complex oscillator, which is a non-Hermitian generalization of the harmonic oscillator Hamiltonian.

It would also be interesting to study the zeroes structure of the $P_{IV}$ and $P_V$ solutions. This could be accomplished because we have analytic solution. On the other hand, in this work we obtained non-singular solutions in the open domain of definitions, i.e., we accept singularities only at the boundary of the domain. Nevertheless, in a recent work \citep{Mor12}, it is studied the harmonic oscillator potential with an infinite barrier at $x=0$, and solutions to $P_{IV}$ with one fixed singularity are obtained. This sort of approach is also possible for $P_{IV}$ and $P_V$.

\appendix
\begin{appendix}
\chapter*{Appendix A}\label{appendixa}
\addcontentsline{toc}{chapter}{Appendix A}
\renewcommand{\theequation}{A.\arabic{equation}}
\rhead[{\bf Appendix A}]{}
In this appendix we are going to derive the orthogonality and completeness relations
\begin{subequations}
\begin{align}
(\psi_E^\sigma,\psi_{E'}^{\sigma'}) = \int_{-\infty}^{\infty} \overline{\psi}_E^{\, \sigma}(x)\psi_{E'}^{\sigma'}(x) \text{d}x 
&=  \delta(E-E') \delta_{\sigma,\sigma'} , \label{a0} \\
\sum_{\sigma = \pm} \int_{-\infty}^{\infty} \psi_E^\sigma(x) \overline\psi_E^{\, \sigma}(x')  \text{d}E&=  \delta(x-x'), \label{a00}
\end{align}\label{a000}
\end{subequations}
\hspace{-1mm}for the eigenfunctions of the inverted oscillator Hamiltonian given in equations \eqref{uplus} and \eqref{uminus}, where
\begin{equation}
N_E = \frac{e^{i(1/2 - iE)\pi/4} \, 2^{iE/2-1} \, \Gamma(1/2-iE)}{\pi^{1/2}\, \Gamma(3/4 - iE/2)} ,
\end{equation}
$\delta_{\sigma,\sigma'}$ is the Kronecker delta function in the indices $\sigma$ and $\sigma'$ and $\delta(y - y')$ is the Dirac delta function in the index $y$. We will point out only the main steps of the derivation given by \citet*[section 8.2]{Wol79}.

First of all, departing from the standard Fourier transform, we introduce the bilateral Mellin transform $f_\sigma^{BM}(\lambda)$ of the function $f(x)$ and its inverse by means of
\begin{subequations}
\begin{align}
f(x) &= (2\pi)^{-1/2} \sum_{\sigma = \pm} \int_{-\infty}^\infty f_\sigma^{BM}(\lambda) \, x_\sigma^{i\lambda - 1/2} \text{d}\lambda,  \label{a1} \\
f_\sigma^{BM}(\lambda) &= (2\pi)^{-1/2} \int_{-\infty}^\infty f(x) \, x_\sigma^{-i\lambda - 1/2} \text{d}x, \qquad \sigma = \pm, \label{a2}
\end{align}
\end{subequations}
where
\begin{subequations}
\begin{align}
x_+ &= 
\begin{cases}
x & {\rm for} \  x>0, \\
0 & {\rm for} \  x<0,
\end{cases} \label{a3} \\
x_- &= \begin{cases}
0 & {\rm for} \  x>0, \\
-x & {\rm for} \  x<0.
\end{cases}  \label{a4}
\end{align}\label{a34}
\end{subequations}
\hspace{-1.8mm}By substituting now equation~\eqref{a1} in equation~\eqref{a2} the following orthogonality relation is obtained
\begin{equation}\label{a5}
(2\pi)^{-1} \int_{-\infty}^\infty x_\sigma^{-i\lambda - 1/2} \, x_{\sigma'}^{i\lambda' - 1/2} \text{d}x= \delta(\lambda - \lambda') \delta_{\sigma,\sigma'} .
\end{equation}

On the other hand, if equation~\eqref{a2} is substituted in equation~\eqref{a1} one arrives at the following completeness relation
\begin{equation}\label{a6}
(2\pi)^{-1}\sum_{\sigma = \pm} \int_{-\infty}^{\infty} x_\sigma^{i\lambda - 1/2} \, {x'}_\sigma^{-i\lambda - 1/2} \text{d}\lambda 
= \delta(x-x'),
\end{equation}
meaning that the set of functions $\{(2\pi)^{-1/2} \, x_\sigma^{i\lambda - 1/2}, \ \sigma = \pm, \ -\infty<\lambda<\infty\}$ constitutes a generalized (Dirac) orthonormal basis for $\mathcal{L}^2(\mathbb{R})$.

Basically, what is left to do now is to express the eigenfunctions $\psi_{E}^{\sigma}(x)$ of the inverted oscillator Hamiltonian in terms of this orthonormal basis, so it is much easier to prove equations~\eqref{a000}. We can accomplish this through a clever use of the Fourier transform \citet*[section 7.5.12]{Wol79}
\begin{equation}\label{a7}
\psi_E^{\sigma}(x) = 2^{iE/2}(2\pi)^{-1} \int_{-\infty}^{\infty} p_\sigma^{-iE - 1/2} \, e^{i(p^2/4 + xp + x^2/2)}\text{d}p,
\end{equation}
where $p_\sigma, \ \sigma=\pm$ are defined similarly as $x_\sigma$ in equations~\eqref{a34}. 

Finally, the substitution of equation~(\ref{a7}) inside the integral of equation~(\ref{a0}) and the use of equation~(\ref{a5}) leads to the right hand side of equation~(\ref{a0}) and hence to the orthogonality relation. In the same way, by substituting equation~(\ref{a7}) into the left hand side of equation~(\ref{a00}) and using equation~(\ref{a6}) one arrives to the right hand side of equation~(\ref{a00}) and thus to the completeness relation.
\end{appendix}
\begin{appendix}
\chapter*{Appendix B}\label{appendixb}
\addcontentsline{toc}{chapter}{Appendix B}
\renewcommand{\theequation}{B.\arabic{equation}}
\rhead[{\bf Appendix B}]{}
\setcounter{equation}{0}
In this appendix we will prove the Wronskian formula for the new states of $H_0$. Let us consider a second-order SUSY transformation
\begin{equation}
H_2B_2^{+}=B_2^{+}H_0,
\end{equation}
generated by two transformation functions $u_j$ such that $H_0 u_j = \epsilon_j u_j$, with $j=1,2$. One can express the action of $B_2^{+}$ over a function $\psi$ in terms of $u_j$ with two different forms, either as an iteration of two first-order SUSY transformations or using Wronskians in the following way
\begin{align}
B_2^{+}\psi&=\frac{W(u_1,u_2,\psi)}{W(u_1,u_2)}\nonumber\\
			&=\frac{1}{2}\left(-\frac{\text{d}}{\text{d}x} + \frac{(u_2^{a})'}{u_2^{a}}\right)\left(-\frac{\text{d}}{\text{d}x} + \frac{u_1'}{u_1}\right)\psi \nonumber\\
			&=\frac{1}{2}\left(-\frac{\text{d}}{\text{d}x} + \frac{(u_1^{a})'}{u_1^{a}}\right)\left(-\frac{\text{d}}{\text{d}x} + \frac{u_2'}{u_2}\right)\psi, 
\end{align}\label{b2mas}
\hspace{-1.8mm}where $u_1^{a}$ is the transformed $u_1$ when we use first $u_2$, similarly for $u_2^{a}$, i.e., we have $u_1^{a}\propto W(u_1,u_2)/u_2$ and $u_2^{a}\propto W(u_1,u_2)/u_1$. From equation~\eqref{b2mas} we can check that $B_2^{+}u_j=0$, $j=1,2$. Now, using $B_2^{-}\equiv (B_2^{+})^{\dagger}$ we can perform the intertwining in the opposite direction, i.e., $H_0B_2^{-}=B_2^{-}H_2$ and we obtain
\begin{equation}
B_2^{-}=\frac{1}{2}\left(\frac{\text{d}}{\text{d}x} + \frac{u_1'}{u_1}\right)\left(\frac{\text{d}}{\text{d}x} + \frac{(u_2^{a})'}{u_2^{a}}\right)
=\frac{1}{2}\left(\frac{\text{d}}{\text{d}x} + \frac{u_2'}{u_2}\right)\left(\frac{\text{d}}{\text{d}x} + \frac{(u_1^{a})'}{u_1^{a}}\right).\label{entreinv}
\end{equation}

From equations~\eqref{entreinv} we can find the eigenfunctions $H_2\tilde{u}_j =\epsilon_j\tilde{u}_j$, which are annihilated by $B_2^{-}$ and they turn out to be
\begin{subequations}
\begin{align}
\tilde{u}_1&=\frac{1}{u_1^{a}}\propto \frac{u_2}{W(u_1,u_2)},\\
\tilde{u}_2&=\frac{1}{u_2^{a}}\propto \frac{u_1}{W(u_1,u_2)}.
\end{align}
\end{subequations}

Repeating these same arguments for the $k$th-order case given by $H_kB_k^{+}=B_k^{+}H_0$ and generated by the $k$ transformation functions $u_j$, such that $H_0u_j=\epsilon_j u_j$, $j=1,\dots ,k$ and $B_k^{+}u_j=0$, the operator $B_k^{-}$ that performs the intertwining in the opposite direction is characterized by the $k$ eigenfunctions $H_k \tilde{u}_j = \epsilon_j \tilde{u}_j$, $j=1,\dots ,k$ such that $B_k^{-}\tilde{u}_j=0$. These eigenfunctions are given by
\begin{subequations}
\begin{align}
\tilde{u}_1&=\frac{1}{u_1^{a}}\propto \frac{W(u_2,\dots , u_{k})}{W(u_1,\dots , u_k)},\\
				&\ \ \vdots \nonumber \\
\tilde{u}_k&=\frac{1}{u_k^{a}}\propto \frac{W(u_1,\dots , u_{k-1})}{W(u_1,\dots , u_k)}.
\end{align}
\end{subequations}
\end{appendix}

\backmatter
\lhead[\bfseries\thepage]{{\bf Bibliography}}
\rhead[{\bf Bibliography}]{\bfseries\thepage}
\bibliography{references}
\addcontentsline{toc}{chapter}{Bibliography}

\printindex
\end{document}